\documentclass{aa}  

\usepackage{graphicx}
\usepackage{hyperref}
\usepackage{setspace}
\usepackage{adjustbox}
\usepackage{balance}
\usepackage[maxfloats=256]{morefloats}
\usepackage{color}
\usepackage{txfonts}
\def\kms{km\,s$^{-1}$}
\def\msun{M$_\odot$}

\begin{document} 

   \title{An interferometric study of B star multiplicity\thanks{Based on observations made with ESO Telescopes at the Paranal Observatory under programmes ID 093.C-0503(A) and 112.2624.}}
    \titlerunning{An interferometric study of B star multiplicity}

   \author{A. J. Frost
          \inst{1},
          H. Sana\inst{2,3},
          J-B Le Bouquin\inst{4},
          H. B. Perets\inst{5},
          J. Bodensteiner\inst{6},
          A. P. Igoshev\inst{7}, 
          G. Banyard\inst{2},
          L. Mahy\inst{8}, \\
          A. M\'erand\inst{9}
        \and
         O. H. Ram\'irez-Agudelo\inst{10}
          }

   \institute{European Southern Observatory, Alonso de Cordova 3107, Vitacura, Santiago, Chile
   \and Institute of Astronomy, KU Leuven, Celestijnlaan 200D, 3001 Leuven, Belgium
   \and Leuven Gravity Institute, KU Leuven, Celestijnenlaan 200D, box
2415, 3001 Leuven, Belgium
   \and Univ. Grenoble Alpes, CNRS, IPAG, F-38000 Grenoble, France
   \and Physics Department, Technion - Israel institute of Technology, Haifa 3200002, Israel
   \and Anton Pannekoek Institute, University of Amsterdam, Science Park 904, 1098 XH Amsterdam, The Netherlands
   \and School of Mathematics, Statistics and Physics, Newcastle University, Newcastle upon Tyne, NE1 7RU, UK
   \and Royal Observatory of Belgium, Avenue Circulaire/Ringlaan 3, 1180 Brussels, Belgium
   \and ESO Headquarters, Karl-Schwarzchild-Str 2 85748 Garching, Germany
   \and German Aerospace Center (DLR), Institute for AI Safety and Security, Rathausallee 12, St. Augustin, 53757, Germany\\
              \email{abigail.frost@eso.org}
             }
   \date{Submitted 1st March 2025. Accepted 5th June 2025.}
    \authorrunning{A. J. Frost et al.}

 
  \abstract
   {Massive stars can have extreme effects on their environments from local to galactic scales. Their multiplicity can affect this influence by altering how they evolve over time by causing dynamical interactions, common-envelope evolution, mergers and more. While O star multiplicity has been studied over a broad separation range (to the point where absolute masses of these systems have been determined and investigations into multiple system formation and interactions have been performed), studies of B star multiplicity are lacking, even though they dominate the production of core-collapse supernovae and neutron stars.}
   {Using interferometry, we investigated the multiplicity of a statistically significant sample of B stars over a range of separations ($\sim$0.5-35 au, given that the average distance to our sample is 412 pc).}
   {We analysed high angular resolution interferometric taken with the PIONIER (Precision Integrated-Optics Near-infrared Imaging ExpeRiment) instrument at the Very Large Telescope Interferometer (VLTI) for a sample of 32 B stars. Using parametric modelling of the closure phases and visibilities, we determined best-fitting models to each of the systems and investigated whether each source was best represented by a single star or a higher-order system. The detection limits were calculated for companions to determine whether they were significant. We then combined our findings from the interferometric data with results from a literature search to determine whether other companions were reported at different separation ranges.}
   {Within the interferometric range 72$\pm$8\% of the B stars are resolved as multiple systems. The most common type of system are binary systems, followed by single stars, triple systems, and quadruple systems. The interferometric companion fraction derived for the sample is 1.88$\pm$0.24. When we accounted for spectroscopic companions that have been confirmed in the literature and wide companions inferred from Gaia data in addition to the companions we found with interferometry, we obtain multiplicity and companion fractions of 0.88$\pm$0.06 and 2.31$\pm$0.27, respectively, for our sample. The number of triple systems increases significantly to the second-most populous system when we account for spectroscopic companions. This suggests that binarity and higher-order multiplicity are as integral to the evolution of B stars as they are for O stars.}
   {}

   \keywords{stars: massive -- techniques: interferometric -- infrared: stars  -- stars: binaries}

   \maketitle
%

\section{Introduction}\label{sec:intro}

Massive stars (M$_{i}\gtrsim$ 8 M$_{\odot}$) are some of the most powerful stars in the Universe. Serving as cosmic engines, their winds, outflows and supernovae (SNe) compress and rarefact molecular clouds and thus affect future generations of stars. Massive stars produce the heaviest elements in the Universe and distribute them into the interstellar medium as they evolve through their end of life explosions. This enriches interstellar chemistry. On galactic scales, the morphology of galactic superwinds depends on the wind activity of massive stars \citep{leith}. If a massive binary system ends its life as a black hole (BH) or neutron star (NS) binary merger, the subsequently released gravitational waves can be felt across galaxies \citep{Abbott2016}. Despite their importance, however, the formation and evolution of massive stars is still not well understood \citep{langrev, pabjulrev}. 

One major uncertainty in the evolution of massive stars is their multiplicity. Close companions ($P_{\rm orb}$<10yrs) in particular can have strong effects on the evolution of massive stars. Mass transfer or stellar mergers could explain apparent evolutionary inconsistencies in some single stars (\citealt{demink11}, \citealt{demink13}, \citealt{fabmerg}, \citealt{frost2024}). A small fraction of these close binary systems result in BH mergers which create gravitational waves \citep{gwtc3}. 

In recent years the role of triples in stellar evolution has also been made evident. Their evolution produces unique evolutionary channels \citep[e.g.][for an overview]{too+16}. The existence of a third companion can affect the evolution of the inner binary through secular \citep{VonZeipel1910,kozai,lidov} and quasi-secular evolution \citep{ant+12}. This can drive the inner binaries to close-pericentre interactions, tidal interactions, migration \citep{maz+79,kis+98,egg+08}, mergers \citep{per+09} and/or mass-transfer. 
Mass loss and mass transfer in triples can also induce instabilities in these systems and change their dynamical evolution. This leads to close encounters and collisions \citep{egg+86,sok04,per+12,Sha+13,mic+14,ham+21,too+21} as well as earlier interactions \citep{kummer23}. The fate of a massive triple system therefore has the potential to be significantly different to that of a massive binary system.

Most recent studies of massive stellar multiplicity thus far have focused on O-type stars. The majority of massive O-type stars form in binary or higher-order multiple systems \citep{masongies09, sana06, sana12, sana13, sana14, moe17, cyprien23, off23} and a large fraction of these systems interact during their lifetimes \citep{Paczynski1967, Podsiadlowski1992, Vanbeveren1994, deMink2013}. 

While they are fainter, less massive and more difficult to characterise, B-type stars create their own set of unique phenomena and play important roles within the Universe. Above $\sim$25-40 M$_{\odot}$, depending on the metallicity, most massive stars directly collapse into BHs. Therefore, most SNe (both core-collapse and thermonuclear type Ia SNe) originate from B-type stars as a result of the initial mass function (IMF, \citealt{kroupa}). B-type stars also create most neutron stars (NS) and pulsars, and they are therefore the source of long-inspiralling NS-NS mergers (e.g. GW170817; \citealt{nsnsmerger}). The merger of two NSs also releases gravitational waves and emission from gamma to radio wavelengths, which allows them to be observed up to cosmological distances. Determining the B star multiplicity fraction is therefore a keystone for understanding a wide range of astrophysical phenomena. 

\citet{abt} studied 109 field B stars in the Galaxy between spectral types B2 and B5 and found an observed spectroscopic binary fraction of 29\%, a bias-corrected spectroscopic binary fraction of 57\%, and a total observed binary fraction of 74\% (including visual companions). \citet{raboud} determined a multiplicity fraction between 52-63\% for the 56 B stars of the cluster NGC 6231. 
\citet{dunstall15} investigated the multiplicity properties of 408 B-type stars in 30 Dor using the VLT-FLAMES\footnote{(Very Large Telescope's Fibre Large Array Multi Element Spectrograph)} Tarantula Survey (VFTS; \citealt{vfts}) and reported a spectroscopic binary fraction of $f_{SB}$(obs) = 25$\pm$2\% for most of the region, with the exception of two older clusters (Hodge 301 and SL639), whose binary fractions were 8$\pm$8\% and 10$\pm$9\%, respectively. Using modelling and synthetic populations, they also constrained the intrinsic multiplicity properties of the less evolved dwarf and giant B-type stars in 30 Dor, and obtained a present-day binary fraction $f_B$(true)=58$\pm$11\% that agreed with the fraction found for the O-type stars in the region. Other recent work by \citet{gazza} used data from VLT-FLAMES to determine a spectroscopic binary fraction of 33$\pm$5\% for 80 B-type stars in the young open cluster NGC 6231, which increases to 52$\pm$8\% when observational biases are considered. While these spectroscopic surveys provided a first probe of the multiplicity of these regions, the sensitivity of spectroscopic observations to multiplicity quickly drops when the period of the orbit is approximately one year and when the stars are of similar mass \citep{gazza}. Beyond this period, interferometry is required to search for companions \citep{sana17}. The $>$1 yr period domain is particularly relevant because most NS+NS mergers are expected to originate from periods in this range \citep{demink16}.

In this paper we determine the multiplicity fraction for 37 B-type stars using high angular resolution data obtained with the Very Large Telescope Interferometer (VLTI; \citealt{vlti}). We describe the observations and the model fitting used to detect companions in Section \ref{sec:meth}. We present and discuss our results in Section \ref{sec:res}. In Section \ref{sec:bias} we simulate artificial populations of B star multiple systems to assess how many companions may have been missed in our work due to observational biases. We conclude in Section \ref{sec:conc}. 
\vspace{-0.4cm}

\section{Method}\label{sec:meth}

\subsection{Observations}

\begin{table*}[h!]
    \centering
    \caption{Information on the B stars of our sample.}
    \label{source_info}
    \begin{adjustbox}{width=\textwidth}
    \begin{tabular}{lllccccc}
        \hline
                \hline
        Name  & RA & DEC & $H$mag & Distance & SpT & Mass & Comments \\
         & (h m s) & ($^{\circ}$ ' ") & & (pc) & &(\msun) & \\
         \hline
          \vspace*{-2mm}\\
      $\gamma$ Peg & 00:13:14.15 & $+$15:11:00.94 & 3.43  &  146$^{+10}_{-11}$   & B2~IV      &  8.8$\pm$0.3\tablefootmark{a} & $\beta$ Cep variable \\ \vspace*{-2mm}\\
        HD\,3379   & 00:36:47.31 & $+$15:13:54.18 & 6.275 &  279$^{+6}_{-6}$     & B2.5~IV    &  5.4$\pm$0.9\tablefootmark{b} & $\beta$ Cep variable \\ \vspace*{-2mm}\\
        HD\,16582  & 02:39:28.96 & $+$00:19:42.63 & 4.74  &  194$^{+9}_{-10}$    & B2~IV      &  8.4$\pm$0.7\tablefootmark{c} & $\beta$ Cep variable \\ \vspace*{-2mm}\\
        HD\,25558  & 04:03:44.60 & $+$05:26:08.23 & 5.58  &  222$^{+12}_{-11}$   & B5~V       &  4.6$\pm$0.5\tablefootmark{d} & \\ \vspace*{-2mm}\\
        HD\,30836  & 04:51:12.36 & $+$05:36:18.37 & 4.09  &  263$^{+20}_{-18}$   & B2~III     & 11.0$\pm$1.0\tablefootmark{e} & \\ \vspace*{-2mm}\\
        HD\,32249  & 05:01:26.35 & $-$07:10:26.27 & 5.30  &  277$^{+10}_{-9}$    & B3~IV      &  7.0$\pm$0.4\tablefootmark{e} &Orion X  \\ \vspace*{-2mm}\\
        HD\,34816  & 05:19:34.52 & $-$13:10:36.44 & 4.98  &  299$^{+18}_{-14}$   & B0.5~V     & 15.0$\pm$3.5\tablefootmark{a} & \\ \vspace*{-2mm}\\
        HD\,35149  & 05:22:50.00 & $+$03:32:39.98 & 5.443 &  575$^{+119}_{-78}$  & B2~V       & 12.5$\pm$0.6\tablefootmark{f} & \\ \vspace*{-2mm}\\     
        HD\,35337  & 05:23:30.15 & $-$13:55:38.46 & 5.80  &  370$^{+15}_{-11}$   & B2~IV      &  9.8$\pm$0.6\tablefootmark{g} & \\ \vspace*{-2mm}\\
        HD\,37017  & 05:35:21.87 & $-$04:29:39.04 & 6.88  &  356$^{+8}_{-9}$     & B1.5~V     &  6.4$\pm$0.5\tablefootmark{g} & \\ \vspace*{-2mm}\\ 
        HD\,51480  & 06:57:09.38 & $-$10:49:28.06 & 5.12  &  1106$^{+31}_{-25}$  & Besh       &  6.4$\pm$0.5\tablefootmark{h} & Be star\\ \vspace*{-2mm}\\
        HD\,66765  & 08:02:55.72 & $-$48:19:29.95 & 7.01  &  372$^{+12}_{-9}$    & B1.5~V     &  6.6$\pm$1.0\tablefootmark{i} & \\ \vspace*{-2mm}\\
        HD\,67621  & 08:06:41.61 & $-$48:29:50.59 & 6.86  &  355$^{+7}_{-6}$     & B2~IV      &  5.5$\pm$0.5\tablefootmark{h} & Vel OB 2 - cluster \\ \vspace*{-2mm}\\
        HD\,105382 & 12:08:05.23 & $-$50:39:40.58 & 4.95  &  109$^{+2}_{-2}$     & B4~III     &  5.7$\pm$0.4\tablefootmark{j} & Lower Cen Crux/Sco OB 2-4, Be star \\ \vspace*{-2mm}\\
        HD\,109026 & 12:32:28.01 & $-$72:07:58.76 & 4.25  &  117$^{+4}_{-4}$     & B3~V       &  5.0$\pm$0.5\tablefootmark{h} & Pulsating variable \\ \vspace*{-2mm}\\ 
        HD\,116658 & 13:25:11.58 & $-$11:09:40.75 & 1.54  & $^*$77$^{+4}_{-3}$   & B1~V       & 11.4$\pm$1.2\tablefootmark{k} & $\beta$ Cep variable \\ \vspace*{-2mm}\\ 
        HD\,121743 & 13:58:16.27 & $-$42:06:02.71 & 4.46  &  140$^{+7}_{-6}$     & B2~IV      &  8.5$\pm$0.3\tablefootmark{f} &  II Sco/Upper Cen Lupus, Variable \\ \vspace*{-2mm}\\
        HD\,132058 & 14:58:31.93 & $-$43:08:02.27 & 3.25  & $^*$117$^{+3}_{-2}$  & B2~III     &  8.8$\pm$0.2\tablefootmark{f} & \\ \vspace*{-2mm}\\
        HD\,133518 & 15:06:55.97 & $-$52:01:47.24 & 6.666 &  602$^{+23}_{-30}$   & B2~VpHe    &  6.3$\pm$0.5\tablefootmark{h} & \\ \vspace*{-2mm}\\ 
        HD\,140008 & 15:42:41.02 & $-$34:42:37.46 & 5.11  &  128$^{+3}_{-3}$     & B5~V       &  4.5$\pm$0.5\tablefootmark{h} & Upper Cen Lupus/ Sco OB 2-3  \\ \vspace*{-2mm}\\
        HD\,147932 & 16:25:35.08 & $-$23:24:18.79 & 5.92  &  122$^{+1}_{-1}$     & B5~V       &  5.0$\pm$0.5\tablefootmark{l} & Ass Sco OB 2-2/Upper Sco/Ophiuchus, Rotationally variable \\ \vspace*{-2mm}\\
        HD\,161701 & 17:47:36.78 & $-$14:43:32.97 & 5.88  &  167$^{+1}_{-2}$     & B9III~pHgMn&3.96$\pm$0.14\tablefootmark{o} & \\ \vspace*{-2mm}\\ 
        HD\,178175 & 19:08:16.70 & $-$19:17:25.05 & 5.39  &  414$^{+24}_{-22}$   & B2~Ve      &  8.8$\pm$0.6\tablefootmark{q} & Be star \\ \vspace*{-2mm}\\
        HD\,189103 & 19:59:44.18 & $-$35:16:34.70 & 4.79  &  208$^{+14}_{-13}$   & B3~IV      &  6.6$\pm$0.1\tablefootmark{f} & \\ \vspace*{-2mm}\\
        HD\,191263 & 20:08:38.28 & $+$10:43:33.11 & 6.656 &  441$^{+11}_{-13}$   & B3~V       &  6.0$\pm$0.5\tablefootmark{h} & \\ \vspace*{-2mm}\\
        HD\,193933 & 20:23:26.26 & $-$14:15:23.17 & 7.068 &  440$^{+10}_{-10}$   & B5~IV      &  5.6$\pm$0.5\tablefootmark{h} & \\ \vspace*{-2mm}\\
        HD\,205637 & 21:37:04.83 & $-$19:27:57.65 & 4.91  &  269$^{+14}_{-13}$   & B3~V       &  8.6$\pm$0.5\tablefootmark{r} & Be star \\ \vspace*{-2mm}\\
        HD\,212076 & 22:21:31.08 & $+$12:12:18.66 & 4.803 & $^*$498$^{+81}_{-61}$& B2~Ve      & 12.5$\pm$0.7\tablefootmark{q} & Be star \\ \vspace*{-2mm}\\
        HD\,212571 & 22:25:16.62 & $+$01:22:38.63 & 5.37  &  330$^{+14}_{-11}$   & B1~III-IVe & 10.7$\pm$0.7\tablefootmark{f} & Be star \\ \vspace*{-2mm}\\
        HD\,224990 & 00:02:19.93 & $-$29:43:13.60 & 5.41  &  207$^{+6}_{-5}$     & B5~V       &  5.5$\pm$0.5\tablefootmark{f} & Cl Blanco 1 (open galactic cluster)  \\ \vspace*{-2mm}\\ 
        MCW 1019   & 22:00:07.93 & $+$06:43:02.81 & 6.212 &  453$^{+27}_{-21}$   & B3~III     &  6.9$\pm$0.7\tablefootmark{s} & \\ \vspace*{-2mm}\\
        $\tau$ Lib & 15:38:39.37 & $-$29:46:39.90 & 4.05  &  112$^{3}_{-2}$      & B2.5~V    &  7.25$\pm$0.49\tablefootmark{e} &  II Sco/Upper Cen Lupus \\ \vspace*{-2mm}\\
        \hline
        \end{tabular}
        \end{adjustbox}
        \tablefoot{ All distances come from Gaia DR3 \citep{dr3}, except for the distances with an asterisk, which were determined by HIPPARCOS  \citep{hipp}. In the penultimate column, we provide masses from the literature (references as footnotes). Any star without information in the final column is a field star without a parent association in the literature (as reported by the SIMBAD Astronomical Database \citep{simbad} and \citealt{cangau}). \newline \textbf{References:}
        \tablefoottext{a}{\citet{2014A&A...566A...7N}},
        \tablefoottext{b}{\citet{2006MNRAS.369L..61H}},
        \tablefoottext{c}{\citet{2009AN....330..317H}},
        \tablefoottext{d}{\citet{sodor14}},
        \tablefoottext{e}{\citet{2010AN....331..349H}},
        \tablefoottext{f}{\citet{2011MNRAS.410..190T}},
        \tablefoottext{g}{\citet{2024A&A...690A.135J}},
        \tablefoottext{h}{\citet{kervbin}},
        \tablefoottext{i}{\citet{2007A&A...468..263C}},
        \tablefoottext{j}{\citet{2004A&A...413..273B}},
        \tablefoottext{k}{\citet{2016MNRAS.458.1964T}},
        \tablefoottext{l}{\citet{2018MNRAS.481.3953A}},
        \tablefoottext{m}{\citet{2010MNRAS.404.1306F}},
        \tablefoottext{n}{\citet{2003PhDT.......251B}},
        \tablefoottext{o}{\citet{2014A&A...561A..63G}},
        \tablefoottext{p}{\citet{1992A&A...264...88V}},
        \tablefoottext{q}{\citet{2016A&A...595A.132Z}},
        \tablefoottext{r}{\citet{silaj14}},
        \tablefoottext{s}{\citet{2016A&A...591L...6I}},
        \tablefoottext{t}{\citet{2022ApJ...937..110L}}
        }
\end{table*}


The data for our sample were taken with the Precision Integrated-Optics Near-infrared Imaging ExpeRiment or PIONIER instrument \citep{pionier}, which combines the beams of four telescopes at the VLTI in the H band (1.66 $\mu$m, $\lambda$/$\Delta\lambda\sim$5, bandwidth$\sim$0.3 $\mu$m). The PIONIER data were reduced using the automated PIONIER pipeline (PNDRS; developed by \citealt{pionier}\footnote{\url{https://www.jmmc.fr/dyn/index.php?m=04&y=15&entry=entry150407-092709}}). PIONIER combines the signal from either the four 8.2 m Unit Telescopes (UTs) or the four 1.8 m Auxiliary Telescopes (ATs). Visibilities and closure phases are therefore obtained over six baselines and three closure phase triangles for both instruments. The observations of our sample were taken with the 1.8 m ATs. Each science observation was bracketed by a calibration observation, which allows the visibilities and closure phases to be calibrated. The projected baseline lengths, $B$, range from $\sim$10-200 m with the ATs, which corresponds to a resolution of $\sim$1-20 mas. PIONIER uses single-mode fibres, and the field of view therefore corresponds to the point spread function (PSF) of the telescope delivered at the fibre injection point, which is $\sim$190 mas for the ATs in the H band. Most of the observations were taken over a sequence of nights in visitor mode as part of ESO programme 093.C-0503(A) (PI: H. Sana). Additional epochs of data were taken from ESO programme 112.2624 (PI: L. Mahy). All the observational data, their model fits, and the bootstrap errors associated with these fits are provided in \href{https://zenodo.org/records/15764971}{this} Zenodo directory. 

\begin{figure*}
   \centering
   \includegraphics[width=180mm]{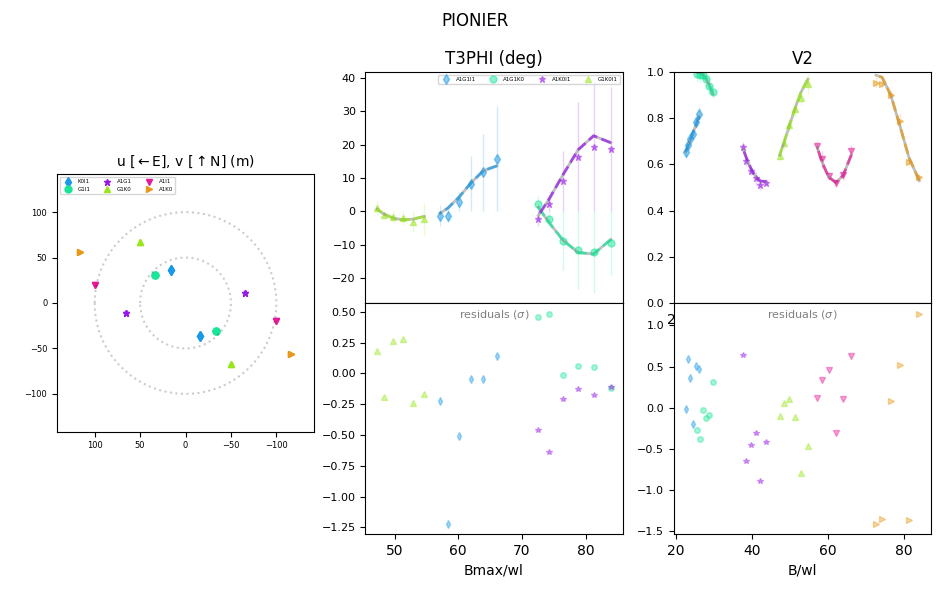}
   \caption{A typical fit to the data of one of our sources, HD\,3379. On the left, we show the $u-v$ coverage of the PIONIER observations. The different colours correspond to the telescope pair with which the particular data were acquired at the VLTI. In the middle, we show the closure phase (T3PHI) fit and residuals in terms of the spatial frequency (B$_{avg}$/$\lambda$; written as Bavg/wl on the axes). On the right, we show the fit to the squared visibilities (V2) and the associated residuals, again in terms of spatial frequency. The data are represented as points in the fits, whilst the model fits are shown as continuous lines.}
              \label{fit}%
    \end{figure*}
    
    \begin{figure*}
   \centering
   \includegraphics[width=150mm]{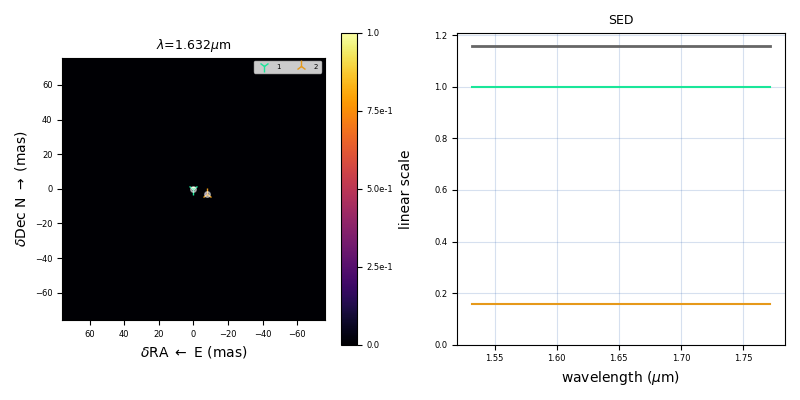}
   \caption{Left: Model image created based on the best-fit model to the data shown in Figure \ref{fit}, showing the primary (1) and secondary (2) stars needed in the model to reproduce the interferometric observables. The colour bar displays the arbitrary flux of the star. Right: The spectral energy distribution displaying the flux ratios of the stars in the model.}
              \label{mod}%
    \end{figure*}

        \begin{figure}
   \centering
   \includegraphics[width=90mm]{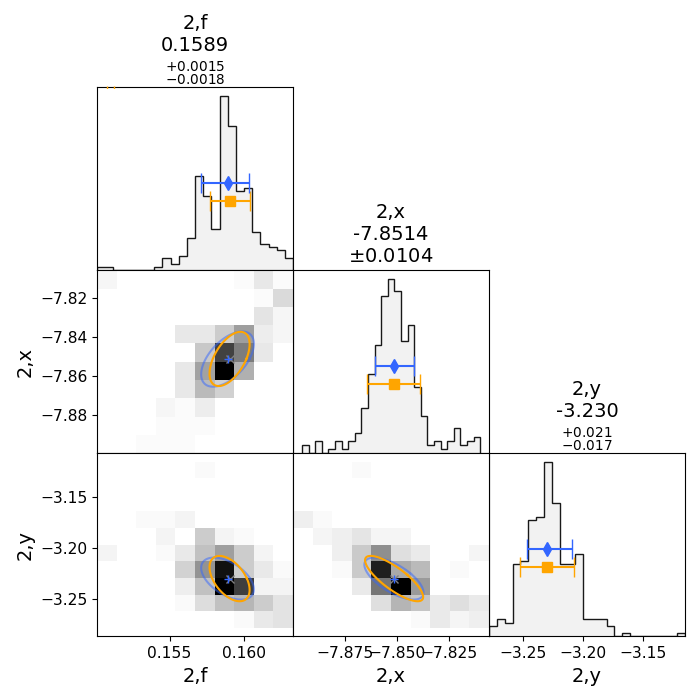}
   \caption{Bootstrapping plot showing the error determination for the dataset fit in Figure \ref{fit}. The fit to all data is shown in orange, and blue corresponds to the bootstrap fit.}
              \label{bs}%
    \end{figure}

        \begin{figure*}
   \centering
   \includegraphics[width=180mm]{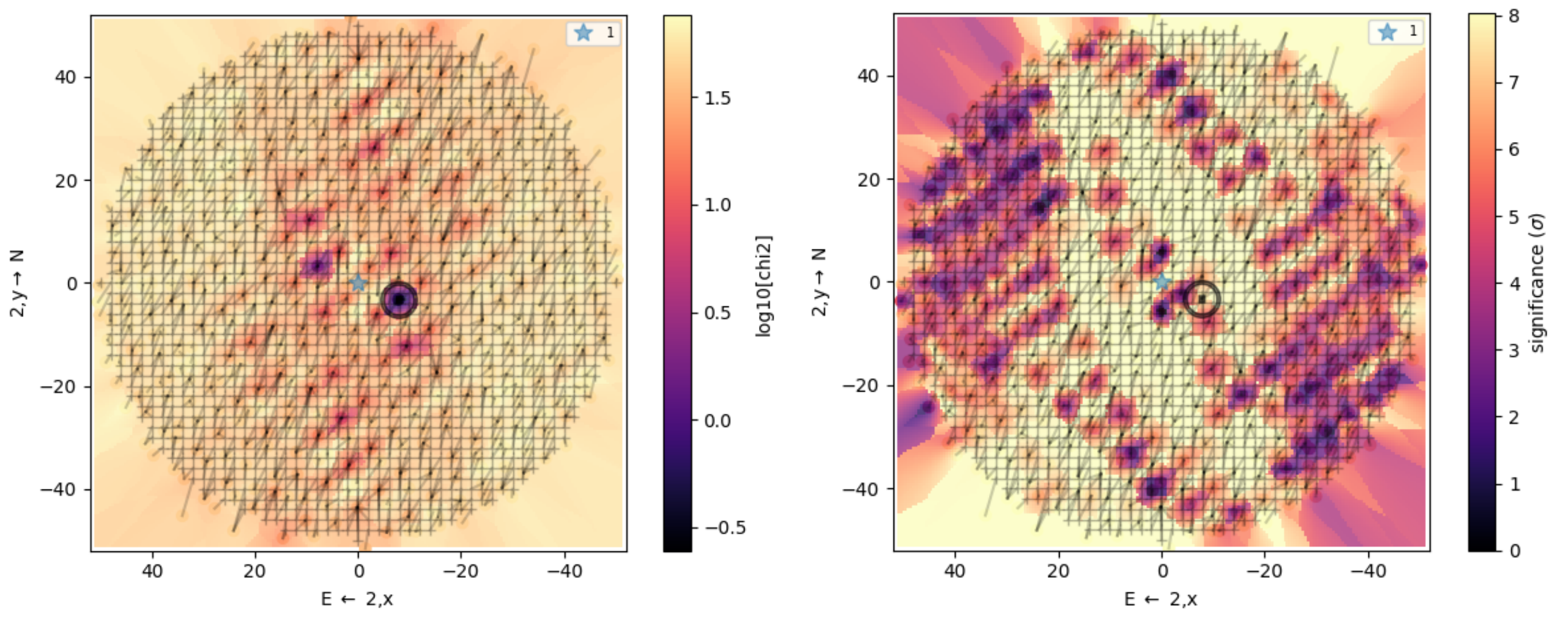}
   \caption{Example of a grid fit, in this case corresponding to Figure \ref{fit}. On the left, we show the grid of potential companion positions that we searched in terms of goodness of fit, as quantified by the $\chi^{2}$. The primary star, which was fixed at position (0,0) during the fit, is shown by a blue star. The black circle shows the (x,y) position of the companion in the best-fitting model. On the right, we show the grid in terms of companion significance. This allowed us to determine whether a companion determined by the best-fitting model was significant in terms of its flux.}
              \label{siggrid}%
    \end{figure*}

\subsection{Sample selection}

The initial sample we analysed was comprised of 37 B-type objects. Many of the data were obtained during visitor-mode observations in 2014 and 2015. As a result, three main selection criteria for the sample existed: their likelihood to be NS progenitors, their visibility in the sky (which depends on the time of the observations), and the observing capabilities of PIONIER. In order to be successfully observed with the instrument based on the typical seeing conditions at Paranal Observatory of 0.6-0.8", objects must have an $H$-band magnitude $\lessapprox$8 mag on average. In comparison to other multiplicity studies such as SMaSH+ \citep{sana14}, which focused on O stars, our sample was limited to closer stars because B-type stars are intrinsically fainter. The majority of the sample are within 500 pc according to the distances determined from Gaia DR3 \citep{dr3} or HIPPARCOS \citep{hipp}. Most of this sample were dwarf stars, followed by subgiants, giants, and supergiants. However, we did not include supergiant stars in this work. This is because B supergiants likely evolved from O-type stars, and they therefore do not contribute to our aim to study B-star multiplicity. With the supergiants removed, the final sample size is 32 stars as listed in Table \ref{source_info}. We searched the literature for dedicated studies that provide mass estimates. We found none, but atmospheric parameters for some sources were retrieved in some papers (allowing us to place the stars in a Hertzprung-Russell diagram). We then used the {\sc bonnsai} tool\footnote{\tt https://www.astro.uni-bonn.de/stars/bonnsai/} \citep{bonnsai} to obtain an evolutionary mass through a comparison with the galactic evolutionary tracks of \citet{brott2011}. When these were not available, we used mass estimates from \citet{kervbin}. Masses $M$ are a useful input in Sect~\ref{sec:bias}, where we evaluate the detection capabilities of our survey. Very precise masses are not needed, however, because the spectroscopic and interferometric detection thresholds only depend on $M$ following $M^{1/3}$.

\subsection{Model fitting}

We fitted the parameters on the PIONIER data using the Python3 module PMOIRED\footnote{\url{https://github.com/amerand/PMOIRED}} \citep{pmoired}, which allowed us to display and model interferometric data stored in the OIFITS format. We used the grid-search capabilities of PMOIRED to search for companions, which are based on a previous tool, CANDID \citep{candid}. In the search for a first companion, the model was composed of a primary star and a companion, which were represented by uniform disks. The position of the primary star was always fixed at (0,0) and the flux of the primary star was fixed at 1, so that the fluxes of any companions were presented in relation to the flux of the primary. The observables we fitted were the squared visibilities (V2) and the closure phases (T3PHI), examples of which are shown in Figure \ref{fit}. The model corresponding to this fit is shown in Figure \ref{mod}. An exploration dictionary was defined that the grid fit iteratively searched over various $x$,$y$ positions. The grid search then fitted the observed data at each point in this defined grid. The number of positions that we probed depended on the number and quality of the data. When a grid is too fine or coarse, an unreliable solution may result. The quality of each of the fits across the grid was assessed using the $\chi^2_{red}$, and this was used to find the best fit of the grid. Additionally, priors were used to ensure that the flux of a companion we found remained greater than 0. At the start of fitting all sources, we assumed that the companion is unresolved, and we therefore fixed the angular diameter to 0.2mas. When models with unresolved stars were not sufficient, models were run for which the diameter of the stars could also be free parameters. Following the grid fitting, we used bootstrapping (e.g. Figure \ref{bs}) to determine the errors on our derived measurements of the sources and to confirm the final values. In the bootstrapping procedure, the data were drawn randomly to create new datasets, and the final parameters and uncertainties were estimated as the average and standard deviation of all the fits that were performed. 

Tests were run on all companions to check the significance of any detected companions (e.g. Figure \ref{siggrid}). As the grid search already computed the $\chi^2_{red}$, the $\chi^2_{red}$ statistics and the number of degrees of freedom were used to define the significance of each of the fits across the grid in terms of $\sigma$. We note that the code saturates numerically for high-significance values. To remain reliably within the numerical accuracy, the maximum significance quoted therefore is 8$\sigma$, which corresponds to a 10$^{-15}$ chance of a false detection. When a binary companion was found to be insignificant, we instead ran single-star fits, where the parameter fit is simply the diameter of the star. 

We also attempted fits of higher-order multiple system during our analysis. We did this using the following method. First, we fitted a binary model. Then, if the fit to the data still appeared to be poor, we ran a grid fit for the next companion. The position of the secondary companion that was found would be fixed during this search, and the grid would iterate on the position of the potential tertiary companion and so on. Running a grid fit first allowed us to determine the level of the total fits and the number of steps across a grid fit that would be appropriate. We then used these values to run a search to determine a detection limit, following the methods of \citet{absil11}. Additionally, the significance of the flux determined at the best-fitting positions for any found companions was tested, and we considered any companion with a flux within a 3$\sigma$ flux distribution to be significant. When the flux of the determined companion was found to be significant and above the detection limit at all data epochs we analysed, we concluded that the system was a higher-order multiple system. We applied some key exceptions. When the position of an additional companion (e.g. tertiary) overlapped with the position of the previously found binary, we did not consider this a valid higher-order multiple. When the first binary grid fit was degenerate as a result of sparse u-v coverage and an additional companion was essentially a mirror of a previously found companion because of this, we did not consider this a valid higher-order multiple either. Finally, when the position of the companion was inconsistent at different data epochs (when they existed) without a significant time difference between epochs (some weeks to months), we did not consider this a valid higher-order multiple. The latter case only occurred for an insignificant fraction of the sample. When the flux varied between epochs, this was accepted for systems that were known variables, but it was not accepted in systems for which variability had not previously been reported. We found interferometric triples in our sample that fulfilled all these criteria, but no higher-order interferometric multiple systems such as quadruples. 

\begin{table*}[h!]
    \centering
    \caption{Details on the multiplicity of the systems derived from the best-fitting datasets.}
\begin{spacing}{1.2} 
    \begin{adjustbox}{width=\textwidth}
    \begin{tabular}{ccccccccccc}
        \hline
                \hline
        Name & Type & $\chi^2_{red}$ & $F_2$ & $\rho_2$ & $F_3$ & $\rho_3$  &  3-$\sigma$ \\
                 &  &  & &  &  &   & detection \\
         & & & & (mas) & & (mas) & distribution \\
         \hline
    HD\,16582 & Single & 0.34 & & & & & 0.002 - 0.011 \\
    HD\,51480* &  & 1.72 &  &  & & & 0.22 - 1.02\\
    HD\,66765 &  & 0.54 & & & & & 0.016 - 0.098 \\
    HD\,67621 &  & 0.34 & & & & & 0.005 - 0.021 \\
    HD\,121743 &  & 0.31 & &  &  & & 0.005 - 0.041 \\
    HD\,189103 &  & 0.56 & &  &  & & 0.004 - 0.023\\
    HD\,205637 &  & 1.21 & & & & & 0.005 - 0.073 \\
    HD\,212571 & & 0.37 & & & & & 0.008 - 0.025 \\
    MCW\,1019 &  & 2.59 & & & & & 0.047 - 0.120 \\
    \hline
$\gamma$ Peg & Binary & 0.11 & 0.036$^{+0.001}_{-0.004}$ & 85.6$\pm$0.1 & & & 0.026 - 0.330\\
HD\,3379 & 	& 0.25 & 0.159$^{+0.002}_{-0.002}$ & 8.49$\pm$0.03 & & & 0.005 - 0.018\\
HD\,25558 &  & 3.15 &	0.793$^{+0.05}_{-0.03}$ & 92.56$\pm$0.20 & & & 0.020 - 0.088\\
HD\,30836 &  & 0.99 & 0.060$^{+0.04}_{-0.02}$ & 3.39$\pm$0.43 & & & 0.010 - 0.073 \\ 
HD\,32249 &  & 16.22 & 0.616$\pm$0.01 & 31.49$\pm$0.04 &  & & 0.071 - 0.278 \\
HD\,34816 &  & 0.77 & 0.049$^{+0.004}_{-0.003}$ & 62.7$\pm$0.1 & & & 0.026 - 0.084 \\
HD\,35337 &  & 0.37 & 0.038$^{+0.002}_{-0.002}$ & 9.84$\pm$0.16 & & & 0.037 - 0.127 \\
HD\,35149 &  & 0.92 & 0.537$\pm$0.008 & 23.34$\pm$0.09 & & & 0.427 - 1.23 \\
HD\,37017 &  & 0.26 & 0.16$^{+0.02}_{-0.01}$ & 1.00$\pm$0.05 & & & 0.157 - 0.756 \\ 
HD\,105382 & & 3.79 & 0.066$^{+0.003}_{-0.005}$ & 41.9$\pm$0.5 & & & 0.102 - 0.397 \\
HD\,109026 & & 0.15 & 0.376$^{+0.002}_{-0.001}$ & 12.94$\pm$0.02 & & & 0.003 - 0.014 \\
HD\,133518 & & 0.45 & 0.085$^{+0.377}_{-0.067}$ & 0.61$\pm$3.74 & & & 0.019 - 0.057 \\
HD\,140008 &  & 0.22 & 0.952$^{+0.002}_{-0.003}$ & 1.130$\pm$0.006 & & & 0.398 - 0.938 \\
HD\,178175 &  & 0.32 & 0.0245$^{+0.0006}_{-0.001}$ & 4.76$\pm$0.06 & & & 0.006 - 0.017 \\
HD\,191263 &  & 0.48 & 0.39$^{+0.14}_{-0.01}$ & 1.67$\pm$0.09 & & & 0.014 - 0.073 \\
HD\,212076 &  & 0.46 & 0.28$^{+0.4}_{-0.3}$ & 0.42$\pm$2.60 & & & 0.013 - 0.092 \\
HD\,224990 &  & 0.62 & 0.270$^{+0.002}_{-0.003}$ & 33.4$\pm$0.1 & & & 0.015 - 0.054 \\
$\tau$ Lib &  & 2.11 & 0.11$^{+0.02}_{-0.01}$ & 7.62$\pm$0.2 & & & 0.066 - 0.693 \\
    \hline 
HD\,116658 & Triple & 2.15 & 0.08$^{+0.03}_{-0.01}$ & 0.99$\pm$0.30 & 0.21$\pm$0.01 & 1.69$\pm$0.20 & 0.047 - 0.178 \\
HD\,132058 &  & 0.21 & 0.0213$\pm$0.002 & 6.5624$\pm$0.1 & 0.012$^{+0.002}_{-0.002}$ & 73.7$\pm$0.2 &  0.009 - 0.050 \\
HD\,147932 & & 0.55 & 0.1122$^{+0.004}_{-0.003}$ & 6.70$\pm$0.1 & 0.028$\pm$0.003 & 83.0$\pm$0.3 & 0.015 - 0.051 \\ 
HD\,161701 &  & 0.11 & 0.219$\pm$0.002 & 1.15$\pm$0.1 & 0.052$\pm$0.003 & 2.89$\pm$0.06 & 0.016 - 0.090 \\
HD\,193933 &  & 0.11 & 0.624$\pm$0.1 & 6.17$\pm$0.02 & 0.021$^{+0.003}_{-0.005}$ & 16.5$\pm$0.6 & 0.007 - 0.037 \\
        \hline
    \end{tabular}
    \end{adjustbox}
    \label{source_params}
    \tablefoot{$F$ represents the flux ratio of the components with relation to the brightest components (whose flux was set to one as a reference point). The subscript number represents whether the companion is a binary (2) or tertiary (3) and so on. The best-fitting datasets are presented, quantified by the reduced chi-square, $\chi^2_{red}$. Systems marked with asterisks are discussed in Section 3.4. We include the distribution of 3$\sigma$ detections in terms of the flux ratio in the last column.}
\end{spacing}
\end{table*}

\section{Results and discussion}\label{sec:res}



\begin{figure}[h!]
   \centering
   \includegraphics[width=70mm]{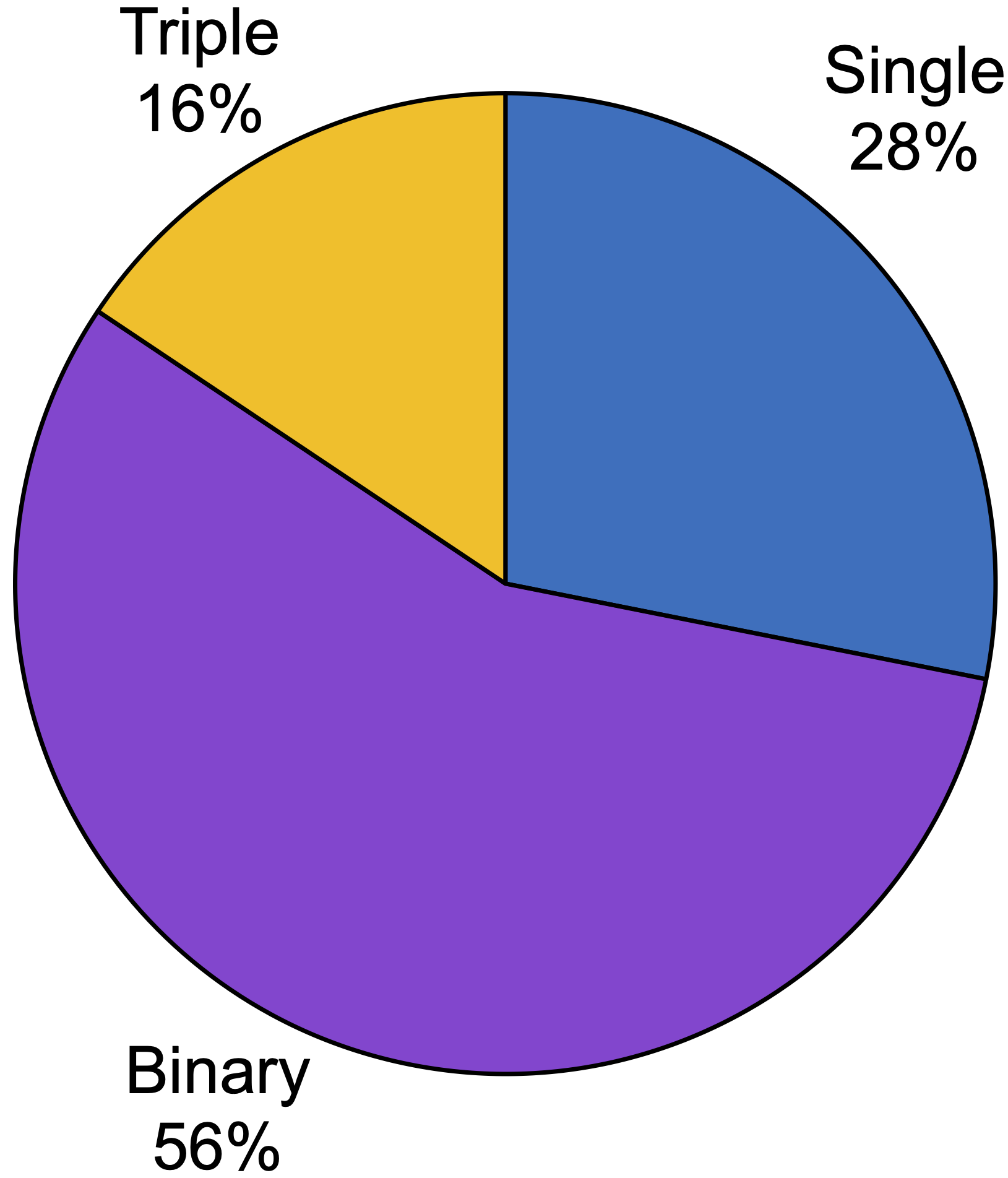}
   \caption{Percentage of the different types of multiple systems detected with interferometry in our B-star sample.}
              \label{interpie}%
    \end{figure}

Following the fitting of the PIONIER data, we detected interferometric companions around 23 of the 32 B stars in our sample. A variety of multiple systems were determined through interferometry and are displayed in Figure \ref{interpie}. Binaries are the most common systems that are detected with interferometry (19), followed by single stars (13) and then triple systems (5). The final fits we obtained to the PIONIER data of our sample are presented in Table \ref{source_params}. We note the sources that are single, and in the case of multiple systems, the fluxes and separations of the sources with respect to the central star. For sources with multiple datasets, the results of the best-fitting epoch are shown. There is no discrepancy between the type of multiple system derived for a source between its datasets at different epochs. We do not discuss the multi-epoch data in detail, but defer this to future work. 

\begin{figure*}[h!]
   \centering
   \includegraphics[width=180mm]{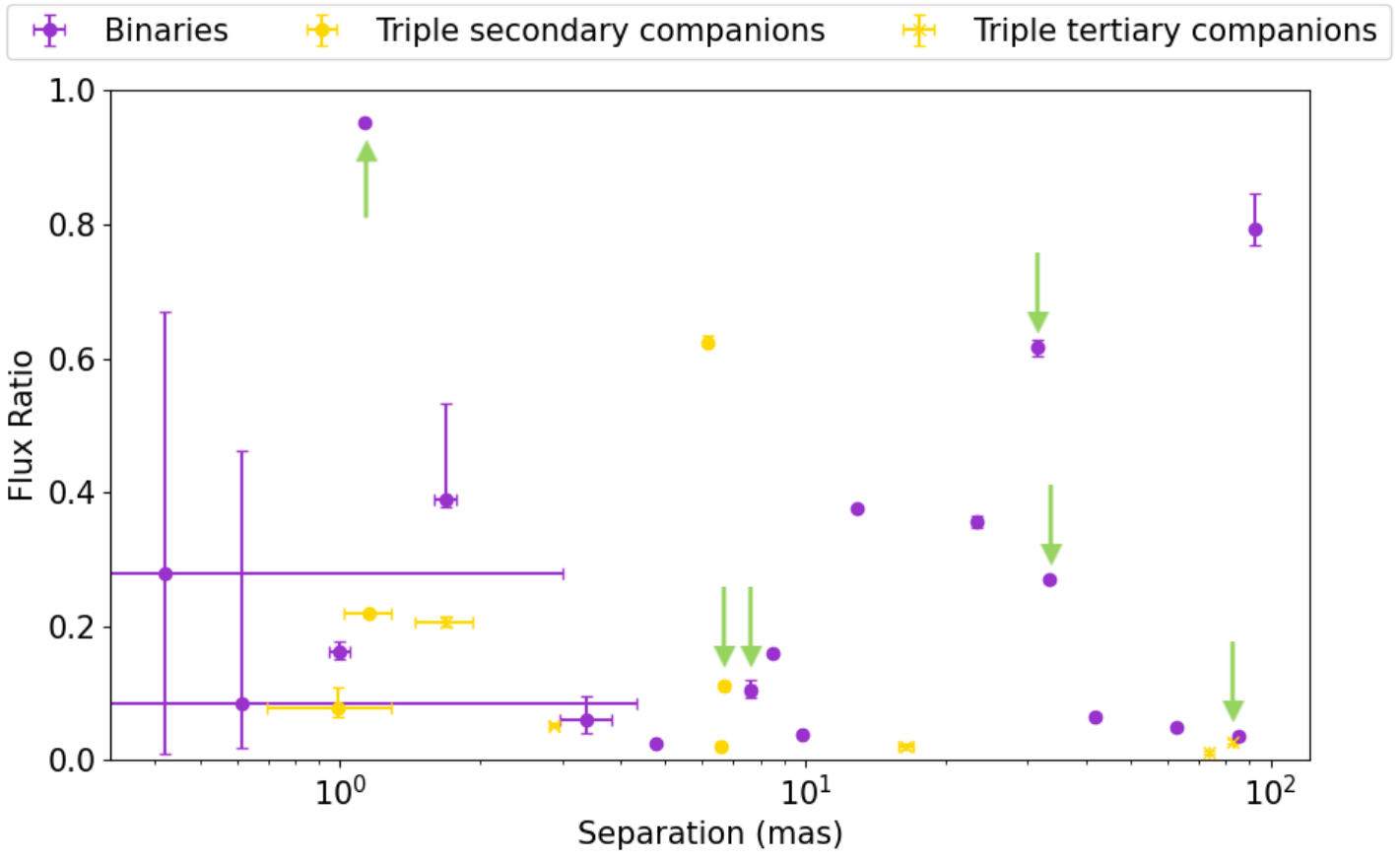}
   \caption{Distribution of companions compared to their flux. Companions belonging to cluster sources are labelled with green arrows, and the remaining sources are field stars. The symbols illustrate the companion that is represented by the point: Binary companions are represented by a circle, and tertiary companions are shown as a triangle.}
              \label{sepflux}%
    \end{figure*}

We calculated the multiplicity fraction ($f_{\mathrm{m}}$) or the ratio of the number of multiple systems ($N_{\mathrm{m}}$) to the sample size ($N$) of our sample following the definitions of \citet{sana14}. Within our interferometric sample, $f_{\mathrm{m}}(N_{\mathrm{m}}/N)$=0.72$\pm$0.08. The statistical error on $f_m$ was calculated using binomial statistics ($\sigma_{f_{\mathrm{m}}}$($f_{\mathrm{m}}$, $N$) = $\sqrt{f_{\mathrm{m}}(1-f_{\mathrm{m}})/N}$, \citealt{sana14}). Similarly, the interferometric companion fraction (the average number of companions per central object or the ratio of the total companions ($N_{\mathrm{c}}$) to $N$) is $f_{\mathrm{c}}(N_{\mathrm{c}}/N)$=1.88$\pm$0.24. The uncertainty on $f_{\mathrm{c}}$ was calculated following Poisson statistics and computed as $\sigma_{f_{\mathrm{c}}}$ = $\sqrt{N_{\mathrm{c}}}$/$N$ \citep{sana14}. Figure \ref{sepflux} shows the separations and fluxes of the detected interferometric companions. For the binaries, the average companion is 27\% of the brightness of the primary star, and the average separation is 23 mas, with a wide range of separations ($\sim$1-90 mas). 

One concern when detecting new companions can be the possibility of chance alignment, especially in clusters. The likelihood that chance alignment causes a false companion detection has been studied in detail for other multiplicity surveys. \citet{sana14} determined the probability of a spurious detection as the result of a chance alignment for their sample of 279 stars. This method has also proven robust in other multiplicity works, such as \citet{madda21}. \citet{sana14} conservatively assumed that all their PIONIER observations were sensitive to separations up 0.2" and found that the probability of spurious detection was always lower than 0.001\%. This means that their interferometric detections were essentially free of spurious detections. It is therefore likely that our detections are also free from contamination by chance alignments. 

               \begin{figure}[b!]
   \centering
   \includegraphics[width=80mm]{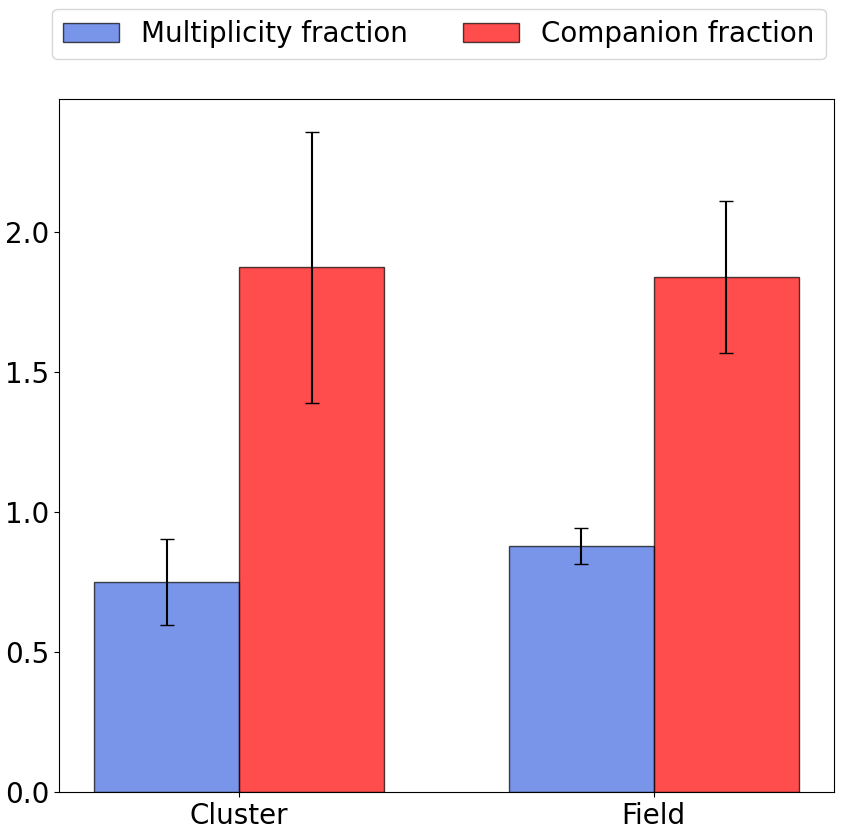}
   \caption{Bar chart displaying the change in interferometric multiplicity and companion fraction for the sources in clusters against those in the open field.}
              \label{fvsc}%
    \end{figure}

\subsection{Trends}

After determining the multiplicity properties of our sample, we investigated whether the characteristics of the stars or their environments had any tangible affect on the number and nature of the companions. Our findings are discussed in the following subsections.

\subsubsection{Trends with source location: Cluster versus field}

First, we investigated the impact of the location on the type of multiple system that is derived. Only 8 of the sample of 32 stars were found to be part of a cluster or stellar association, meaning that these are small-number statistics. For completeness, we note that all but 2 of the cluster systems are multiples, which corresponds to $f_m$=0.75$\pm$0.15, with a companion fraction of 1.88$\pm$0.5. The multiplicity fraction for the cluster sources is larger than for the field sources, where $f_m$=0.71$\pm$0.09. All these fractions are shown in Figure \ref{fvsc}. This is consistent with the hypothesis that a cluster environment could give rise to the disruption of wide systems through dynamical encounters, which leads to more single stars. The field sources have a companion fraction of $f_c$=1.88$\pm$0.28, however, which is equivalent to that of the cluster sources (within the errors). We therefore conclude that a sample that includes more cluster sources is required for real conclusions to be made about the differences or lack thereof between the multiplicity and companions of B stars in cluster versus field environments.

Stellar evolution and age might also affect the multiplicity. Some of our older systems might have had more massive companions that already evolved to explode as supernovae that either disrupted the system or led to a compact companion beyond our detection limits. Therefore, some potential bias remains for older systems to be seen having lower multiplicity, which would also propagate to dependences on both stellar type (see next section) and the environment. Since field stars might be older on average than cluster stars, this might then also decrease the multiplicity fraction among field stars, as seen in our sample. Much work has been done in the literature to determine the ages of stellar clusters, and as a result, we can speculate on the ages of our cluster sources. The majority of our cluster sources come from the Sco association. This cluster has been postulated to be between 4 and 8 million years old (\citealt{gazza} and references within). Our Sco sources contain a mix of single, binary, and triple systems. The remaining systems are in the Orion X association (HD\,32249), the open galactic cluster Blanco 1 (HD\,224990), and the Vel OB 2 cluster (HD\,67621). The first two are both interferometric binaries, and the last appears to be a single star. Blanco 1 is estimated to be 100-150Myr \citep{blanco} old, which is much older than the Sco association. Because our sample is small and we detect no trends, we again refrain from drawing conclusions regarding the multiplicity we detect with age.

           \begin{figure}
   \centering
   \includegraphics[width=90mm]{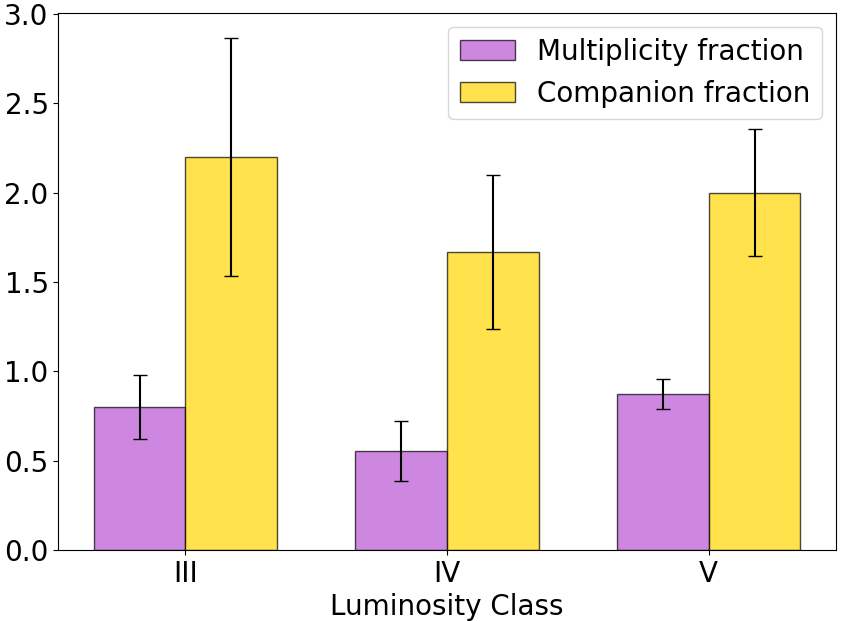}
   \caption{Interferometric multiplicity and companion fractions for the B stars based on their luminosity class.}
    \label{lclassfracs}
    \end{figure}
    
    \begin{figure}
   \centering
   \includegraphics[width=90mm]{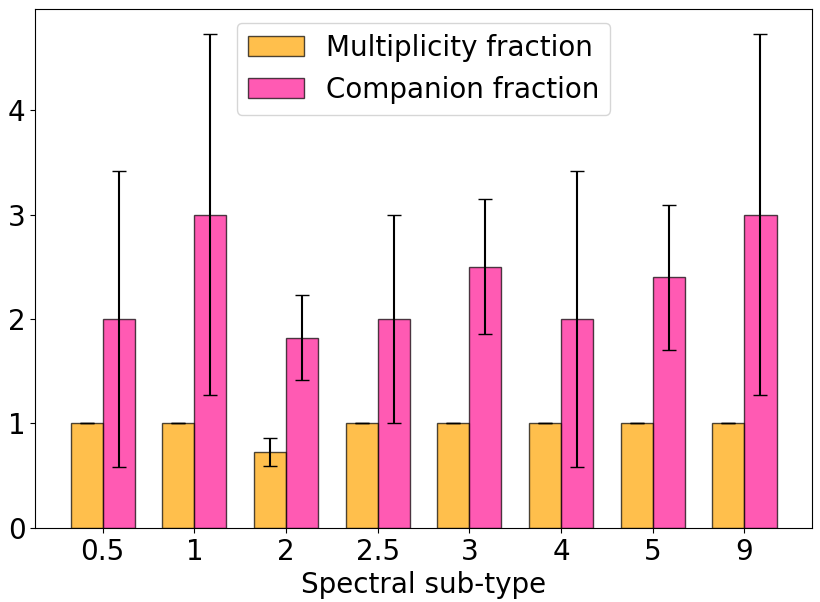}
   \caption{Interferometric multiplicity and companion fractions for the B stars based on their spectral sub-type.}
    \label{sptfracs}%
    \end{figure}

\subsubsection{Trends with luminosity class}

The first stellar characteristic we investigated is the luminosity class, or in other words, the evolutionary stage of the source. We only considered sources with a clear luminosity class and therefore ignored any sources with spectral types such as IV/V. Clear luminosity classes were determined for 30 out of 32 of the stars in our sample. Most of the stars in this group ($\sim$53\%) are in the luminosity class V and are therefore dwarfs. Sub-giants (IV) were the second most common class of star ($\sim$30\%), followed by giants (III, $\sim$17\%). Figure \ref{lclassfracs} shows the variation in multiplicity and companion fraction with luminosity class. The dwarfs have the most statistically significant results because they make up the majority of the sample and have $f_m$=0.88$\pm$0.08 and $f_c$ = 2.00$\pm$0.35. Only 5 sources are giants, and their fractions are accordingly much less reliable, but we note that they have a multiplicity fraction of 0.80$\pm$0.18 and a companion fraction of 2.20$\pm$0.66. Nine sources are sub-giants and have a multiplicity fraction of 0.56$\pm$0.17 and a companion fraction of 1.67$\pm$0.43.


\subsubsection{Trends with spectral sub-type}
The second stellar property we probed is the spectral sub-type. We calculated the multiplicity and companion fractions of each sub-type and present them in Fig. \ref{sptfracs}. The temperature, radius, and mass decrease significantly in dwarfs from type B0.5 to B9, and the surface gravity (log$g$) is the only quantity to stay more or less constant at $\sim$4 \citep{coxbook}. Spectral types B0.5, B1, B2.5, B4, and B9 apply to two or fewer stars, and it is therefore not possible to search for trends. Between spectral types B2 and B3, multiplicity and companion fraction increase as the spectral type increases, but we note that all our B3 sources are multiples. All our B5 sources are also multiple systems, although the companion fraction is slightly lower than that of the B3 sources.

\subsection{Mass estimates}\label{sect:mass}

We estimated the masses of the interferometric companions using the flux ratios of the companions obtained from the fitting. The mass ratio $q$ can be defined as $q$=M$_{\mathrm{comp}}$/M$_{\mathrm{prim}}$, or the companion mass over the mass of the primary star. Following \citet{cyprien23} and using the relations of \citet{jb17} and \citet{martins05}, we approximated the mass ratio of MS stars as $q$ = $F_H^{0.7}$, where $F_{H}$ is the H-band flux. This method is unreliable when the central component of the multiple system is an unresolved binary because the mass ratio uses the combined flux of both components, and hence, the estimated mass ratio will not be accurate. Excluding the sources with known spectroscopic companions, a lack of data, or conflicting results for this reason, we provide the mass ratio estimates for the remaining sources in Table \ref{masstab}. The error estimates in these calculations are based solely on the systematic errors derived for the $H$-band flux ratios with PMOIRED.  Long-term monitoring of all systems in our sample with both spectroscopy and interferometry would enable us to derive results on a more statistically significant sample.

\begin{table*}
    \centering
    \caption{Mass ratio estimates for the sources and their interferometrically detected companions.}
    \begin{tabular}{ccc}
        \hline
                \hline
         Name & \multicolumn{2}{c}{Mass ratio} \\
          & Companion 1 & Companion 2 \\ \\
         \hline
HD\,3379 & 0.276$^{+0.002}_{-0.002}$ & \\ \\ 
HD\,34816 & 0.121$^{+0.006}_{-0.005}$ & \\ \\ 
*HD\,116658 & 0.331$^{+0.01}_{-0.007}$ & 0.168$^{+0.05}_{-0.020}$ \\ \\ 
HD\,132058 & 0.0676$\pm$0.004 & 0.045$^{+0.006}_{-0.004}$ \\ \\
HD\,147932 & 0.216$^{+0.006}_{-0.004}$ & 0.083$\pm$0.006\\ \\
HD\,178175 & 0.075$^{+0.001}_{-0.003}$ & \\ \\
HD\,191263 & 0.517$^{+0.1}_{-0.01}$ & \\ \\
HD\,212076 & 0.41$^{+0.4}_{-0.3}$  & \\ \\
    \hline 
    \end{tabular}
    \label{masstab}
        \tablefoot{The masses for sources marked with asterisks are lower limits because they might be contaminated by spectroscopic sources, as detailed in Section 3.4.}
\end{table*}

\subsection{Estimating a complete multiplicity fraction}

A literature search was performed for companions within the inner working angle (IWA) and beyond the outer working angle (OWA) of PIONIER, including the use of the ninth catalogue of spectroscopic binary orbits by \citet{specbins}. In addition to the interferometric companions we detected, we found that 15 systems have confirmed spectroscopic companions (listed in Table \ref{spec_comps}) and that a further 10 systems were studied previously and were determined not to harbour a spectroscopic companion. The remaining 11 sources in the sample either have conflicting reports of companions/non-detections or lack the data. Eight of the spectroscopic companions, given the length of their periods and their distances, likely constitute one of the companions we detected with interferometry.

We also searched for wide companions using the methods of \citet{andhag} and \citet{elbadryrix}. They computed the angular separation on the sky ($\Delta\Theta$), the difference in parallax ($\Delta\omega$) and its error ($\sigma_{\Delta\omega}$), the difference in proper motion ($\Delta\mu$) and its error ($\sigma_{\Delta\mu}$), and the potentially expected difference in proper motion due to orbital motion ($\mu_{orb}$) to calculate the likelihood of whether a Gaia DR3 source orbited another. When these methods were applied to our sample, we found that two sources, HD\,121743 and HD\,37017, have potential Gaia companions (described in Table \ref{wide_comps}). While this may appear low for a sample of 32 stars, \citet{andhag} reported a rate of ultra-wide companions of 0.044, which would mean that only 1.7 stars from this sample are expected to have wide companions.

\begin{figure}[b]
   \centering
   \includegraphics[width=75mm]{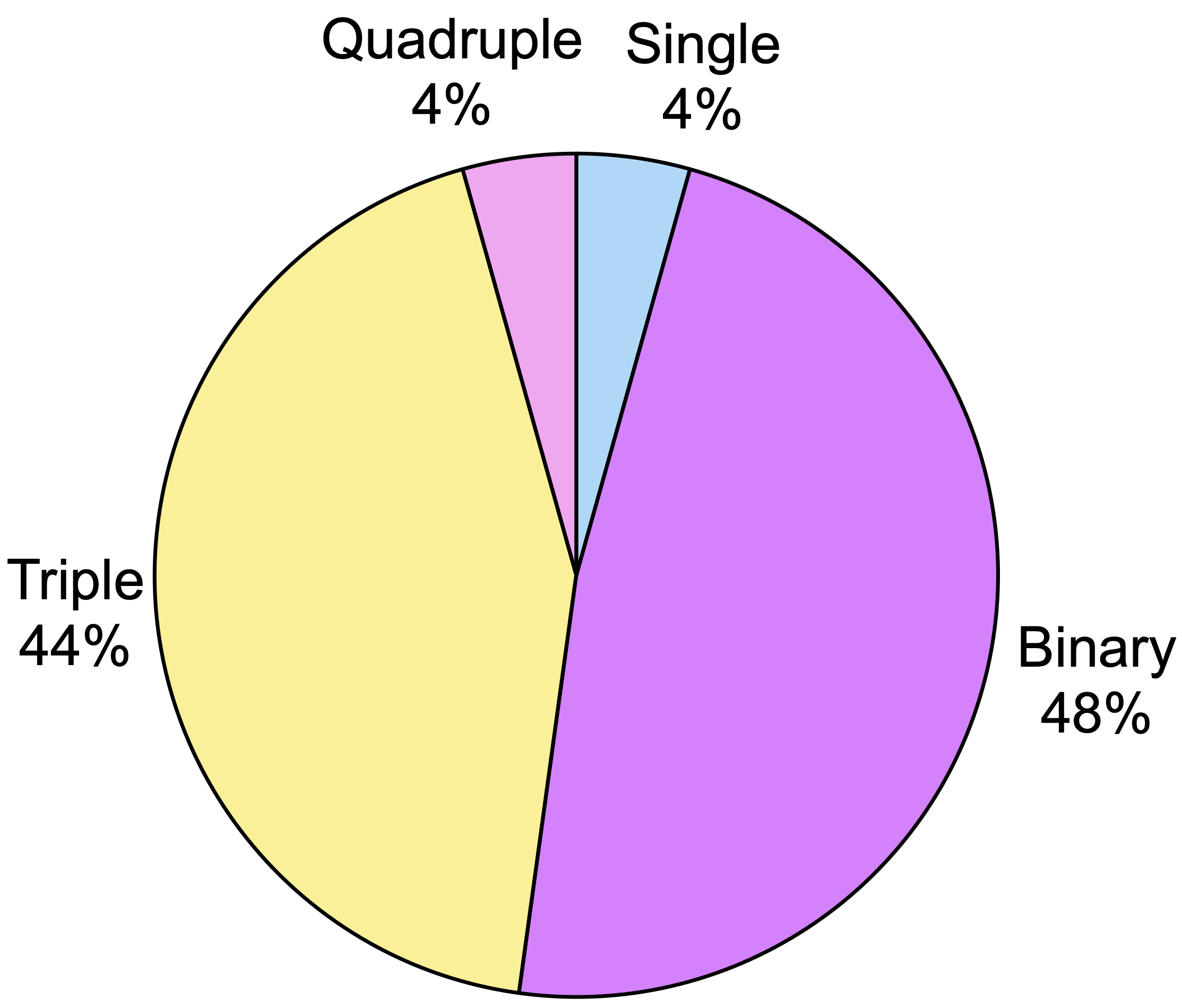}
\caption{Pie chart showing the multiplicity of our entire B star sample, taking into account spectroscopic, interferometric, and Gaia DR3 companions. This chart excludes stars for which the results in the literature are unclear for the presence of a spectroscopic companion or those that lack spectroscopic data for a search for such a companion, meaning that these percentages are calculated for a sub-sample of 23 sources and not for the total 32.}
              \label{compie}%
    \end{figure}

We calculated a complete multiplicity and companion fraction for the sources in our sample for which a clear report of companions/non-detections for spectroscopic companions is available. This sub-set of our original interferometrically studied sample constitutes 23 sources. We find the multiplicity fraction for this sub-sample to be 0.96$\pm$0.04 and the companion fraction to be 2.65$\pm$0.34. The distribution of the different companions is illustrated in Figure \ref{compie}. Notably, a significant number of the systems that appear in the interferometric data as binaries are in fact hierarchical triples because they have an inner spectroscopic companion. The higher-order multiplicity fraction for this sub-sample is $\sim$47\%. 

\begin{figure}[b]
   \centering
   \includegraphics[width=75mm]{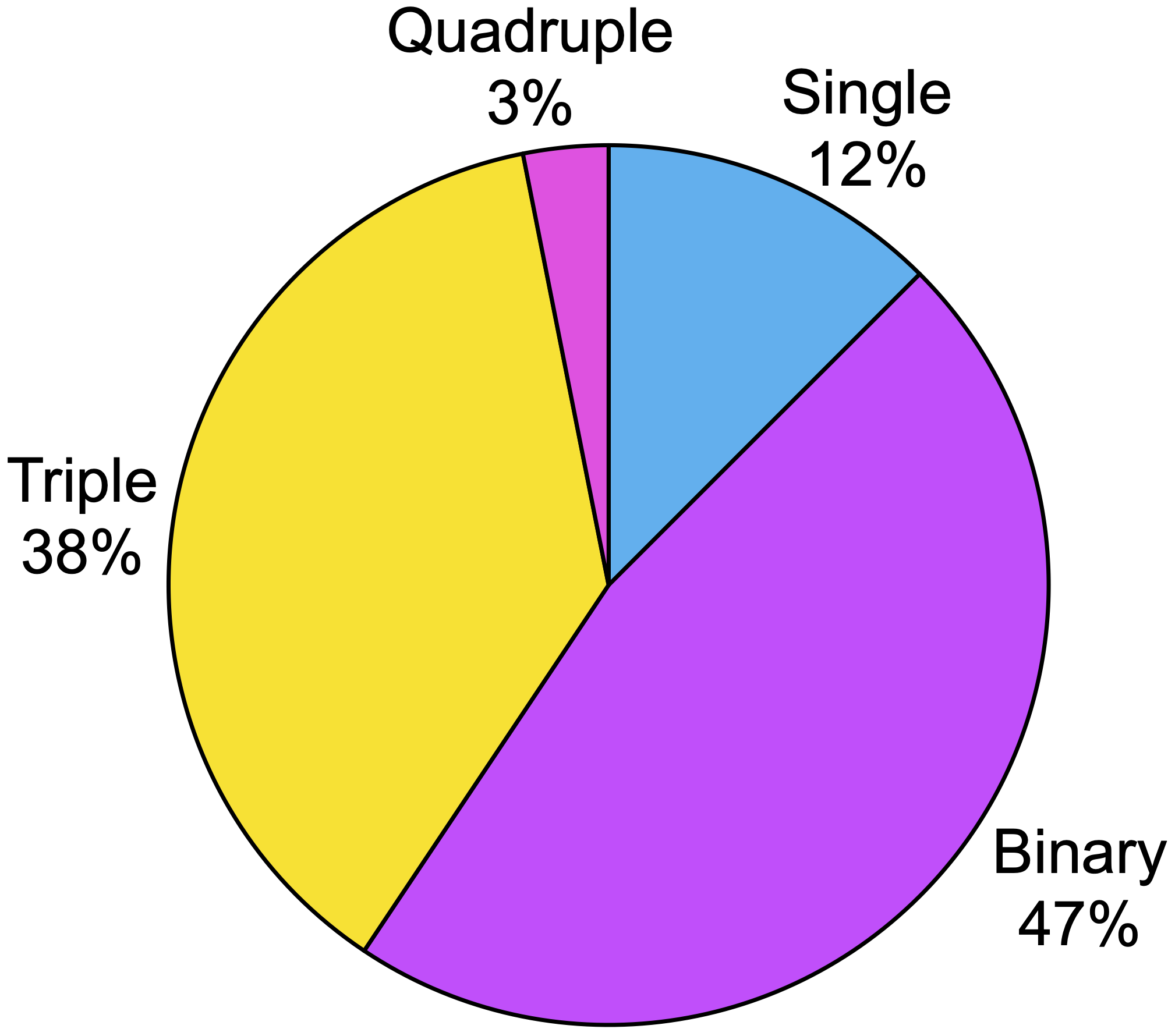}
   \caption{Pie chart showing the multiplicity of our entire B-star sample of 32 sources, taking into account spectroscopic, interferometric, and Gaia DR3 companions.}
              \label{compie_total}%
    \end{figure}

We note that most of the interferometric single stars in our sample lack data or show conflicting results for spectroscopic companions, which we did not include in the complete multiplicity analysis. For transparency, we also include Figure \ref{compie_total}, which does not exclude stars based on unclear spectroscopic detections or non-detections. In this case, the multiplicity fraction is 0.88$\pm$0.06 and the companion fraction is 2.31$\pm$0.27. The higher-order multiplicity fraction in this case is $\sim$40\%. 

\begin{table*}
    \centering
    \caption{Results of the literature search for spectroscopic and/or long-period companions.}\label{t:spectro}
    \begin{tabular}{cccc}
        \hline
                \hline
         & Name & Reference & Notes \\
         \hline
 Clear detections & $^*$HD\,25558 & \citet{sodor14} & $P$=8.9$\pm$0.5yr \\
      & HD\,30836 & \citet{sbhd30836}, \citet{mahy22} & $P$=9.5191d, P=9.519999$\pm$0.000409d \\
      & $^*$HD\,35337 & \citet{abt} & $P$=106.7$\pm$0.4d \\
      & HD\,37017 & \citet{sbhd37017} & $P$=18.6556$\pm$0.0017d \\
      & $^*$HD\,105382 & \citet{kervbin} & $R_{orbit}$ from Gaia DR2 - 2.469 au \\
      & HD\,116658 & \citet{alfvir} & $P$=4.0145d\\
     & HD\,140008 & \citet{sbhd140008} & $P$=12.26d \\
     & HD\,161701 & \citet{sbHD161701} & $P$=12.4520d \\
     & HD\,189103 & \citet{sbHD189103} & $P$=2.1051d \\
     & $^*$HD\,205637 & \citet{rivi06} & $P$=128.5d \\ 
     & $^*$HD\,212571 & \citet{bjork2002} & $P$=84.07$\pm$0.02d \\
     & $^*$HD\,224990 & \citet{sbHD224990} & $P$=1740$\pm$22d \\ 
     & $^*$MCW 1019 & \citet{sbmcw1019} & $P$=2245$^{+25}_{-30}$d \\
    & $\tau$ Lib & \citet{specbins} & $P$=3.29d \\
             \hline
    Non-detections & HD\,3379 & \citet{abt84} & \\
        & & \citet{abt} & \\
    & & \citet{telt06} & \\
    & HD\,16582 & \citet{abt} & \\
    & HD\,34816 & \citet{telt06} & \\
    & HD\,121743 & \citet{brown97} & \\
    & & \citet{shat02} & \\
    & & \citet{telt06} & \\
    & HD\,132058 & \citet{brown97} &  \\
     & & \citet{telt06} &  \\ 
    & HD\,147932 & \citet{brown97} & \\
    & & \citet{rosslowe18} & \\
  & HD\,178175 & \citet{abt84} & \\
    & HD\,191263 & \citet{abt84} & \\
    & HD\,212076 & \citet{percy77} & \\
    & & \citet{hanu87} & \\
    & & \citet{chauville01} & \\
    \hline 
    Conflicting results &  HD\,109026 & & \\ 
     & $\gamma$ Peg & & \\
     & HD\,66765 & & \\
     & HD\,35149 & & \\
     & HD\,51480 & & \\
     & HD\,32249 & & \\
    \hline 
    Lack of data & HD\,133518 & & \\
      & HD\,67621  & & \\
      & HD\,193933 & & \\
    \hline 
    \end{tabular}
    \label{spec_comps}
    \tablefoot{Sources with an asterisk are those whose companions are on long enough orbits to overlap with the interferometric detection range.}
\end{table*}

\begin{table*}
    \centering
        \caption{Wide binaries found in the Gaia DR3 database for the entire sample.} 
 \begin{tabular}{llccrcccrcccc}
        \hline
        Name        &  Gaia primary & Gaia companion & $\Delta \theta$  & $\Delta\varpi \pm \sigma_{\Delta\varpi}$ & $\Delta \mu\pm \sigma_{\Delta\mu}$ & $\rho$ \\
                    & (Gaia DR3 name) & (Gaia DR3 name)  & (arcsec)           & (mas)                                          & (mas year$^{-1}$)                        & (a.u.)\\
        \hline
HD\,121743  &  6110381256839901440  &  6110394073027855360  &  293.3  & $ 0.049  \pm  0.279 $ & $ 0.972  \pm  0.424  $ &  41302.0  \\
  &   &  6110385929769619328  &  634.1  & $ 0.067  \pm  0.276 $ & $ 0.702  \pm  0.404  $ &  89314.1  \\
HD\,37017  &  3209634905754969856  &  3209634905754971136  &  22.8  & $ 0.178  \pm  0.155 $ & $ 0.67  \pm  0.116  $ &  8175.8  \\
    \hline
    \end{tabular}
    \label{wide_comps}
    \tablefoot{$\rho$ is the projected orbital separation.}
\end{table*}

\subsubsection{The prevalence of triple systems}

The most likely expected configuration of stars in a bound triple system is a close inner binary and a tertiary companion at a much larger separation that orbits the centre of mass of the inner binary. Such a system is referred to as a hierarchical triple system \citep{hier}. This is expected to be the most common form of triple system because when the inner and outer orbits in a triple have similar radii, the system can become dynamically unstable, causing a star to be ejected from the system \citep{kiseleva}. One of the most commonly used tests to determine whether a triple system is stable is that of \citet{marda}. This calculation requires inclination and eccentricity information for the systems, however, which we do not possess for most of our sample.

The separations of our sample of interferometric candidate triples alone yield separation ratios of 58\%, 9\%, 9\%, 40\%, and 37\% between the separation of the binary companion and the primary and the separation of tertiary and the primary. This means that two systems have ratios lower than 15\%, which makes them likely hierarchical triples. The remaining three interferometric triples are systems with separation ratios between 15-55\%, which makes their fates more uncertain without more orbital information. The timescale for an unstable system is about 30 times the inner binary period \citep{grishin}, which is in turn much shorter than the expected lifetimes of these systems. It is statistically highly improbable that the majority of the comparable-separation triple systems we observed are actually unstable. Instead, it is likely that apparent non-hierarchical triples are a result of projection effects, and that most and likely all of the systems are in a stable, hierarchical configuration. Orbital monitoring of all these triple systems will allow us to separate the systems that may be on the precipice of collapse from those that are more stable and to eliminate any projection effects, but this is beyond the scope of this paper.

By observing a sample of 165 spectroscopic binaries with follow-up NACO imaging, \citet{toko06} found that short-period binaries often have outer companions (63\%$\pm$5\% after correction for incompleteness), with 96\% of their sample having an outer companion when the period of the inner binary was shorter than 3d and 34\% for periods longer than 12d. This was for a sample of solar-type stars, whose observed multiplicity and companion fractions are expected to be lower on average than that of B-type stars. We used the information we compiled to calculate our complete multiplicity fraction to compare our results with their work. Of our sample, only HD\,189103 and $\tau$ Lib have spectroscopic companions with periods of 3d or shorter (the former has no outer companion, and the latter has one interferometric companion), but this is extreme low-number statistics, and we therefore do not expect agreement with \citet{toko06}. 

\subsection{Notes on specific sources and types of star}

\subsubsection{HD\,51480}

When we fitted single-star models to our datasets, we fitted the diameter of the uniform disk instead of fixing it such that it could be modelled as an unresolved point source (0.2 mas). The majority of the sources that we fitted with single-star models converged on uniform disk values of 0.5 mas or lower. HD\,51480 is much more extended, however, with a fit diameter of 1.38$^{+0.03}_{-0.08}$mas, corresponding to $\sim$1 au in diameter given the distance to the source. This is consistent because a decreasing visibility profile is typical of an extended structure. Additionally, the spectra from \citet{murphy20} show P-Cygni profiles, which might imply the presence of a hypergiant star. The deviations from the smooth curve typical of a disk caused us to also try multiple companion models for the system, however. In these cases, we were not succesful when we left the diameter of the primary free because the model was unable to converge. A binary model formed of two point sources can also fit the data well, with the flux of the secondary companion comparable to that the primary within the errors (0.92$\pm$0.16) and a very small separation (0.6995$\pm$0.046). This did not improve the fit to the data recorded across the K0G1 baseline, however (see Figure A.3). The system was classified as a Be star by \citet{hd51480}. \citet{wang18} included the source in their candidates for Be- and sub-dwarf O-star systems, but the signal associated with the system did not pass their selection criterion for further study. We also performed a triple fit to determine whether there were any hints of a faint, more distant companion. A triple model in which the secondary was fixed as described improved the fit, but the tertiary companion was below our calculated detection limit as well. We therefore chose a single-star fit as the final model.

\subsubsection{Be stars}

Classical Be stars, as their name suggests, show strong emission lines that are associated with rapid rotation circumstellar decretion disks that are generated as a result of this rotation and non-radial pulsations (e.g. \citealt{struve}, \citealt{rivi13}). The Be-star phase is observed to be transient, and $\sim$20\% at least of the B-type stars are observed in this form. The multiplicity of Be stars is important for understanding their nature because one major potential cause of their rapid rotation is binary interaction, although the exact percentage is still debated (e.g. \citealt{vanbev}, \citealt{pols}, \citealt{hastings21}). The binary formation channel assumes that two MS stars are in a close binary orbit, with one star being more massive than the other. The more massive star will swell first, overfill its Roche lobe, and transfer mass to the other star. This transfer spins the second star up, and eventually, only the He core of the (originally) more massive stars remains. This He star is then likely to go through a SNe event, which may result in the formation of a Be X-ray binary. Suggestions have also been made that triple systems may also play a role in the creation of Be systems (e.g. \citealt{dodd}).

The sample includes six stars that have been reported to be Be stars. The Be stars in our sample show a range of multiplicities. When we consider the complete multiplicity fraction, five of the systems are binaries, and the remaining system is a single star. Two of the binaries, HD\,105382 and HD\,178175, are interferometric, whilst the remaining systems are binaries because they have spectroscopic companions. We measured a separation for the companion of HD\,105382 of $\sim$37 mas, corresponding to $\sim$4 au given the source distance, and $\sim$5 mas/$\sim$2 au for HD\,178175. 

The remaining binary Be systems are HD\,205637, HD\,212571, and HD\,212076. \citet{klem19} reported a downturn in the spectral energy distribution of HD\,205637, which indicates a binary companion, and \citet{rivi06} also reported it to be a binary. BeSS spectra of this system show strong double-peaked emission and narrow lines, implying that if this is a classical Be system, we view it close to edge-on.

HD\,212571 (also known as pi Aqr) is one of the most famous Be stars. The source is also included in the BeSS catalogue \citep{bess}. A number of comments on its multiplicity have been made based on spectroscopic data. \citet{bjork2002} proposed that based on trailing H-$\alpha$ emission and radial velocity variations, the source might be a binary system with a period of $\sim$84 days, and \citet{langer2020} note that this companion might be a stripped star. If this is indeed a signal from the companion, the orbit is relatively long. However, \citet{klem19} reported no evidence of a companion in the study of the source SED, and \citet{wang18} and \citet{horch} failed to find an indication of a companion in their spectra. During the fitting process, we found that a model consisting of a single extended disk can fit the data of the source. When fitting the data of HD\,212571, we were able to fit a binary model. The separation of the companion was easy to constrain and was very small ($\sim$0.15 mas/0.05 au on average). This separation might correspond to a companion in the spectroscopic range. The errors on the flux were large (>80\%), however, and varied greatly between epochs. We therefore deemed a binary model unreliable and settled on a single-star model, in agreement with \citet{klem19}, \citet{wang18} and \citet{horch}. 

The final Be-star system is HD\,51480, which we discussed in the previous subsection. Because the system appears to have a large extended component, it is unclear where the Be signature comes from in the system because the predicted size of this extended component is much larger than the predicted sizes of Be-star disks (1 au vs. orders of stellar radii). Since the nature of the system is not clear, we do not further discuss it as a Be star.

\subsubsection{Variable stars}

Another common type of star in our B-star sample is variable stars. A number of comments have been made in the literature between variability and multiplicity. Eight of the stars in our sample are noted as variable in the literature. Two systems appear to be single, four are binaries, and two are triple systems.

One particularly discussed group of variable stars are classical Cepheid stars, which radially pulsate on rigid periods. \citet{evanscep} postulated based on Hubble UV spectra that the massive stellar system Y Car, which was previously classified as a binary classic Cepheid, was in fact a triple system. Similarly, a study of a sample of Galactic Cepheids with interferometric data by \citet{galcep} detected several companions.

The most prolific type of variable in the sample are the $\beta$-Cep variables. The stars have masses $\gtrapprox$7\,M$_{\odot}$ and exhibit minor rapid variations in brightness \citep{bceprev}. \citet{lefevre} studied the variability of 30\% of the OB stars observed with HIPPARCOS and found that OBe stars were more variable than the average, OB MS stars are less variable than the average, and that OB supergiants show average variability. This variation is thought to be due to pulsations at their surface driven by the $\kappa$-mechanism (whereby high-opacity regions of partly ionised H and He in the stellar surface sink and rise in a cycle) and p-mode pulsations (radial pulsations caused by pressure waves travelling longitudinally that disturb the H and He material in the stellar envelope; \citealt{lefevre}).

A type of binary that can result in variability are eclipsing binaries, in which stars within the system orbit on a plane that obstructs our line of sight. We searched the literature for evidence of eclipsing binaries within our sample. We find that three of our triple systems, HD\,116658, HD\,25558 (both from \citealt{lefevre}) and HD\,35149 \citep{luc21}, were categorised as eclipsing binaries. \citet{lefevre} only tentatively described HD\,116658 as an eclipsing binary, stating a possible period of 4.014\,d. This is almost identical to the period of the spectroscopic companion detected around the system by \citet{shob}, and we tentatively assume that this spectroscopic companion is the secondary star we detected with interferometry. \citet{lefevre} defined a period of 1.532d for HD\,25558. A spectroscopic companion again exists for this system, but it has a long period, and it is thus more likely our interferometrically detected companion. \citet{luc21} noted the presence of an eclipsing binary with a 2.28d period for the system HD\,35149. {\citet{sodor14} found a long-period spectroscopic companion, which we suggest is the companion we detected interferometrically.
When the eclipsing companion is taken into account, the system might therefore be in fact a triple, and we included it as such in our complete multiplicity fraction plots and calculations (Section 3.3).

\subsection{Comparison with other work}

Using the results of \citet{silaj14}, we attributed masses to the dwarf stars of our sample using their spectral sub-type. We omitted the giants we observed because of the large uncertainties associated with the determination of their stellar properties. \citet{moe17} collected the multiplicity statistics of a variety of O, B, and A stars by studying the data of a large number of sources using observations that covered much of the separation range of companions. They determined that at low masses, the most likely form of a stellar system is a single-star system ($\sim$0.6), followed by a binary system ($\sim$0.3), then a triple system ($\sim$0.07), and a quadruple system ($\sim$0.01). As the mass of the system increases, these ratios change. Above $\sim$2\,M$_{\odot}$, the binary system becomes the most probable system. After $\sim$12\,M$_{\odot}$, the triple becomes the most likely system, and after $\sim$23\,M$_{\odot}$, the quadruple system is most probable. We find that the triples range in mass from $\sim$6-15\,M$_{\odot}$. The binary mass range is 8-11\,M$_{\odot}$. Only one dwarf system is a quadruple (with an estimated primary mass of 8\,M$_{\odot}$). Of the dwarf star systems that we can compare with \citealt{silaj14}, the binaries have masses between 5.9-9.11\,M$_{\odot}$ and the triples have masses between 5.9-13.21\,M$_{\odot}$. 

In order to compare the multiplicity fraction we derived with interferometry for B stars to a similar survey for O stars, we considered the SMaSH+ survey \citep{sana14}. After combining their results from PIONIER data with known spectroscopic companions, they concluded that 91$\pm$3\% of the O-star systems they observed were multiple systems. This agrees with the complete multiplicity fraction we derived and the conclusion of the two works is the same: The majority of massive stars form in at least a binary system. 

 \citet{gazza} recently studied the multiplicity properties of B stars specifically in one open cluster, NGC 6231. They determined a spectroscopic binary fraction for the cluster of 52$\pm$8\% after bias correction, which is significantly lower than the fraction we derived for the cluster sources of our sample. However we noted that they probed a very different physical separation range for their stars.

An interesting difference between our study and those we discussed above is the fact that the vast majority (SMaSH+) or all of the stars \citep{gazza} are in clusters or associations, whilst the majority of our B-star sample is field stars. One previous study of binary systems in clusters, \citet{hu}, determined an early constraint on the variation of the binary fraction in a young star cluster. They determined that more binaries should be found towards the centre of clusters. \citet{deacon20} recently used Gaia DR2 data to investigate the multiplicity of stars in the Alpha Per, Pleiades, and Praesepe clusters. They reported that the average separation of the wide binary systems is smaller than that of the field binaries, implying that while some systems are able to persist, it is likely that the dynamically highly processing cluster environment disrupts most wide binary systems. The increased gravitational potential wells of O-type stars may mean that for the SMaSH+ sample with its many cluster members, the outer companions were disrupted in the cluster environments and were removed. This would lower their companion fraction.

\begin{figure*}
\includegraphics[width=90mm]{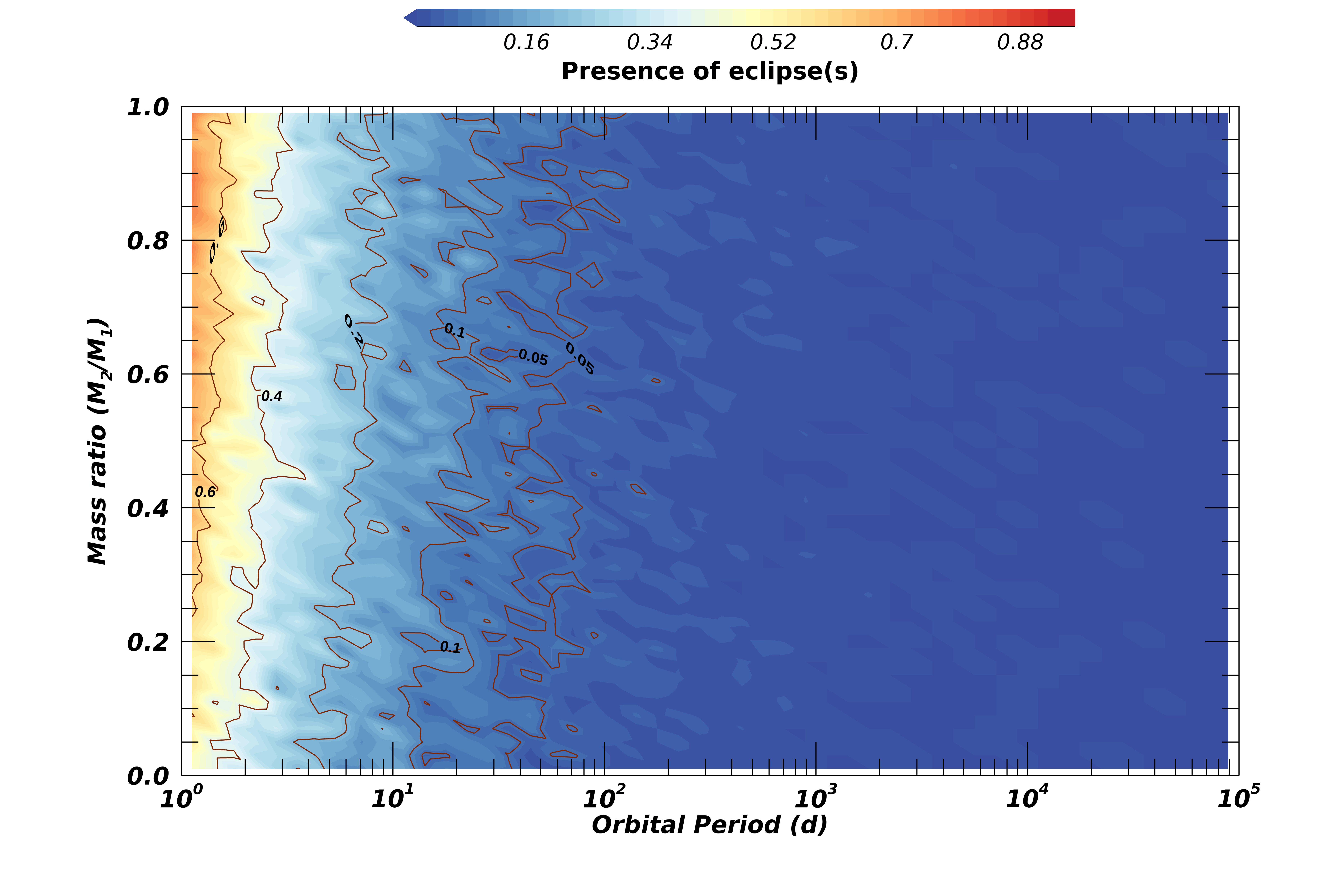}
\includegraphics[width=90mm]{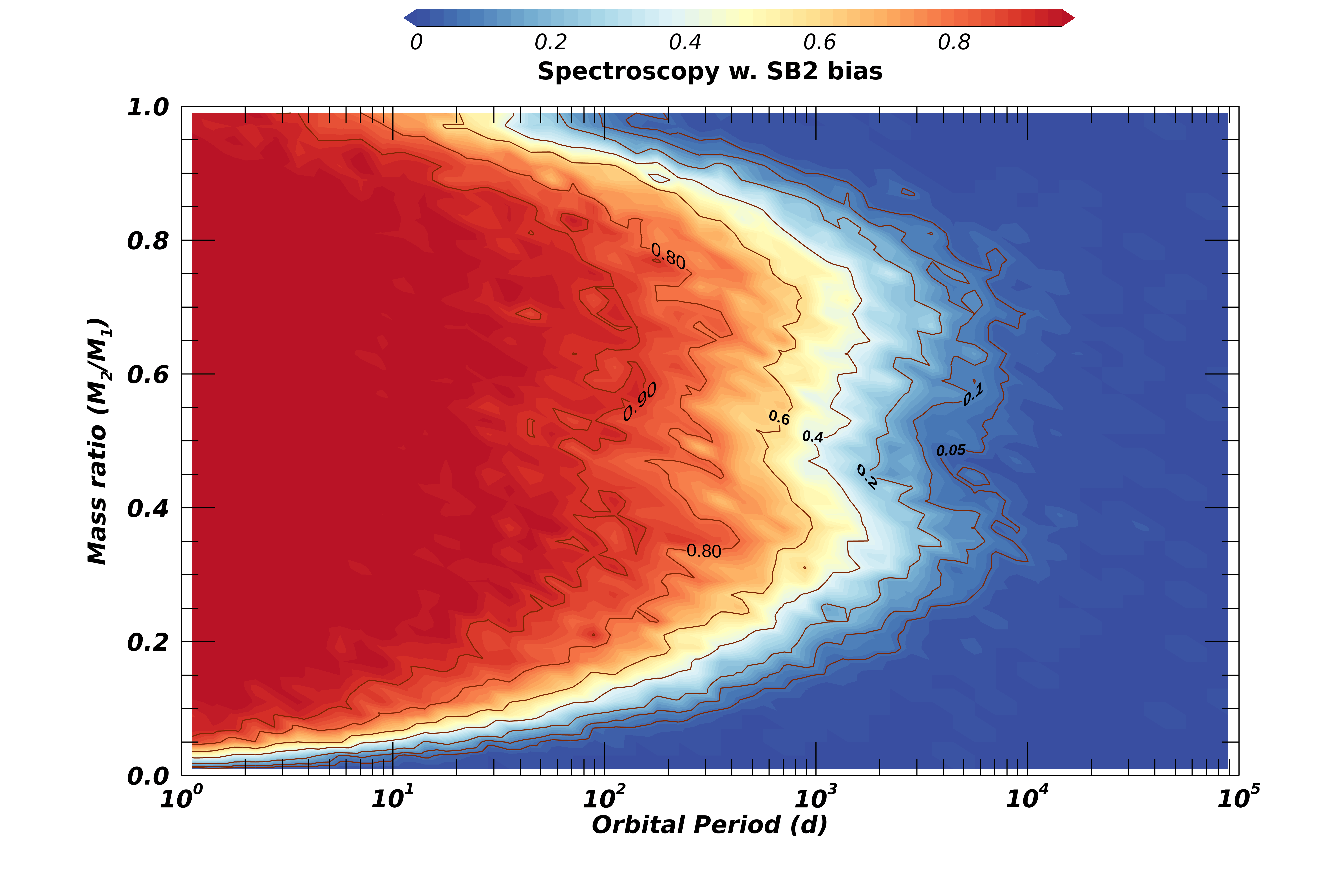}\\
\includegraphics[width=90mm]{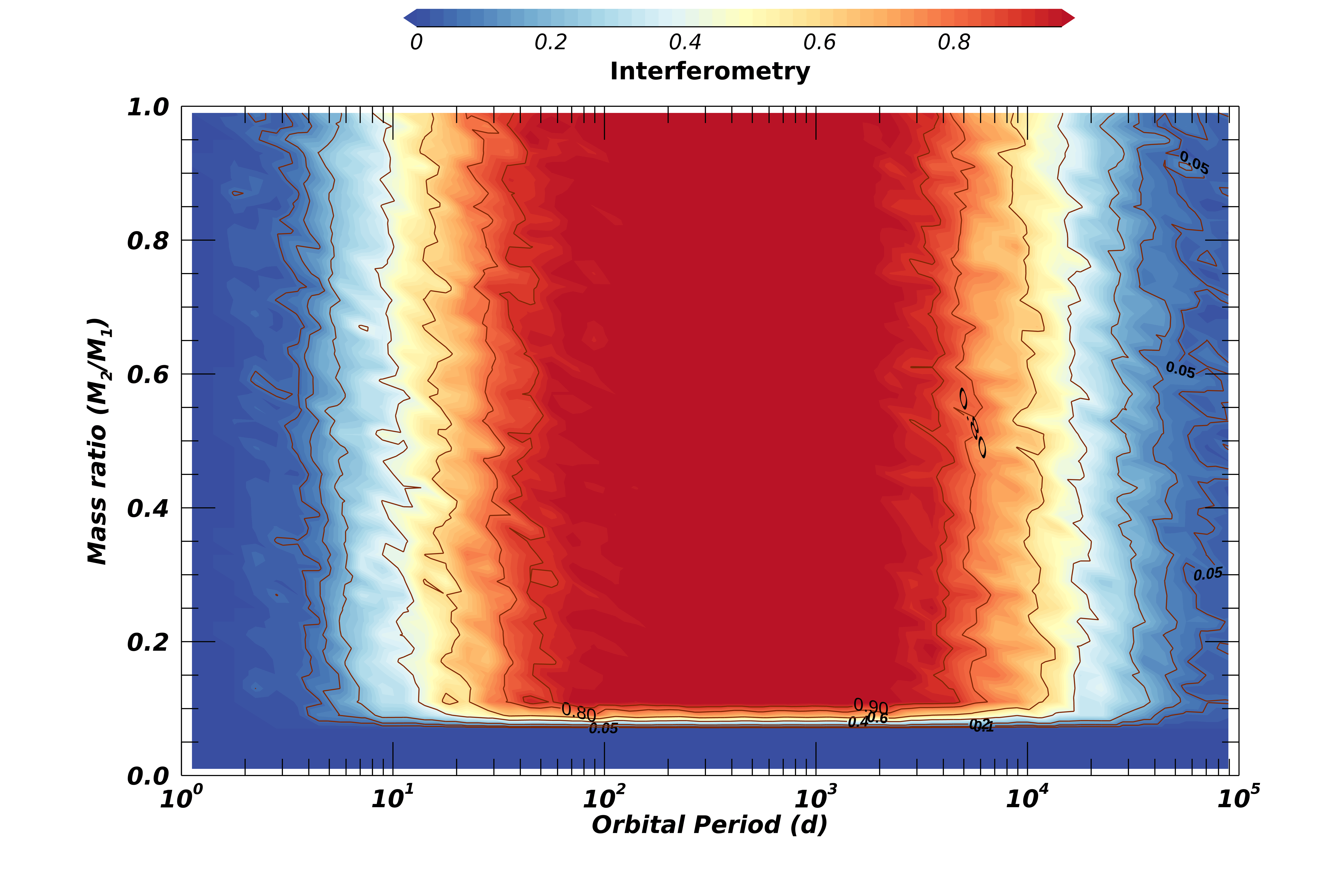}
\includegraphics[width=90mm]{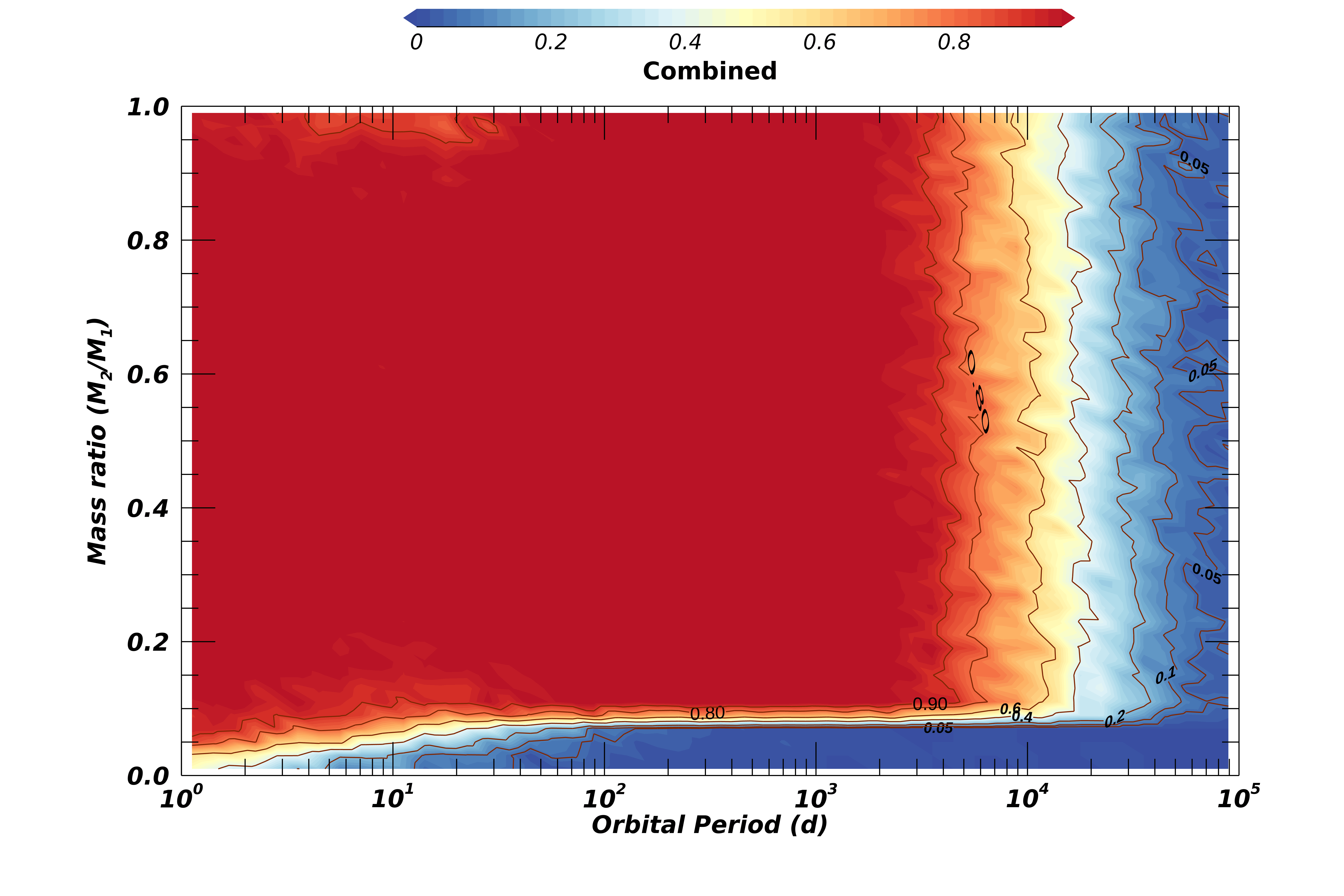}
\caption{Detection probability of a binary system as a function of its orbital period $P$ and mass ratio $q=M_2/M_1$.}
\label{f:detect}
\end{figure*}
\section{Observational biases and our capability of detecting companions}\label{sec:bias}

Our results are purely observational so far. In this section, we evaluate the sensitivity of different techniques to detecting companions to our sample of B-type stars. In this exercise, we ignore the supergiant stars, whose mass-luminosity and mass-radius relation is different from that of the remaining stars of the sample. This allows us to focus on the near-ZAMS multiplicity properties of B stars. 

We used a Monte Carlo approach to simulate artificial populations of B-type binaries \citep{sana13,bodensteiner2021}, and we applied detection criteria specific to various binary detection techniques to evaluate the likelihood that a companion is detected as a function of various physical and orbital properties. We adopted the following input distributions: an \"Opik-law for orbital periods, covering the range $\log P=0...5$, a uniform distribution for mass ratios in the range $q=M_2/M_1=0.01...1.0$, and a power-law distribution $f_e \propto e^{-0.5}$ for eccentricities \citep{sana12}. The highest eccentricities depend on the orbital period and were computed such that the separation at periastron was larger than 0.1 au. We also adopted random orientations of the orbital planes in the three-dimensional space and random times of periastron passage. We used a global mass-radius relation to evaluate the presence of eclipses  \citep[$R = M^{0.72}$;][]{eker2018}  and the relation of the mass ratio to the flux ratio $q=F_H^{0.7}$ of Sect.~\ref{sect:mass} to estimate the $H$-band flux ratio.  
For our simulations to closely represent our sample properties, we drew the distances and  masses from normal distributions centred on the values listed in Table~\ref{source_info} and used the respective uncertainties as 1$\sigma$ dispersions. We drew 10\,000 populations of 33 main-sequence B-type binaries and investigated the detection of companions by simulated observing campaigns using three different techniques: interferometry, spectroscopy, and photometry. The specific equations are given in \citet{SanaVrancken2025}. We limit our discussion to the detectability as a function of orbital periods and mass ratios as these are the dominant parameters in determining the detectability of a binary system. The sensitivity of the various detection methods and a combination thereof is summarised in Fig.~\ref{f:detect}.  

For interferometry, we simulated a single observational epoch, randomly placed along the orbit, and we compared the projected angular separation $\rho$ between the two components of the binary systems to the inner and outer working angles of PIONIER (IWA and OWA). We considered a companion to be detectable when the projected separation satisfies $1 \leq \rho/\mathrm{mas} \leq 100$ and the H-band flux ratio is higher than 3\%\ as representative of the upper envelope of our 3$\sigma$ detection limit (Table~\ref{source_params}). Figure~\ref{f:detect} shows that the detection probability is highly independent of the mass ratio down to the flux-contrast threshold of 3\%, which sets a sharp limit at $q=0.085$. The period sensitivity range is  mostly set by the distance of our targets and the IWA and OWA of PIONIER, resulting in  a very homogenous high ($>90$\%)  detection probability between 50 and 3000~days. The detection probability remains significant ($>50$\%) from 15 to 10\,000~days. 

For spectroscopy, we simulated a representative observing campaign consisting of ten observational epochs spread over 1000 days ($\approx$ 3 years) with a radial velocity (RV) precision of 3~\kms. We simulated measured RVs accounting for the blending of spectral lines using the dedicated line-blending simulations for medium-resolution spectroscopy of \citet{gazza}. We adopted a detection threshold at a peak-to-peak RV variation of 20~\kms\ at least and with a variability confidence higher than 4$\sigma$ (e.g. \citealt{gazza}). Spectroscopy is very sensitive at short periods ($>90$\%\ for $P<90$~d), except for low- and high-mass ratios, where the detection probability is impacted by insufficient reflex motion of the primary and by the SB2 line-blending bias, respectively. Spectroscopy remains sensitive up to periods of $\sim 1000$ days in favourable configurations. 

While photometry is not the main source of information of this study, we still included it in our simulations because following \citet{sana2025}, it is a straightforward addition for estimating the occurrence of eclipses for a given orbit and pair of stellar radii. We conservatively assumed that any eclipse would lead to a detection, regardless of  considerations on the quality of the photometric campaign or the number of epochs. Our simulations confirmed that the bulk of eclipsing binaries is limited to orbital periods of a few days, but also that a small fraction of systems still show eclipses up to periods of some months.

The combination of these various techniques leads to a  detection probability of 99\%\ up to periods of 1000~days (96\%\ up to 30 yr) and a mass ratio down to $\approx 0.1$. For the shortest period ($P<10$), 40\%\ of the lowest mass-ratio systems ($q<0.1$) avoid detection, which is the main limitation to the overall 95\%-detection probability in that period range. In the range 10-1000~d, the overall detection is 91\%, rising to 99\%\ when systems with $q<0.1$ are ignored. Based on this  high detection probability, we consider that multiplicity fractions based on systems that have both interferometry and archival spectroscopy do not need a bias correction down to $q=0.1$ and periods of a few years.  

For systems with insufficient spectroscopy, we thus conclude that one to three companions with $q<0.1$ could have been missed (assuming a uniform mass-ratio distribution) in our sample of 32 targets. The set of 12 systems with insufficient or inconclusive spectroscopy (see Table~\ref{t:spectro}) deserves further discussion because they may have undetected companions in the short-period range. When we again ignore the supergiants, 7 of the 21 objects with complete detections (i.e. considering both spectroscopy and interferometry) have companions with orbital periods of $P<20$~days. When we apply this occurrence rate of 0.3 to the 10 non-supergiants with missing or conflicting spectroscopic results, we could have missed three short period systems. The sample of main-sequence stars with insufficient spectroscopy is dominated by single (4) and binary (5) stars. These non-detection are therefore likely to decrease the fraction of single stars from 19\%\ to 15\% and to increase the triple fraction by about the same amount. These corrections remain smaller than the statistical uncertainties, however, which are $\sim$9\%. We therefore conclude that our determined multiplicity and companion fractions are reliable and that, at worst, we slightly underestimate the number of companions around the stars in our sample.

\section{Conclusions}\label{sec:conc}

Using high-resolution H-band interferometric data from PIONIER/VLTI, we have probed the multiplicity of a sample of 32 B-type stars in the interferometric range. The sample mostly consisted of field stars, and 8 stars of the sample belong to clusters or associations. We used parametric models created with the code PMOIRED to fit the synthetic closure phases and squared visibilities from these models to the observed closure phases and visibilities obtained with PIONIER to determine the number of stars and their characteristics, primarily, their flux and separation. 

Following this analysis, we determined an interferometric multiplicity fraction of 0.72$\pm$0.08 for this sample of stars. The most common form of multiple system we detected is a triple system, with an overall interferometric companion fraction of 1.88$\pm$0.24. In our sample, the majority of the detected companions have separations of 1-20 mas and fluxes that are $<$30\% than the flux of the brightest star in the system. This implies that most of the companions to these stars have a lower mass, which agrees with the observed IMF. 

We combined our interferometric results with spectroscopic companions and eclipsing binaries detected in the literature and with wide companions that were statistically derived from Gaia DR3 using the proper motions and distances of the stars surrounding our sources. Using simulations of B-star multiples, we determined that our interferometric results and those from combined techniques are not strongly affected by observational biases. These `complete' statistics result in multiplicity and companion fractions of 0.88$\pm$0.06 and 2.31$\pm$0.27, respectively for our sample, with many interferometric binaries becoming hierarchical triples because they have a spectroscopic companion. The multiplicity of B stars therefore likely dominates their evolution, and the role of tertiary companions in this process is likely also significant. 


\section{Data availability}

The observational data, model fits and error calculations associated with the results found in this paper can be found at \href{https://zenodo.org/records/15764971}{this} Zenodo directory (DOI: 10.5281/zenodo.15764971).

\begin{acknowledgements}

This work was published with support from the European Research Council under European Union’s Horizon 2020 research programmes (grant agreements No 772225 and No 865932-ERC-SNeX) and the FWO Odysseus program under project G0F8H6N. We also thank the referee for their time and effort in the consideration of our manuscript and their useful feedback. 
    
\end{acknowledgements}

\bibliographystyle{aa}
\bibliography{bstar}

@preamble{ " \newcommand{\noop}[1]{} " }

@ARTICLE{sana2025,
    author  = {{Sana}, H. and {Shenar}, T. and {Bodensteiner}, J. and {et al.}},
    year    = "\noop{3001}2025",
    journal = {Nature Astronomy},
    }

@ARTICLE{VonZeipel1910,
       author = {{von Zeipel}, H.},
        title = "{Sur l'application des s{\'e}ries de M. Lindstedt {\`a} l'{\'e}tude du mouvement des com{\`e}tes p{\'e}riodiques}",
      journal = {Astronomische Nachrichten},
         year = 1910,
        month = mar,
       volume = {183},
       number = {22},
        pages = {345},
          doi = {10.1002/asna.19091832202},
       adsurl = {https://ui.adsabs.harvard.edu/abs/1910AN....183..345V}
}

@ARTICLE{dodd,
       author = {{Dodd}, Jonathan M. and {Oudmaijer}, Ren{\'e} D. and {Radley}, Isaac C. and {Vioque}, Miguel and {Frost}, Abigail J.},
        title = "{Gaia uncovers difference in B and Be star binarity at small scales: evidence for mass transfer causing the Be phenomenon}",
      journal = {\mnras},
         year = 2024,
        month = jan,
       volume = {527},
       number = {2},
        pages = {3076-3086},
          doi = {10.1093/mnras/stad3105},
archivePrefix = {arXiv},
       eprint = {2310.05653},
 primaryClass = {astro-ph.SR},
       adsurl = {https://ui.adsabs.harvard.edu/abs/2024MNRAS.527.3076D}
}

@ARTICLE{absil11,
       author = {{Absil}, O. and {Le Bouquin}, J. -B. and {Berger}, J. -P. and {Lagrange}, A. -M. and {Chauvin}, G. and {Lazareff}, B. and {Zins}, G. and {Haguenauer}, P. and {Jocou}, L. and {Kern}, P. and {Millan-Gabet}, R. and {Rochat}, S. and {Traub}, W.},
        title = "{Searching for faint companions with VLTI/PIONIER. I. Method and first results}",
      journal = {\aap},
         year = 2011,
        month = nov,
       volume = {535},
          eid = {A68},
        pages = {A68},
          doi = {10.1051/0004-6361/201117719},
archivePrefix = {arXiv},
       eprint = {1110.1178},
 primaryClass = {astro-ph.EP},
       adsurl = {https://ui.adsabs.harvard.edu/abs/2011A&A...535A..68A}
}

@ARTICLE{candid,
       author = {{Gallenne}, A. and {M{\'e}rand}, A. and {Kervella}, P. and {Monnier}, J.~D. and {Schaefer}, G.~H. and {Baron}, F. and {Breitfelder}, J. and {Le Bouquin}, J.~B. and {Roettenbacher}, R.~M. and {Gieren}, W. and {Pietrzy{\'n}ski}, G. and {McAlister}, H. and {ten Brummelaar}, T. and {Sturmann}, J. and {Sturmann}, L. and {Turner}, N. and {Ridgway}, S. and {Kraus}, S.},
        title = "{Robust high-contrast companion detection from interferometric observations. The CANDID algorithm and an application to six binary Cepheids}",
      journal = {\aap},
         year = 2015,
        month = jul,
       volume = {579},
          eid = {A68},
        pages = {A68},
          doi = {10.1051/0004-6361/201525917},
archivePrefix = {arXiv},
       eprint = {1505.02715},
 primaryClass = {astro-ph.SR},
       adsurl = {https://ui.adsabs.harvard.edu/abs/2015A&A...579A..68G}
}

@ARTICLE{kummer23,
       author = {{Kummer}, F. and {Toonen}, S. and {de Koter}, A.},
        title = "{The main evolutionary pathways of massive hierarchical triple stars}",
      journal = {\aap},
         year = 2023,
        month = oct,
       volume = {678},
          eid = {A60},
        pages = {A60},
          doi = {10.1051/0004-6361/202347179},
archivePrefix = {arXiv},
       eprint = {2306.09400},
 primaryClass = {astro-ph.SR},
       adsurl = {https://ui.adsabs.harvard.edu/abs/2023A&A...678A..60K}
}

@ARTICLE{frost2024,
       author = {{Frost}, A.~J. and {Sana}, H. and {Mahy}, L. and {Wade}, G. and {Barron}, J. and {Le Bouquin}, J. -B. and {M{\'e}rand}, A. and {Schneider}, F.~R.~N. and {Shenar}, T. and {Barb{\'a}}, R.~H. and {Bowman}, D.~M. and {Fabry}, M. and {Farhang}, A. and {Marchant}, P. and {Morrell}, N.~I. and {Smoker}, J.~V.},
        title = "{A magnetic massive star has experienced a stellar merger}",
      journal = {Science},
         year = 2024,
        month = apr,
       volume = {384},
       number = {6692},
        pages = {214-217},
          doi = {10.1126/science.adg7700},
archivePrefix = {arXiv},
       eprint = {2404.10167},
 primaryClass = {astro-ph.SR},
       adsurl = {https://ui.adsabs.harvard.edu/abs/2024Sci...384..214F}
}

@ARTICLE{alfvir,
       author = {{Harrington}, David and {Koenigsberger}, Gloria and {Olgu{\'\i}n}, Enrique and {Ilyin}, Ilya and {Berdyugina}, Svetlana V. and {Lara}, Bruno and {Moreno}, Edmundo},
        title = "{Alpha Virginis: line-profile variations and orbital elements}",
      journal = {\aap},
         year = 2016,
        month = may,
       volume = {590},
          eid = {A54},
        pages = {A54},
          doi = {10.1051/0004-6361/201526507},
archivePrefix = {arXiv},
       eprint = {1604.02057},
 primaryClass = {astro-ph.SR},
       adsurl = {https://ui.adsabs.harvard.edu/abs/2016A&A...590A..54H}
}

@INPROCEEDINGS{leith,
   author = {{Leitherer}, C.},
    title = "{Massive Stars in Starburst Galaxies and the Origin of Galactic Superwinds.}",
booktitle = {Reviews in Modern Astronomy},
     year = 1994,
   series = {Reviews in Modern Astronomy},
   volume = 7,
   editor = {{Klare}, G.},
    pages = {73-102},
   adsurl = {http://adsabs.harvard.edu/abs/1994RvMA....7...73L}
}

@ARTICLE{hier,
       author = {{Evans}, David S.},
        title = "{Stars of Higher Multiplicity}",
      journal = {\qjras},
         year = 1968,
        month = dec,
       volume = {9},
        pages = {388},
       adsurl = {https://ui.adsabs.harvard.edu/abs/1968QJRAS...9..388E}
}

@INPROCEEDINGS{marda,
       author = {{Mardling}, R. and {Aarseth}, S.},
        title = "{Dynamics and Stability of Three-Body Systems}",
    booktitle = {The Dynamics of Small Bodies in the Solar System, A Major Key to Solar System Studies},
         year = 1999,
       editor = {{Steves}, Bonnie A. and {Roy}, Archie E.},
       series = {NATO Advanced Study Institute (ASI) Series C},
       volume = {522},
        month = jan,
        pages = {385},
       adsurl = {https://ui.adsabs.harvard.edu/abs/1999ASIC..522..385M}
}

@ARTICLE{kiseleva,
       author = {{Kiseleva}, G. and {Eggleton}, P.~P. and {Anosova}, J.~P.},
        title = "{A note on the stability of hierarchical triple stars with initially circular orbits.}",
      journal = {\mnras},
         year = 1994,
        month = mar,
       volume = {267},
        pages = {161-166},
          doi = {10.1093/mnras/267.1.161},
       adsurl = {https://ui.adsabs.harvard.edu/abs/1994MNRAS.267..161K}
}

@ARTICLE{SanaVrancken2025,
       author = {{Sana}, Hugues and {Vrancken}, Jasmine},
        title = "{Observing Binaries}",
      journal = {arXiv e-prints},
         year = 2025,
        month = apr,
          eid = {arXiv:2504.00548},
        pages = {arXiv:2504.00548},
          doi = {10.48550/arXiv.2504.00548},
archivePrefix = {arXiv},
       eprint = {2504.00548},
 primaryClass = {astro-ph.SR},
       adsurl = {https://ui.adsabs.harvard.edu/abs/2025arXiv250400548S}
}

@ARTICLE{2024A&A...690A.135J,
       author = {{Jin}, Harim and {Langer}, Norbert and {Lennon}, Daniel J. and {Proffitt}, Charles R.},
        title = "{Boron depletion in Galactic early B-type stars reveals two different main sequence star populations}",
      journal = {\aap},
         year = 2024,
        month = oct,
       volume = {690},
          eid = {A135},
        pages = {A135},
          doi = {10.1051/0004-6361/202450896},
archivePrefix = {arXiv},
       eprint = {2405.18266},
 primaryClass = {astro-ph.SR},
       adsurl = {https://ui.adsabs.harvard.edu/abs/2024A&A...690A.135J}
}

@ARTICLE{2022ApJ...937..110L,
       author = {{Liu}, Zhicun and {Cui}, Wenyuan and {Liu}, Chao and {Alexeeva}, Sofya and {Shi}, Jianrong and {Zhao}, Gang},
        title = "{Chemical Composition of B-type Stars from LAMOST DR5}",
      journal = {\apj},
         year = 2022,
        month = oct,
       volume = {937},
       number = {2},
          eid = {110},
        pages = {110},
          doi = {10.3847/1538-4357/ac8cf5},
       adsurl = {https://ui.adsabs.harvard.edu/abs/2022ApJ...937..110L}
}

@ARTICLE{2018MNRAS.481.3953A,
       author = {{Allen}, Christine and {Ruelas-Mayorga}, Alex and {S{\'a}nchez}, Leonardo J. and {Costero}, Rafael},
        title = "{The dynamical evolution of multiple systems of trapezium type}",
      journal = {\mnras},
         year = 2018,
        month = dec,
       volume = {481},
       number = {3},
        pages = {3953-3965},
          doi = {10.1093/mnras/sty2502},
archivePrefix = {arXiv},
       eprint = {1809.03537},
 primaryClass = {astro-ph.SR},
       adsurl = {https://ui.adsabs.harvard.edu/abs/2018MNRAS.481.3953A}
}

@ARTICLE{2016A&A...595A.132Z,
       author = {{Zorec}, J. and {Fr{\'e}mat}, Y. and {Domiciano de Souza}, A. and {Royer}, F. and {Cidale}, L. and {Hubert}, A. -M. and {Semaan}, T. and {Martayan}, C. and {Cochetti}, Y.~R. and {Arias}, M.~L. and {Aidelman}, Y. and {Stee}, P.},
        title = "{Critical study of the distribution of rotational velocities of Be stars. I. Deconvolution methods, effects due to gravity darkening, macroturbulence, and binarity}",
      journal = {\aap},
         year = 2016,
        month = nov,
       volume = {595},
          eid = {A132},
        pages = {A132},
          doi = {10.1051/0004-6361/201628760},
       adsurl = {https://ui.adsabs.harvard.edu/abs/2016A&A...595A.132Z}
}

@ARTICLE{2016A&A...591L...6I,
       author = {{Irrgang}, A. and {Desphande}, A. and {Moehler}, S. and {Mugrauer}, M. and {Janousch}, D.},
        title = "{The slowly pulsating B-star 18 Pegasi: A testbed for upper main sequence stellar evolution}",
      journal = {\aap},
         year = 2016,
        month = jun,
       volume = {591},
          eid = {L6},
        pages = {L6},
          doi = {10.1051/0004-6361/201628844},
archivePrefix = {arXiv},
       eprint = {1605.09267},
 primaryClass = {astro-ph.SR},
       adsurl = {https://ui.adsabs.harvard.edu/abs/2016A&A...591L...6I}
}

@ARTICLE{2016MNRAS.458.1964T,
       author = {{Tkachenko}, A. and {Matthews}, J.~M. and {Aerts}, C. and {Pavlovski}, K. and {P{\'a}pics}, P.~I. and {Zwintz}, K. and {Cameron}, C. and {Walker}, G.~A.~H. and {Kuschnig}, R. and {Degroote}, P. and {Debosscher}, J. and {Moravveji}, E. and {Kolbas}, V. and {Guenther}, D.~B. and {Moffat}, A.~F.~J. and {Rowe}, J.~F. and {Rucinski}, S.~M. and {Sasselov}, D. and {Weiss}, W.~W.},
        title = "{Stellar modelling of Spica, a high-mass spectroscopic binary with a {\ensuremath{\beta}} Cep variable primary component}",
      journal = {\mnras},
         year = 2016,
        month = may,
       volume = {458},
       number = {2},
        pages = {1964-1976},
          doi = {10.1093/mnras/stw255},
archivePrefix = {arXiv},
       eprint = {1601.08069},
 primaryClass = {astro-ph.SR},
       adsurl = {https://ui.adsabs.harvard.edu/abs/2016MNRAS.458.1964T}
}

@ARTICLE{2014A&A...566A...7N,
       author = {{Nieva}, Mar{\'\i}a-Fernanda and {Przybilla}, Norbert},
        title = "{Fundamental properties of nearby single early B-type stars}",
      journal = {\aap},
         year = 2014,
        month = jun,
       volume = {566},
          eid = {A7},
        pages = {A7},
          doi = {10.1051/0004-6361/201423373},
archivePrefix = {arXiv},
       eprint = {1412.1418},
 primaryClass = {astro-ph.SR},
       adsurl = {https://ui.adsabs.harvard.edu/abs/2014A&A...566A...7N}
}

@ARTICLE{2014A&A...561A..63G,
       author = {{Gonz{\'a}lez}, J.~F. and {Saffe}, C. and {Castelli}, F. and {Hubrig}, S. and {Ilyin}, I. and {Sch{\"o}ller}, M. and {Carroll}, T.~A. and {Leone}, F. and {Giarrusso}, M.},
        title = "{HD 161701, a chemically peculiar binary with a HgMn primary and an Ap secondary}",
      journal = {\aap},
         year = 2014,
        month = jan,
       volume = {561},
          eid = {A63},
        pages = {A63},
          doi = {10.1051/0004-6361/201322327},
       adsurl = {https://ui.adsabs.harvard.edu/abs/2014A&A...561A..63G}
}

@ARTICLE{2011MNRAS.410..190T,
       author = {{Tetzlaff}, N. and {Neuh{\"a}user}, R. and {Hohle}, M.~M.},
        title = "{A catalogue of young runaway Hipparcos stars within 3 kpc from the Sun}",
      journal = {\mnras},
         year = 2011,
        month = jan,
       volume = {410},
       number = {1},
        pages = {190-200},
          doi = {10.1111/j.1365-2966.2010.17434.x},
archivePrefix = {arXiv},
       eprint = {1007.4883},
 primaryClass = {astro-ph.GA},
       adsurl = {https://ui.adsabs.harvard.edu/abs/2011MNRAS.410..190T}
}

@ARTICLE{2010MNRAS.404.1306F,
       author = {{Fraser}, M. and {Dufton}, P.~L. and {Hunter}, I. and {Ryans}, R.~S.~I.},
        title = "{Atmospheric parameters and rotational velocities for a sample of Galactic B-type supergiants}",
      journal = {\mnras},
         year = 2010,
        month = may,
       volume = {404},
       number = {3},
        pages = {1306-1320},
          doi = {10.1111/j.1365-2966.2010.16392.x},
archivePrefix = {arXiv},
       eprint = {1001.3337},
 primaryClass = {astro-ph.SR},
       adsurl = {https://ui.adsabs.harvard.edu/abs/2010MNRAS.404.1306F}
}

@ARTICLE{2010AN....331..349H,
       author = {{Hohle}, M.~M. and {Neuh{\"a}user}, R. and {Schutz}, B.~F.},
        title = "{Masses and luminosities of O- and B-type stars and red supergiants}",
      journal = {Astronomische Nachrichten},
         year = 2010,
        month = apr,
       volume = {331},
       number = {4},
        pages = {349},
          doi = {10.1002/asna.200911355},
archivePrefix = {arXiv},
       eprint = {1003.2335},
 primaryClass = {astro-ph.SR},
       adsurl = {https://ui.adsabs.harvard.edu/abs/2010AN....331..349H}
}

@ARTICLE{2009AN....330..317H,
       author = {{Hubrig}, S. and {Briquet}, M. and {De Cat}, P. and {Sch{\"o}ller}, M. and {Morel}, T. and {Ilyin}, I.},
        title = "{New magnetic field measurements of {\ensuremath{\beta}} Cephei stars and slowly pulsating B stars}",
      journal = {Astronomische Nachrichten},
         year = 2009,
        month = apr,
       volume = {330},
       number = {4},
        pages = {317},
          doi = {10.1002/asna.200811187},
archivePrefix = {arXiv},
       eprint = {0902.1314},
 primaryClass = {astro-ph.SR},
       adsurl = {https://ui.adsabs.harvard.edu/abs/2009AN....330..317H}
}

@ARTICLE{2007A&A...468..263C,
       author = {{Cidale}, L.~S. and {Arias}, M.~L. and {Torres}, A.~F. and {Zorec}, J. and {Fr{\'e}mat}, Y. and {Cruzado}, A.},
        title = "{Fundamental parameters of He-weak and He-strong stars}",
      journal = {\aap},
         year = 2007,
        month = jun,
       volume = {468},
       number = {1},
        pages = {263-272},
          doi = {10.1051/0004-6361:20066454},
archivePrefix = {arXiv},
       eprint = {0705.0541},
 primaryClass = {astro-ph},
       adsurl = {https://ui.adsabs.harvard.edu/abs/2007A&A...468..263C}
}

@ARTICLE{2006MNRAS.369L..61H,
       author = {{Hubrig}, S. and {Briquet}, M. and {Sch{\"o}ller}, M. and {De Cat}, P. and {Mathys}, G. and {Aerts}, C.},
        title = "{Discovery of magnetic fields in the {\ensuremath{\beta}}Cephei star {\ensuremath{\xi}}$^{1}$ CMa and in several slowly pulsating B stars$^{*}$}",
      journal = {\mnras},
         year = 2006,
        month = jun,
       volume = {369},
       number = {1},
        pages = {L61-L65},
          doi = {10.1111/j.1745-3933.2006.00175.x},
archivePrefix = {arXiv},
       eprint = {astro-ph/0604283},
 primaryClass = {astro-ph},
       adsurl = {https://ui.adsabs.harvard.edu/abs/2006MNRAS.369L..61H}
}

@ARTICLE{2004A&A...413..273B,
       author = {{Briquet}, M. and {Aerts}, C. and {L{\"u}ftinger}, T. and {De Cat}, P. and {Piskunov}, N.~E. and {Scuflaire}, R.},
        title = "{He and Si surface inhomogeneities of four Bp variable stars}",
      journal = {\aap},
         year = 2004,
        month = jan,
       volume = {413},
        pages = {273-283},
          doi = {10.1051/0004-6361:20031450},
       adsurl = {https://ui.adsabs.harvard.edu/abs/2004A&A...413..273B}
}

@PHDTHESIS{2003PhDT.......251B,
       author = {{Burnley}, Adam Warwick},
        title = "{Mass loss from hot, luminous stars}",
       school = {University College London, UK},
         year = 2003,
        month = jan,
       adsurl = {https://ui.adsabs.harvard.edu/abs/2003PhDT.......251B}
}

@ARTICLE{1992A&A...264...88V,
       author = {{van Genderen}, A.~M. and {van den Bosch}, F.~C. and {Dessing}, F. and {Fehmers}, G.~C. and {van Grunsven}, J. and {van der Heiden}, R. and {Janssens}, A.~M. and {Kalter}, R. and {van der Meer}, R.~L.~J. and {van Ojik}, R. and {Smit}, J.~M. and {Zijderveld}, M.~J.},
        title = "{Light variations of massive stars (alpha Cygni variables). XIII. The B-type hypergiants R 81 (LBV), HD 80077 (LBV?), HD 168607 = V 4029 Sagittarii (LBV) and HD 168625 = V 4030 Sagittarii.}",
      journal = {\aap},
         year = 1992,
        month = oct,
       volume = {264},
        pages = {88-104},
       adsurl = {https://ui.adsabs.harvard.edu/abs/1992A&A...264...88V}
}

@ARTICLE{eker2018,
       author = {{Eker}, Z. and {Bak{\i}{\c{s}}}, V. and {Bilir}, S. and {Soydugan}, F. and {Steer}, I. and {Soydugan}, E. and {Bak{\i}{\c{s}}}, H. and {Ali{\c{c}}avu{\c{s}}}, F. and {Aslan}, G. and {Alpsoy}, M.},
        title = "{Interrelated main-sequence mass-luminosity, mass-radius, and mass-effective temperature relations}",
      journal = {\mnras},
         year = 2018,
        month = oct,
       volume = {479},
       number = {4},
        pages = {5491-5511},
          doi = {10.1093/mnras/sty1834},
archivePrefix = {arXiv},
       eprint = {1807.02568},
 primaryClass = {astro-ph.SR},
       adsurl = {https://ui.adsabs.harvard.edu/abs/2018MNRAS.479.5491E}
}

@ARTICLE{bodensteiner2021,
       author = {{Bodensteiner}, J. and {Sana}, H. and {Wang}, C. and {Langer}, N. and {Mahy}, L. and {Banyard}, G. and {de Koter}, A. and {de Mink}, S.~E. and {Evans}, C.~J. and {G{\"o}tberg}, Y. and {Patrick}, L.~R. and {Schneider}, F.~R.~N. and {Tramper}, F.},
        title = "{The young massive SMC cluster NGC 330 seen by MUSE. II. Multiplicity properties of the massive-star population}",
      journal = {\aap},
         year = 2021,
        month = aug,
       volume = {652},
          eid = {A70},
        pages = {A70},
          doi = {10.1051/0004-6361/202140507},
archivePrefix = {arXiv},
       eprint = {2104.13409},
 primaryClass = {astro-ph.SR},
       adsurl = {https://ui.adsabs.harvard.edu/abs/2021A&A...652A..70B}
}

@ARTICLE{gazza,
       author = {{Banyard}, G. and {Sana}, H. and {Mahy}, L. and {Bodensteiner}, J. and {Villase{\~n}or}, J.~I. and {Evans}, C.~J.},
        title = "{The observed multiplicity properties of B-type stars in the Galactic young open cluster NGC 6231}",
      journal = {\aap},
         year = 2022,
        month = feb,
       volume = {658},
          eid = {A69},
        pages = {A69},
          doi = {10.1051/0004-6361/202141037},
archivePrefix = {arXiv},
       eprint = {2108.07814},
 primaryClass = {astro-ph.SR},
       adsurl = {https://ui.adsabs.harvard.edu/abs/2022A&A...658A..69B}
}

@ARTICLE{demink11,
       author = {{de Mink}, S.~E. and {Langer}, N. and {Izzard}, R.~G.},
        title = "{Binaries are the best single stars}",
      journal = {Bulletin de la Societe Royale des Sciences de Liege},
         year = 2011,
        month = jan,
       volume = {80},
        pages = {543-548},
archivePrefix = {arXiv},
       eprint = {1010.2200},
 primaryClass = {astro-ph.SR},
       adsurl = {https://ui.adsabs.harvard.edu/abs/2011BSRSL..80..543D}
}

@ARTICLE{pionier,
       author = {{Le Bouquin}, J. -B. and {Berger}, J. -P. and {Lazareff}, B. and
         {Zins}, G. and {Haguenauer}, P. and {Jocou}, L. and {Kern}, P. and
         {Millan-Gabet}, R. and {Traub}, W. and {Absil}, O. and
         {Augereau}, J. -C. and {Benisty}, M. and {Blind}, N. and {Bonfils}, X. and
         {Bourget}, P. and {Delboulbe}, A. and {Feautrier}, P. and
         {Germain}, M. and {Gitton}, P. and {Gillier}, D. and {Kiekebusch}, M. and
         {Kluska}, J. and {Knudstrup}, J. and {Labeye}, P. and {Lizon}, J. -L. and
         {Monin}, J. -L. and {Magnard}, Y. and {Malbet}, F. and {Maurel}, D. and
         {M{\'e}nard}, F. and {Micallef}, M. and {Michaud}, L. and
         {Montagnier}, G. and {Morel}, S. and {Moulin}, T. and {Perraut}, K. and
         {Popovic}, D. and {Rabou}, P. and {Rochat}, S. and {Rojas}, C. and
         {Roussel}, F. and {Roux}, A. and {Stadler}, E. and {Stefl}, S. and
         {Tatulli}, E. and {Ventura}, N.},
        title = "{PIONIER: a 4-telescope visitor instrument at VLTI}",
      journal = {\aap},
         year = 2011,
        month = nov,
       volume = {535},
          eid = {A67},
        pages = {A67},
          doi = {10.1051/0004-6361/201117586},
archivePrefix = {arXiv},
       eprint = {1109.1918},
 primaryClass = {astro-ph.IM},
       adsurl = {https://ui.adsabs.harvard.edu/abs/2011A&A...535A..67L}
}

@ARTICLE{vfts,
       author = {{Evans}, C.~J. and {Taylor}, W.~D. and {H{\'e}nault-Brunet}, V. and
         {Sana}, H. and {de Koter}, A. and {Sim{\'o}n-D{\'\i}az}, S. and
         {Carraro}, G. and {Bagnoli}, T. and {Bastian}, N. and
         {Bestenlehner}, J.~M. and {Bonanos}, A.~Z. and {Bressert}, E. and
         {Brott}, I. and {Campbell}, M.~A. and {Cantiello}, M. and
         {Clark}, J.~S. and {Costa}, E. and {Crowther}, P.~A. and
         {de Mink}, S.~E. and {Doran}, E. and {Dufton}, P.~L. and
         {Dunstall}, P.~R. and {Friedrich}, K. and {Garcia}, M. and
         {Gieles}, M. and {Gr{\"a}fener}, G. and {Herrero}, A. and
         {Howarth}, I.~D. and {Izzard}, R.~G. and {Langer}, N. and
         {Lennon}, D.~J. and {Ma{\'\i}z Apell{\'a}niz}, J. and {Markova}, N. and
         {Najarro}, F. and {Puls}, J. and {Ramirez}, O.~H. and
         {Sab{\'\i}n-Sanjuli{\'a}n}, C. and {Smartt}, S.~J. and {Stroud}, V.~E. and
         {van Loon}, J. Th. and {Vink}, J.~S. and {Walborn}, N.~R.},
        title = "{The VLT-FLAMES Tarantula Survey. I. Introduction and observational overview}",
      journal = {\aap},
         year = 2011,
        month = jun,
       volume = {530},
          eid = {A108},
        pages = {A108},
          doi = {10.1051/0004-6361/201116782},
archivePrefix = {arXiv},
       eprint = {1103.5386},
 primaryClass = {astro-ph.SR},
       adsurl = {https://ui.adsabs.harvard.edu/abs/2011A&A...530A.108E}
}

@ARTICLE{demink13,
       author = {{de Mink}, S.~E. and {Langer}, N. and {Izzard}, R.~G. and {Sana}, H. and
         {de Koter}, A.},
        title = "{The Rotation Rates of Massive Stars: The Role of Binary Interaction through Tides, Mass Transfer, and Mergers}",
      journal = {\apj},
         year = 2013,
        month = feb,
       volume = {764},
       number = {2},
          eid = {166},
        pages = {166},
          doi = {10.1088/0004-637X/764/2/166},
archivePrefix = {arXiv},
       eprint = {1211.3742},
 primaryClass = {astro-ph.SR},
       adsurl = {https://ui.adsabs.harvard.edu/abs/2013ApJ...764..166D}
}

@ARTICLE{fabmerg,
       author = {{Schneider}, Fabian R.~N. and {Ohlmann}, Sebastian T. and
         {Podsiadlowski}, Philipp and {R{\"o}pke}, Friedrich K. and
         {Balbus}, Steven A. and {Pakmor}, R{\"u}diger and {Springel}, Volker},
        title = "{Stellar mergers as the origin of magnetic massive stars}",
      journal = {\nat},
         year = 2019,
        month = oct,
       volume = {574},
       number = {7777},
        pages = {211-214},
          doi = {10.1038/s41586-019-1621-5},
archivePrefix = {arXiv},
       eprint = {1910.14058},
 primaryClass = {astro-ph.SR},
       adsurl = {https://ui.adsabs.harvard.edu/abs/2019Natur.574..211S}
}

@ARTICLE{dunstall15,
       author = {{Dunstall}, P.~R. and {Dufton}, P.~L. and {Sana}, H. and {Evans}, C.~J. and
         {Howarth}, I.~D. and {Sim{\'o}n-D{\'\i}az}, S. and {de Mink}, S.~E. and
         {Langer}, N. and {Ma{\'\i}z Apell{\'a}niz}, J. and {Taylor}, W.~D.},
        title = "{The VLT-FLAMES Tarantula Survey. XXII. Multiplicity properties of the B-type stars}",
      journal = {\aap},
         year = 2015,
        month = aug,
       volume = {580},
          eid = {A93},
        pages = {A93},
          doi = {10.1051/0004-6361/201526192},
archivePrefix = {arXiv},
       eprint = {1505.07121},
 primaryClass = {astro-ph.SR},
       adsurl = {https://ui.adsabs.harvard.edu/abs/2015A&A...580A..93D}
}

@ARTICLE{sana14,
       author = {{Sana}, H. and {Le Bouquin}, J. -B. and {Lacour}, S. and {Berger}, J. -P. and {Duvert}, G. and {Gauchet}, L. and {Norris}, B. and {Olofsson}, J. and {Pickel}, D. and {Zins}, G. and {Absil}, O. and {de Koter}, A. and {Kratter}, K. and {Schnurr}, O. and {Zinnecker}, H.},
        title = "{Southern Massive Stars at High Angular Resolution: Observational Campaign and Companion Detection}",
      journal = {\apjs},
         year = 2014,
        month = nov,
       volume = {215},
       number = {1},
          eid = {15},
        pages = {15},
          doi = {10.1088/0067-0049/215/1/15},
archivePrefix = {arXiv},
       eprint = {1409.6304},
 primaryClass = {astro-ph.SR},
       adsurl = {https://ui.adsabs.harvard.edu/abs/2014ApJS..215...15S}
}

@ARTICLE{hipp,
       author = {{Perryman}, M.~A.~C. and {Lindegren}, L. and {Kovalevsky}, J. and {Hog}, E. and {Bastian}, U. and {Bernacca}, P.~L. and {Creze}, M. and {Donati}, F. and {Grenon}, M. and {Grewing}, M. and {van Leeuwen}, F. and {van der Marel}, H. and {Mignard}, F. and {Murray}, C.~A. and {Le Poole}, R.~S. and {Schrijver}, H. and {Turon}, C. and {Arenou}, F. and {Froeschle}, M. and {Petersen}, C.~S.},
        title = "{The Hipparcos Catalogue.}",
      journal = {\aap},
         year = 1997,
        month = jul,
       volume = {500},
        pages = {501-504},
       adsurl = {https://ui.adsabs.harvard.edu/abs/1997A&A...323L..49P}
}

@ARTICLE{sbmcw1019,
       author = {{Irrgang}, A. and {Desphande}, A. and {Moehler}, S. and {Mugrauer}, M. and {Janousch}, D.},
        title = "{The slowly pulsating B-star 18 Pegasi: A testbed for upper main sequence stellar evolution}",
      journal = {\aap},
         year = 2016,
        month = jun,
       volume = {591},
          eid = {L6},
        pages = {L6},
          doi = {10.1051/0004-6361/201628844},
archivePrefix = {arXiv},
       eprint = {1605.09267},
 primaryClass = {astro-ph.SR},
       adsurl = {https://ui.adsabs.harvard.edu/abs/2016A&A...591L...6I}
}

@ARTICLE{sbhd37017,
       author = {{Leone}, F. and {Catanzaro}, G.},
        title = "{Orbital elements of binary systems with a chemically peculiar star}",
      journal = {\aap},
         year = 1999,
        month = mar,
       volume = {343},
        pages = {273-280},
       adsurl = {https://ui.adsabs.harvard.edu/abs/1999A&A...343..273L}
}

@ARTICLE{kroupa,
       author = {{Kroupa}, Pavel},
        title = "{The Initial Mass Function of Stars: Evidence for Uniformity in Variable Systems}",
      journal = {Science},
         year = 2002,
        month = jan,
       volume = {295},
       number = {5552},
        pages = {82-91},
          doi = {10.1126/science.1067524},
archivePrefix = {arXiv},
       eprint = {astro-ph/0201098},
 primaryClass = {astro-ph},
       adsurl = {https://ui.adsabs.harvard.edu/abs/2002Sci...295...82K}
}

@ARTICLE{madda21,
       author = {{Reggiani}, M. and {Rainot}, A. and {Sana}, H. and {Almeida}, L.~A. and {Caballero-Nieves}, S. and {Kratter}, K. and {Lacour}, S. and {Le Bouquin}, J. -B. and {Zinnecker}, H.},
        title = "{Probing the low-mass end of the companion mass function for O-type stars}",
      journal = {\aap},
         year = 2022,
        month = apr,
       volume = {660},
          eid = {A122},
        pages = {A122},
          doi = {10.1051/0004-6361/202142418},
archivePrefix = {arXiv},
       eprint = {2112.10831},
 primaryClass = {astro-ph.SR},
       adsurl = {https://ui.adsabs.harvard.edu/abs/2022A&A...660A.122R}
}

@ARTICLE{grishin,
       author = {{Grishin}, Evgeni and {Perets}, Hagai B. and {Zenati}, Yossef and {Michaely}, Erez},
        title = "{Generalized Hill-stability criteria for hierarchical three-body systems at arbitrary inclinations}",
      journal = {\mnras},
         year = 2017,
        month = apr,
       volume = {466},
       number = {1},
        pages = {276-285},
          doi = {10.1093/mnras/stw3096},
archivePrefix = {arXiv},
       eprint = {1609.05912},
 primaryClass = {astro-ph.EP},
       adsurl = {https://ui.adsabs.harvard.edu/abs/2017MNRAS.466..276G}
}

@ARTICLE{sbHD189103,
       author = {{Wilson}, Ralph E.},
        title = "{On the period-eccentricity relation in binary systems}",
      journal = {\aj},
         year = 1921,
        month = may,
       volume = {33},
        pages = {147-150},
          doi = {10.1086/104448},
       adsurl = {https://ui.adsabs.harvard.edu/abs/1921AJ.....33..147W}
}

@ARTICLE{sbhd30836,
       author = {{Luyten}, W.~J.},
        title = "{A Rediscussion of the Orbits of Seventy-Seven Spectroscopic Binaries}",
      journal = {\apj},
         year = 1936,
        month = jul,
       volume = {84},
        pages = {85},
          doi = {10.1086/143751},
       adsurl = {https://ui.adsabs.harvard.edu/abs/1936ApJ....84...85L}
}

@ARTICLE{sodor14,
       author = {{S{\'o}dor}, {\'A}. and {De Cat}, P. and {Wright}, D.~J. and {Neiner}, C. and {Briquet}, M. and {Lampens}, P. and {Dukes}, R.~J. and {Henry}, G.~W. and {Williamson}, M.~H. and {Brunsden}, E. and {Pollard}, K.~R. and {Cottrell}, P.~L. and {Maisonneuve}, F. and {Kilmartin}, P.~M. and {Matthews}, J. and {Kallinger}, T. and {Beck}, P.~G. and {Kambe}, E. and {Engelbrecht}, C.~A. and {Czanik}, R.~J. and {Yang}, S. and {Hashimoto}, O. and {Honda}, S. and {Fu}, J.~N. and {Castanheira}, B. and {Lehmann}, H. and {Bogn{\'a}r}, Zs. and {Behara}, N. and {Scaringi}, S. and {Van Winckel}, H. and {Menu}, J. and {Lobel}, A. and {Mathias}, P. and {Saesen}, S. and {Vu{\v{c}}kovi{\'c}}, M. and {MiMeS Collaboration}},
        title = "{Extensive study of HD 25558, a long-period double-lined binary with two SPB components}",
      journal = {\mnras},
         year = 2014,
        month = mar,
       volume = {438},
       number = {4},
        pages = {3535-3556},
          doi = {10.1093/mnras/stt2466},
archivePrefix = {arXiv},
       eprint = {1312.6307},
 primaryClass = {astro-ph.SR},
       adsurl = {https://ui.adsabs.harvard.edu/abs/2014MNRAS.438.3535S}
}

@ARTICLE{sbHD224990,
       author = {{Gonz{\'a}lez}, J.~F. and {Levato}, H.},
        title = "{Spectroscopic study of the open cluster Blanco 1}",
      journal = {\aap},
         year = 2009,
        month = nov,
       volume = {507},
       number = {1},
        pages = {541-547},
          doi = {10.1051/0004-6361/200912772},
       adsurl = {https://ui.adsabs.harvard.edu/abs/2009A&A...507..541G}
}

@ARTICLE{sbHD161701,
       author = {{Hube}, Douglas P.},
        title = "{The Spectrographic Orbit of H.D. 161701}",
      journal = {\jrasc},
         year = 1969,
        month = oct,
       volume = {63},
        pages = {229},
       adsurl = {https://ui.adsabs.harvard.edu/abs/1969JRASC..63..229H}
}

@ARTICLE{sbhd140008,
       author = {{Thackeray}, A.~D. and {Hutchings}, F.~B.},
        title = "{Orbits of two double-lined binaries HD 140008 and 178322}",
      journal = {\mnras},
         year = 1965,
        month = jan,
       volume = {129},
        pages = {191},
          doi = {10.1093/mnras/129.2.191},
       adsurl = {https://ui.adsabs.harvard.edu/abs/1965MNRAS.129..191T}
}

@ARTICLE{brott2011,
       author = {{Brott}, I. and {de Mink}, S.~E. and {Cantiello}, M. and {Langer}, N. and {de Koter}, A. and {Evans}, C.~J. and {Hunter}, I. and {Trundle}, C. and {Vink}, J.~S.},
        title = "{Rotating massive main-sequence stars. I. Grids of evolutionary models and isochrones}",
      journal = {\aap},
         year = 2011,
        month = jun,
       volume = {530},
          eid = {A115},
        pages = {A115},
          doi = {10.1051/0004-6361/201016113},
archivePrefix = {arXiv},
       eprint = {1102.0530},
 primaryClass = {astro-ph.SR},
       adsurl = {https://ui.adsabs.harvard.edu/abs/2011A&A...530A.115B}
}

@ARTICLE{bonnsai,
       author = {{Schneider}, F.~R.~N. and {Langer}, N. and {de Koter}, A. and {Brott}, I. and {Izzard}, R.~G. and {Lau}, H.~H.~B.},
        title = "{Bonnsai: a Bayesian tool for comparing stars with stellar evolution models}",
      journal = {\aap},
         year = 2014,
        month = oct,
       volume = {570},
          eid = {A66},
        pages = {A66},
          doi = {10.1051/0004-6361/201424286},
archivePrefix = {arXiv},
       eprint = {1408.3409},
 primaryClass = {astro-ph.SR},
       adsurl = {https://ui.adsabs.harvard.edu/abs/2014A&A...570A..66S}
}

@ARTICLE{kervbin,
       author = {{Kervella}, Pierre and {Arenou}, Fr{\'e}d{\'e}ric and {Mignard}, Fran{\c{c}}ois and {Th{\'e}venin}, Fr{\'e}d{\'e}ric},
        title = "{Stellar and substellar companions of nearby stars from Gaia DR2. Binarity from proper motion anomaly}",
      journal = {\aap},
         year = 2019,
        month = mar,
       volume = {623},
          eid = {A72},
        pages = {A72},
          doi = {10.1051/0004-6361/201834371},
archivePrefix = {arXiv},
       eprint = {1811.08902},
 primaryClass = {astro-ph.SR},
       adsurl = {https://ui.adsabs.harvard.edu/abs/2019A&A...623A..72K}
}

@ARTICLE{shob,
       author = {{Shobbrook}, R.~R. and {Lomb}, N.~R. and {Herbison-Evans}, D.},
        title = "{The short period light and velocity variations in Alpha Virginis.}",
      journal = {\mnras},
         year = 1972,
        month = jan,
       volume = {156},
        pages = {165},
          doi = {10.1093/mnras/156.2.165},
       adsurl = {https://ui.adsabs.harvard.edu/abs/1972MNRAS.156..165S}
}

@ARTICLE{andhag,
       author = {{Igoshev}, Andrei P. and {Perets}, Hagai B.},
        title = "{Wide binary companions to massive stars and their use in constraining natal kicks}",
      journal = {\mnras},
         year = 2019,
        month = jul,
       volume = {486},
       number = {3},
        pages = {4098-4113},
          doi = {10.1093/mnras/stz1024},
archivePrefix = {arXiv},
       eprint = {1901.05972},
 primaryClass = {astro-ph.HE},
       adsurl = {https://ui.adsabs.harvard.edu/abs/2019MNRAS.486.4098I}
}

@ARTICLE{specbins,
       author = {{Pourbaix}, D. and {Tokovinin}, A.~A. and {Batten}, A.~H. and {Fekel}, F.~C. and {Hartkopf}, W.~I. and {Levato}, H. and {Morrell}, N.~I. and {Torres}, G. and {Udry}, S.},
        title = "{S$_{B$^{9}$}$: The ninth catalogue of spectroscopic binary orbits}",
      journal = {\aap},
         year = 2004,
        month = sep,
       volume = {424},
        pages = {727-732},
          doi = {10.1051/0004-6361:20041213},
archivePrefix = {arXiv},
       eprint = {astro-ph/0406573},
 primaryClass = {astro-ph},
       adsurl = {https://ui.adsabs.harvard.edu/abs/2004A&A...424..727P}
}

@ARTICLE{moe17,
       author = {{Moe}, Maxwell and {Di Stefano}, Rosanne},
        title = "{Mind Your Ps and Qs: The Interrelation between Period (P) and Mass-ratio (Q) Distributions of Binary Stars}",
      journal = {\apjs},
         year = 2017,
        month = jun,
       volume = {230},
       number = {2},
          eid = {15},
        pages = {15},
          doi = {10.3847/1538-4365/aa6fb6},
archivePrefix = {arXiv},
       eprint = {1606.05347},
 primaryClass = {astro-ph.SR},
       adsurl = {https://ui.adsabs.harvard.edu/abs/2017ApJS..230...15M}
}

@ARTICLE{murphy20,
       author = {{Murphy}, Simon J. and {Gray}, Richard O. and {Corbally}, Christopher J. and {Kuehn}, Charles and {Bedding}, Timothy R. and {Killam}, Josiah},
        title = "{The discovery of lambda Bootis stars - the Southern Survey II}",
      journal = {\mnras},
         year = 2020,
        month = dec,
       volume = {499},
       number = {2},
        pages = {2701-2713},
          doi = {10.1093/mnras/staa2347},
archivePrefix = {arXiv},
       eprint = {2008.02392},
 primaryClass = {astro-ph.SR},
       adsurl = {https://ui.adsabs.harvard.edu/abs/2020MNRAS.499.2701M}
}

@ARTICLE{wang18,
       author = {{Wang}, Luqian and {Gies}, Douglas R. and {Peters}, Geraldine J.},
        title = "{Detection of Additional Be+sdO Systems from IUE Spectroscopy}",
      journal = {\apj},
         year = 2018,
        month = feb,
       volume = {853},
       number = {2},
          eid = {156},
        pages = {156},
          doi = {10.3847/1538-4357/aaa4b8},
archivePrefix = {arXiv},
       eprint = {1801.01066},
 primaryClass = {astro-ph.SR},
       adsurl = {https://ui.adsabs.harvard.edu/abs/2018ApJ...853..156W}
}

@ARTICLE{hd51480,
       author = {{Bidelman}, W.~P. and {MacConnell}, D.~J.},
        title = "{The brighter stars astrophysical interest in the southern sky.}",
      journal = {\aj},
         year = 1973,
        month = oct,
       volume = {78},
        pages = {687-733},
          doi = {10.1086/111475},
       adsurl = {https://ui.adsabs.harvard.edu/abs/1973AJ.....78..687B}
}

@ARTICLE{dr3,
       author = {{Bailer-Jones}, C.~A.~L. and {Rybizki}, J. and {Fouesneau}, M. and {Demleitner}, M. and {Andrae}, R.},
        title = "{Estimating Distances from Parallaxes. V. Geometric and Photogeometric Distances to 1.47 Billion Stars in Gaia Early Data Release 3}",
      journal = {\aj},
         year = 2021,
        month = mar,
       volume = {161},
       number = {3},
          eid = {147},
        pages = {147},
          doi = {10.3847/1538-3881/abd806},
archivePrefix = {arXiv},
       eprint = {2012.05220},
 primaryClass = {astro-ph.SR},
       adsurl = {https://ui.adsabs.harvard.edu/abs/2021AJ....161..147B}
}

@ARTICLE{bceprev,
       author = {{Lesh}, J.~R. and {Aizenman}, M.~L.},
        title = "{The observational status of the beta Cephei stars.}",
      journal = {\araa},
         year = 1978,
        month = jan,
       volume = {16},
        pages = {215},
          doi = {10.1146/annurev.aa.16.090178.001243},
       adsurl = {https://ui.adsabs.harvard.edu/abs/1978ARA&A..16..215L}
}

@ARTICLE{klem19,
       author = {{Klement}, Robert and {Carciofi}, A.~C. and {Rivinius}, T. and {Ignace}, R. and {Matthews}, L.~D. and {Torstensson}, K. and {Gies}, D. and {Vieira}, R.~G. and {Richardson}, N.~D. and {Domiciano de Souza}, A. and {Bjorkman}, J.~E. and {Hallinan}, G. and {Faes}, D.~M. and {Mota}, B. and {Gullingsrud}, A.~D. and {de Breuck}, C. and {Kervella}, P. and {Cur{\'e}}, M. and {Gunawan}, D.},
        title = "{Prevalence of SED Turndown among Classical Be Stars: Are All Be Stars Close Binaries?}",
      journal = {\apj},
         year = 2019,
        month = nov,
       volume = {885},
       number = {2},
          eid = {147},
        pages = {147},
          doi = {10.3847/1538-4357/ab48e7},
archivePrefix = {arXiv},
       eprint = {1909.12413},
 primaryClass = {astro-ph.SR},
       adsurl = {https://ui.adsabs.harvard.edu/abs/2019ApJ...885..147K}
}

@ARTICLE{horch,
       author = {{Horch}, Elliott P. and {Tokovinin}, Andrei and {Weiss}, Samuel A. and {L{\"o}bb}, J{\'a}nos and {Casetti-Dinescu}, Dana I. and {Granucci}, Nicole M. and {Hess}, Nicole M. and {Everett}, Mark E. and {van Belle}, Gerard T. and {Winters}, Jennifer G. and {Nusdeo}, Daniel A. and {Henry}, Todd J. and {Howell}, Steve B. and {Teske}, Johanna K. and {Hirsch}, Lea A. and {Scott}, Nicholas J. and {Matson}, Rachel A. and {Kane}, Stephen R.},
        title = "{Observations of Binary Stars with the Differential Speckle Survey Instrument. VIII. Measures of Metal-poor and Triple Stars from 2015 to 2018}",
      journal = {\aj},
         year = 2019,
        month = feb,
       volume = {157},
       number = {2},
          eid = {56},
        pages = {56},
          doi = {10.3847/1538-3881/aaf87e},
archivePrefix = {arXiv},
       eprint = {1812.05178},
 primaryClass = {astro-ph.SR},
       adsurl = {https://ui.adsabs.harvard.edu/abs/2019AJ....157...56H}
}

@ARTICLE{langer2020,
       author = {{Langer}, N. and {Baade}, D. and {Bodensteiner}, J. and {Greiner}, J. and {Rivinius}, Th. and {Martayan}, Ch. and {Borre}, C.~C.},
        title = "{{\ensuremath{\gamma}} Cas stars: Normal Be stars with discs impacted by the wind of a helium-star companion?}",
      journal = {\aap},
         year = 2020,
        month = jan,
       volume = {633},
          eid = {A40},
        pages = {A40},
          doi = {10.1051/0004-6361/201936736},
archivePrefix = {arXiv},
       eprint = {1911.06508},
 primaryClass = {astro-ph.SR},
       adsurl = {https://ui.adsabs.harvard.edu/abs/2020A&A...633A..40L}
}

@ARTICLE{shat02,
       author = {{Shatsky}, N. and {Tokovinin}, A.},
        title = "{The mass ratio distribution of B-type visual binaries in the Sco OB2 association}",
      journal = {\aap},
         year = 2002,
        month = jan,
       volume = {382},
        pages = {92-103},
          doi = {10.1051/0004-6361:20011542},
archivePrefix = {arXiv},
       eprint = {astro-ph/0109456},
 primaryClass = {astro-ph},
       adsurl = {https://ui.adsabs.harvard.edu/abs/2002A&A...382...92S}
}

@ARTICLE{brown97,
       author = {{Brown}, A.~G.~A. and {Verschueren}, W.},
        title = "{High S/N Echelle spectroscopy in young stellar groups. II. Rotational velocities of early-type stars in SCO OB2.}",
      journal = {\aap},
         year = 1997,
        month = mar,
       volume = {319},
        pages = {811-838},
archivePrefix = {arXiv},
       eprint = {astro-ph/9608089},
 primaryClass = {astro-ph},
       adsurl = {https://ui.adsabs.harvard.edu/abs/1997A&A...319..811B}
}

@ARTICLE{telt06,
       author = {{Telting}, J.~H. and {Schrijvers}, C. and {Ilyin}, I.~V. and {Uytterhoeven}, K. and {De Ridder}, J. and {Aerts}, C. and {Henrichs}, H.~F.},
        title = "{A high-resolution spectroscopy survey of {\ensuremath{\beta}} Cephei pulsations in bright stars}",
      journal = {\aap},
         year = 2006,
        month = jun,
       volume = {452},
       number = {3},
        pages = {945-953},
          doi = {10.1051/0004-6361:20054730},
       adsurl = {https://ui.adsabs.harvard.edu/abs/2006A&A...452..945T}
}

@ARTICLE{rivi06,
       author = {{Rivinius}, Th. and {{\v{S}}tefl}, S. and {Baade}, D.},
        title = "{Bright Be-shell stars}",
      journal = {\aap},
         year = 2006,
        month = nov,
       volume = {459},
       number = {1},
        pages = {137-145},
          doi = {10.1051/0004-6361:20053008},
       adsurl = {https://ui.adsabs.harvard.edu/abs/2006A&A...459..137R}
}

@ARTICLE{bjork2002,
       author = {{Bjorkman}, Karen S. and {Miroshnichenko}, Anatoly S. and {McDavid}, David and {Pogrosheva}, Tatiana M.},
        title = "{A Study of {\ensuremath{\pi}} Aquarii during a Quasi-normal Star Phase: Refined Fundamental Parameters and Evidence for Binarity}",
      journal = {\apj},
         year = 2002,
        month = jul,
       volume = {573},
       number = {2},
        pages = {812-824},
          doi = {10.1086/340751},
archivePrefix = {arXiv},
       eprint = {astro-ph/0203357},
 primaryClass = {astro-ph},
       adsurl = {https://ui.adsabs.harvard.edu/abs/2002ApJ...573..812B}
}

@ARTICLE{bess,
       author = {{Neiner}, C. and {de Batz}, B. and {Cochard}, F. and {Floquet}, M. and {Mekkas}, A. and {Desnoux}, V.},
        title = "{The Be Star Spectra (BeSS) Database}",
      journal = {\aj},
         year = 2011,
        month = nov,
       volume = {142},
       number = {5},
          eid = {149},
        pages = {149},
          doi = {10.1088/0004-6256/142/5/149},
       adsurl = {https://ui.adsabs.harvard.edu/abs/2011AJ....142..149N}
}

@ARTICLE{sana06,
       author = {{Sana}, H. and {Rauw}, G. and {Naz{\'e}}, Y. and {Gosset}, E. and {Vreux}, J. -M.},
        title = "{An XMM-Newton view of the young open cluster NGC 6231 - II. The OB star population}",
      journal = {\mnras},
         year = 2006,
        month = oct,
       volume = {372},
       number = {2},
        pages = {661-678},
          doi = {10.1111/j.1365-2966.2006.10847.x},
archivePrefix = {arXiv},
       eprint = {astro-ph/0607486},
 primaryClass = {astro-ph},
       adsurl = {https://ui.adsabs.harvard.edu/abs/2006MNRAS.372..661S}
}

@ARTICLE{sana13,
       author = {{Sana}, H. and {de Koter}, A. and {de Mink}, S.~E. and {Dunstall}, P.~R. and {Evans}, C.~J. and {H{\'e}nault-Brunet}, V. and {Ma{\'\i}z Apell{\'a}niz}, J. and {Ram{\'\i}rez-Agudelo}, O.~H. and {Taylor}, W.~D. and {Walborn}, N.~R. and {Clark}, J.~S. and {Crowther}, P.~A. and {Herrero}, A. and {Gieles}, M. and {Langer}, N. and {Lennon}, D.~J. and {Vink}, J.~S.},
        title = "{The VLT-FLAMES Tarantula Survey. VIII. Multiplicity properties of the O-type star population}",
      journal = {\aap},
         year = 2013,
        month = feb,
       volume = {550},
          eid = {A107},
        pages = {A107},
          doi = {10.1051/0004-6361/201219621},
archivePrefix = {arXiv},
       eprint = {1209.4638},
 primaryClass = {astro-ph.SR},
       adsurl = {https://ui.adsabs.harvard.edu/abs/2013A&A...550A.107S}
}

@ARTICLE{deacon20,
       author = {{Deacon}, N.~R. and {Kraus}, A.~L.},
        title = "{Wide binaries are rare in open clusters}",
      journal = {\mnras},
         year = 2020,
        month = aug,
       volume = {496},
       number = {4},
        pages = {5176-5200},
          doi = {10.1093/mnras/staa1877},
archivePrefix = {arXiv},
       eprint = {2006.06679},
 primaryClass = {astro-ph.SR},
       adsurl = {https://ui.adsabs.harvard.edu/abs/2020MNRAS.496.5176D}
}

@ARTICLE{hu,
       author = {{Hu}, Y. and {Deng}, L. and {de Grijs}, Richard and {Liu}, Q. and {Goodwin}, Simon P.},
        title = "{The Binary Fraction of the Young Cluster NGC 1818 in the Large Magellanic Cloud}",
      journal = {\apj},
         year = 2010,
        month = nov,
       volume = {724},
       number = {1},
        pages = {649-656},
          doi = {10.1088/0004-637X/724/1/649},
archivePrefix = {arXiv},
       eprint = {0801.2814},
 primaryClass = {astro-ph},
       adsurl = {https://ui.adsabs.harvard.edu/abs/2010ApJ...724..649H}
}

@ARTICLE{cangau,
       author = {{Cantat-Gaudin}, T. and {Anders}, F.},
        title = "{Clusters and mirages: cataloguing stellar aggregates in the Milky Way}",
      journal = {\aap},
         year = 2020,
        month = jan,
       volume = {633},
          eid = {A99},
        pages = {A99},
          doi = {10.1051/0004-6361/201936691},
archivePrefix = {arXiv},
       eprint = {1911.07075},
 primaryClass = {astro-ph.SR},
       adsurl = {https://ui.adsabs.harvard.edu/abs/2020A&A...633A..99C}
}

@ARTICLE{simbad,
       author = {{Wenger}, M. and {Ochsenbein}, F. and {Egret}, D. and {Dubois}, P. and {Bonnarel}, F. and {Borde}, S. and {Genova}, F. and {Jasniewicz}, G. and {Lalo{\"e}}, S. and {Lesteven}, S. and {Monier}, R.},
        title = "{The SIMBAD astronomical database. The CDS reference database for astronomical objects}",
      journal = {\aaps},
         year = 2000,
        month = apr,
       volume = {143},
        pages = {9-22},
          doi = {10.1051/aas:2000332},
archivePrefix = {arXiv},
       eprint = {astro-ph/0002110},
 primaryClass = {astro-ph},
       adsurl = {https://ui.adsabs.harvard.edu/abs/2000A&AS..143....9W}
}

@ARTICLE{mahy22,
       author = {{Mahy}, L. and {Sana}, H. and {Shenar}, T. and {Sen}, K. and {Langer}, N. and {Marchant}, P. and {Abdul-Masih}, M. and {Banyard}, G. and {Bodensteiner}, J. and {Bowman}, D.~M. and {Dsilva}, K. and {Fabry}, M. and {Hawcroft}, C. and {Janssens}, S. and {Van Reeth}, T. and {Eldridge}, C.},
        title = "{Identifying quiescent compact objects in massive Galactic single-lined spectroscopic binaries}",
      journal = {\aap},
         year = 2022,
        month = aug,
       volume = {664},
          eid = {A159},
        pages = {A159},
          doi = {10.1051/0004-6361/202243147},
archivePrefix = {arXiv},
       eprint = {2207.07752},
 primaryClass = {astro-ph.SR},
       adsurl = {https://ui.adsabs.harvard.edu/abs/2022A&A...664A.159M}
}

@INPROCEEDINGS{off23,
       author = {{Offner}, S.~S.~R. and {Moe}, M. and {Kratter}, K.~M. and {Sadavoy}, S.~I. and {Jensen}, E.~L.~N. and {Tobin}, J.~J.},
        title = "{The Origin and Evolution of Multiple Star Systems}",
    booktitle = {Protostars and Planets VII},
         year = 2023,
       editor = {{Inutsuka}, S. and {Aikawa}, Y. and {Muto}, T. and {Tomida}, K. and {Tamura}, M.},
       series = {Astronomical Society of the Pacific Conference Series},
       volume = {534},
        month = jul,
        pages = {275},
          doi = {10.48550/arXiv.2203.10066},
archivePrefix = {arXiv},
       eprint = {2203.10066},
 primaryClass = {astro-ph.SR},
       adsurl = {https://ui.adsabs.harvard.edu/abs/2023ASPC..534..275O}
}

@ARTICLE{too+21,
       author = {{Toonen}, S. and {Boekholt}, T.~C.~N. and {Portegies Zwart}, S.},
        title = "{Stellar triples on the edge. Comprehensive overview of the evolution of destabilised triples leading to stellar and binary exotica}",
      journal = {\aap},
         year = 2022,
        month = may,
       volume = {661},
          eid = {A61},
        pages = {A61},
          doi = {10.1051/0004-6361/202141991},
archivePrefix = {arXiv},
       eprint = {2108.04272},
 primaryClass = {astro-ph.SR},
       adsurl = {https://ui.adsabs.harvard.edu/abs/2022A&A...661A..61T}
}

@ARTICLE{ant+12,
       author = {{Antonini}, Fabio and {Perets}, Hagai B.},
        title = "{Secular Evolution of Compact Binaries near Massive Black Holes: Gravitational Wave Sources and Other Exotica}",
      journal = {\apj},
         year = 2012,
        month = sep,
       volume = {757},
       number = {1},
          eid = {27},
        pages = {27},
          doi = {10.1088/0004-637X/757/1/27},
archivePrefix = {arXiv},
       eprint = {1203.2938},
 primaryClass = {astro-ph.GA},
       adsurl = {https://ui.adsabs.harvard.edu/abs/2012ApJ...757...27A}
}

@ARTICLE{ham+21,
       author = {{Hamers}, Adrian S. and {Glanz}, Hila and {Neunteufel}, Patrick},
        title = "{A Statistical View of the Stable and Unstable Roche Lobe Overflow of a Tertiary Star onto the Inner Binary in Triple Systems}",
      journal = {\apjs},
         year = 2022,
        month = mar,
       volume = {259},
       number = {1},
          eid = {25},
        pages = {25},
          doi = {10.3847/1538-4365/ac49e7},
archivePrefix = {arXiv},
       eprint = {2110.00024},
 primaryClass = {astro-ph.SR},
       adsurl = {https://ui.adsabs.harvard.edu/abs/2022ApJS..259...25H}
}

@ARTICLE{kis+98,
       author = {{Kiseleva}, L.~G. and {Eggleton}, P.~P. and {Mikkola}, S.},
        title = "{Tidal friction in triple stars}",
      journal = {\mnras},
         year = 1998,
        month = oct,
       volume = {300},
       number = {1},
        pages = {292-302},
          doi = {10.1046/j.1365-8711.1998.01903.x},
       adsurl = {https://ui.adsabs.harvard.edu/abs/1998MNRAS.300..292K}
}

@ARTICLE{egg+86,
       author = {{Eggleton}, P.~P. and {Verbunt}, F.},
        title = "{Triple star evolution and the formation of short-period, low-mass X-ray binaries.}",
      journal = {\mnras},
         year = 1986,
        month = may,
       volume = {220},
        pages = {13P-18},
          doi = {10.1093/mnras/220.1.13P},
       adsurl = {https://ui.adsabs.harvard.edu/abs/1986MNRAS.220P..13E}
}

@ARTICLE{maz+79,
       author = {{Mazeh}, T. and {Shaham}, J.},
        title = "{The orbital evolution of close triple systems: the binary eccentricity.}",
      journal = {\aap},
         year = 1979,
        month = aug,
       volume = {77},
        pages = {145},
       adsurl = {https://ui.adsabs.harvard.edu/abs/1979A&A....77..145M}
}

@ARTICLE{kozai,
       author = {{Kozai}, Yoshihide},
        title = "{Secular perturbations of asteroids with high inclination and eccentricity}",
      journal = {\aj},
         year = 1962,
        month = nov,
       volume = {67},
        pages = {591-598},
          doi = {10.1086/108790},
       adsurl = {https://ui.adsabs.harvard.edu/abs/1962AJ.....67..591K}
}

@ARTICLE{lidov,
       author = {{Lidov}, M.~L.},
        title = "{The evolution of orbits of artificial satellites of planets under the action of gravitational perturbations of external bodies}",
      journal = {\planss},
         year = 1962,
        month = oct,
       volume = {9},
       number = {10},
        pages = {719-759},
          doi = {10.1016/0032-0633(62)90129-0},
       adsurl = {https://ui.adsabs.harvard.edu/abs/1962P&SS....9..719L}
}

@ARTICLE{rosslowe18,
       author = {{Rosslowe}, C.~K. and {Crowther}, Paul A.},
        title = "{A deep near-infrared spectroscopic survey of the Scutum-Crux arm for Wolf-Rayet stars}",
      journal = {\mnras},
         year = 2018,
        month = jan,
       volume = {473},
       number = {3},
        pages = {2853-2870},
          doi = {10.1093/mnras/stx2103},
archivePrefix = {arXiv},
       eprint = {1708.03582},
 primaryClass = {astro-ph.SR},
       adsurl = {https://ui.adsabs.harvard.edu/abs/2018MNRAS.473.2853R}
}

@ARTICLE{chauville01,
       author = {{Chauville}, J. and {Zorec}, J. and {Ballereau}, D. and {Morrell}, N. and {Cidale}, L. and {Garcia}, A.},
        title = "{High and intermediate-resolution spectroscopy of Be stars 4481 lines}",
      journal = {\aap},
         year = 2001,
        month = nov,
       volume = {378},
        pages = {861-882},
          doi = {10.1051/0004-6361:20011202},
       adsurl = {https://ui.adsabs.harvard.edu/abs/2001A&A...378..861C}
}

@ARTICLE{hanu87,
       author = {{Hanuschik}, R.~W.},
        title = "{High-resolution emission-line spectroscopy of Be stars. II. Fe II andother weak emission lines.}",
      journal = {\aap},
         year = 1987,
        month = feb,
       volume = {173},
        pages = {299-314},
       adsurl = {https://ui.adsabs.harvard.edu/abs/1987A&A...173..299H}
}

@ARTICLE{percy77,
       author = {{Percy}, J.~R. and {Lane}, M.~C.},
        title = "{Search for beta Cephei stars. I: Photometric and spectroscopic studies of northern B-type stars.}",
      journal = {\aj},
         year = 1977,
        month = may,
       volume = {82},
        pages = {353-359},
          doi = {10.1086/112057},
       adsurl = {https://ui.adsabs.harvard.edu/abs/1977AJ.....82..353P}
}

@ARTICLE{abt84,
       author = {{Abt}, H.~A. and {Cardona}, O.},
        title = "{Be stars in binaries.}",
      journal = {\apj},
         year = 1984,
        month = oct,
       volume = {285},
        pages = {190-194},
          doi = {10.1086/162490},
       adsurl = {https://ui.adsabs.harvard.edu/abs/1984ApJ...285..190A}
}

@ARTICLE{abt,
       author = {{Abt}, Helmut A. and {Gomez}, Ana E. and {Levy}, Saul G.},
        title = "{The Frequency and Formation Mechanism of B2--B5 Main-Sequence Binaries}",
      journal = {\apjs},
         year = 1990,
        month = oct,
       volume = {74},
        pages = {551},
          doi = {10.1086/191508},
       adsurl = {https://ui.adsabs.harvard.edu/abs/1990ApJS...74..551A}
}

@ARTICLE{Sha+13,
       author = {{Shappee}, Benjamin J. and {Thompson}, Todd A.},
        title = "{The Mass-loss-induced Eccentric Kozai Mechanism: A New Channel for the Production of Close Compact Object-Stellar Binaries}",
      journal = {\apj},
         year = 2013,
        month = mar,
       volume = {766},
       number = {1},
          eid = {64},
        pages = {64},
          doi = {10.1088/0004-637X/766/1/64},
archivePrefix = {arXiv},
       eprint = {1204.1053},
 primaryClass = {astro-ph.SR},
       adsurl = {https://ui.adsabs.harvard.edu/abs/2013ApJ...766...64S}
}

@ARTICLE{sok04,
       author = {{Soker}, Noam},
        title = "{Wind accretion by a binary stellar system and disc formation}",
      journal = {\mnras},
         year = 2004,
        month = jun,
       volume = {350},
       number = {4},
        pages = {1366-1372},
          doi = {10.1111/j.1365-2966.2004.07731.x},
archivePrefix = {arXiv},
       eprint = {astro-ph/0402364},
 primaryClass = {astro-ph},
       adsurl = {https://ui.adsabs.harvard.edu/abs/2004MNRAS.350.1366S}
}

@ARTICLE{per+12,
       author = {{Perets}, Hagai B. and {Kratter}, Kaitlin M.},
        title = "{The Triple Evolution Dynamical Instability: Stellar Collisions in the Field and the Formation of Exotic Binaries}",
      journal = {\apj},
         year = 2012,
        month = dec,
       volume = {760},
       number = {2},
          eid = {99},
        pages = {99},
          doi = {10.1088/0004-637X/760/2/99},
archivePrefix = {arXiv},
       eprint = {1203.2914},
 primaryClass = {astro-ph.SR},
       adsurl = {https://ui.adsabs.harvard.edu/abs/2012ApJ...760...99P}
}

@ARTICLE{mic+14,
       author = {{Michaely}, Erez and {Perets}, Hagai B.},
        title = "{Secular Dynamics in Hierarchical Three-body Systems with Mass Loss and Mass Transfer}",
      journal = {\apj},
         year = 2014,
        month = oct,
       volume = {794},
       number = {2},
          eid = {122},
        pages = {122},
          doi = {10.1088/0004-637X/794/2/122},
archivePrefix = {arXiv},
       eprint = {1406.3035},
 primaryClass = {astro-ph.SR},
       adsurl = {https://ui.adsabs.harvard.edu/abs/2014ApJ...794..122M}
}

@ARTICLE{per+09,
       author = {{Perets}, Hagai B. and {Fabrycky}, Daniel C.},
        title = "{On the Triple Origin of Blue Stragglers}",
      journal = {\apj},
         year = 2009,
        month = jun,
       volume = {697},
       number = {2},
        pages = {1048-1056},
          doi = {10.1088/0004-637X/697/2/1048},
archivePrefix = {arXiv},
       eprint = {0901.4328},
 primaryClass = {astro-ph.SR},
       adsurl = {https://ui.adsabs.harvard.edu/abs/2009ApJ...697.1048P}
}

@ARTICLE{too+16,
       author = {{Toonen}, Silvia and {Hamers}, Adrian and {Portegies Zwart}, Simon},
        title = "{The evolution of hierarchical triple star-systems}",
      journal = {Computational Astrophysics and Cosmology},
         year = 2016,
        month = dec,
       volume = {3},
       number = {1},
          eid = {6},
        pages = {6},
          doi = {10.1186/s40668-016-0019-0},
archivePrefix = {arXiv},
       eprint = {1612.06172},
 primaryClass = {astro-ph.SR},
       adsurl = {https://ui.adsabs.harvard.edu/abs/2016ComAC...3....6T}
}

@ARTICLE{toko06,
       author = {{Tokovinin}, A. and {Thomas}, S. and {Sterzik}, M. and {Udry}, S.},
        title = "{Tertiary companions to close spectroscopic binaries}",
      journal = {\aap},
         year = 2006,
        month = may,
       volume = {450},
       number = {2},
        pages = {681-693},
          doi = {10.1051/0004-6361:20054427},
archivePrefix = {arXiv},
       eprint = {astro-ph/0601518},
 primaryClass = {astro-ph},
       adsurl = {https://ui.adsabs.harvard.edu/abs/2006A&A...450..681T}
}

@ARTICLE{luc21,
       author = {{IJspeert}, L.~W. and {Tkachenko}, A. and {Johnston}, C. and {Garcia}, S. and {De Ridder}, J. and {Van Reeth}, T. and {Aerts}, C.},
        title = "{An all-sky sample of intermediate- to high-mass OBA-type eclipsing binaries observed by TESS}",
      journal = {\aap},
         year = 2021,
        month = aug,
       volume = {652},
          eid = {A120},
        pages = {A120},
          doi = {10.1051/0004-6361/202141489},
archivePrefix = {arXiv},
       eprint = {2107.10005},
 primaryClass = {astro-ph.SR},
       adsurl = {https://ui.adsabs.harvard.edu/abs/2021A&A...652A.120I}
}

@ARTICLE{blanco,
       author = {{Moraux}, E. and {Bouvier}, J. and {Stauffer}, J.~R. and {Barrado y Navascu{\'e}s}, D. and {Cuillandre}, J. -C.},
        title = "{The lower mass function of the young open cluster Blanco 1: from 30 M$_{Jup}$ to 3 M$_{{\ensuremath{\odot}}}$}",
      journal = {\aap},
         year = 2007,
        month = aug,
       volume = {471},
       number = {2},
        pages = {499-513},
          doi = {10.1051/0004-6361:20066308},
archivePrefix = {arXiv},
       eprint = {0706.2102},
 primaryClass = {astro-ph},
       adsurl = {https://ui.adsabs.harvard.edu/abs/2007A&A...471..499M}
}

@ARTICLE{egg+08,
       author = {{Eggleton}, P.~P. and {Tokovinin}, A.~A.},
        title = "{A catalogue of multiplicity among bright stellar systems}",
      journal = {\mnras},
         year = 2008,
        month = sep,
       volume = {389},
       number = {2},
        pages = {869-879},
          doi = {10.1111/j.1365-2966.2008.13596.x},
archivePrefix = {arXiv},
       eprint = {0806.2878},
 primaryClass = {astro-ph},
       adsurl = {https://ui.adsabs.harvard.edu/abs/2008MNRAS.389..869E}
}

@ARTICLE{galcep,
       author = {{Gallenne}, A. and {Kervella}, P. and {Borgniet}, S. and {M{\'e}rand}, A. and {Pietrzy{\'n}ski}, G. and {Gieren}, W. and {Monnier}, J.~D. and {Schaefer}, G.~H. and {Evans}, N.~R. and {Anderson}, R.~I. and {Baron}, F. and {Roettenbacher}, R.~M. and {Karczmarek}, P.},
        title = "{Multiplicity of Galactic Cepheids from long-baseline interferometry. IV. New detected companions from MIRC and PIONIER observations}",
      journal = {\aap},
         year = 2019,
        month = feb,
       volume = {622},
          eid = {A164},
        pages = {A164},
          doi = {10.1051/0004-6361/201834614},
archivePrefix = {arXiv},
       eprint = {1812.09989},
 primaryClass = {astro-ph.SR},
       adsurl = {https://ui.adsabs.harvard.edu/abs/2019A&A...622A.164G}
}

@ARTICLE{evanscep,
       author = {{Evans}, Nancy Remage and {Carpenter}, Kenneth G. and {Robinson}, Richard and {Kienzle}, Francesco and {Dekas}, Anne E.},
        title = "{High-Mass Triple Systems: The Classical Cepheid Y Carinae}",
      journal = {\aj},
         year = 2005,
        month = aug,
       volume = {130},
       number = {2},
        pages = {789-793},
          doi = {10.1086/430458},
archivePrefix = {arXiv},
       eprint = {astro-ph/0504169},
 primaryClass = {astro-ph},
       adsurl = {https://ui.adsabs.harvard.edu/abs/2005AJ....130..789E}
}

@ARTICLE{lefevre,
       author = {{Lef{\`e}vre}, L. and {Marchenko}, S.~V. and {Moffat}, A.~F.~J. and {Acker}, A.},
        title = "{A systematic study of variability among OB-stars based on HIPPARCOS photometry}",
      journal = {\aap},
         year = 2009,
        month = nov,
       volume = {507},
       number = {2},
        pages = {1141-1201},
          doi = {10.1051/0004-6361/200912304},
       adsurl = {https://ui.adsabs.harvard.edu/abs/2009A&A...507.1141L}
}

@ARTICLE{pols,
       author = {{Pols}, O.~R. and {Cote}, J. and {Waters}, L.~B.~F.~M. and {Heise}, J.},
        title = "{The formation of Be stars through close binary evolution.}",
      journal = {\aap},
         year = 1991,
        month = jan,
       volume = {241},
        pages = {419},
       adsurl = {https://ui.adsabs.harvard.edu/abs/1991A&A...241..419P}
}

@ARTICLE{vanbev,
       author = {{van Bever}, J. and {Vanbeveren}, D.},
        title = "{The number of B-type binary mass gainers in general, binary Be stars in particular, predicted by close binary evolution.}",
      journal = {\aap},
         year = 1997,
        month = jun,
       volume = {322},
        pages = {116-126},
       adsurl = {https://ui.adsabs.harvard.edu/abs/1997A&A...322..116V}
}

@ARTICLE{struve,
       author = {{Struve}, Otto},
        title = "{On the Origin of Bright Lines in Spectra of Stars of Class B}",
      journal = {\apj},
         year = 1931,
        month = mar,
       volume = {73},
        pages = {94},
          doi = {10.1086/143298},
       adsurl = {https://ui.adsabs.harvard.edu/abs/1931ApJ....73...94S}
}

@ARTICLE{sana17,
       author = {{Sana}, H. and {Ram{\'\i}rez-Tannus}, M.~C. and {de Koter}, A. and {Kaper}, L. and {Tramper}, F. and {Bik}, A.},
        title = "{A dearth of short-period massive binaries in the young massive star forming region M 17. Evidence for a large orbital separation at birth?}",
      journal = {\aap},
         year = 2017,
        month = mar,
       volume = {599},
          eid = {L9},
        pages = {L9},
          doi = {10.1051/0004-6361/201630087},
archivePrefix = {arXiv},
       eprint = {1702.02153},
 primaryClass = {astro-ph.SR},
       adsurl = {https://ui.adsabs.harvard.edu/abs/2017A&A...599L...9S}
}

@ARTICLE{coxbook,
       author = {{Cox}, Arthur N. and {Pilachowski}, Catherine A.},
        title = "{Allen's Astrophysical Quantities}",
      journal = {Physics Today},
         year = 2000,
        month = oct,
       volume = {53},
       number = {10},
        pages = {77},
          doi = {10.1063/1.1325201},
       adsurl = {https://ui.adsabs.harvard.edu/abs/2000PhT....53j..77C}
}

@ARTICLE{silaj14,
       author = {{Silaj}, J. and {Jones}, C.~E. and {Sigut}, T.~A.~A. and {Tycner}, C.},
        title = "{The H{\ensuremath{\alpha}} Profiles of Be Shell Stars}",
      journal = {\apj},
         year = 2014,
        month = nov,
       volume = {795},
       number = {1},
          eid = {82},
        pages = {82},
          doi = {10.1088/0004-637X/795/1/82},
       adsurl = {https://ui.adsabs.harvard.edu/abs/2014ApJ...795...82S}
}

@INPROCEEDINGS{pmoired,
       author = {{M{\'e}rand}, Antoine},
        title = "{Flexible spectro-interferometric modelling of OIFITS data with PMOIRED}",
    booktitle = {Optical and Infrared Interferometry and Imaging VIII},
         year = 2022,
       editor = {{M{\'e}rand}, Antoine and {Sallum}, Stephanie and {Sanchez-Bermudez}, Joel},
       series = {Society of Photo-Optical Instrumentation Engineers (SPIE) Conference Series},
       volume = {12183},
        month = aug,
          eid = {121831N},
        pages = {121831N},
          doi = {10.1117/12.2626700},
archivePrefix = {arXiv},
       eprint = {2207.11047},
 primaryClass = {astro-ph.IM},
       adsurl = {https://ui.adsabs.harvard.edu/abs/2022SPIE12183E..1NM}
}

@ARTICLE{rivi13,
       author = {{Rivinius}, Thomas and {Carciofi}, Alex C. and {Martayan}, Christophe},
        title = "{Classical Be stars. Rapidly rotating B stars with viscous Keplerian decretion disks}",
      journal = {\aapr},
         year = 2013,
        month = oct,
       volume = {21},
          eid = {69},
        pages = {69},
          doi = {10.1007/s00159-013-0069-0},
archivePrefix = {arXiv},
       eprint = {1310.3962},
 primaryClass = {astro-ph.SR},
       adsurl = {https://ui.adsabs.harvard.edu/abs/2013A&ARv..21...69R}
}

@ARTICLE{deMink2013,
       author = {{de Mink}, S.~E. and {Langer}, N. and {Izzard}, R.~G. and {Sana}, H. and {de Koter}, A.},
        title = "{The Rotation Rates of Massive Stars: The Role of Binary Interaction through Tides, Mass Transfer, and Mergers}",
      journal = {\apj},
         year = 2013,
        month = feb,
       volume = {764},
       number = {2},
          eid = {166},
        pages = {166},
          doi = {10.1088/0004-637X/764/2/166},
archivePrefix = {arXiv},
       eprint = {1211.3742},
 primaryClass = {astro-ph.SR},
       adsurl = {https://ui.adsabs.harvard.edu/abs/2013ApJ...764..166D}
}

@ARTICLE{Vanbeveren1994,
       author = {{Vanbeveren}, D. and {De Loore}, C.},
        title = "{The evolution of the mass gainer in massive close binaries.}",
      journal = {\aap},
         year = 1994,
        month = oct,
       volume = {290},
        pages = {129-132},
       adsurl = {https://ui.adsabs.harvard.edu/abs/1994A&A...290..129V}
}

@ARTICLE{Podsiadlowski1992,
       author = {{Podsiadlowski}, Ph. and {Joss}, P.~C. and {Hsu}, J.~J.~L.},
        title = "{Presupernova Evolution in Massive Interacting Binaries}",
      journal = {\apj},
         year = 1992,
        month = may,
       volume = {391},
        pages = {246},
          doi = {10.1086/171341},
       adsurl = {https://ui.adsabs.harvard.edu/abs/1992ApJ...391..246P}
}

@ARTICLE{Paczynski1967,
       author = {{Paczy{\'n}ski}, B.},
        title = "{Gravitational Waves and the Evolution of Close Binaries}",
      journal = {\actaa},
         year = 1967,
        month = jan,
       volume = {17},
        pages = {287},
       adsurl = {https://ui.adsabs.harvard.edu/abs/1967AcA....17..287P}
}

@ARTICLE{elbadryrix,
       author = {{El-Badry}, Kareem and {Rix}, Hans-Walter},
        title = "{Imprints of white dwarf recoil in the separation distribution of Gaia wide binaries}",
      journal = {\mnras},
         year = 2018,
        month = nov,
       volume = {480},
       number = {4},
        pages = {4884-4902},
          doi = {10.1093/mnras/sty2186},
archivePrefix = {arXiv},
       eprint = {1807.06011},
 primaryClass = {astro-ph.SR},
       adsurl = {https://ui.adsabs.harvard.edu/abs/2018MNRAS.480.4884E}
}

@ARTICLE{cyprien23,
       author = {{Lanthermann}, C. and {Le Bouquin}, J. -B. and {Sana}, H. and {M{\'e}rand}, A. and {Monnier}, J.~D. and {Perraut}, K. and {Frost}, A.~J. and {Mahy}, L. and {Gosset}, E. and {De Becker}, M. and {Kraus}, S. and {Anugu}, N. and {Davies}, C.~L. and {Ennis}, J. and {Gardner}, T. and {Labdon}, A. and {Setterholm}, B. and {ten Brummelaar}, T. and {Schaefer}, G.~H.},
        title = "{Multiplicity of northern bright O-type stars with optical long baseline interferometry. Results of the pilot survey}",
      journal = {\aap},
         year = 2023,
        month = apr,
       volume = {672},
          eid = {A6},
        pages = {A6},
          doi = {10.1051/0004-6361/202245364},
archivePrefix = {arXiv},
       eprint = {2302.03168},
 primaryClass = {astro-ph.SR},
       adsurl = {https://ui.adsabs.harvard.edu/abs/2023A&A...672A...6L}
}

@INPROCEEDINGS{vlti,
       author = {{Haubois}, Xavier and {M{\'e}rand}, Antoine and {Abuter}, Roberto and {Araneda}, Juan Pablo and {Bian}, Fuyan and {Bourget}, Pierre and {Bristow}, Paul and {Burgos}, Pablo and {Delplancke-Str{\"o}bele}, Fran{\c{c}}oise and {Dembet}, Roderick and {Gil}, Juan Pablo and {Glindemann}, Andreas and {Gont{\'e}}, Fr{\'e}d{\'e}ric and {Guajardo}, Patricia and {Hubin}, Norbert and {Hummel}, Christian and {Korhonen}, Heidi and {Labdon}, Aaron and {Kolb}, Johann and {Kosmalski}, Johan and {Lacour}, Sylvestre and {Paladini}, Claudia and {Pallanca}, Laurent and {Pasquini}, Luca and {Percheron}, Isabelle and {Riquelme}, Miguel and {Rivinius}, Thomas and {Sani}, Eleonora and {Schmidtobreick}, Linda and {Scicluna}, Peter and {Sch{\"o}ller}, Markus and {Schuhler}, Nicolas and {Tristram}, Konrad and {Wittkowski}, Markus and {Woillez}, Julien and {Zins}, G{\'e}rard},
        title = "{VLTI status update}",
    booktitle = {Optical and Infrared Interferometry and Imaging VIII},
         year = 2022,
       editor = {{M{\'e}rand}, Antoine and {Sallum}, Stephanie and {Sanchez-Bermudez}, Joel},
       series = {Society of Photo-Optical Instrumentation Engineers (SPIE) Conference Series},
       volume = {12183},
        month = aug,
          eid = {1218306},
        pages = {1218306},
          doi = {10.1117/12.2635405},
       adsurl = {https://ui.adsabs.harvard.edu/abs/2022SPIE12183E..06H}
}

@ARTICLE{gwtc3,
       author = {{Abbott}, R. and {Abbott}, T.~D. and {Acernese}, F. and {Ackley}, K. and {Adams}, C. and {Adhikari}, N. and {Adhikari}, R.~X. and {Adya}, V.~B. and {Affeldt}, C. and {Agarwal}, D. and {Agathos}, M. and {Agatsuma}, K. and {Aggarwal}, N. and {Aguiar}, O.~D. and {Aiello}, L. and {Ain}, A. and {Ajith}, P. and {Akcay}, S. and {Akutsu}, T. and {Albanesi}, S. and {Allocca}, A. and {Altin}, P.~A. and {Amato}, A. and {Anand}, C. and {Anand}, S. and {Ananyeva}, A. and {Anderson}, S.~B. and {Anderson}, W.~G. and {Ando}, M. and {Andrade}, T. and {Andres}, N. and {Andri{\'c}}, T. and {Angelova}, S.~V. and {Ansoldi}, S. and {Antelis}, J.~M. and {Antier}, S. and {Appert}, S. and {Arai}, Koji and {Arai}, Koya and {Arai}, Y. and {Araki}, S. and {Araya}, A. and {Araya}, M.~C. and {Areeda}, J.~S. and {Ar{\`e}ne}, M. and {Aritomi}, N. and {Arnaud}, N. and {Arogeti}, M. and {Aronson}, S.~M. and {Arun}, K.~G. and {Asada}, H. and {Asali}, Y. and {Ashton}, G. and {Aso}, Y. and {Assiduo}, M. and {Aston}, S.~M. and {Astone}, P. and {Aubin}, F. and {Austin}, C. and {Babak}, S. and {Badaracco}, F. and {Bader}, M.~K.~M. and {Badger}, C. and {Bae}, S. and {Bae}, Y. and {Baer}, A.~M. and {Bagnasco}, S. and {Bai}, Y. and {Baiotti}, L. and {Baird}, J. and {Bajpai}, R. and {Ball}, M. and {Ballardin}, G. and {Ballmer}, S.~W. and {Balsamo}, A. and {Baltus}, G. and {Banagiri}, S. and {Bankar}, D. and {Barayoga}, J.~C. and {Barbieri}, C. and {Barish}, B.~C. and {Barker}, D. and {Barneo}, P. and {Barone}, F. and {Barr}, B. and {Barsotti}, L. and {Barsuglia}, M. and {Barta}, D. and {Bartlett}, J. and {Barton}, M.~A. and {Bartos}, I. and {Bassiri}, R. and {Basti}, A. and {Bawaj}, M. and {Bayley}, J.~C. and {Baylor}, A.~C. and {Bazzan}, M. and {B{\'e}csy}, B. and {Bedakihale}, V.~M. and {Bejger}, M. and {Belahcene}, I. and {Benedetto}, V. and {Beniwal}, D. and {Bennett}, T.~F. and {Bentley}, J.~D. and {Benyaala}, M. and {Bergamin}, F. and {Berger}, B.~K. and {Bernuzzi}, S. and {Berry}, C.~P.~L. and {Bersanetti}, D. and {Bertolini}, A. and {Betzwieser}, J. and {Beveridge}, D. and {Bhandare}, R. and {Bhardwaj}, U. and {Bhattacharjee}, D. and {Bhaumik}, S. and {Bilenko}, I.~A. and {Billingsley}, G. and {Bini}, S. and {Birney}, R. and {Birnholtz}, O. and {Biscans}, S. and {Bischi}, M. and {Biscoveanu}, S. and {Bisht}, A. and {Biswas}, B. and {Bitossi}, M. and {Bizouard}, M. -A. and {Blackburn}, J.~K. and {Blair}, C.~D. and {Blair}, D.~G. and {Blair}, R.~M. and {Bobba}, F. and {Bode}, N. and {Boer}, M. and {Bogaert}, G. and {Boldrini}, M. and {Bonavena}, L.~D. and {Bondu}, F. and {Bonilla}, E. and {Bonnand}, R. and {Booker}, P. and {Boom}, B.~A. and {Bork}, R. and {Boschi}, V. and {Bose}, N. and {Bose}, S. and {Bossilkov}, V. and {Boudart}, V. and {Bouffanais}, Y. and {Bozzi}, A. and {Bradaschia}, C. and {Brady}, P.~R. and {Bramley}, A. and {Branch}, A. and {Branchesi}, M. and {Brandt}, J. and {Brau}, J.~E. and {Breschi}, M. and {Briant}, T. and {Briggs}, J.~H. and {Brillet}, A. and {Brinkmann}, M. and {Brockill}, P. and {Brooks}, A.~F. and {Brooks}, J. and {Brown}, D.~D. and {Brunett}, S. and {Bruno}, G. and {Bruntz}, R. and {Bryant}, J. and {Bulik}, T. and {Bulten}, H.~J. and {Buonanno}, A. and {Buscicchio}, R. and {Buskulic}, D. and {Buy}, C. and {Byer}, R.~L. and {Davies}, G.~S. Cabourn and {Cadonati}, L. and {Cagnoli}, G. and {Cahillane}, C. and {Bustillo}, J. Calder{\'o}n and {Callaghan}, J.~D. and {Callister}, T.~A. and {Calloni}, E. and {Cameron}, J. and {Camp}, J.~B. and {Canepa}, M. and {Canevarolo}, S. and {Cannavacciuolo}, M. and {Cannon}, K.~C. and {Cao}, H. and {Cao}, Z. and {Capocasa}, E. and {Capote}, E. and {Carapella}, G. and {Carbognani}, F.},
        title = "{GWTC-3: Compact Binary Coalescences Observed by LIGO and Virgo during the Second Part of the Third Observing Run}",
      journal = {Physical Review X},
         year = 2023,
        month = oct,
       volume = {13},
       number = {4},
          eid = {041039},
        pages = {041039},
          doi = {10.1103/PhysRevX.13.041039},
archivePrefix = {arXiv},
       eprint = {2111.03606},
 primaryClass = {gr-qc},
       adsurl = {https://ui.adsabs.harvard.edu/abs/2023PhRvX..13d1039A}
}

@ARTICLE{masongies09,
       author = {{Mason}, Brian D. and {Hartkopf}, William I. and {Gies}, Douglas R. and {Henry}, Todd J. and {Helsel}, John W.},
        title = "{The High Angular Resolution Multiplicity of Massive Stars}",
      journal = {\aj},
         year = 2009,
        month = feb,
       volume = {137},
       number = {2},
        pages = {3358-3377},
          doi = {10.1088/0004-6256/137/2/3358},
archivePrefix = {arXiv},
       eprint = {0811.0492},
 primaryClass = {astro-ph},
       adsurl = {https://ui.adsabs.harvard.edu/abs/2009AJ....137.3358M}
}

@ARTICLE{pabjulrev,
       author = {{Marchant}, Pablo and {Bodensteiner}, Julia},
        title = "{The Evolution of Massive Binary Stars}",
      journal = {\araa},
         year = 2024,
        month = sep,
       volume = {62},
       number = {1},
        pages = {21-61},
          doi = {10.1146/annurev-astro-052722-105936},
archivePrefix = {arXiv},
       eprint = {2311.01865},
 primaryClass = {astro-ph.SR},
       adsurl = {https://ui.adsabs.harvard.edu/abs/2024ARA&A..62...21M}
}

@ARTICLE{raboud,
       author = {{Raboud}, D.},
        title = "{Binarity among B-stars in NGC 6231.}",
      journal = {\aap},
         year = 1996,
        month = nov,
       volume = {315},
        pages = {384-395},
       adsurl = {https://ui.adsabs.harvard.edu/abs/1996A&A...315..384R}
}

@ARTICLE{langrev,
       author = {{Langer}, N.},
        title = "{Presupernova Evolution of Massive Single and Binary Stars}",
      journal = {\araa},
         year = 2012,
        month = sep,
       volume = {50},
        pages = {107-164},
          doi = {10.1146/annurev-astro-081811-125534},
archivePrefix = {arXiv},
       eprint = {1206.5443},
 primaryClass = {astro-ph.SR},
       adsurl = {https://ui.adsabs.harvard.edu/abs/2012ARA&A..50..107L}
}

@ARTICLE{sana12,
       author = {{Sana}, H. and {de Mink}, S.~E. and {de Koter}, A. and {Langer}, N. and
         {Evans}, C.~J. and {Gieles}, M. and {Gosset}, E. and {Izzard}, R.~G. and
         {Le Bouquin}, J. -B. and {Schneider}, F.~R.~N.},
        title = "{Binary Interaction Dominates the Evolution of Massive Stars}",
      journal = {Science},
         year = "2012",
        month = "Jul",
       volume = {337},
       number = {6093},
        pages = {444},
          doi = {10.1126/science.1223344},
archivePrefix = {arXiv},
       eprint = {1207.6397},
 primaryClass = {astro-ph.SR},
       adsurl = {https://ui.adsabs.harvard.edu/abs/2012Sci...337..444S}
}

@ARTICLE{hastings21,
       author = {{Hastings}, B. and {Langer}, N. and {Wang}, C. and {Schootemeijer}, A. and {Milone}, A.~P.},
        title = "{Stringent upper limit on Be star fractions produced by binary interaction}",
      journal = {\aap},
         year = 2021,
        month = sep,
       volume = {653},
          eid = {A144},
        pages = {A144},
          doi = {10.1051/0004-6361/202141269},
archivePrefix = {arXiv},
       eprint = {2106.12263},
 primaryClass = {astro-ph.SR},
       adsurl = {https://ui.adsabs.harvard.edu/abs/2021A&A...653A.144H}
}

@ARTICLE{demink16,
       author = {{de Mink}, S.~E. and {Mandel}, I.},
        title = "{The chemically homogeneous evolutionary channel for binary black hole mergers: rates and properties of gravitational-wave events detectable by advanced LIGO}",
      journal = {\mnras},
         year = 2016,
        month = aug,
       volume = {460},
       number = {4},
        pages = {3545-3553},
          doi = {10.1093/mnras/stw1219},
archivePrefix = {arXiv},
       eprint = {1603.02291},
 primaryClass = {astro-ph.HE},
       adsurl = {https://ui.adsabs.harvard.edu/abs/2016MNRAS.460.3545D}
}

@ARTICLE{Abbott2016,
       author = {{Abbott}, B.~P. and {Abbott}, R. and {Abbott}, T.~D. and
         {Abernathy}, M.~R. and {Acernese}, F. and {Ackley}, K. and {Adams}, C. and
         {Adams}, T. and {Addesso}, P. and {Adhikari}, R.~X. and {Adya}, V.~B. and
         {Affeldt}, C. and {Agathos}, M. and {Agatsuma}, K. and {Aggarwal}, N. and
         {Aguiar}, O.~D. and {Aiello}, L. and {Ain}, A. and {Ajith}, P. and
         {Allen}, B. and {Allocca}, A. and {Altin}, P.~A. and {Anderson}, S.~B. and
         {Anderson}, W.~G. and {Arai}, K. and {Araya}, M.~C. and
         {Arceneaux}, C.~C. and {Areeda}, J.~S. and {Arnaud}, N. and
         {Arun}, K.~G. and {Ascenzi}, S. and {Ashton}, G. and {Ast}, M. and
         {Aston}, S.~M. and {Astone}, P. and {Aufmuth}, P. and {Aulbert}, C. and
         {Babak}, S. and {Bacon}, P. and {Bader}, M.~K.~M. and {Baker}, P.~T. and
         {Baldaccini}, F. and {Ballardin}, G. and {Ballmer}, S.~W. and
         {Barayoga}, J.~C. and {Barclay}, S.~E. and {Barish}, B.~C. and
         {Barker}, D. and {Barone}, F. and {Barr}, B. and {Barsotti}, L. and
         {Barsuglia}, M. and {Barta}, D. and {Bartlett}, J. and {Bartos}, I. and
         {Bassiri}, R. and {Basti}, A. and {Batch}, J.~C. and {Baune}, C. and
         {Bavigadda}, V. and {Bazzan}, M. and {Bejger}, M. and {Bell}, A.~S. and
         {Berger}, B.~K. and {Bergmann}, G. and {Berry}, C.~P.~L. and
         {Bersanetti}, D. and {Bertolini}, A. and {Betzwieser}, J. and
         {Bhagwat}, S. and {Bhandare}, R. and {Bilenko}, I.~A. and
         {Billingsley}, G. and {Birch}, J. and {Birney}, R. and {Birnholtz}, O. and
         {Biscans}, S. and {Bisht}, A. and {Bitossi}, M. and {Biwer}, C. and
         {Bizouard}, M.~A. and {Blackburn}, J.~K. and {Blair}, C.~D. and
         {Blair}, D.~G. and {Blair}, R.~M. and {Bloemen}, S. and {Bock}, O. and
         {Boer}, M. and {Bogaert}, G. and {Bogan}, C. and {Bohe}, A. and
         {Bond}, C. and {Bondu}, F. and {Bonnand}, R. and {Boom}, B.~A. and
         {Bork}, R. and {Boschi}, V. and {Bose}, S. and {Bouffanais}, Y. and
         {Bozzi}, A. and {Bradaschia}, C. and {Brady}, P.~R. and
         {Braginsky}, V.~B. and {Branchesi}, M. and {Brau}, J.~E. and
         {Briant}, T. and {Brillet}, A. and {Brinkmann}, M. and {Brisson}, V. and
         {Brockill}, P. and {Broida}, J.~E. and {Brooks}, A.~F. and
         {Brown}, D.~A. and {Brown}, D.~D. and {Brown}, N.~M. and {Brunett}, S. and
         {Buchanan}, C.~C. and {Buikema}, A. and {Bulik}, T. and
         {Bulten}, H.~J. and {Buonanno}, A. and {Buskulic}, D. and {Buy}, C. and
         {Byer}, R.~L. and {Cabero}, M. and {Cadonati}, L. and {Cagnoli}, G. and
         {Cahillane}, C. and {Calder{\'o}n Bustillo}, J. and {Callister}, T. and
         {Calloni}, E. and {Camp}, J.~B. and {Cannon}, K.~C. and {Cao}, J. and
         {Capano}, C.~D. and {Capocasa}, E. and {Carbognani}, F. and
         {Caride}, S. and {Casanueva Diaz}, J. and {Casentini}, C. and
         {Caudill}, S. and {Cavagli{\`a}}, M. and {Cavalier}, F. and
         {Cavalieri}, R. and {Cella}, G. and {Cepeda}, C.~B. and
         {Cerboni Baiardi}, L. and {Cerretani}, G. and {Cesarini}, E. and
         {Chamberlin}, S.~J. and {Chan}, M. and {Chao}, S. and {Charlton}, P. and
         {Chassande-Mottin}, E. and {Cheeseboro}, B.~D. and {Chen}, H.~Y. and
         {Chen}, Y. and {Cheng}, C. and {Chincarini}, A. and {Chiummo}, A. and
         {Cho}, H.~S. and {Cho}, M. and {Chow}, J.~H. and {Christensen}, N. and
         {Chu}, Q. and {Chua}, S. and {Chung}, S. and {Ciani}, G. and
         {Clara}, F. and {Clark}, J.~A. and {Cleva}, F. and {Coccia}, E. and
         {Cohadon}, P. -F. and {Colla}, A. and {Collette}, C.~G. and
         {Cominsky}, L. and {Constancio}, M. and {Conte}, A. and {Conti}, L. and
         {Cook}, D. and {Corbitt}, T.~R. and {Cornish}, N. and {Corsi}, A. and
         {Cortese}, S. and {Costa}, C.~A. and {Coughlin}, M.~W. and
         {Coughlin}, S.~B. and {Coulon}, J. -P. and {Countryman}, S.~T. and
         {Couvares}, P. and {Cowan}, E.~E. and {Coward}, D.~M. and
         {Cowart}, M.~J. and {Coyne}, D.~C. and {Coyne}, R. and {Craig}, K. and
         {Creighton}, J.~D.~E. and {Cripe}, J. and {Crowder}, S.~G. and
         {Cumming}, A. and {Cunningham}, L. and {Cuoco}, E. and
         {Dal Canton}, T. and {Danilishin}, S.~L. and {D'Antonio}, S. and
         {Danzmann}, K. and {Darman}, N.~S. and {Dasgupta}, A. and
         {Da Silva Costa}, C.~F. and {Dattilo}, V. and {Dave}, I. and
         {Davier}, M. and {Davies}, G.~S. and {Daw}, E.~J. and {Day}, R. and
         {De}, S. and {DeBra}, D. and {Debreczeni}, G. and {Degallaix}, J. and
         {De Laurentis}, M. and {Del{\'e}glise}, S. and {Del Pozzo}, W. and
         {Denker}, T. and {Dent}, T. and {Dergachev}, V. and {De Rosa}, R. and
         {DeRosa}, R.~T. and {DeSalvo}, R. and {Devine}, R.~C. and {Dhurand
        har}, S. and {D{\'\i}az}, M.~C. and {Di Fiore}, L. and
         {Di Giovanni}, M. and {Di Girolamo}, T. and {Di Lieto}, A. and
         {Di Pace}, S. and {Di Palma}, I. and {Di Virgilio}, A. and
         {Dolique}, V. and {Donovan}, F. and {Dooley}, K.~L. and {Doravari}, S. and
         {Douglas}, R. and {Downes}, T.~P. and {Drago}, M. and
         {Drever}, R.~W.~P. and {Driggers}, J.~C. and {Ducrot}, M. and
         {Dwyer}, S.~E. and {Edo}, T.~B. and {Edwards}, M.~C. and {Effler}, A. and
         {Eggenstein}, H. -B. and {Ehrens}, P. and {Eichholz}, J. and
         {Eikenberry}, S.~S. and {Engels}, W. and {Essick}, R.~C. and
         {Etzel}, T. and {Evans}, M. and {Evans}, T.~M. and {Everett}, R. and
         {Factourovich}, M. and {Fafone}, V. and {Fair}, H. and {Fairhurst}, S. and
         {Fan}, X. and {Fang}, Q. and {Farinon}, S. and {Farr}, B. and
         {Farr}, W.~M. and {Favata}, M. and {Fays}, M. and {Fehrmann}, H. and
         {Fejer}, M.~M. and {Fenyvesi}, E. and {Ferrante}, I. and
         {Ferreira}, E.~C. and {Ferrini}, F. and {Fidecaro}, F. and {Fiori}, I. and
         {Fiorucci}, D. and {Fisher}, R.~P. and {Flaminio}, R. and
         {Fletcher}, M. and {Fong}, H. and {Fournier}, J. -D. and {Frasca}, S. and
         {Frasconi}, F. and {Frei}, Z. and {Freise}, A. and {Frey}, R. and
         {Frey}, V. and {Fritschel}, P. and {Frolov}, V.~V. and {Fulda}, P. and
         {Fyffe}, M. and {Gabbard}, H.~A.~G. and {Gaebel}, S. and {Gair}, J.~R. and
         {Gammaitoni}, L. and {Gaonkar}, S.~G. and {Garufi}, F. and {Gaur}, G. and
         {Gehrels}, N. and {Gemme}, G. and {Geng}, P. and {Genin}, E. and
         {Gennai}, A. and {George}, J. and {Gergely}, L. and {Germain}, V. and
         {Ghosh}, Abhirup and {Ghosh}, Archisman and {Ghosh}, S. and
         {Giaime}, J.~A. and {Giardina}, K.~D. and {Giazotto}, A. and
         {Gill}, K. and {Glaefke}, A. and {Goetz}, E. and {Goetz}, R. and
         {Gondan}, L. and {Gonz{\'a}lez}, G. and {Gonzalez Castro}, J.~M. and
         {Gopakumar}, A. and {Gordon}, N.~A. and {Gorodetsky}, M.~L. and
         {Gossan}, S.~E. and {Gosselin}, M. and {Gouaty}, R. and {Grado}, A. and
         {Graef}, C. and {Graff}, P.~B. and {Granata}, M. and {Grant}, A. and
         {Gras}, S. and {Gray}, C. and {Greco}, G. and {Green}, A.~C. and
         {Groot}, P. and {Grote}, H. and {Grunewald}, S. and {Guidi}, G.~M. and
         {Guo}, X. and {Gupta}, A. and {Gupta}, M.~K. and {Gushwa}, K.~E. and
         {Gustafson}, E.~K. and {Gustafson}, R. and {Hacker}, J.~J. and
         {Hall}, B.~R. and {Hall}, E.~D. and {Hamilton}, H. and {Hammond}, G. and
         {Haney}, M. and {Hanke}, M.~M. and {Hanks}, J. and {Hanna}, C. and
         {Hannam}, M.~D. and {Hanson}, J. and {Hardwick}, T. and {Harms}, J. and
         {Harry}, G.~M. and {Harry}, I.~W. and {Hart}, M.~J. and
         {Hartman}, M.~T. and {Haster}, C. -J. and {Haughian}, K. and
         {Healy}, J. and {Heidmann}, A. and {Heintze}, M.~C. and {Heitmann}, H. and
         {Hello}, P. and {Hemming}, G. and {Hendry}, M. and {Heng}, I.~S. and
         {Hennig}, J. and {Henry}, J. and {Heptonstall}, A.~W. and {Heurs}, M. and
         {Hild}, S. and {Hoak}, D. and {Hofman}, D. and {Holt}, K. and
         {Holz}, D.~E. and {Hopkins}, P. and {Hough}, J. and {Houston}, E.~A. and
         {Howell}, E.~J. and {Hu}, Y.~M. and {Huang}, S. and {Huerta}, E.~A. and
         {Huet}, D. and {Hughey}, B. and {Husa}, S. and {Huttner}, S.~H. and
         {Huynh-Dinh}, T. and {Indik}, N. and {Ingram}, D.~R. and {Inta}, R. and
         {Isa}, H.~N. and {Isac}, J. -M. and {Isi}, M. and {Isogai}, T. and
         {Iyer}, B.~R. and {Izumi}, K. and {Jacqmin}, T. and {Jang}, H. and
         {Jani}, K. and {Jaranowski}, P. and {Jawahar}, S. and {Jian}, L. and
         {Jim{\'e}nez-Forteza}, F. and {Johnson}, W.~W. and
         {Johnson-McDaniel}, N.~K. and {Jones}, D.~I. and {Jones}, R. and
         {Jonker}, R.~J.~G. and {Ju}, L. and {K}, Haris and {Kalaghatgi}, C.~V. and
         {Kalogera}, V. and {Kandhasamy}, S. and {Kang}, G. and {Kanner}, J.~B. and
         {Kapadia}, S.~J. and {Karki}, S. and {Karvinen}, K.~S. and
         {Kasprzack}, M. and {Katsavounidis}, E. and {Katzman}, W. and
         {Kaufer}, S. and {Kaur}, T. and {Kawabe}, K. and
         {K{\'e}f{\'e}lian}, F. and {Kehl}, M.~S. and {Keitel}, D. and
         {Kelley}, D.~B. and {Kells}, W. and {Kennedy}, R. and {Key}, J.~S. and
         {Khalili}, F.~Y. and {Khan}, I. and {Khan}, S. and {Khan}, Z. and
         {Khazanov}, E.~A. and {Kijbunchoo}, N. and {Kim}, Chi-Woong and
         {Kim}, Chunglee and {Kim}, J. and {Kim}, K. and {Kim}, N. and
         {Kim}, W. and {Kim}, Y. -M. and {Kimbrell}, S.~J. and {King}, E.~J. and
         {King}, P.~J. and {Kissel}, J.~S. and {Klein}, B. and {Kleybolte}, L. and
         {Klimenko}, S. and {Koehlenbeck}, S.~M. and {Koley}, S. and
         {Kondrashov}, V. and {Kontos}, A. and {Korobko}, M. and {Korth}, W.~Z. and
         {Kowalska}, I. and {Kozak}, D.~B. and {Kringel}, V. and {Krishnan}, B. and
         {Kr{\'o}lak}, A. and {Krueger}, C. and {Kuehn}, G. and {Kumar}, P. and
         {Kumar}, R. and {Kuo}, L. and {Kutynia}, A. and {Lackey}, B.~D. and {Land
        ry}, M. and {Lange}, J. and {Lantz}, B. and {Lasky}, P.~D. and
         {Laxen}, M. and {Lazzarini}, A. and {Lazzaro}, C. and {Leaci}, P. and
         {Leavey}, S. and {Lebigot}, E.~O. and {Lee}, C.~H. and {Lee}, H.~K. and
         {Lee}, H.~M. and {Lee}, K. and {Lenon}, A. and {Leonardi}, M. and
         {Leong}, J.~R. and {Leroy}, N. and {Letendre}, N. and {Levin}, Y. and
         {Lewis}, J.~B. and {Li}, T.~G.~F. and {Libson}, A. and
         {Littenberg}, T.~B. and {Lockerbie}, N.~A. and {Lombardi}, A.~L. and
         {London}, L.~T. and {Lord}, J.~E. and {Lorenzini}, M. and
         {Loriette}, V. and {Lormand}, M. and {Losurdo}, G. and {Lough}, J.~D. and
         {Lousto}, C. and {L{\"u}ck}, H. and {Lundgren}, A.~P. and {Lynch}, R. and
         {Ma}, Y. and {Machenschalk}, B. and {MacInnis}, M. and
         {Macleod}, D.~M. and {Maga{\~n}a-Sandoval}, F. and
         {Maga{\~n}a Zertuche}, L. and {Magee}, R.~M. and {Majorana}, E. and
         {Maksimovic}, I. and {Malvezzi}, V. and {Man}, N. and {Mandel}, I. and
         {Mandic}, V. and {Mangano}, V. and {Mansell}, G.~L. and {Manske}, M. and
         {Mantovani}, M. and {Marchesoni}, F. and {Marion}, F. and
         {M{\'a}rka}, S. and {M{\'a}rka}, Z. and {Markosyan}, A.~S. and
         {Maros}, E. and {Martelli}, F. and {Martellini}, L. and
         {Martin}, I.~W. and {Martynov}, D.~V. and {Marx}, J.~N. and
         {Mason}, K. and {Masserot}, A. and {Massinger}, T.~J. and
         {Masso-Reid}, M. and {Mastrogiovanni}, S. and {Matichard}, F. and
         {Matone}, L. and {Mavalvala}, N. and {Mazumder}, N. and {McCarthy}, R. and
         {McClelland}, D.~E. and {McCormick}, S. and {McGuire}, S.~C. and
         {McIntyre}, G. and {McIver}, J. and {McManus}, D.~J. and {McRae}, T. and
         {McWilliams}, S.~T. and {Meacher}, D. and {Meadors}, G.~D. and
         {Meidam}, J. and {Melatos}, A. and {Mendell}, G. and {Mercer}, R.~A. and
         {Merilh}, E.~L. and {Merzougui}, M. and {Meshkov}, S. and
         {Messenger}, C. and {Messick}, C. and {Metzdorff}, R. and
         {Meyers}, P.~M. and {Mezzani}, F. and {Miao}, H. and {Michel}, C. and
         {Middleton}, H. and {Mikhailov}, E.~E. and {Milano}, L. and
         {Miller}, A.~L. and {Miller}, A. and {Miller}, B.~B. and {Miller}, J. and
         {Millhouse}, M. and {Minenkov}, Y. and {Ming}, J. and {Mirshekari}, S. and
         {Mishra}, C. and {Mitra}, S. and {Mitrofanov}, V.~P. and
         {Mitselmakher}, G. and {Mittleman}, R. and {Moggi}, A. and {Mohan}, M. and
         {Mohapatra}, S.~R.~P. and {Montani}, M. and {Moore}, B.~C. and
         {Moore}, C.~J. and {Moraru}, D. and {Moreno}, G. and {Morriss}, S.~R. and
         {Mossavi}, K. and {Mours}, B. and {Mow-Lowry}, C.~M. and {Mueller}, G. and
         {Muir}, A.~W. and {Mukherjee}, Arunava and {Mukherjee}, D. and
         {Mukherjee}, S. and {Mukund}, N. and {Mullavey}, A. and {Munch}, J. and
         {Murphy}, D.~J. and {Murray}, P.~G. and {Mytidis}, A. and
         {Nardecchia}, I. and {Naticchioni}, L. and {Nayak}, R.~K. and
         {Nedkova}, K. and {Nelemans}, G. and {Nelson}, T.~J.~N. and {Neri}, M. and
         {Neunzert}, A. and {Newton}, G. and {Nguyen}, T.~T. and
         {Nielsen}, A.~B. and {Nissanke}, S. and {Nitz}, A. and {Nocera}, F. and
         {Nolting}, D. and {Normandin}, M.~E.~N. and {Nuttall}, L.~K. and
         {Oberling}, J. and {Ochsner}, E. and {O'Dell}, J. and {Oelker}, E. and
         {Ogin}, G.~H. and {Oh}, J.~J. and {Oh}, S.~H. and {Ohme}, F. and
         {Oliver}, M. and {Oppermann}, P. and {Oram}, Richard J. and
         {O'Reilly}, B. and {O'Shaughnessy}, R. and {Ottaway}, D.~J. and
         {Overmier}, H. and {Owen}, B.~J. and {Pai}, A. and {Pai}, S.~A. and
         {Palamos}, J.~R. and {Palashov}, O. and {Palomba}, C. and
         {Pal-Singh}, A. and {Pan}, H. and {Pan}, Y. and {Pankow}, C. and
         {Pannarale}, F. and {Pant}, B.~C. and {Paoletti}, F. and {Paoli}, A. and
         {Papa}, M.~A. and {Paris}, H.~R. and {Parker}, W. and {Pascucci}, D. and
         {Pasqualetti}, A. and {Passaquieti}, R. and {Passuello}, D. and
         {Patricelli}, B. and {Patrick}, Z. and {Pearlstone}, B.~L. and
         {Pedraza}, M. and {Pedurand}, R. and {Pekowsky}, L. and {Pele}, A. and
         {Penn}, S. and {Perreca}, A. and {Perri}, L.~M. and {Pfeiffer}, H.~P. and
         {Phelps}, M. and {Piccinni}, O.~J. and {Pichot}, M. and
         {Piergiovanni}, F. and {Pierro}, V. and {Pillant}, G. and {Pinard}, L. and
         {Pinto}, I.~M. and {Pitkin}, M. and {Poe}, M. and {Poggiani}, R. and
         {Popolizio}, P. and {Porter}, E. and {Post}, A. and {Powell}, J. and
         {Prasad}, J. and {Predoi}, V. and {Prestegard}, T. and {Price}, L.~R. and
         {Prijatelj}, M. and {Principe}, M. and {Privitera}, S. and {Prix}, R. and
         {Prodi}, G.~A. and {Prokhorov}, L. and {Puncken}, O. and {Punturo}, M. and
         {Puppo}, P. and {P{\"u}rrer}, M. and {Qi}, H. and {Qin}, J. and
         {Qiu}, S. and {Quetschke}, V. and {Quintero}, E.~A. and
         {Quitzow-James}, R. and {Raab}, F.~J. and {Rabeling}, D.~S. and
         {Radkins}, H. and {Raffai}, P. and {Raja}, S. and {Rajan}, C. and
         {Rakhmanov}, M. and {Rapagnani}, P. and {Raymond}, V. and
         {Razzano}, M. and {Re}, V. and {Read}, J. and {Reed}, C.~M. and
         {Regimbau}, T. and {Rei}, L. and {Reid}, S. and {Reitze}, D.~H. and
         {Rew}, H. and {Reyes}, S.~D. and {Ricci}, F. and {Riles}, K. and
         {Rizzo}, M. and {Robertson}, N.~A. and {Robie}, R. and {Robinet}, F. and
         {Rocchi}, A. and {Rolland}, L. and {Rollins}, J.~G. and {Roma}, V.~J. and
         {Romano}, J.~D. and {Romano}, R. and {Romanov}, G. and {Romie}, J.~H. and
         {Rosi{\'n}ska}, D. and {Rowan}, S. and {R{\"u}diger}, A. and
         {Ruggi}, P. and {Ryan}, K. and {Sachdev}, S. and {Sadecki}, T. and
         {Sadeghian}, L. and {Sakellariadou}, M. and {Salconi}, L. and
         {Saleem}, M. and {Salemi}, F. and {Samajdar}, A. and {Sammut}, L. and
         {Sanchez}, E.~J. and {Sandberg}, V. and {Sandeen}, B. and {Sand
        ers}, J.~R. and {Sassolas}, B. and {Sathyaprakash}, B.~S. and
         {Saulson}, P.~R. and {Sauter}, O.~E.~S. and {Savage}, R.~L. and
         {Sawadsky}, A. and {Schale}, P. and {Schilling}, R. and {Schmidt}, J. and
         {Schmidt}, P. and {Schnabel}, R. and {Schofield}, R.~M.~S. and
         {Sch{\"o}nbeck}, A. and {Schreiber}, E. and {Schuette}, D. and
         {Schutz}, B.~F. and {Scott}, J. and {Scott}, S.~M. and {Sellers}, D. and
         {Sengupta}, A.~S. and {Sentenac}, D. and {Sequino}, V. and
         {Sergeev}, A. and {Setyawati}, Y. and {Shaddock}, D.~A. and
         {Shaffer}, T. and {Shahriar}, M.~S. and {Shaltev}, M. and
         {Shapiro}, B. and {Shawhan}, P. and {Sheperd}, A. and
         {Shoemaker}, D.~H. and {Shoemaker}, D.~M. and {Siellez}, K. and
         {Siemens}, X. and {Sieniawska}, M. and {Sigg}, D. and {Silva}, A.~D. and
         {Singer}, A. and {Singer}, L.~P. and {Singh}, A. and {Singh}, R. and
         {Singhal}, A. and {Sintes}, A.~M. and {Slagmolen}, B.~J.~J. and
         {Smith}, J.~R. and {Smith}, N.~D. and {Smith}, R.~J.~E. and
         {Son}, E.~J. and {Sorazu}, B. and {Sorrentino}, F. and {Souradeep}, T. and
         {Srivastava}, A.~K. and {Staley}, A. and {Steinke}, M. and
         {Steinlechner}, J. and {Steinlechner}, S. and {Steinmeyer}, D. and
         {Stephens}, B.~C. and {Stevenson}, S. and {Stone}, R. and
         {Strain}, K.~A. and {Straniero}, N. and {Stratta}, G. and
         {Strauss}, N.~A. and {Strigin}, S. and {Sturani}, R. and
         {Stuver}, A.~L. and {Summerscales}, T.~Z. and {Sun}, L. and
         {Sunil}, S. and {Sutton}, P.~J. and {Swinkels}, B.~L. and
         {Szczepa{\'n}czyk}, M.~J. and {Tacca}, M. and {Talukder}, D. and
         {Tanner}, D.~B. and {T{\'a}pai}, M. and {Tarabrin}, S.~P. and
         {Taracchini}, A. and {Taylor}, R. and {Theeg}, T. and {Thirugnanasamband
        am}, M.~P. and {Thomas}, E.~G. and {Thomas}, M. and {Thomas}, P. and
         {Thorne}, K.~A. and {Thrane}, E. and {Tiwari}, S. and {Tiwari}, V. and
         {Tokmakov}, K.~V. and {Toland}, K. and {Tomlinson}, C. and
         {Tonelli}, M. and {Tornasi}, Z. and {Torres}, C.~V. and
         {Torrie}, C.~I. and {T{\"o}yr{\"a}}, D. and {Travasso}, F. and
         {Traylor}, G. and {Trifir{\`o}}, D. and {Tringali}, M.~C. and
         {Trozzo}, L. and {Tse}, M. and {Turconi}, M. and {Tuyenbayev}, D. and
         {Ugolini}, D. and {Unnikrishnan}, C.~S. and {Urban}, A.~L. and
         {Usman}, S.~A. and {Vahlbruch}, H. and {Vajente}, G. and {Valdes}, G. and
         {Vallisneri}, M. and {van Bakel}, N. and {van Beuzekom}, M. and
         {van den Brand}, J.~F.~J. and {Van Den Broeck}, C. and {Vand
        er-Hyde}, D.~C. and {van der Schaaf}, L. and {van Heijningen}, J.~V. and
         {van Veggel}, A.~A. and {Vardaro}, M. and {Vass}, S. and
         {Vas{\'u}th}, M. and {Vaulin}, R. and {Vecchio}, A. and {Vedovato}, G. and
         {Veitch}, J. and {Veitch}, P.~J. and {Venkateswara}, K. and
         {Verkindt}, D. and {Vetrano}, F. and {Vicer{\'e}}, A. and
         {Vinciguerra}, S. and {Vine}, D.~J. and {Vinet}, J. -Y. and
         {Vitale}, S. and {Vo}, T. and {Vocca}, H. and {Vorvick}, C. and
         {Voss}, D.~V. and {Vousden}, W.~D. and {Vyatchanin}, S.~P. and
         {Wade}, A.~R. and {Wade}, L.~E. and {Wade}, M. and {Walker}, M. and
         {Wallace}, L. and {Walsh}, S. and {Wang}, G. and {Wang}, H. and
         {Wang}, M. and {Wang}, X. and {Wang}, Y. and {Ward}, R.~L. and
         {Warner}, J. and {Was}, M. and {Weaver}, B. and {Wei}, L. -W. and
         {Weinert}, M. and {Weinstein}, A.~J. and {Weiss}, R. and {Wen}, L. and
         {We{\ss}els}, P. and {Westphal}, T. and {Wette}, K. and
         {Whelan}, J.~T. and {Whitcomb}, S.~E. and {Whiting}, B.~F. and
         {Williams}, R.~D. and {Williamson}, A.~R. and {Willis}, J.~L. and
         {Willke}, B. and {Wimmer}, M.~H. and {Winkler}, W. and {Wipf}, C.~C. and
         {Wittel}, H. and {Woan}, G. and {Woehler}, J. and {Worden}, J. and
         {Wright}, J.~L. and {Wu}, D.~S. and {Wu}, G. and {Yablon}, J. and
         {Yam}, W. and {Yamamoto}, H. and {Yancey}, C.~C. and {Yu}, H. and
         {Yvert}, M. and {Zadro{\.Z}ny}, A. and {Zangrando}, L. and
         {Zanolin}, M. and {Zendri}, J. -P. and {Zevin}, M. and {Zhang}, L. and
         {Zhang}, M. and {Zhang}, Y. and {Zhao}, C. and {Zhou}, M. and
         {Zhou}, Z. and {Zhu}, X.~J. and {Zucker}, M.~E. and {Zuraw}, S.~E. and
         {Zweizig}, J. and {LIGO Scientific Collaboration} and
         {Virgo Collaboration}},
        title = "{Binary Black Hole Mergers in the First Advanced LIGO Observing Run}",
      journal = {Physical Review X},
         year = 2016,
        month = oct,
       volume = {6},
       number = {4},
          eid = {041015},
        pages = {041015},
          doi = {10.1103/PhysRevX.6.041015},
archivePrefix = {arXiv},
       eprint = {1606.04856},
 primaryClass = {gr-qc},
       adsurl = {https://ui.adsabs.harvard.edu/abs/2016PhRvX...6d1015A}
}

@ARTICLE{jb17,
       author = {{Le Bouquin}, J. -B. and {Sana}, H. and {Gosset}, E. and {De Becker}, M. and {Duvert}, G. and {Absil}, O. and {Anthonioz}, F. and {Berger}, J. -P. and {Ertel}, S. and {Grellmann}, R. and {Guieu}, S. and {Kervella}, P. and {Rabus}, M. and {Willson}, M.},
        title = "{Resolved astrometric orbits of ten O-type binaries}",
      journal = {\aap},
         year = 2017,
        month = may,
       volume = {601},
          eid = {A34},
        pages = {A34},
          doi = {10.1051/0004-6361/201629260},
archivePrefix = {arXiv},
       eprint = {1608.03525},
 primaryClass = {astro-ph.SR},
       adsurl = {https://ui.adsabs.harvard.edu/abs/2017A&A...601A..34L}
}

@ARTICLE{martins05,
       author = {{Martins}, F. and {Schaerer}, D. and {Hillier}, D.~J.},
        title = "{A new calibration of stellar parameters of Galactic O stars}",
      journal = {\aap},
         year = 2005,
        month = jun,
       volume = {436},
       number = {3},
        pages = {1049-1065},
          doi = {10.1051/0004-6361:20042386},
archivePrefix = {arXiv},
       eprint = {astro-ph/0503346},
 primaryClass = {astro-ph},
       adsurl = {https://ui.adsabs.harvard.edu/abs/2005A&A...436.1049M}
}

\begin{appendix}
\section{Plots and fits for each source} \label{a:plotsandfits}
Below we present the observations, model fits and images and bootstrap error calculations for the entire sample of sources. Note that we do not include model images for the single star models, just their fits to the data and the bootstrap plots for the size of the uniform disks. The order of the sources matches Table 2. 




\begin{figure*}[t!]
   \centering
   \includegraphics[width=180mm]{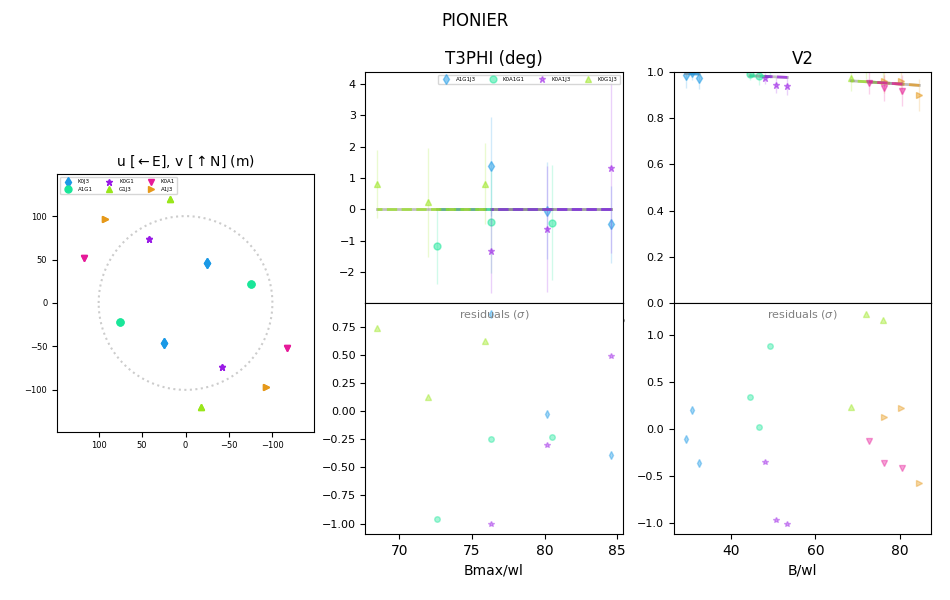}
   \caption{The best-fit data for HD\,16582. On the left, the $u-v$ coverage from the observations with PIONIER can be seen. The different colours correspond to the telescope pair with which that particular data was acquired at the VLTI. In the middle, the closure phase (T3PHI) fit and residuals are shown in terms of the spatial frequency (B$_{avg}$/$\lambda$, written as Bavg/wl on the axes). On the right, the fit to the squared visibilities (V2) and the associated residuals are shown, again in terms of spatial frequency. Across the two fits, the data are represented as points whilst the fit as a continuous line.}
    \label{HD16582fit}%
    \end{figure*}

        \begin{figure}
   \centering
   \includegraphics[width=70mm]{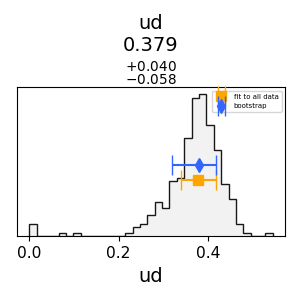}
   \caption{Bootstrapping plot showing the error determination for the dataset fit in Figure \ref{HD16582fit}.}
    \label{HD16582bs}%
    \end{figure}

    \begin{figure*}
   \centering
   \includegraphics[width=180mm]{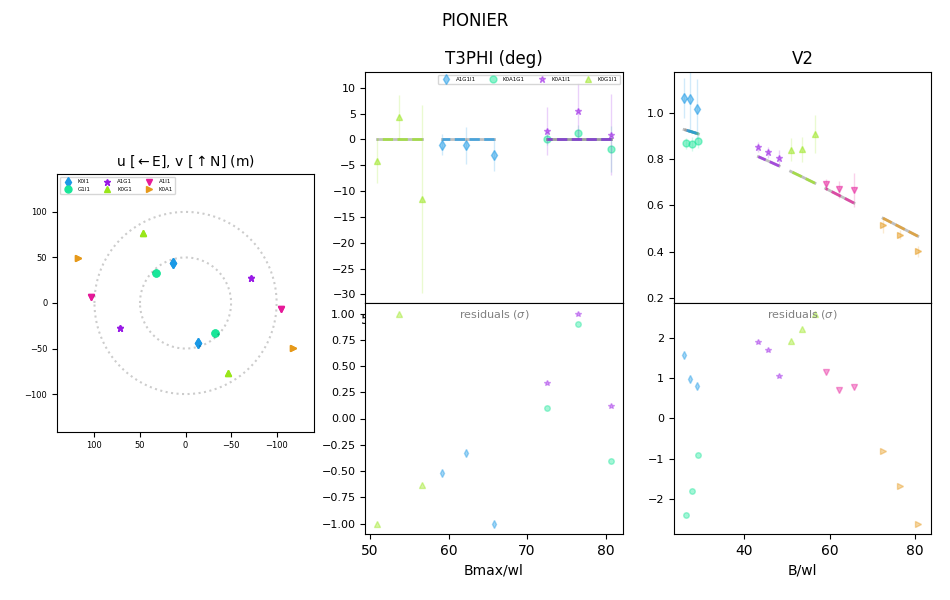}
   \caption{The best-fit data for HD\,51480.}
    \label{HD51480fit}%
    \end{figure*}

        \begin{figure}
   \centering
   \includegraphics[width=70mm]{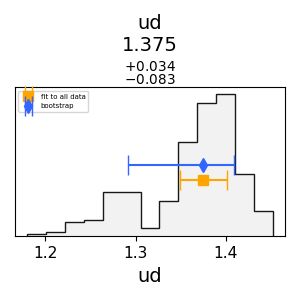}
   \caption{Bootstrapping plot showing the error determination for the dataset fit in Figure \ref{HD51480fit}.}
    \label{HD51480bs}
    \end{figure}



            \begin{figure*}
   \centering
   \includegraphics[width=180mm]{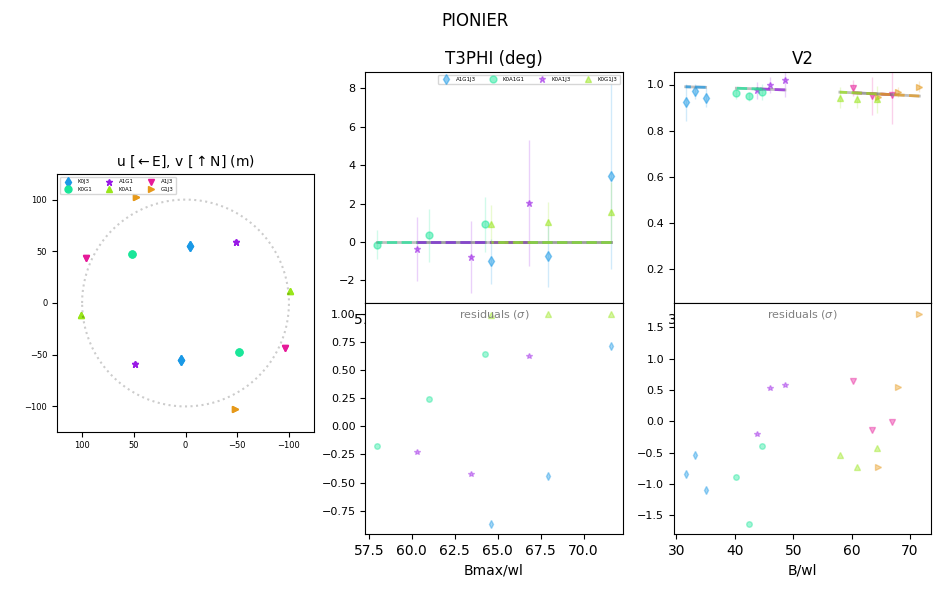}
   \caption{The best-fit data for HD\,66765.}
    \label{HD66765fit}
    \end{figure*}

        \begin{figure}
   \centering
   \includegraphics[width=70mm]{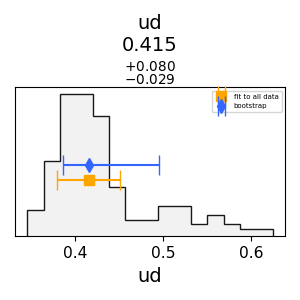}
   \caption{Bootstrapping plot showing the error determination for the dataset fit in Figure \ref{HD66765fit}.}
    \label{HD64760bs}%
    \end{figure}

                \begin{figure*}
   \centering
   \includegraphics[width=180mm]{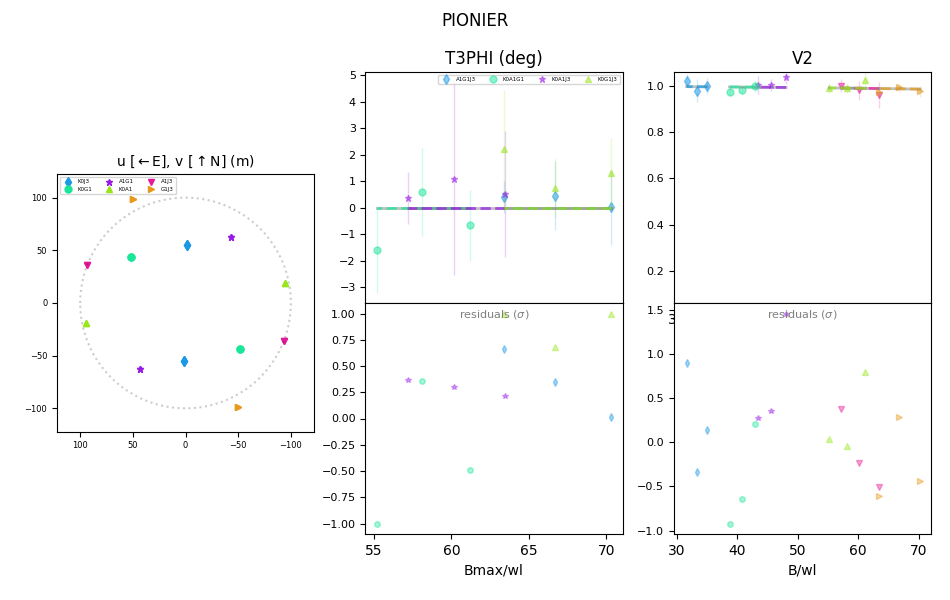}
   \caption{The best-fit data for HD\,67621.}
    \label{HD67621fit}
    \end{figure*}

        \begin{figure}
   \centering
   \includegraphics[width=70mm]{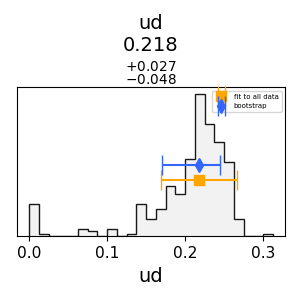}
   \caption{Bootstrapping plot showing the error determination for the dataset fit in Figure \ref{HD67621fit}.}
    \label{HD67621bs}%
    \end{figure}

                    \begin{figure*}
   \centering
   \includegraphics[width=180mm]{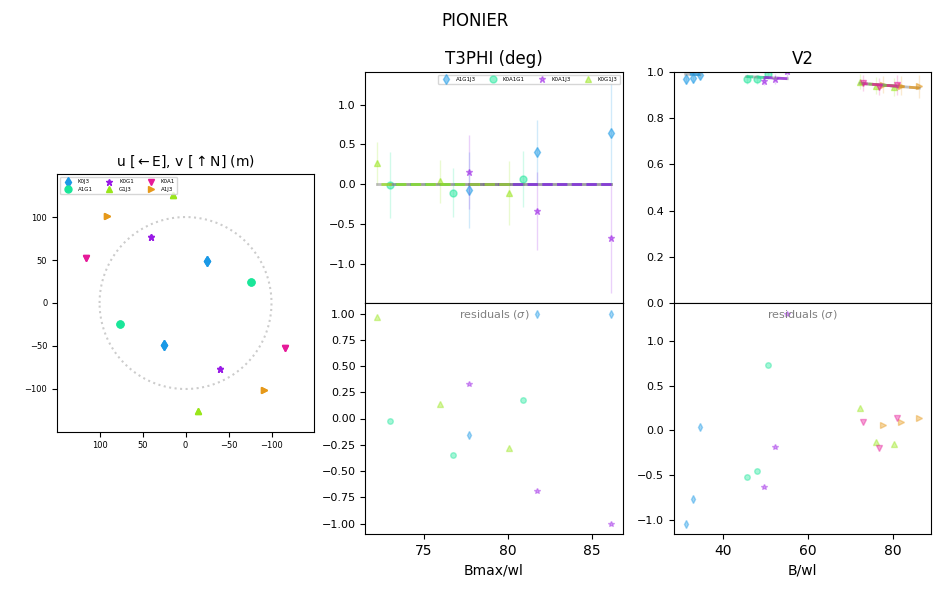}
   \caption{The best-fit data for HD\,121743.}
    \label{HD121743fit}
    \end{figure*}

        \begin{figure}
   \centering
   \includegraphics[width=70mm]{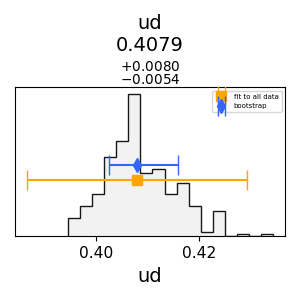}
   \caption{Bootstrapping plot showing the error determination for the dataset fit in Figure \ref{HD121743fit}.}
    \label{HD121743bs}%
    \end{figure}





\begin{figure*}
   \centering
   \includegraphics[width=180mm]{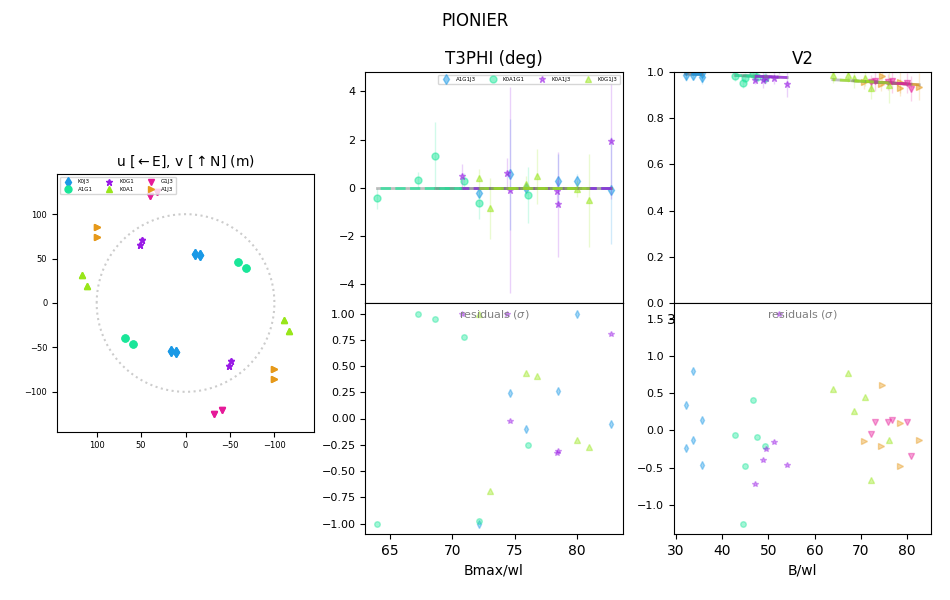}
   \caption{The best-fit data for HD\,189103.}
    \label{HD189103fit}
    \end{figure*}

        \begin{figure}
   \centering
   \includegraphics[width=70mm]{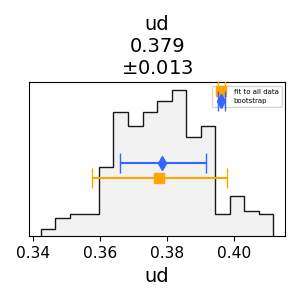}
   \caption{Bootstrapping plot showing the error determination for the dataset fit in Figure \ref{HD189103fit}.}
    \label{HD189103bs}%
    \end{figure}

\begin{figure*}
   \centering
   \includegraphics[width=180mm]{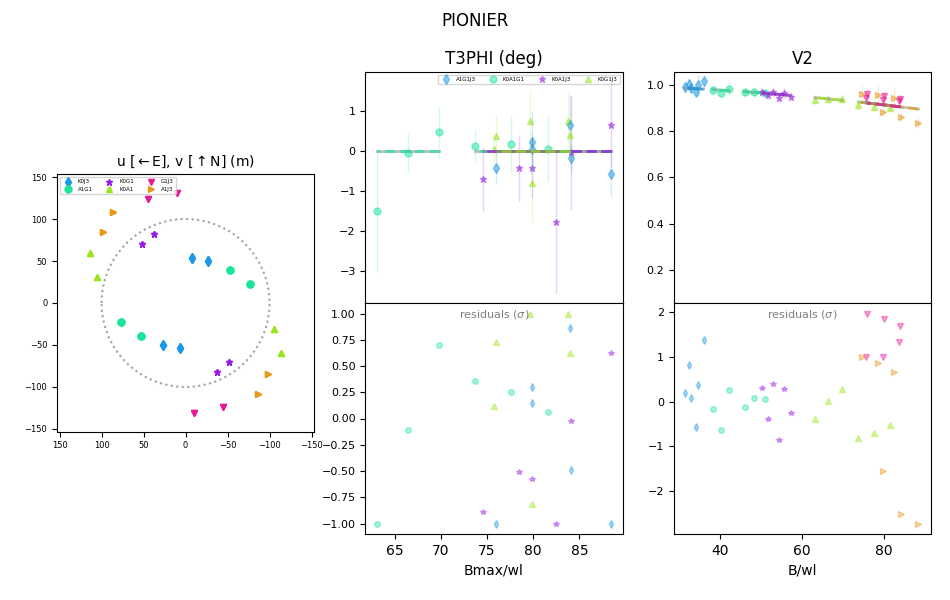}
   \caption{The best-fit data for HD\,205637.}
    \label{HD205637fit}
    \end{figure*}

        \begin{figure}
   \centering
   \includegraphics[width=70mm]{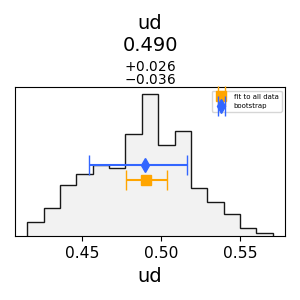}
   \caption{Bootstrapping plot showing the error determination for the dataset fit in Figure \ref{HD205637fit}.}
    \label{HD205637bs}%
    \end{figure}

\begin{figure*}
   \centering
   \includegraphics[width=180mm]{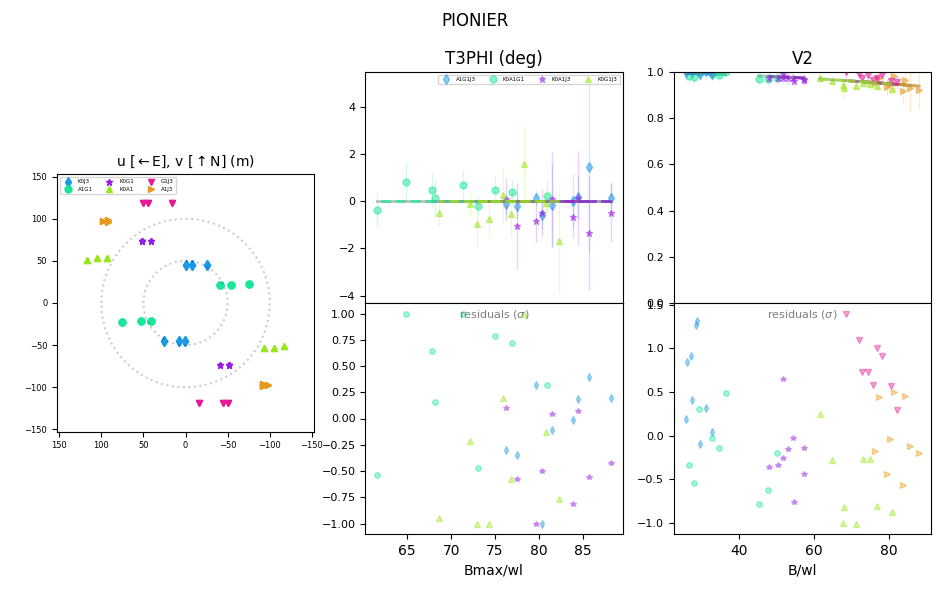}
   \caption{The best-fit data for HD\,212571.}
    \label{HD212571fit}
    \end{figure*}

        \begin{figure}
   \centering
   \includegraphics[width=70mm]{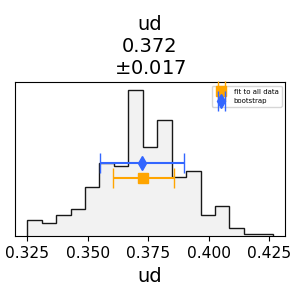}
   \caption{Bootstrapping plot showing the error determination for the dataset fit in Figure \ref{HD212571fit}.}
    \label{HD212571bs}%
    \end{figure}

\begin{figure*}
   \centering
   \includegraphics[width=180mm]{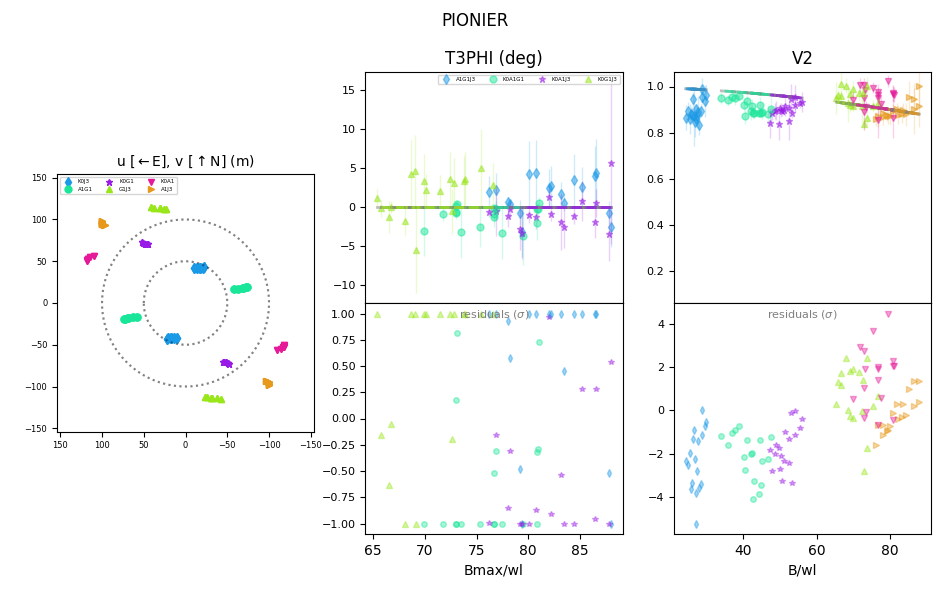}
   \caption{The best-fit data for MCW\,1019.}
    \label{MCW1019fit}
    \end{figure*}

        \begin{figure}
   \centering
   \includegraphics[width=70mm]{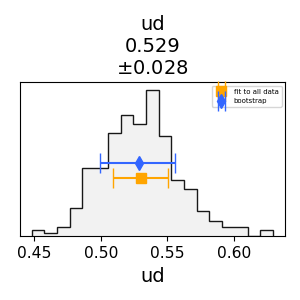}
   \caption{Bootstrapping plot showing the error determination for the dataset fit in Figure \ref{MCW1019fit}.}
    \label{MCW10191bs}%
    \end{figure}

\begin{figure*}
   \centering
   \includegraphics[width=180mm]{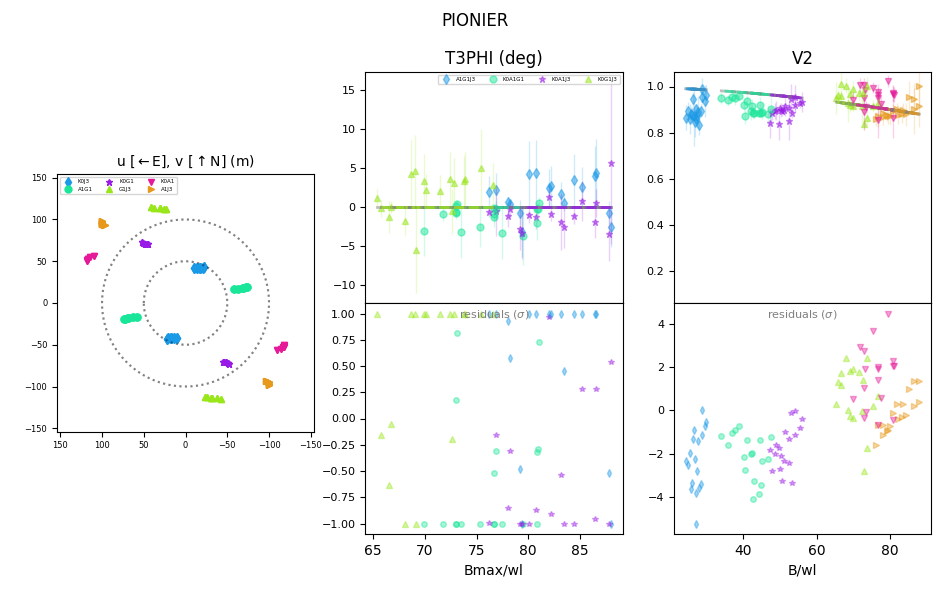}
   \caption{The best-fit data for MCW\,1019.}
    \label{MCW1019fit}
    \end{figure*}

        \begin{figure}
   \centering
   \includegraphics[width=70mm]{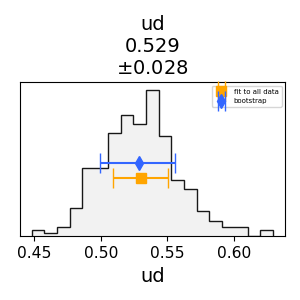}
   \caption{Bootstrapping plot showing the error determination for the dataset fit in Figure \ref{MCW1019fit}.}
    \label{MCW10191bs}%
    \end{figure}




             \begin{figure*}
   \centering
   \includegraphics[width=170mm]{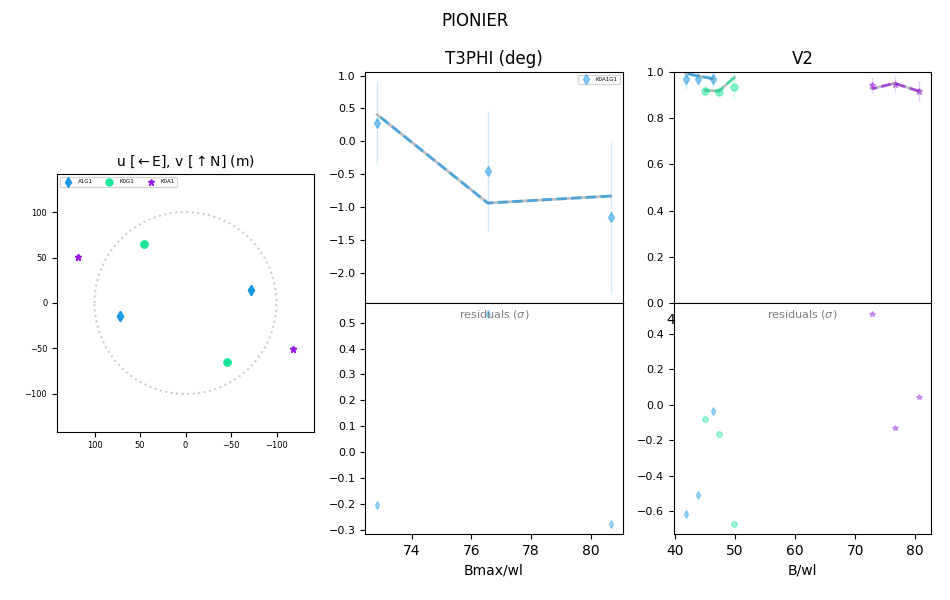}
   \caption{The best-fit data for $\gamma$ Peg. On the left, the $u-v$ coverage from the observations with PIONIER can be seen. The different colours correspond to the telescope pair with which that particular data was acquired at the VLTI. In the middle, the closure phase (T3PHI) fit and residuals are shown in terms of the spatial frequency (B$_{avg}$/$\lambda$, written as Bavg/wl on the axes). On the right, the fit to the squared visibilities (V2) and the associated residuals are shown, again in terms of spatial frequency. Across the two fits, the data are represented as points whilst the fit as a continuous line.}
    \end{figure*}
    
    \begin{figure*}
   \centering
   \includegraphics[width=140mm]{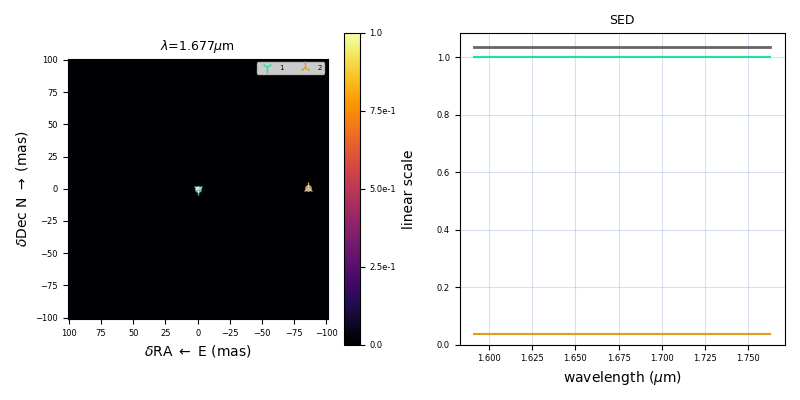}
   \caption{Model image created based on the best-fitting model of $\gamma$ Peg.}
    \end{figure*}

    \begin{figure*}
   \centering
   \includegraphics[width=120mm]{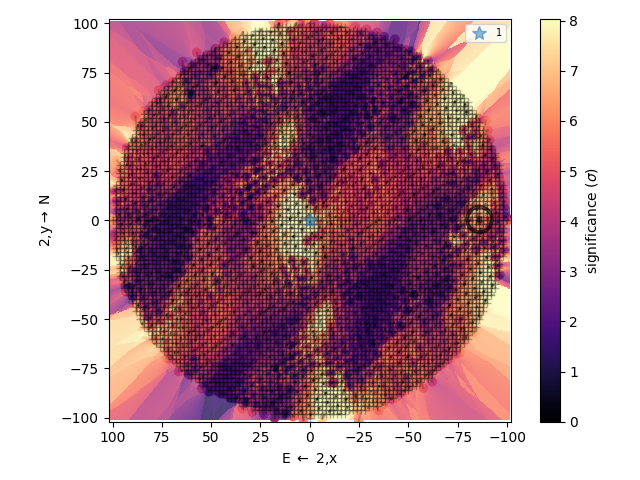}
   \caption{Plot showing the grid search used to search for the companion around $\gamma$ Peg and the companion's significance.}
    \end{figure*}
    
    \begin{figure*}
   \centering
   \includegraphics[width=120mm]{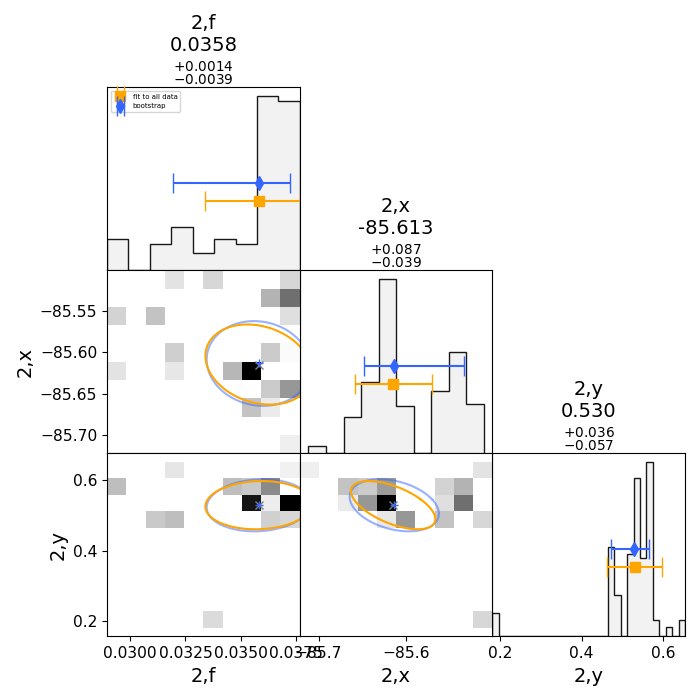}
   \caption{Bootstrapping plot showing the error determination for the dataset of $\gamma$ Peg.}
    \end{figure*}


             \begin{figure*}
   \centering
   \includegraphics[width=170mm]{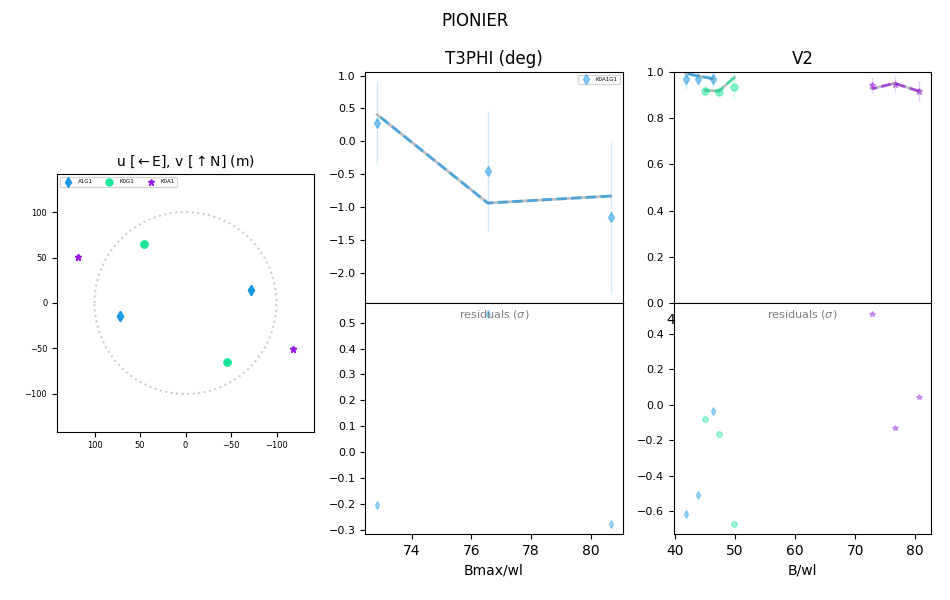}
   \caption{The best-fit data for $\gamma$ Peg. On the left, the $u-v$ coverage from the observations with PIONIER can be seen. The different colours correspond to the telescope pair with which that particular data was acquired at the VLTI. In the middle, the closure phase (T3PHI) fit and residuals are shown in terms of the spatial frequency (B$_{avg}$/$\lambda$, written as Bavg/wl on the axes). On the right, the fit to the squared visibilities (V2) and the associated residuals are shown, again in terms of spatial frequency. Across the two fits, the data are represented as points whilst the fit as a continuous line.}
    \end{figure*}
    
    \begin{figure*}
   \centering
   \includegraphics[width=140mm]{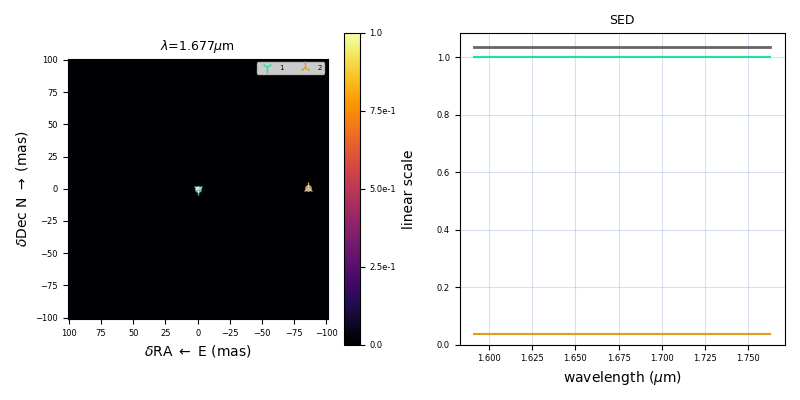}
   \caption{Model image created based on the best-fitting model of $\gamma$ Peg.}
    \end{figure*}

    \begin{figure*}
   \centering
   \includegraphics[width=120mm]{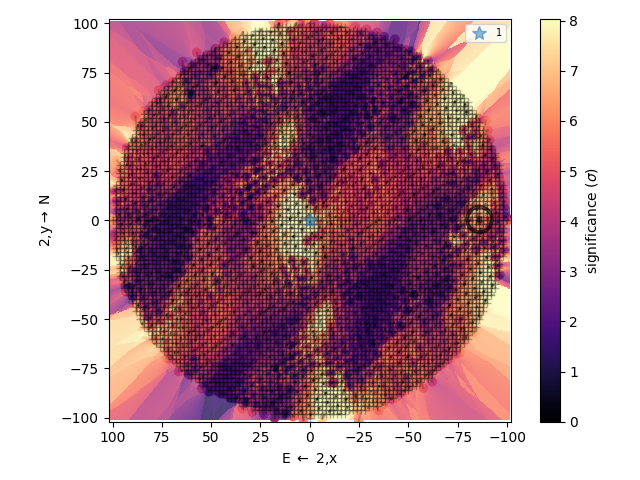}
   \caption{Plot showing the grid search used to search for the companion around $\gamma$ Peg and the companion's significance.}
    \end{figure*}
    
    \begin{figure*}
   \centering
   \includegraphics[width=120mm]{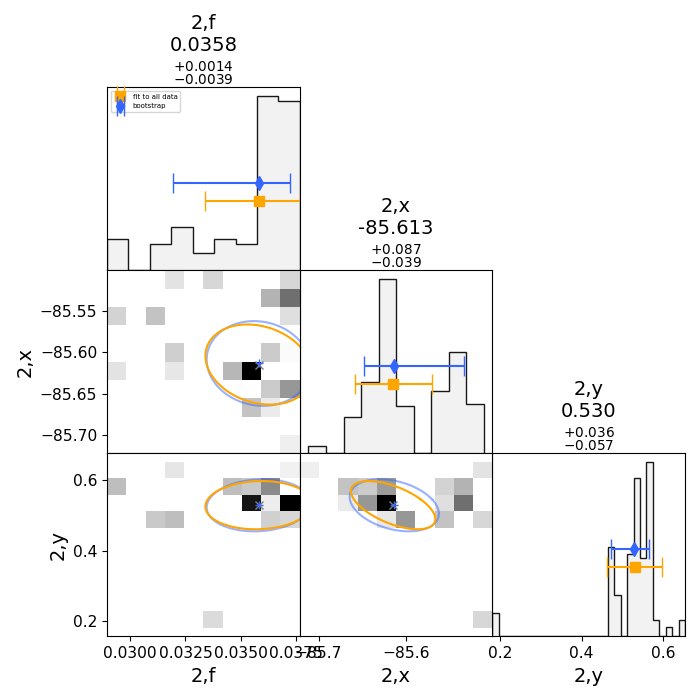}
   \caption{Bootstrapping plot showing the error determination for the dataset of $\gamma$ Peg.}
    \end{figure*}

             \begin{figure*}
   \centering
   \includegraphics[width=170mm]{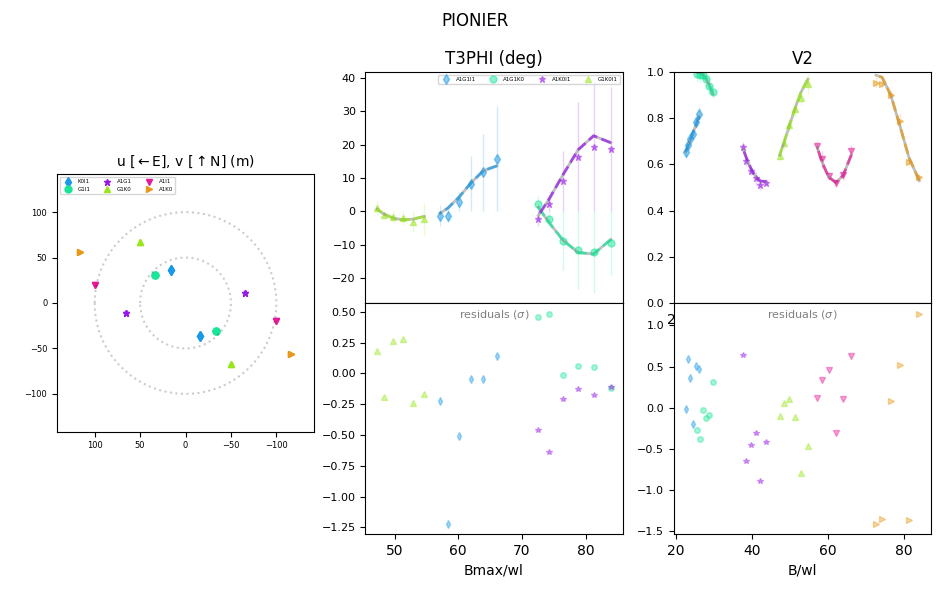}
   \caption{The best-fit data for HD\,3379. On the left, the $u-v$ coverage from the observations with PIONIER can be seen. The different colours correspond to the telescope pair with which that particular data was acquired at the VLTI. In the middle, the closure phase (T3PHI) fit and residuals are shown in terms of the spatial frequency (B$_{avg}$/$\lambda$, written as Bavg/wl on the axes). On the right, the fit to the squared visibilities (V2) and the associated residuals are shown, again in terms of spatial frequency. Across the two fits, the data are represented as points whilst the fit as a continuous line.}
    \end{figure*}
    
    \begin{figure*}
   \centering
   \includegraphics[width=140mm]{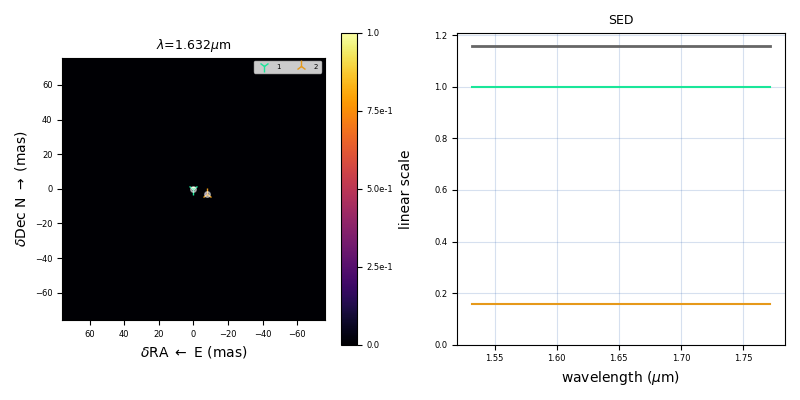}
   \caption{Model image created based on the best-fitting model of HD\,3379.}
    \end{figure*}

    \begin{figure*}
   \centering
   \includegraphics[width=120mm]{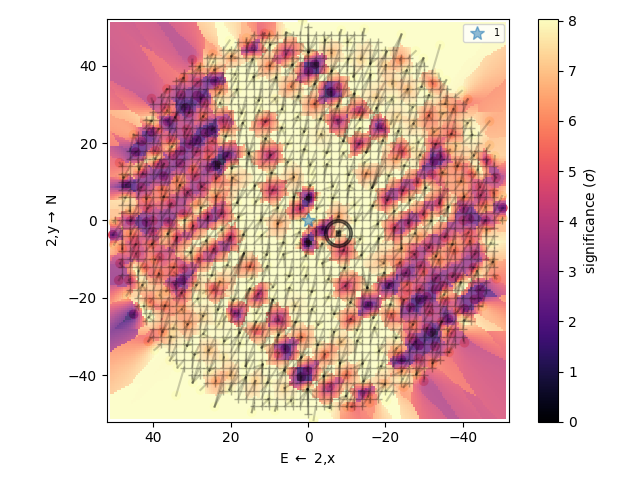}
   \caption{Plot showing the grid search used to search for the companion around HD\,3379 and the companion's significance.}
    \end{figure*}
    
    \begin{figure*}
   \centering
   \includegraphics[width=120mm]{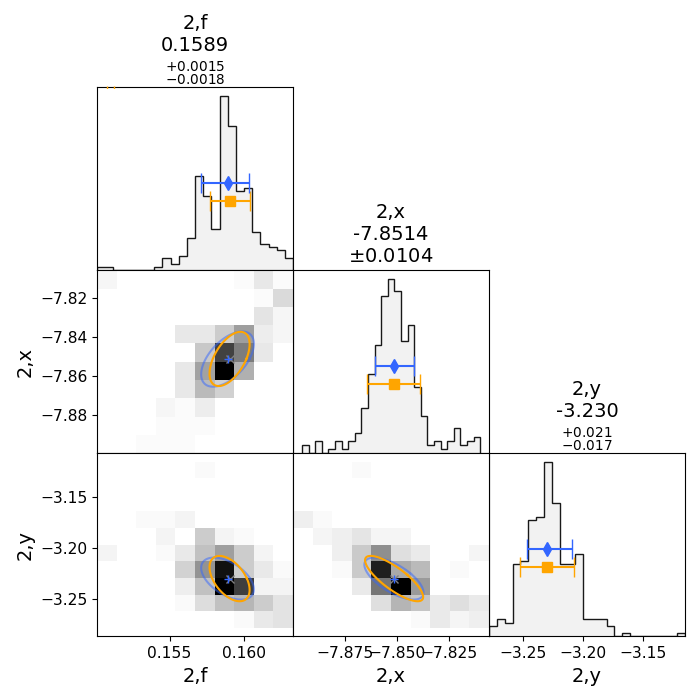}
   \caption{Bootstrapping plot showing the error determination for the dataset of HD\,3379.}
    \end{figure*}

             \begin{figure*}
   \centering
   \includegraphics[width=170mm]{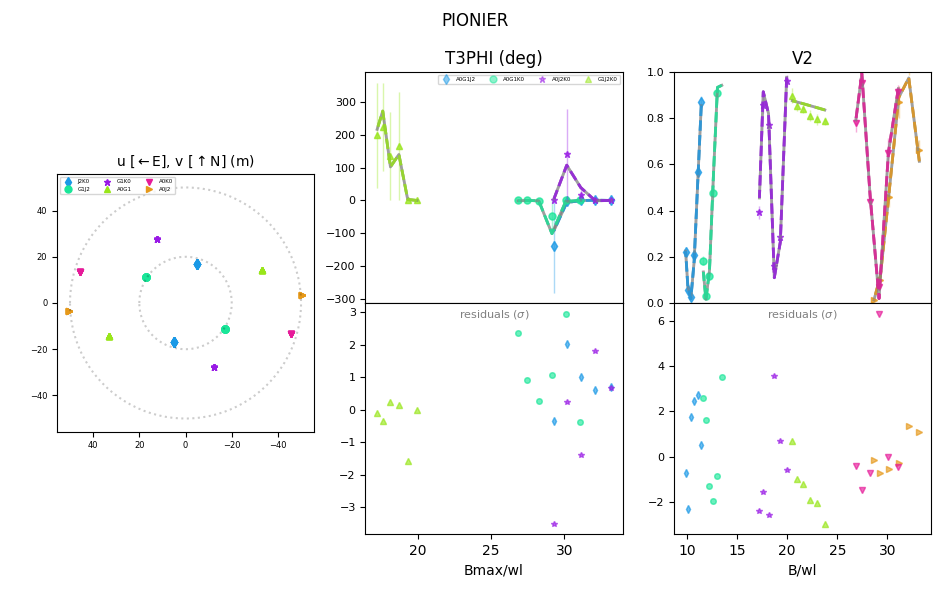}
   \caption{The best-fit data for HD\,25558. On the left, the $u-v$ coverage from the observations with PIONIER can be seen. The different colours correspond to the telescope pair with which that particular data was acquired at the VLTI. In the middle, the closure phase (T3PHI) fit and residuals are shown in terms of the spatial frequency (B$_{avg}$/$\lambda$, written as Bavg/wl on the axes). On the right, the fit to the squared visibilities (V2) and the associated residuals are shown, again in terms of spatial frequency. Across the two fits, the data are represented as points whilst the fit as a continuous line.}
    \end{figure*}
    
    \begin{figure*}
   \centering
   \includegraphics[width=140mm]{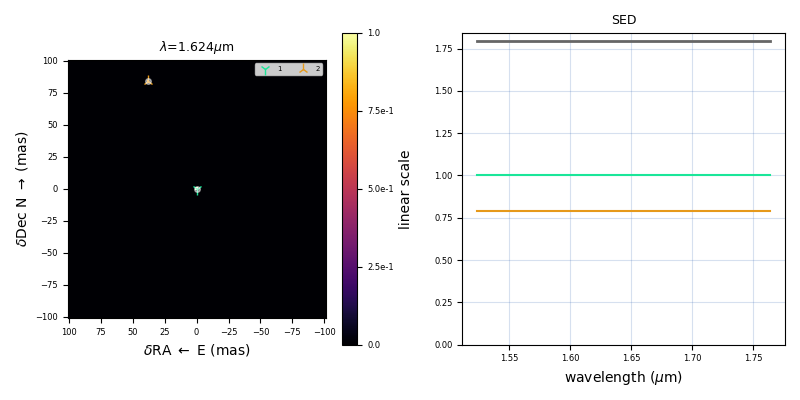}
   \caption{Model image created based on the best-fitting model of HD\,25558.}
    \end{figure*}

    \begin{figure*}
   \centering
   \includegraphics[width=120mm]{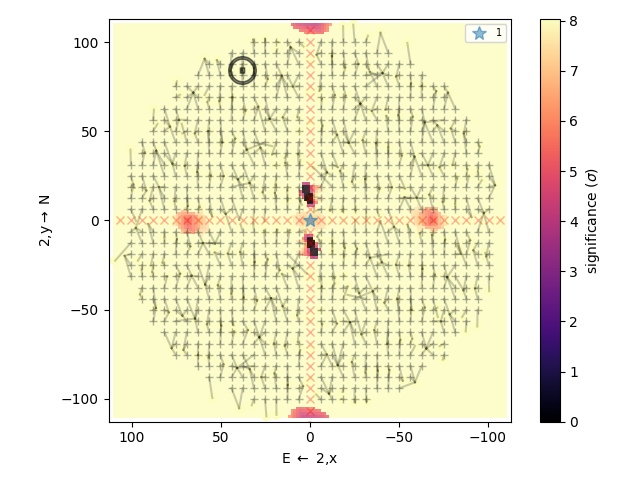}
   \caption{Plot showing the grid search used to search for the companion around HD\,25558 and the companion's significance.}
    \end{figure*}
    
    \begin{figure*}
   \centering
   \includegraphics[width=120mm]{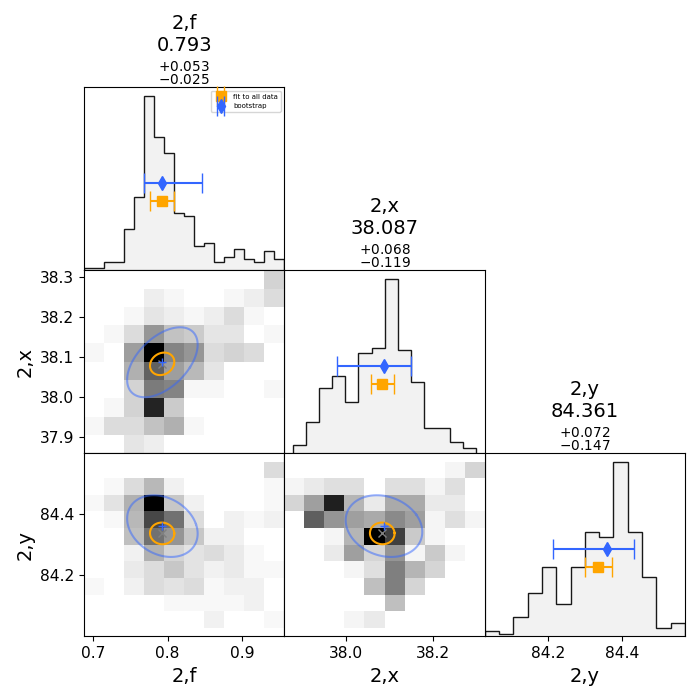}
   \caption{Bootstrapping plot showing the error determination for the dataset of HD\,25558.}
    \end{figure*}

                 \begin{figure*}
   \centering
   \includegraphics[width=170mm]{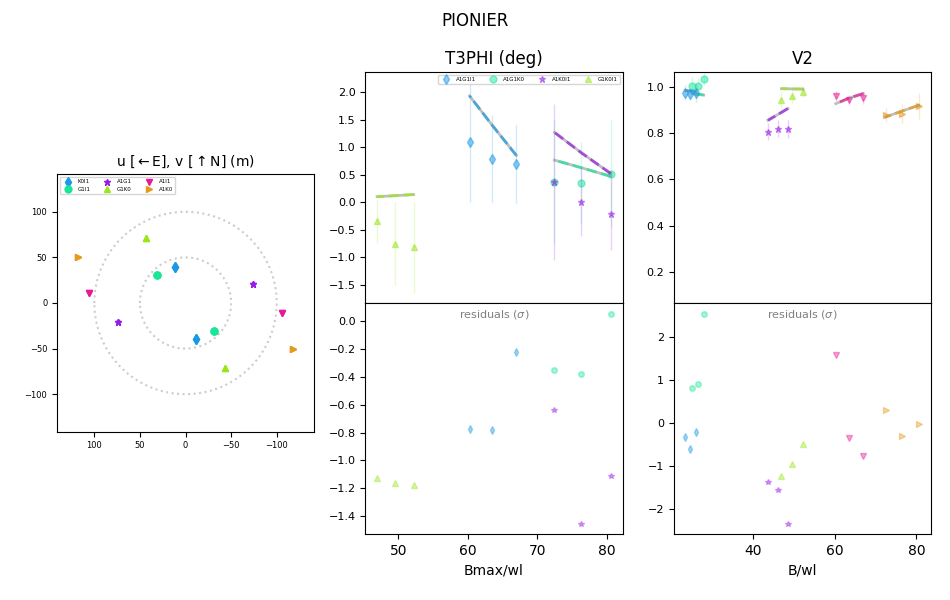}
   \caption{The best-fit data for HD\,30836. On the left, the $u-v$ coverage from the observations with PIONIER can be seen. The different colours correspond to the telescope pair with which that particular data was acquired at the VLTI. In the middle, the closure phase (T3PHI) fit and residuals are shown in terms of the spatial frequency (B$_{avg}$/$\lambda$, written as Bavg/wl on the axes). On the right, the fit to the squared visibilities (V2) and the associated residuals are shown, again in terms of spatial frequency. Across the two fits, the data are represented as points whilst the fit as a continuous line.}
    \end{figure*}
    
    \begin{figure*}
   \centering
   \includegraphics[width=140mm]{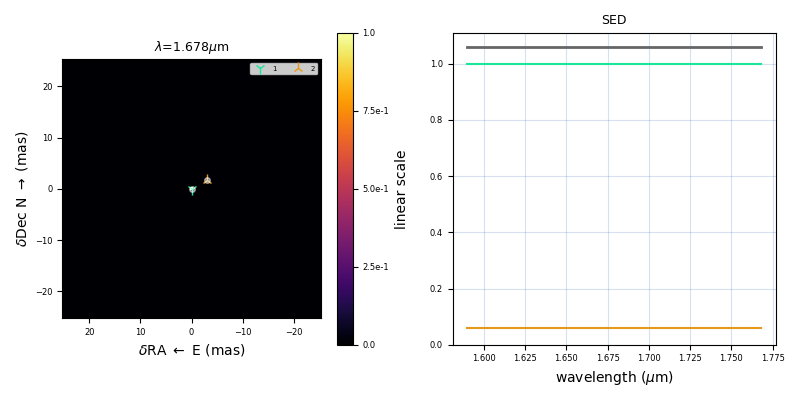}
   \caption{Model image created based on the best-fitting model of HD\,30836.}
    \end{figure*}

    \begin{figure*}
   \centering
   \includegraphics[width=120mm]{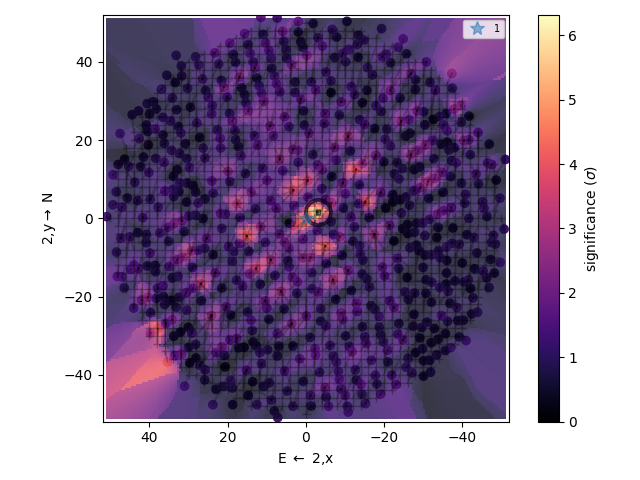}
   \caption{Plot showing the grid search used to search for the companion around HD\,30836 and the companion's significance.}
    \end{figure*}
    
    \begin{figure*}
   \centering
   \includegraphics[width=120mm]{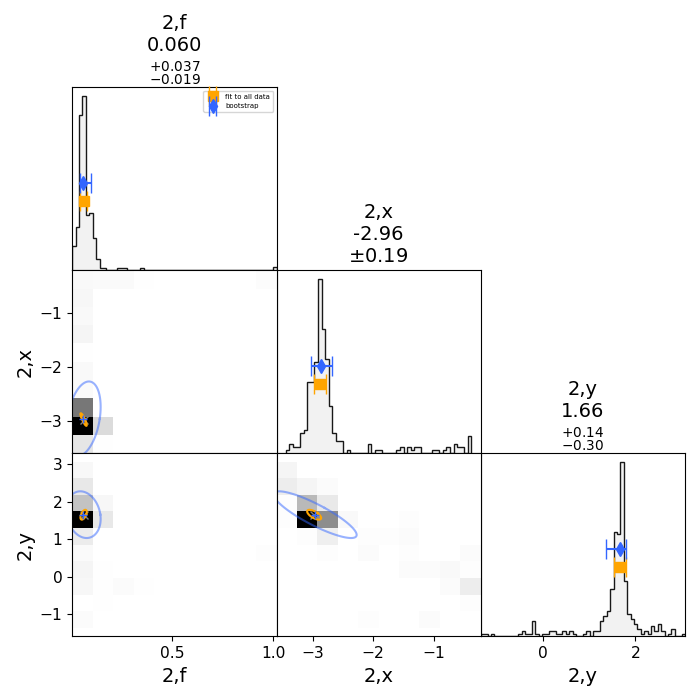}
   \caption{Bootstrapping plot showing the error determination for the dataset of HD\,30836.}
    \end{figure*}

                     \begin{figure*}
   \centering
   \includegraphics[width=170mm]{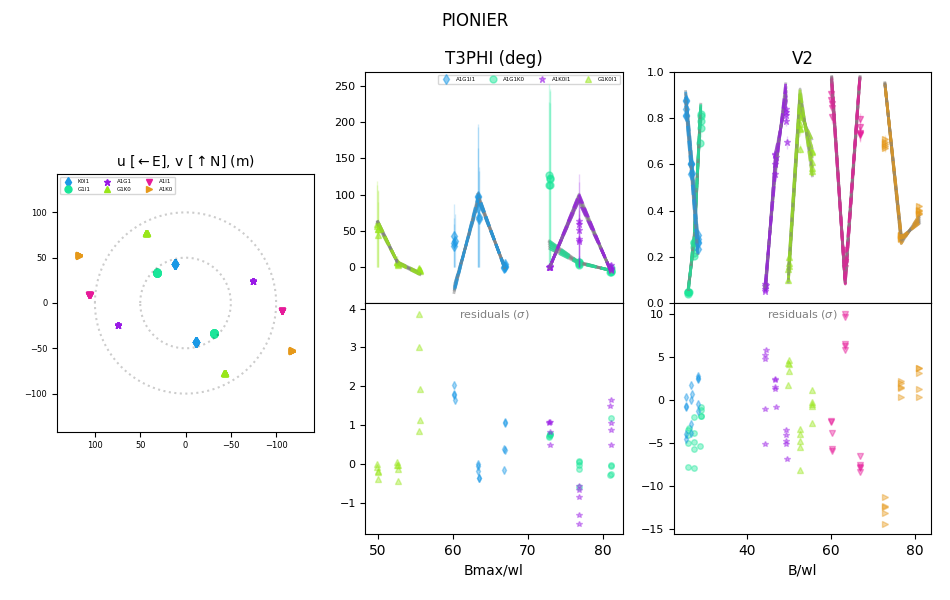}
   \caption{The best-fit data for HD\,32249. On the left, the $u-v$ coverage from the observations with PIONIER can be seen. The different colours correspond to the telescope pair with which that particular data was acquired at the VLTI. In the middle, the closure phase (T3PHI) fit and residuals are shown in terms of the spatial frequency (B$_{avg}$/$\lambda$, written as Bavg/wl on the axes). On the right, the fit to the squared visibilities (V2) and the associated residuals are shown, again in terms of spatial frequency. Across the two fits, the data are represented as points whilst the fit as a continuous line.}
    \end{figure*}
    
    \begin{figure*}
   \centering
   \includegraphics[width=140mm]{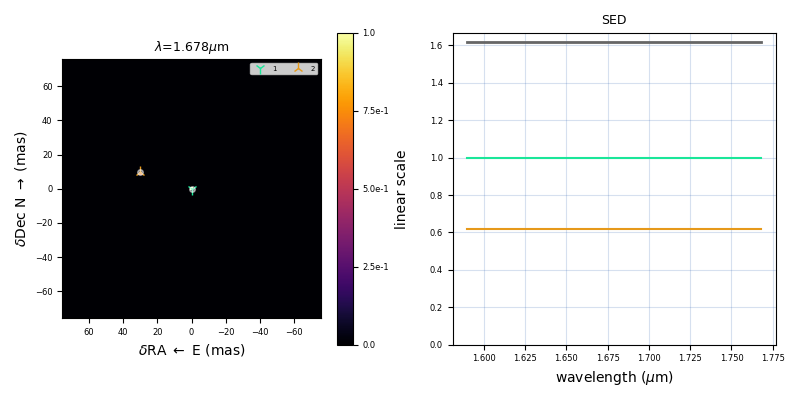}
   \caption{Model image created based on the best-fitting model of HD\,32249.}
    \end{figure*}

    \begin{figure*}
   \centering
   \includegraphics[width=120mm]{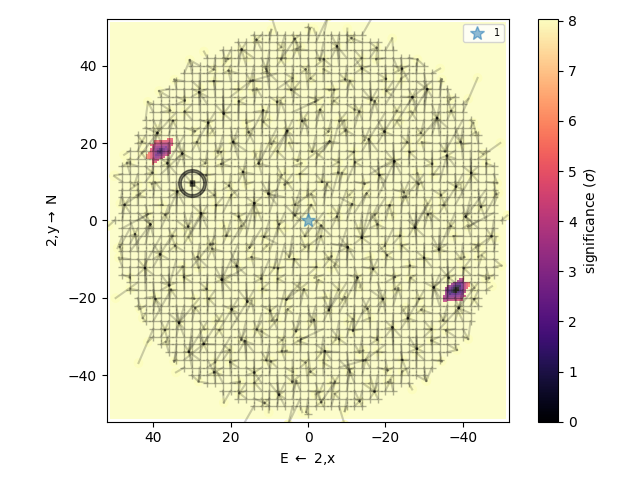}
   \caption{Plot showing the grid search used to search for the companion around HD\,32249 and the companion's significance.}
    \end{figure*}
    
    \begin{figure*}
   \centering
   \includegraphics[width=120mm]{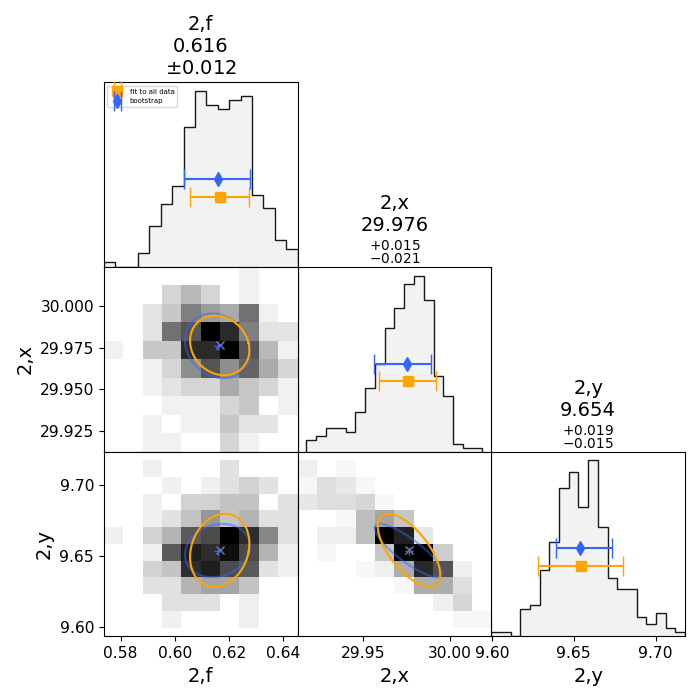}
   \caption{Bootstrapping plot showing the error determination for the dataset of HD\,32249.}
    \end{figure*}

                     \begin{figure*}
   \centering
   \includegraphics[width=170mm]{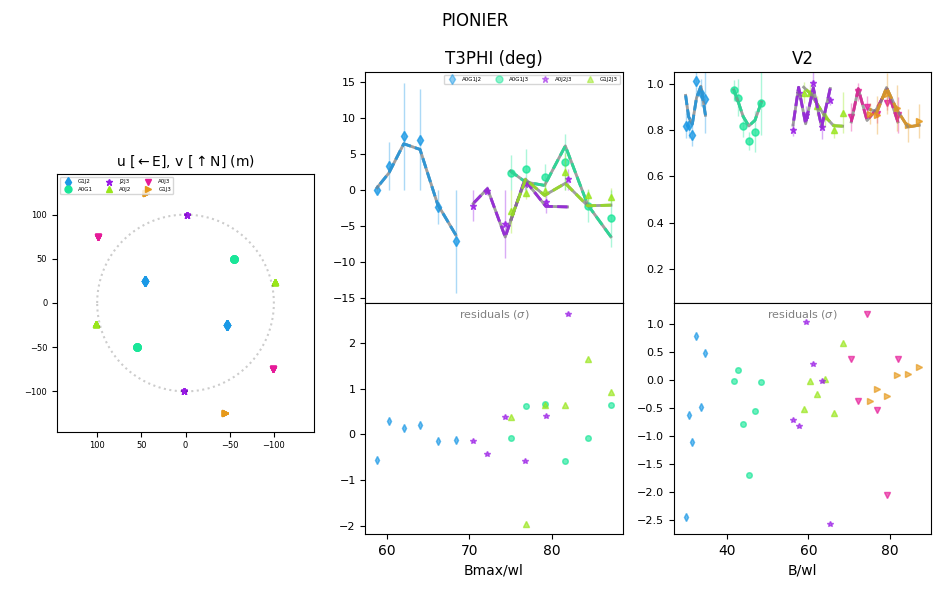}
   \caption{The best-fit data for HD\,34816. On the left, the $u-v$ coverage from the observations with PIONIER can be seen. The different colours correspond to the telescope pair with which that particular data was acquired at the VLTI. In the middle, the closure phase (T3PHI) fit and residuals are shown in terms of the spatial frequency (B$_{avg}$/$\lambda$, written as Bavg/wl on the axes). On the right, the fit to the squared visibilities (V2) and the associated residuals are shown, again in terms of spatial frequency. Across the two fits, the data are represented as points whilst the fit as a continuous line.}
    \end{figure*}
    
    \begin{figure*}
   \centering
   \includegraphics[width=140mm]{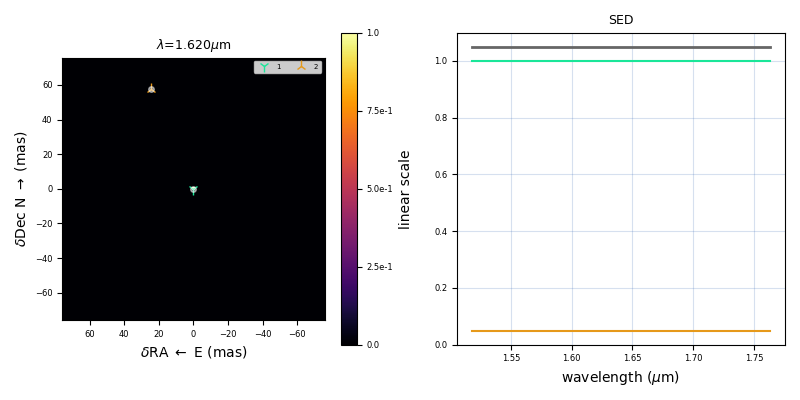}
   \caption{Model image created based on the best-fitting model of HD\,34816.}
    \end{figure*}

    \begin{figure*}
   \centering
   \includegraphics[width=120mm]{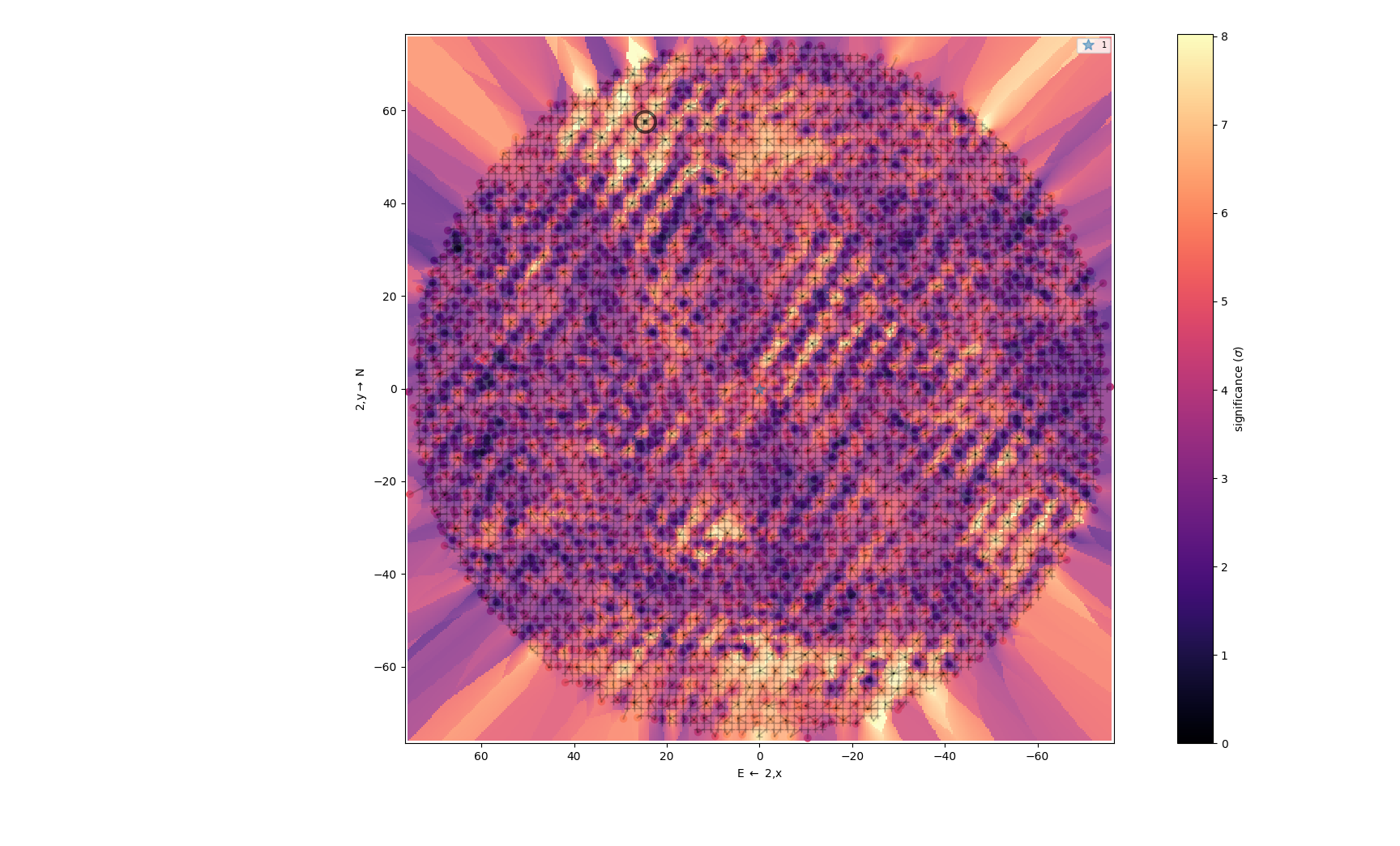}
   \caption{Plot showing the grid search used to search for the companion around HD\,34816 and the companion's significance.}
    \end{figure*}
    
    \begin{figure*}
   \centering
   \includegraphics[width=120mm]{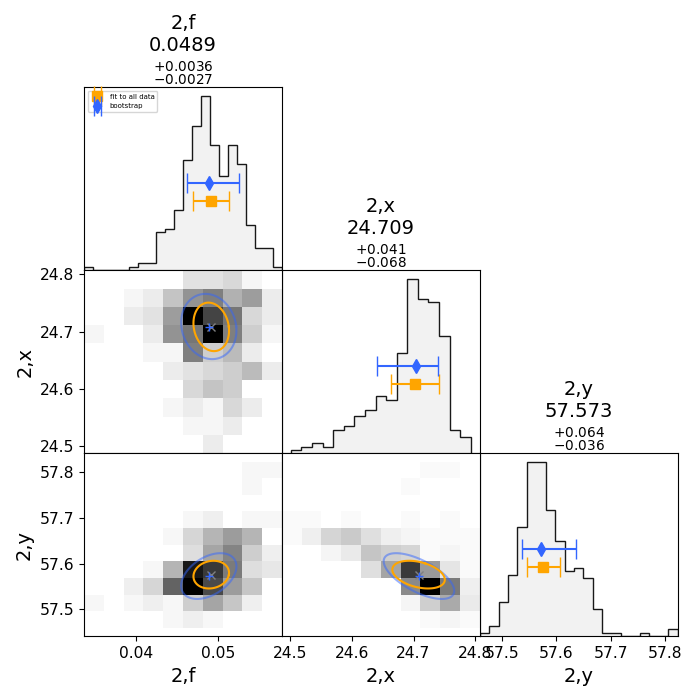}
   \caption{Bootstrapping plot showing the error determination for the dataset of HD\,34816.}
    \end{figure*}

                         \begin{figure*}
   \centering
   \includegraphics[width=170mm]{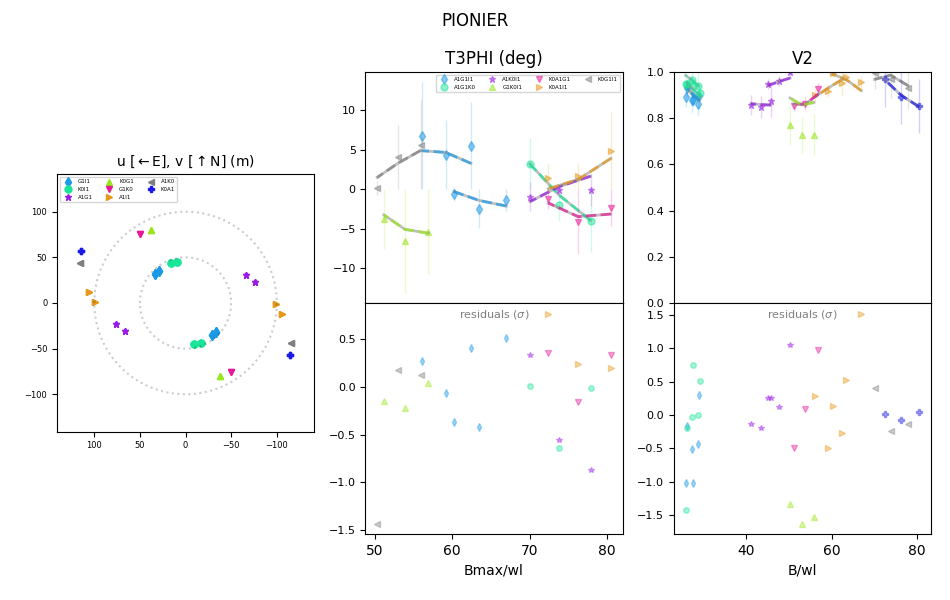}
   \caption{The best-fit data for HD\,35337. On the left, the $u-v$ coverage from the observations with PIONIER can be seen. The different colours correspond to the telescope pair with which that particular data was acquired at the VLTI. In the middle, the closure phase (T3PHI) fit and residuals are shown in terms of the spatial frequency (B$_{avg}$/$\lambda$, written as Bavg/wl on the axes). On the right, the fit to the squared visibilities (V2) and the associated residuals are shown, again in terms of spatial frequency. Across the two fits, the data are represented as points whilst the fit as a continuous line.}
    \end{figure*}
    
    \begin{figure*}
   \centering
   \includegraphics[width=140mm]{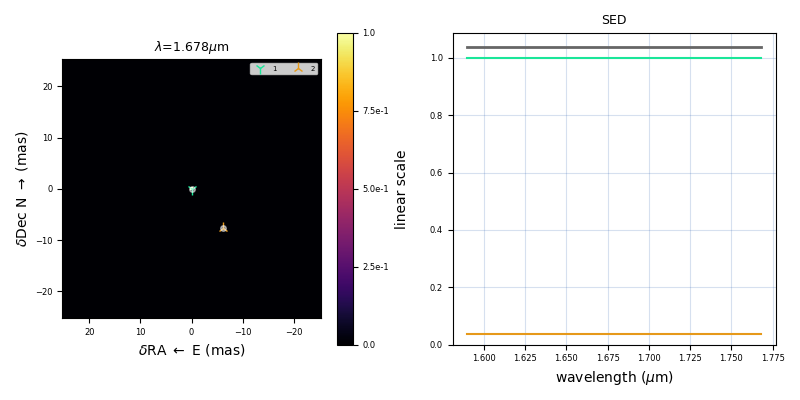}
   \caption{Model image created based on the best-fitting model of HD\,35337.}
    \end{figure*}

    \begin{figure*}
   \centering
   \includegraphics[width=120mm]{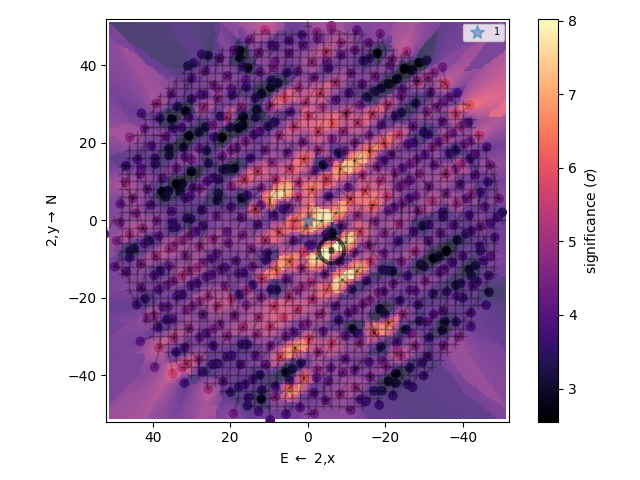}
   \caption{Plot showing the grid search used to search for the companion around HD\,35337 and the companion's significance.}
    \end{figure*}
    
    \begin{figure*}
   \centering
   \includegraphics[width=120mm]{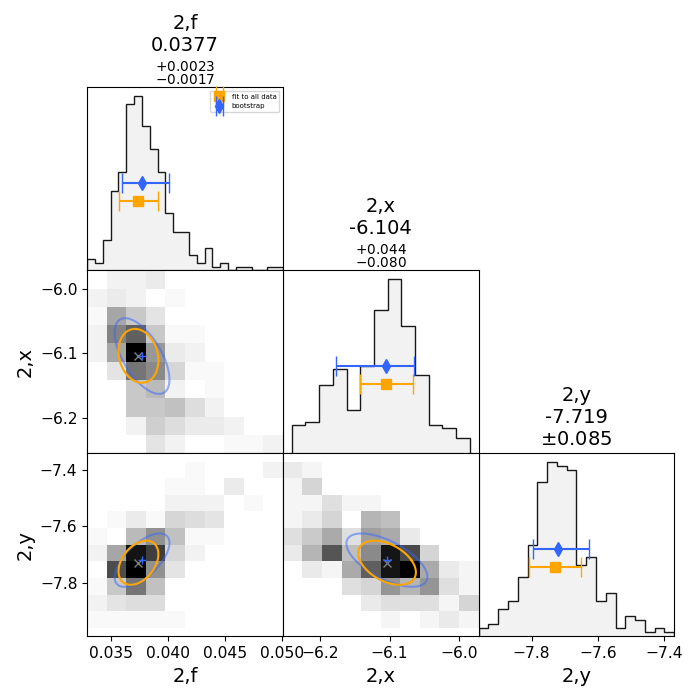}
   \caption{Bootstrapping plot showing the error determination for the dataset of HD\,35337.}
    \end{figure*}

                         \begin{figure*}
   \centering
   \includegraphics[width=170mm]{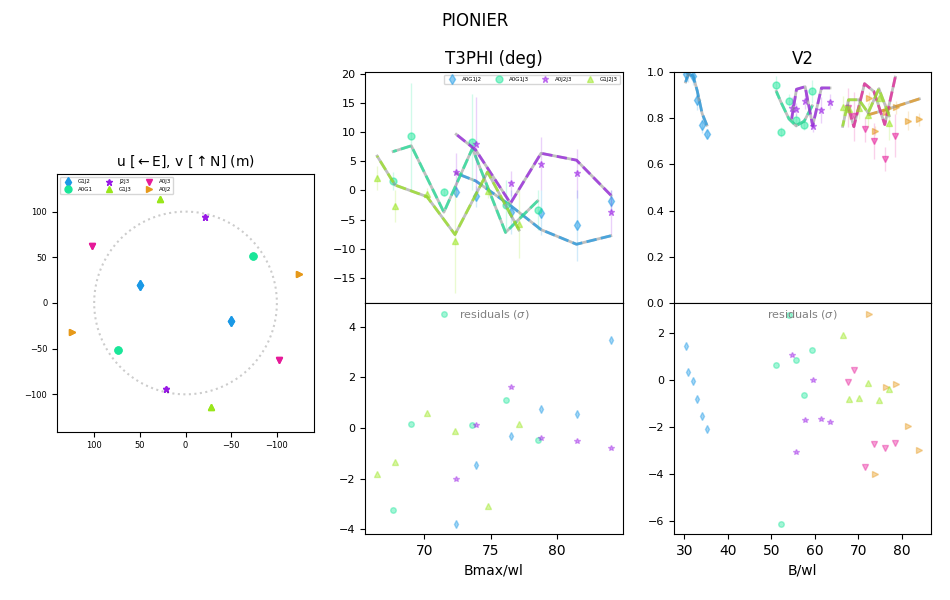}
   \caption{The best-fit data for HD\,105382. On the left, the $u-v$ coverage from the observations with PIONIER can be seen. The different colours correspond to the telescope pair with which that particular data was acquired at the VLTI. In the middle, the closure phase (T3PHI) fit and residuals are shown in terms of the spatial frequency (B$_{avg}$/$\lambda$, written as Bavg/wl on the axes). On the right, the fit to the squared visibilities (V2) and the associated residuals are shown, again in terms of spatial frequency. Across the two fits, the data are represented as points whilst the fit as a continuous line.}
    \end{figure*}
    
    \begin{figure*}
   \centering
   \includegraphics[width=140mm]{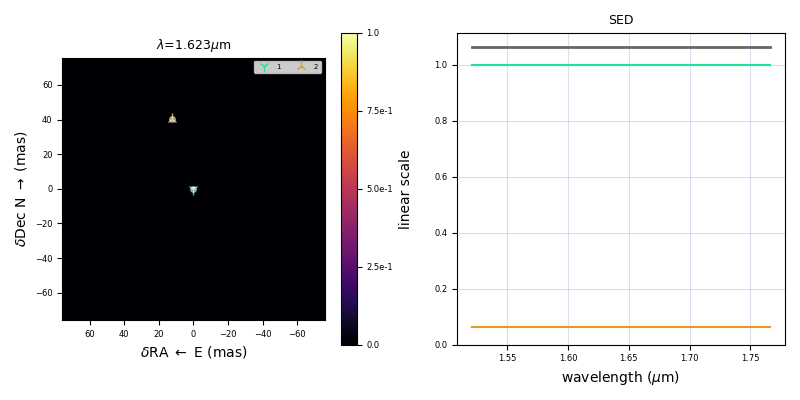}
   \caption{Model image created based on the best-fitting model of HD\,105382.}
    \end{figure*}

    \begin{figure*}
   \centering
   \includegraphics[width=120mm]{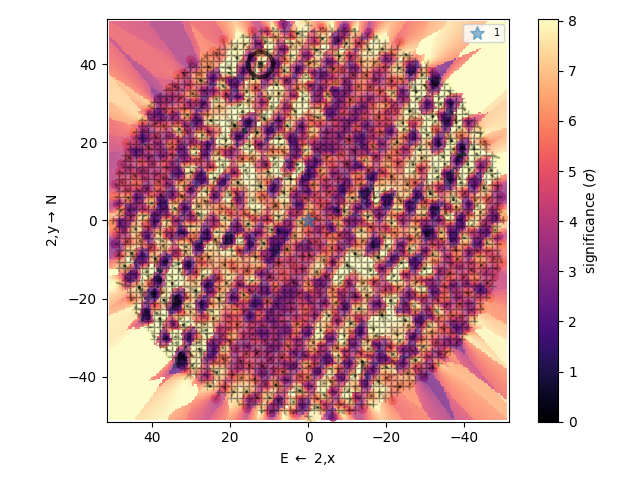}
   \caption{Plot showing the grid search used to search for the companion around HD\,105382 and the companion's significance.}
    \end{figure*}
    
    \begin{figure*}
   \centering
   \includegraphics[width=120mm]{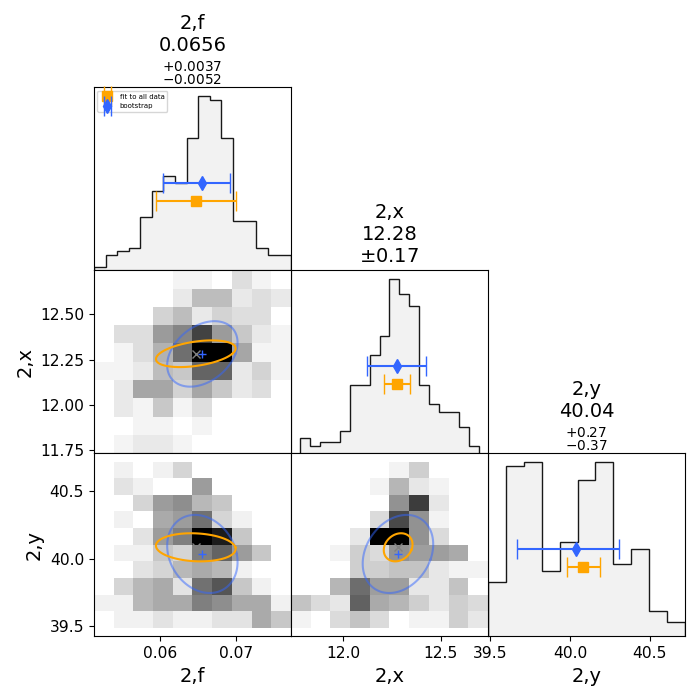}
   \caption{Bootstrapping plot showing the error determination for the dataset of HD\,105382.}
    \end{figure*}

                         \begin{figure*}
   \centering
   \includegraphics[width=170mm]{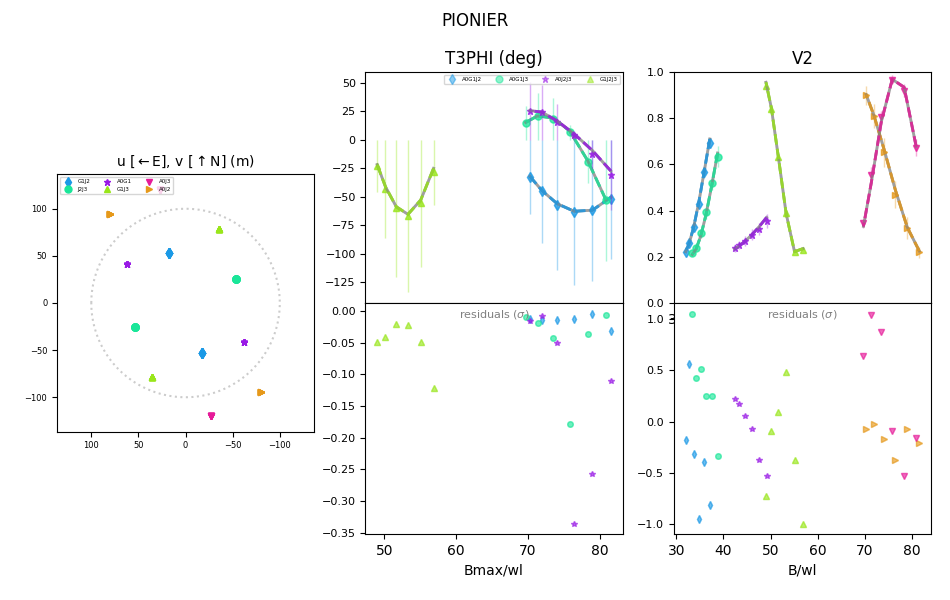}
   \caption{The best-fit data for HD\,109026. On the left, the $u-v$ coverage from the observations with PIONIER can be seen. The different colours correspond to the telescope pair with which that particular data was acquired at the VLTI. In the middle, the closure phase (T3PHI) fit and residuals are shown in terms of the spatial frequency (B$_{avg}$/$\lambda$, written as Bavg/wl on the axes). On the right, the fit to the squared visibilities (V2) and the associated residuals are shown, again in terms of spatial frequency. Across the two fits, the data are represented as points whilst the fit as a continuous line.}
    \end{figure*}
    
    \begin{figure*}
   \centering
   \includegraphics[width=140mm]{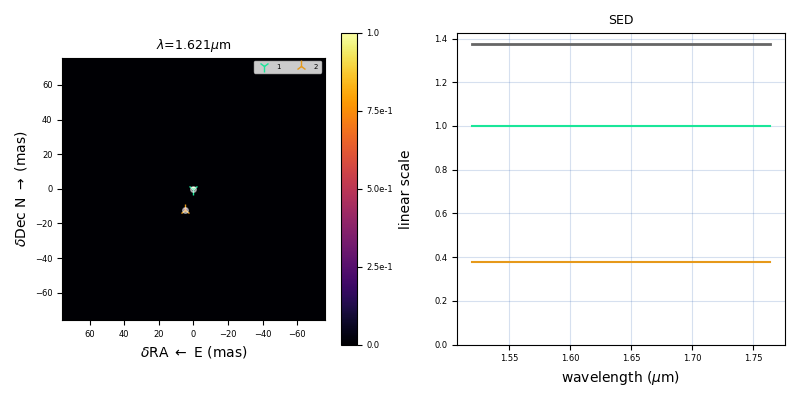}
   \caption{Model image created based on the best-fitting model of HD\,109026.}
    \end{figure*}

    \begin{figure*}
   \centering
   \includegraphics[width=120mm]{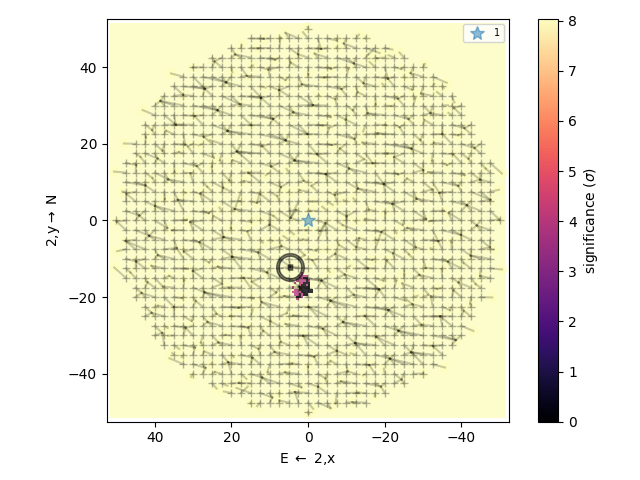}
   \caption{Plot showing the grid search used to search for the companion around HD\,109026 and the companion's significance.}
    \end{figure*}
    
    \begin{figure*}
   \centering
   \includegraphics[width=120mm]{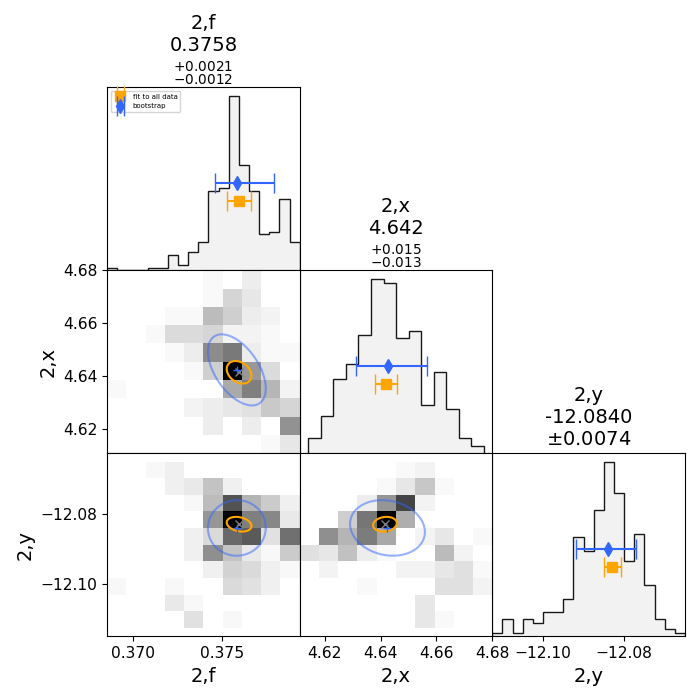}
   \caption{Bootstrapping plot showing the error determination for the dataset of HD\,109026.}
    \end{figure*}

                         \begin{figure*}
   \centering
   \includegraphics[width=170mm]{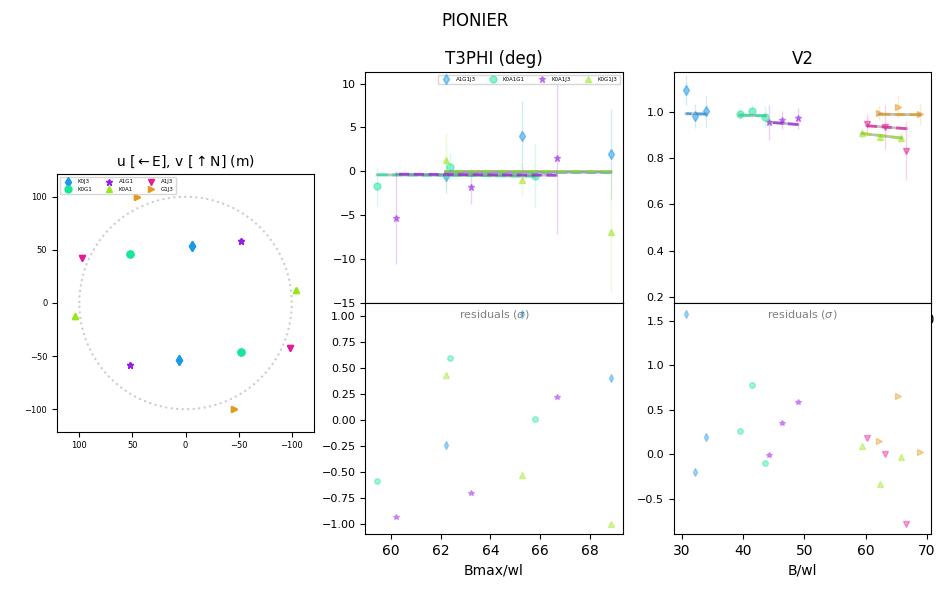}
   \caption{The best-fit data for HD\,133518. On the left, the $u-v$ coverage from the observations with PIONIER can be seen. The different colours correspond to the telescope pair with which that particular data was acquired at the VLTI. In the middle, the closure phase (T3PHI) fit and residuals are shown in terms of the spatial frequency (B$_{avg}$/$\lambda$, written as Bavg/wl on the axes). On the right, the fit to the squared visibilities (V2) and the associated residuals are shown, again in terms of spatial frequency. Across the two fits, the data are represented as points whilst the fit as a continuous line.}
    \end{figure*}
    
    \begin{figure*}
   \centering
   \includegraphics[width=140mm]{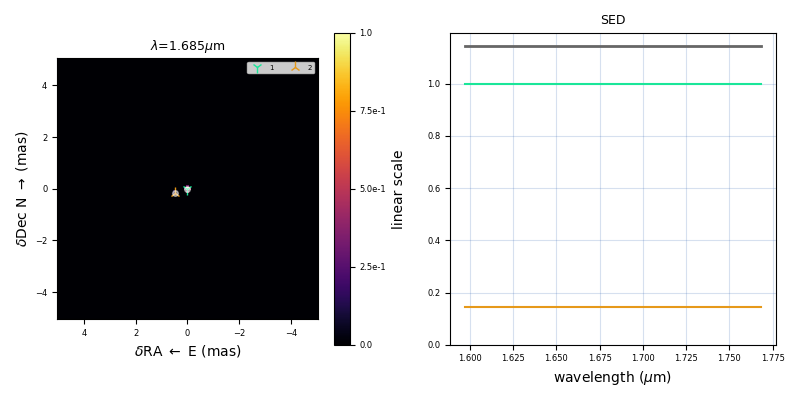}
   \caption{Model image created based on the best-fitting model of HD\,133518.}
    \end{figure*}

    \begin{figure*}
   \centering
   \includegraphics[width=120mm]{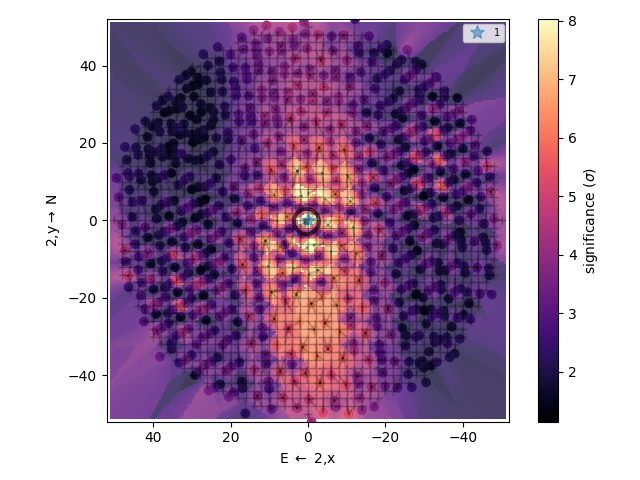}
   \caption{Plot showing the grid search used to search for the companion around HD\,133518 and the companion's significance.}
    \end{figure*}
    
    \begin{figure*}
   \centering
   \includegraphics[width=120mm]{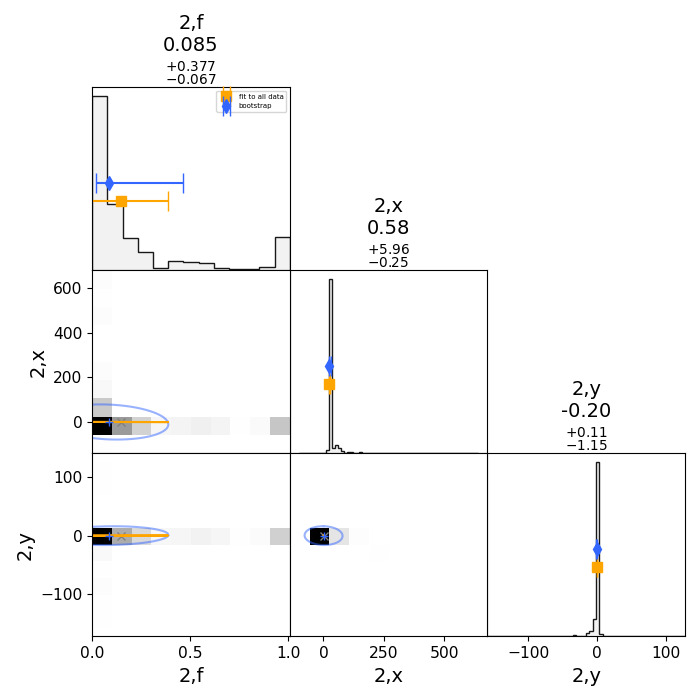}
   \caption{Bootstrapping plot showing the error determination for the dataset of HD\,133518.}
    \end{figure*}

                         \begin{figure*}
   \centering
   \includegraphics[width=170mm]{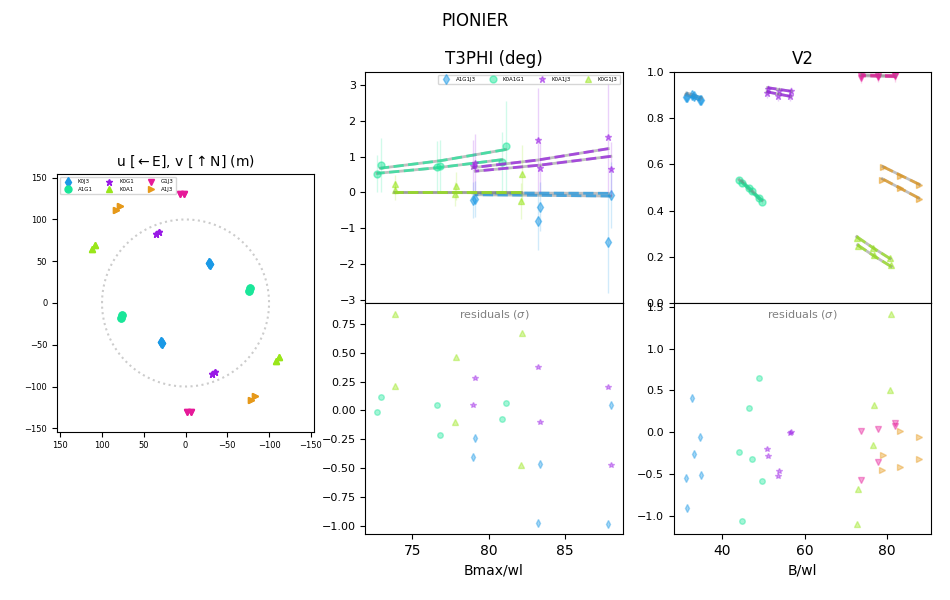}
   \caption{The best-fit data for HD\,140008. On the left, the $u-v$ coverage from the observations with PIONIER can be seen. The different colours correspond to the telescope pair with which that particular data was acquired at the VLTI. In the middle, the closure phase (T3PHI) fit and residuals are shown in terms of the spatial frequency (B$_{avg}$/$\lambda$, written as Bavg/wl on the axes). On the right, the fit to the squared visibilities (V2) and the associated residuals are shown, again in terms of spatial frequency. Across the two fits, the data are represented as points whilst the fit as a continuous line.}
    \end{figure*}
    
    \begin{figure*}
   \centering
   \includegraphics[width=140mm]{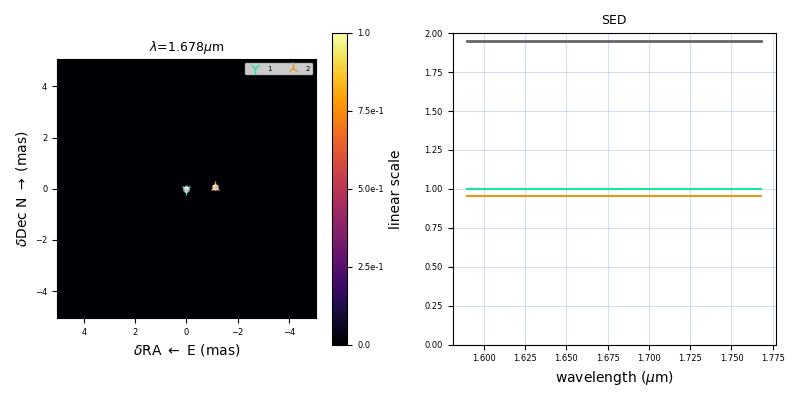}
   \caption{Model image created based on the best-fitting model of HD\,140008.}
    \end{figure*}

    \begin{figure*}
   \centering
   \includegraphics[width=120mm]{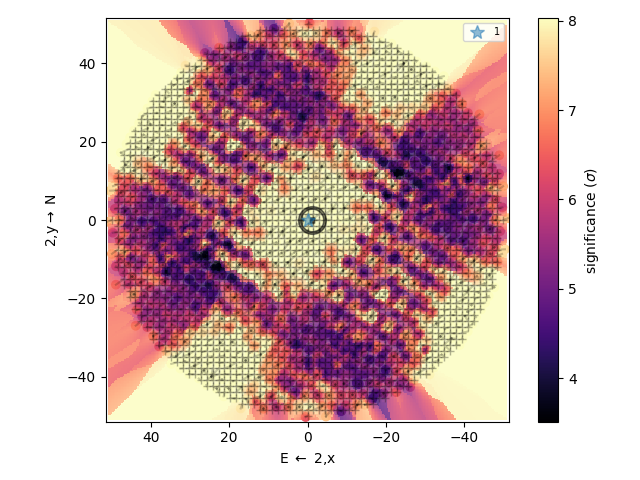}
   \caption{Plot showing the grid search used to search for the companion around HD\,140008 and the companion's significance.}
    \end{figure*}
    
    \begin{figure*}
   \centering
   \includegraphics[width=120mm]{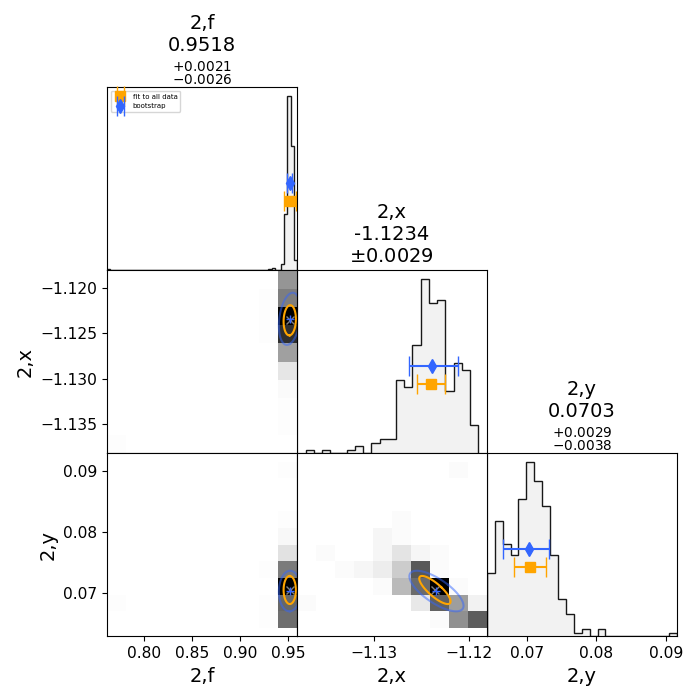}
   \caption{Bootstrapping plot showing the error determination for the dataset of HD\,140008.}
    \end{figure*}

    

    

                         \begin{figure*}
   \centering
   \includegraphics[width=170mm]{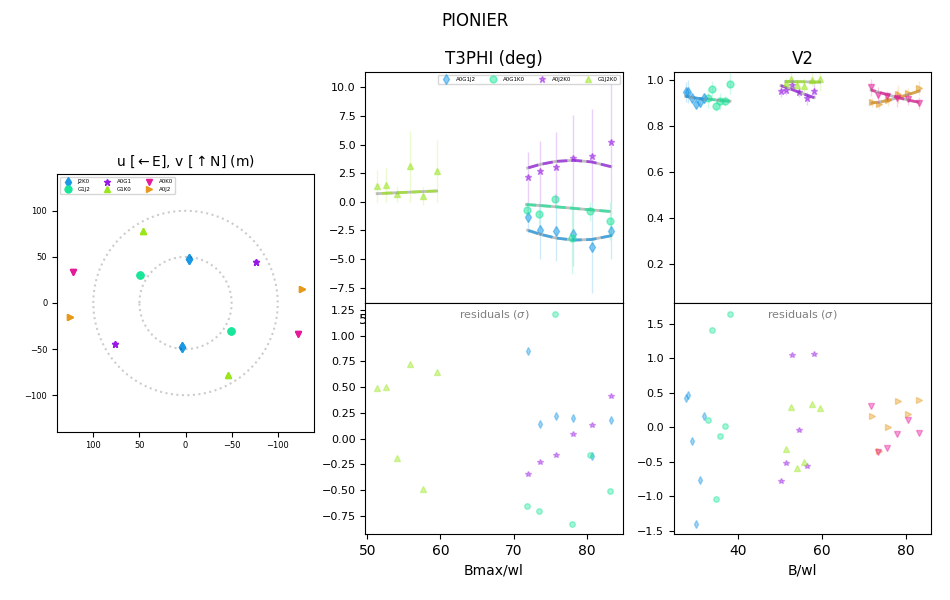}
   \caption{The best-fit data for HD\,178175. On the left, the $u-v$ coverage from the observations with PIONIER can be seen. The different colours correspond to the telescope pair with which that particular data was acquired at the VLTI. In the middle, the closure phase (T3PHI) fit and residuals are shown in terms of the spatial frequency (B$_{avg}$/$\lambda$, written as Bavg/wl on the axes). On the right, the fit to the squared visibilities (V2) and the associated residuals are shown, again in terms of spatial frequency. Across the two fits, the data are represented as points whilst the fit as a continuous line.}
    \end{figure*}
    
    \begin{figure*}
   \centering
   \includegraphics[width=140mm]{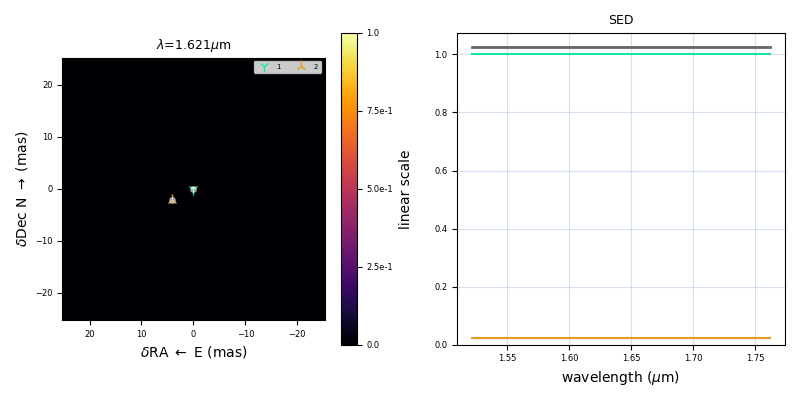}
   \caption{Model image created based on the best-fitting model of HD\,178175.}
    \end{figure*}

    \begin{figure*}
   \centering
   \includegraphics[width=120mm]{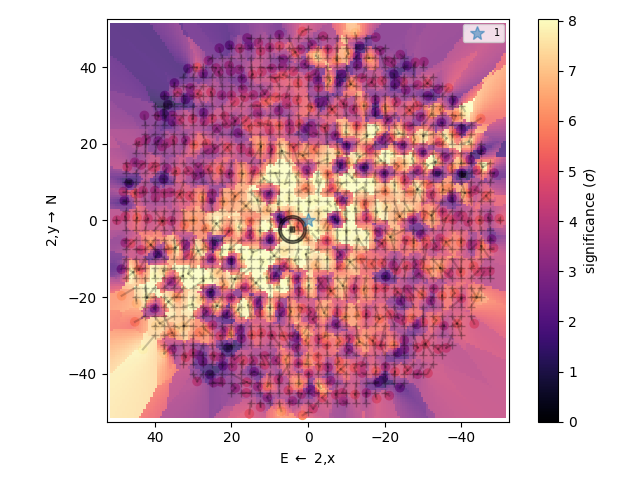}
   \caption{Plot showing the grid search used to search for the companion around HD\,178175 and the companion's significance.}
    \end{figure*}
    
    \begin{figure*}
   \centering
   \includegraphics[width=120mm]{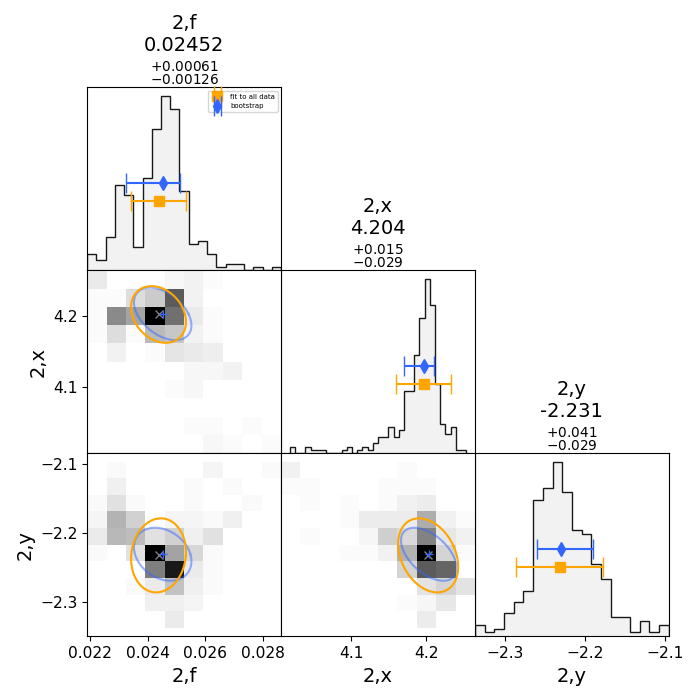}
   \caption{Bootstrapping plot showing the error determination for the dataset of HD\,178175.}
    \end{figure*}

                         \begin{figure*}
   \centering
   \includegraphics[width=170mm]{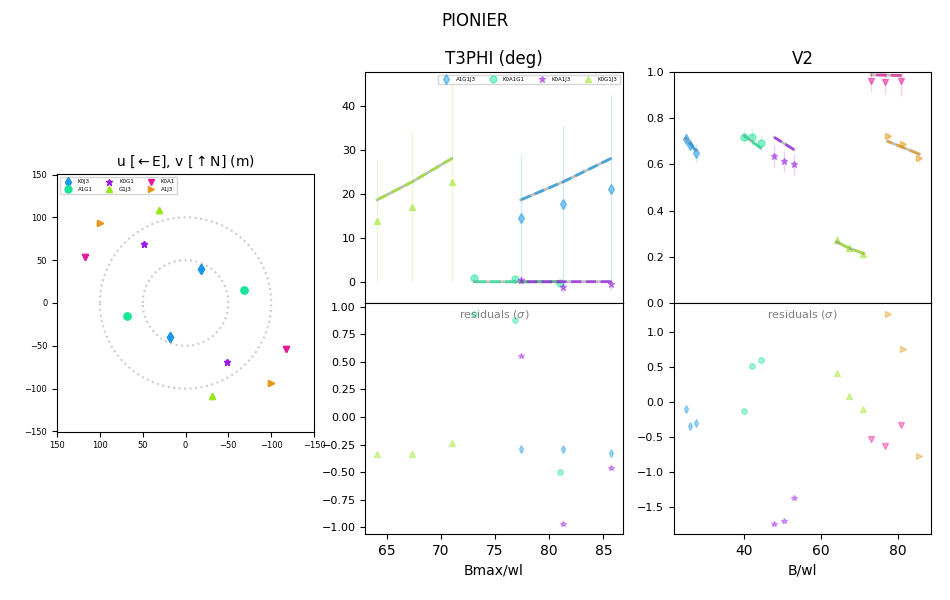}
   \caption{The best-fit data for HD\,191263. On the left, the $u-v$ coverage from the observations with PIONIER can be seen. The different colours correspond to the telescope pair with which that particular data was acquired at the VLTI. In the middle, the closure phase (T3PHI) fit and residuals are shown in terms of the spatial frequency (B$_{avg}$/$\lambda$, written as Bavg/wl on the axes). On the right, the fit to the squared visibilities (V2) and the associated residuals are shown, again in terms of spatial frequency. Across the two fits, the data are represented as points whilst the fit as a continuous line.}
    \end{figure*}
    
    \begin{figure*}
   \centering
   \includegraphics[width=140mm]{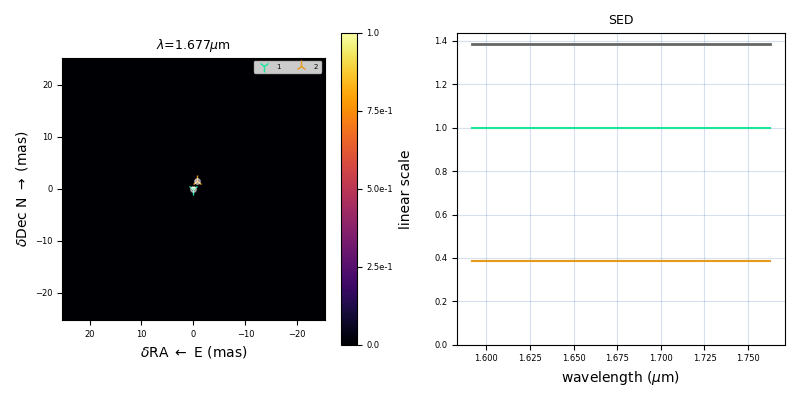}
   \caption{Model image created based on the best-fitting model of HD\,191263.}
    \end{figure*}

    \begin{figure*}
   \centering
   \includegraphics[width=120mm]{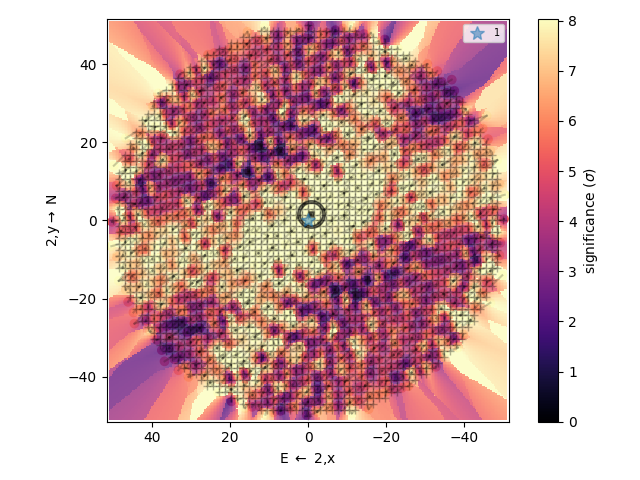}
   \caption{Plot showing the grid search used to search for the companion around HD\,191263 and the companion's significance.}
    \end{figure*}
    
    \begin{figure*}
   \centering
   \includegraphics[width=120mm]{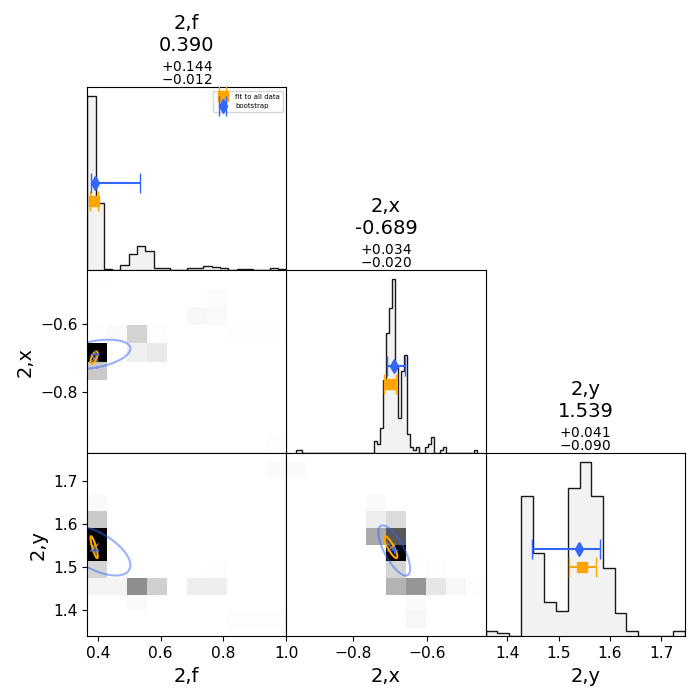}
   \caption{Bootstrapping plot showing the error determination for the dataset of HD\,191263.}
    \end{figure*}

                         \begin{figure*}
   \centering
   \includegraphics[width=170mm]{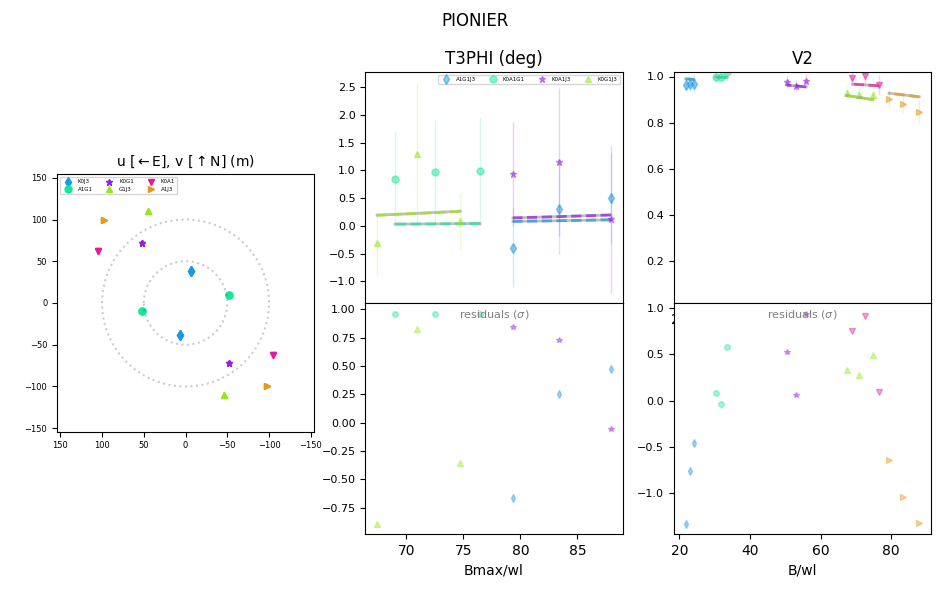}
   \caption{The best-fit data for HD\,212076. On the left, the $u-v$ coverage from the observations with PIONIER can be seen. The different colours correspond to the telescope pair with which that particular data was acquired at the VLTI. In the middle, the closure phase (T3PHI) fit and residuals are shown in terms of the spatial frequency (B$_{avg}$/$\lambda$, written as Bavg/wl on the axes). On the right, the fit to the squared visibilities (V2) and the associated residuals are shown, again in terms of spatial frequency. Across the two fits, the data are represented as points whilst the fit as a continuous line.}
    \end{figure*}
    
    \begin{figure*}
   \centering
   \includegraphics[width=140mm]{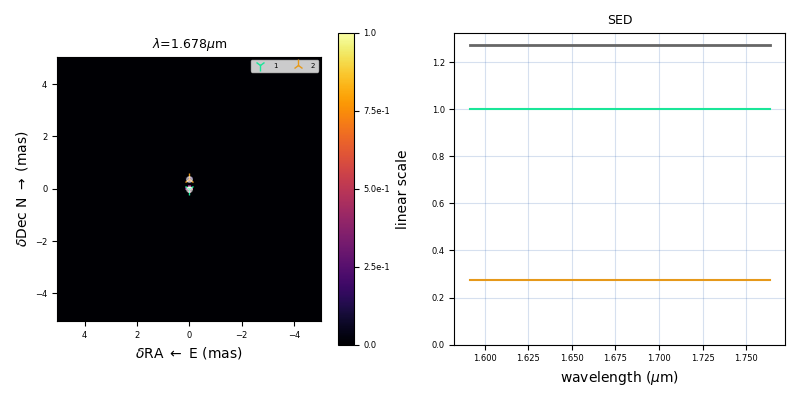}
   \caption{Model image created based on the best-fitting model of HD\,212076.}
    \end{figure*}

    \begin{figure*}
   \centering
   \includegraphics[width=120mm]{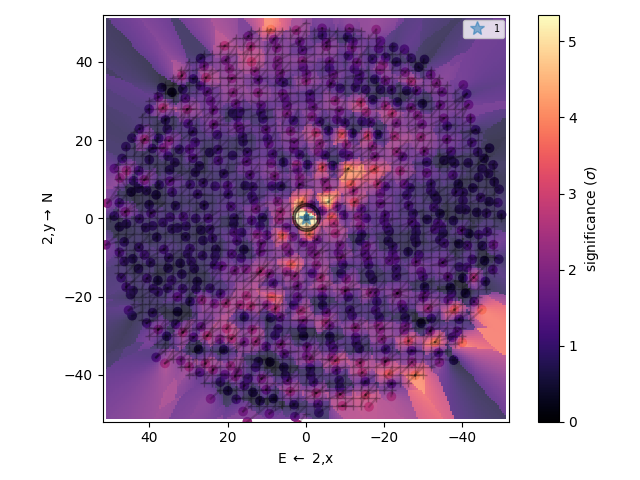}
   \caption{Plot showing the grid search used to search for the companion around HD\,212076 and the companion's significance.}
    \end{figure*}
    
    \begin{figure*}
   \centering
   \includegraphics[width=120mm]{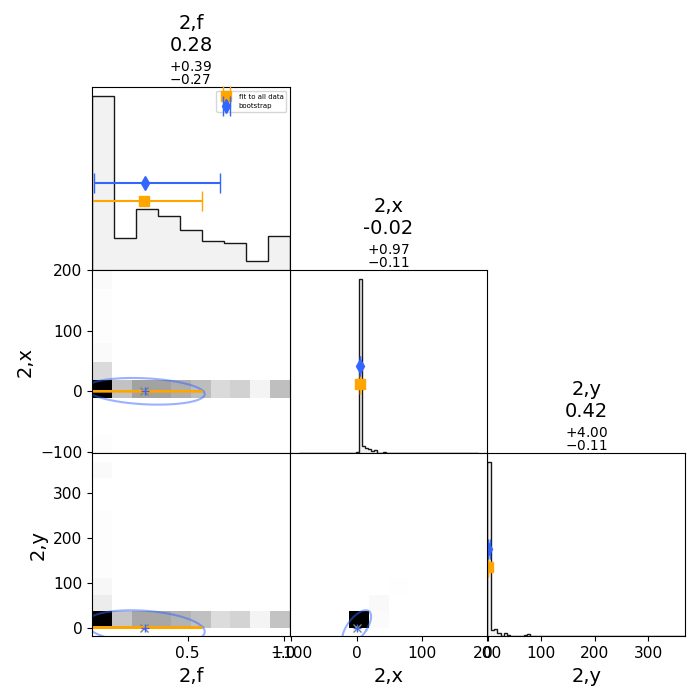}
   \caption{Bootstrapping plot showing the error determination for the dataset of HD\,212076.}
    \end{figure*}

                         \begin{figure*}
   \centering
   \includegraphics[width=170mm]{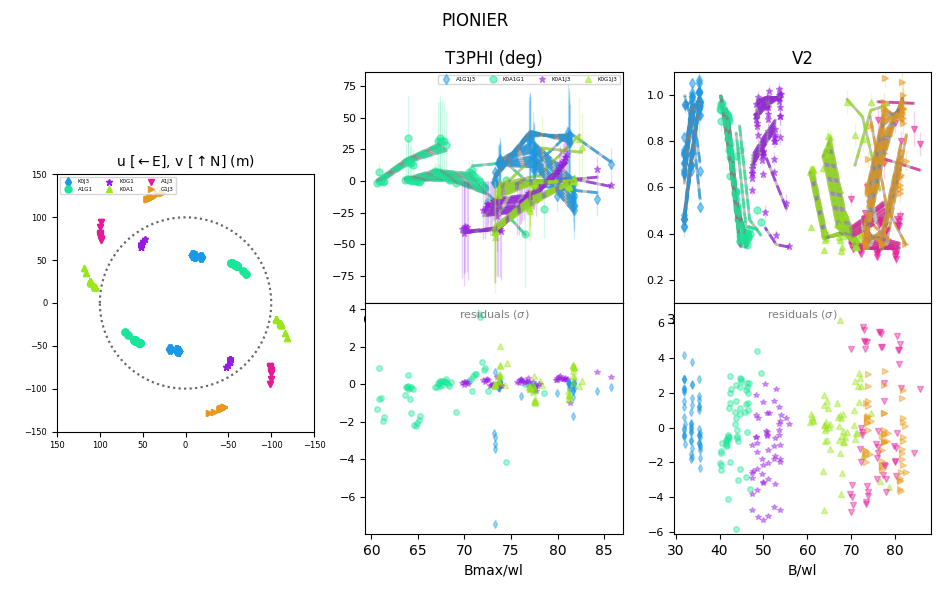}
   \caption{The best-fit data for HD\,224990. On the left, the $u-v$ coverage from the observations with PIONIER can be seen. The different colours correspond to the telescope pair with which that particular data was acquired at the VLTI. In the middle, the closure phase (T3PHI) fit and residuals are shown in terms of the spatial frequency (B$_{avg}$/$\lambda$, written as Bavg/wl on the axes). On the right, the fit to the squared visibilities (V2) and the associated residuals are shown, again in terms of spatial frequency. Across the two fits, the data are represented as points whilst the fit as a continuous line.}
    \end{figure*}
    
    \begin{figure*}
   \centering
   \includegraphics[width=140mm]{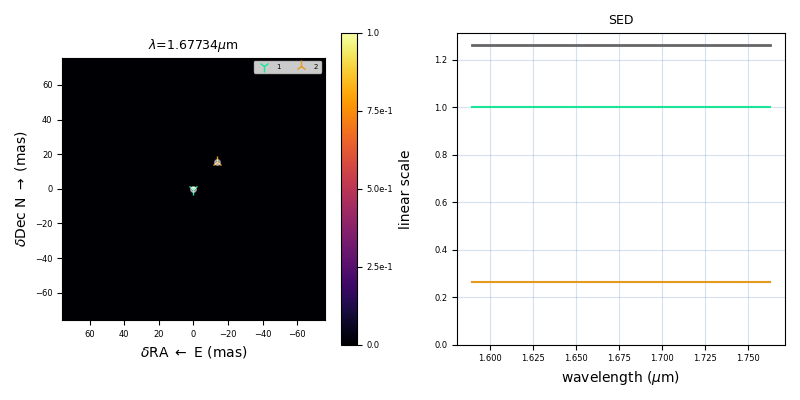}
   \caption{Model image created based on the best-fitting model of HD\,224990.}
    \end{figure*}

    \begin{figure*}
   \centering
   \includegraphics[width=120mm]{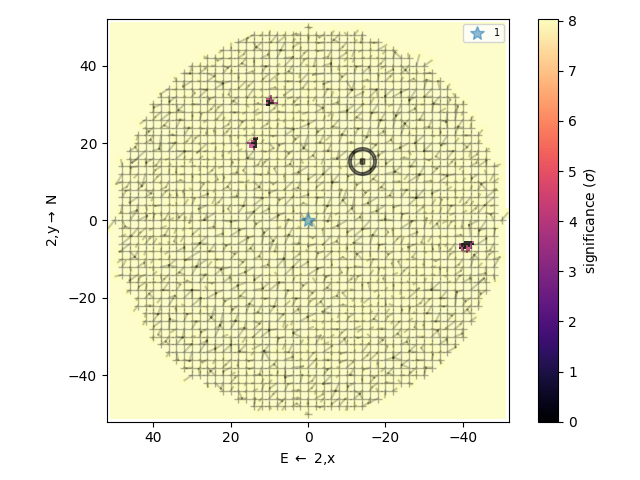}
   \caption{Plot showing the grid search used to search for the companion around HD\,224990 and the companion's significance.}
    \end{figure*}
    
    \begin{figure*}
   \centering
   \includegraphics[width=120mm]{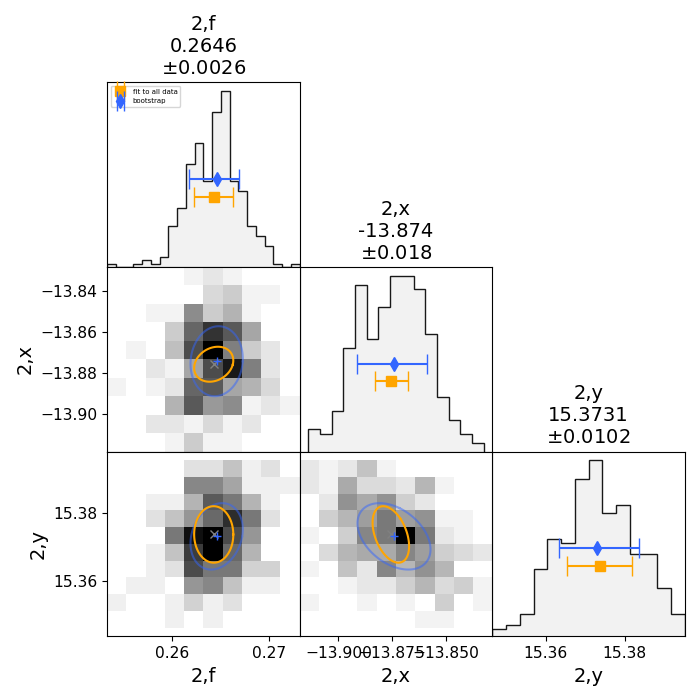}
   \caption{Bootstrapping plot showing the error determination for the dataset of HD\,224990.}
    \end{figure*}

                         \begin{figure*}
   \centering
   \includegraphics[width=170mm]{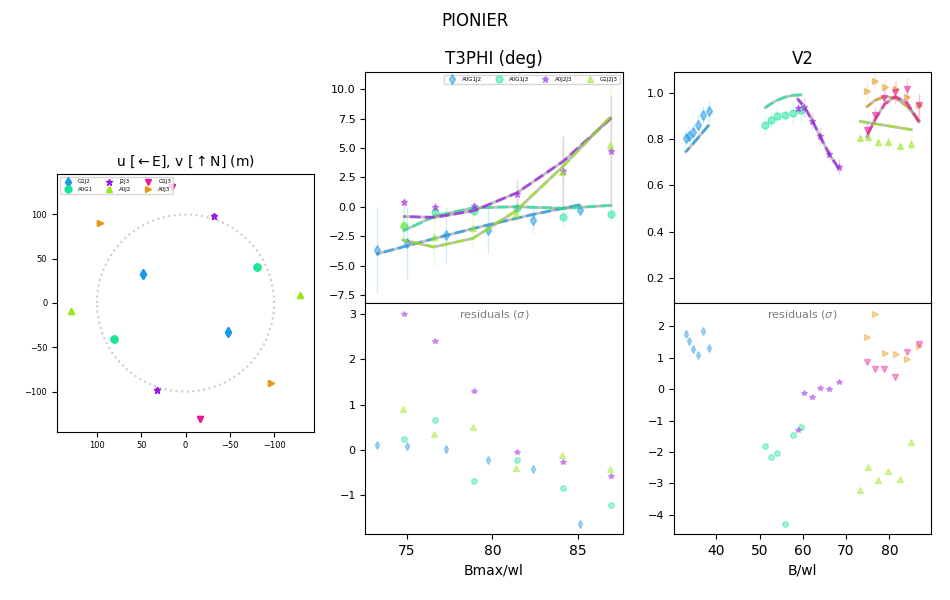}
   \caption{The best-fit data for HD\,224990. On the left, the $u-v$ coverage from the observations with PIONIER can be seen. The different colours correspond to the telescope pair with which that particular data was acquired at the VLTI. In the middle, the closure phase (T3PHI) fit and residuals are shown in terms of the spatial frequency (B$_{avg}$/$\lambda$, written as Bavg/wl on the axes). On the right, the fit to the squared visibilities (V2) and the associated residuals are shown, again in terms of spatial frequency. Across the two fits, the data are represented as points whilst the fit as a continuous line.}
    \end{figure*}
    
    \begin{figure*}
   \centering
   \includegraphics[width=140mm]{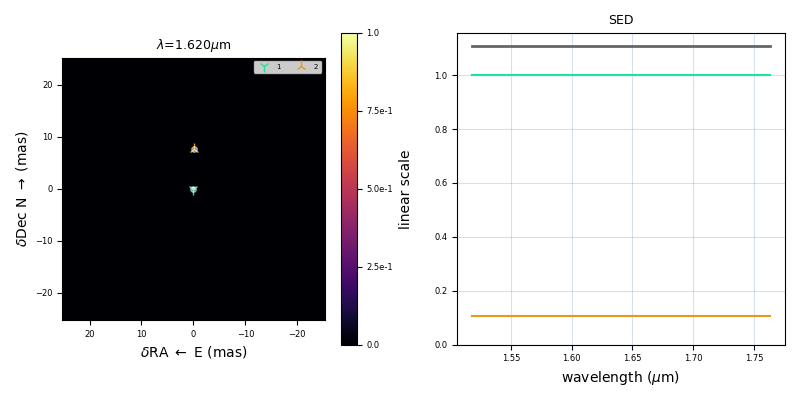}
   \caption{Model image created based on the best-fitting model of HD\,224990.}
    \end{figure*}

    \begin{figure*}
   \centering
   \includegraphics[width=120mm]{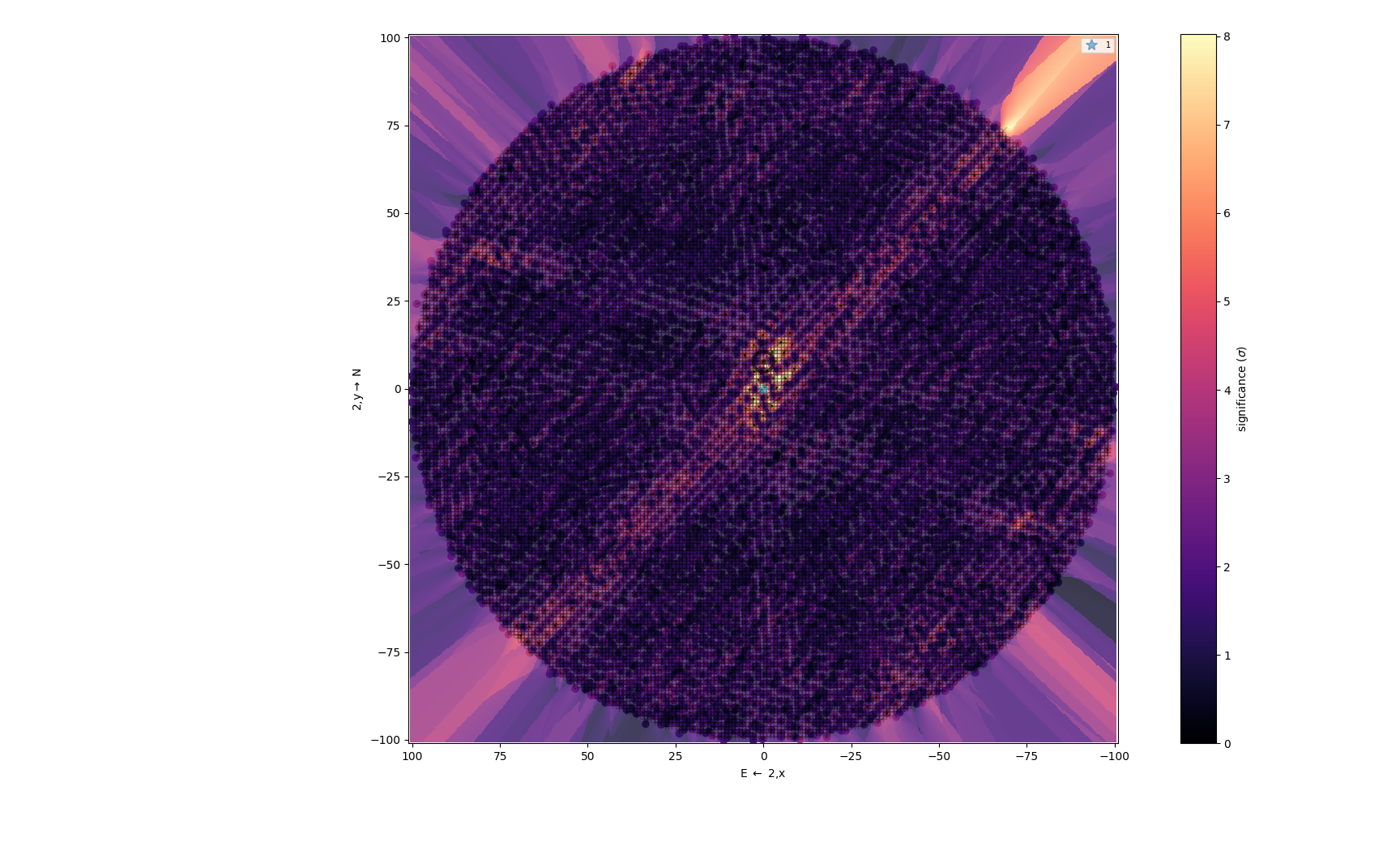}
   \caption{Plot showing the grid search used to search for the companion around HD\,224990 and the companion's significance.}
    \end{figure*}
    
    \begin{figure*}
   \centering
   \includegraphics[width=120mm]{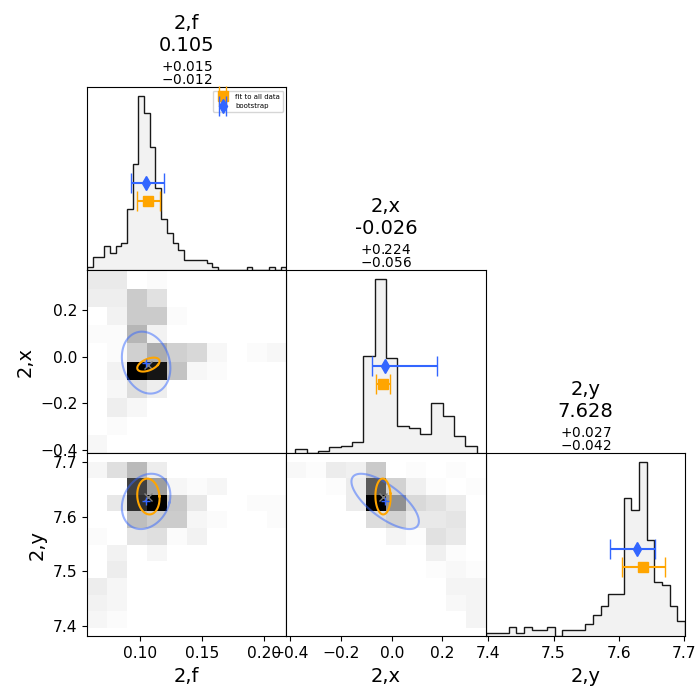}
   \caption{Bootstrapping plot showing the error determination for the dataset of HD\,224990.}
    \end{figure*}

                        \begin{figure*}
   \centering
   \includegraphics[width=170mm]{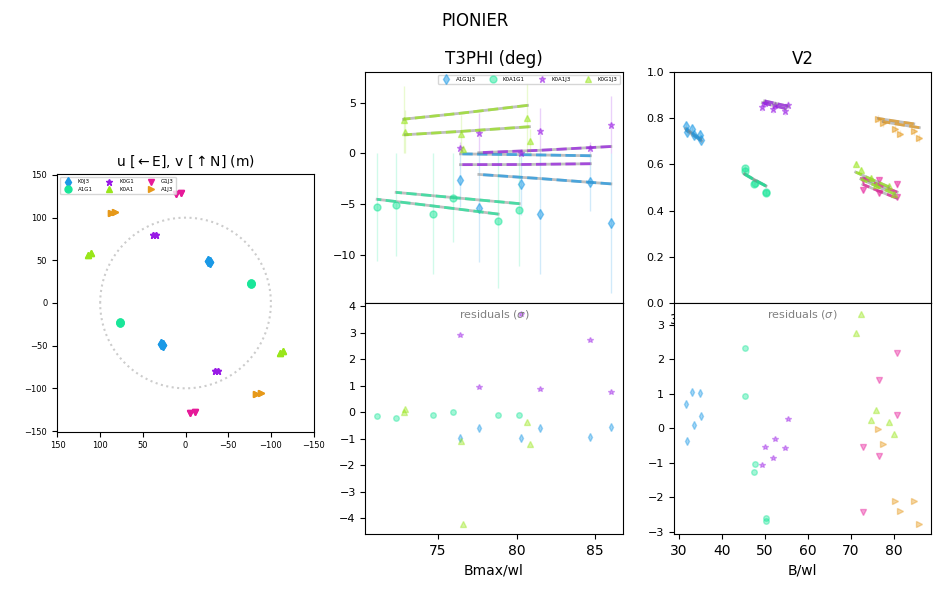}
   \caption{The best-fit data for HD\,116658. On the left, the $u-v$ coverage from the observations with PIONIER can be seen. The different colours correspond to the telescope pair with which that particular data was acquired at the VLTI. In the middle, the closure phase (T3PHI) fit and residuals are shown in terms of the spatial frequency (B$_{avg}$/$\lambda$, written as Bavg/wl on the axes). On the right, the fit to the squared visibilities (V2) and the associated residuals are shown, again in terms of spatial frequency. Across the two fits, the data are represented as points whilst the fit as a continuous line.}
    \end{figure*}
    
    \begin{figure*}
   \centering
   \includegraphics[width=140mm]{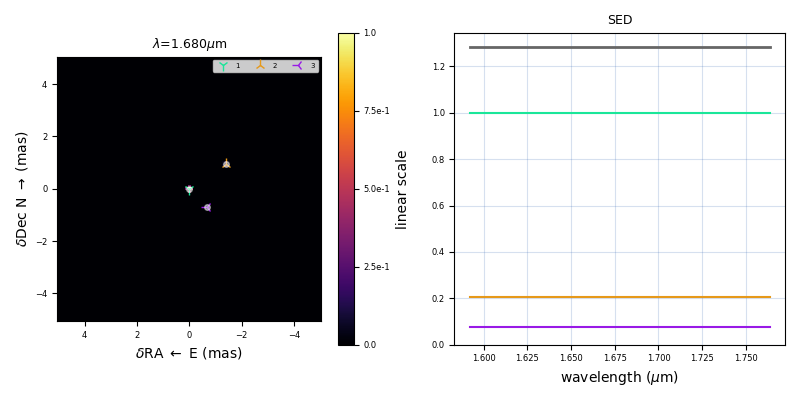}
   \caption{Model image created based on the best-fitting model of HD\,116658.}
    \end{figure*}

    \begin{figure*}
   \centering
   \includegraphics[width=120mm]{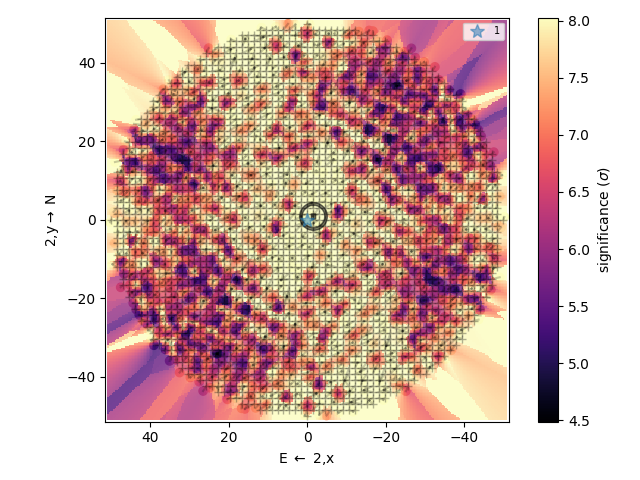}
   \caption{Plot showing the grid search used to search for the secondary companion around HD\,116658 and the companion's significance.}
    \end{figure*}
    
    \begin{figure*}
   \centering
   \includegraphics[width=120mm]{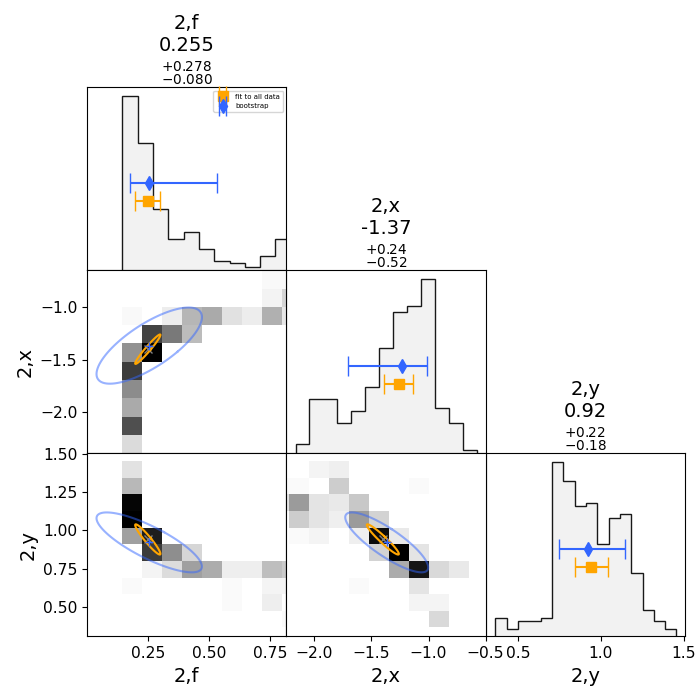}
   \caption{Bootstrapping plot showing the error determination for the parameters of the secondary star of HD\,116658.}
    \end{figure*}

    \begin{figure*}
   \centering
   \includegraphics[width=120mm]{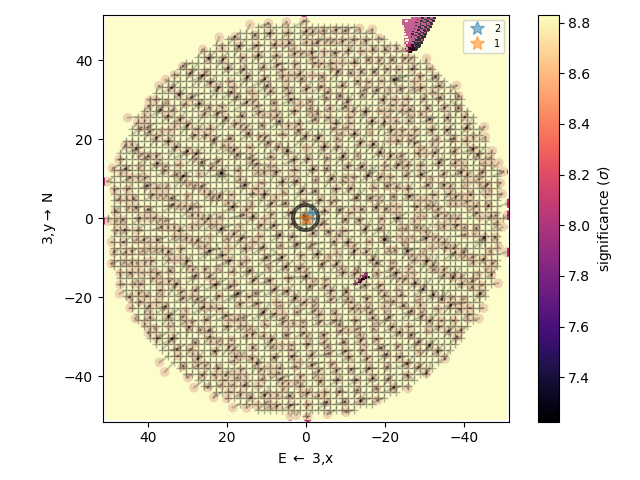}
   \caption{Plot showing the grid search used to search for the tertiary companion around HD\,116658 and the companion's significance.}
    \end{figure*}
    
    \begin{figure*}
   \centering
   \includegraphics[width=120mm]{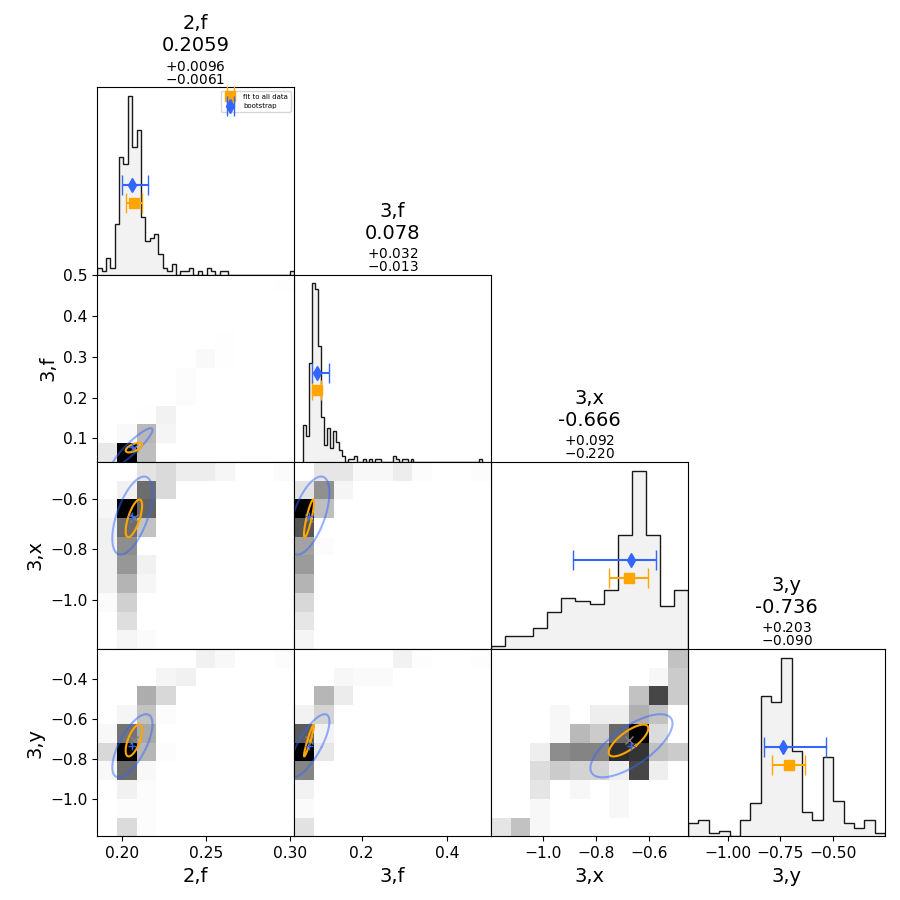}
   \caption{Bootstrapping plot showing the error determination of the tertiary star of HD\,116658.}
    \end{figure*}

                        \begin{figure*}
   \centering
   \includegraphics[width=170mm]{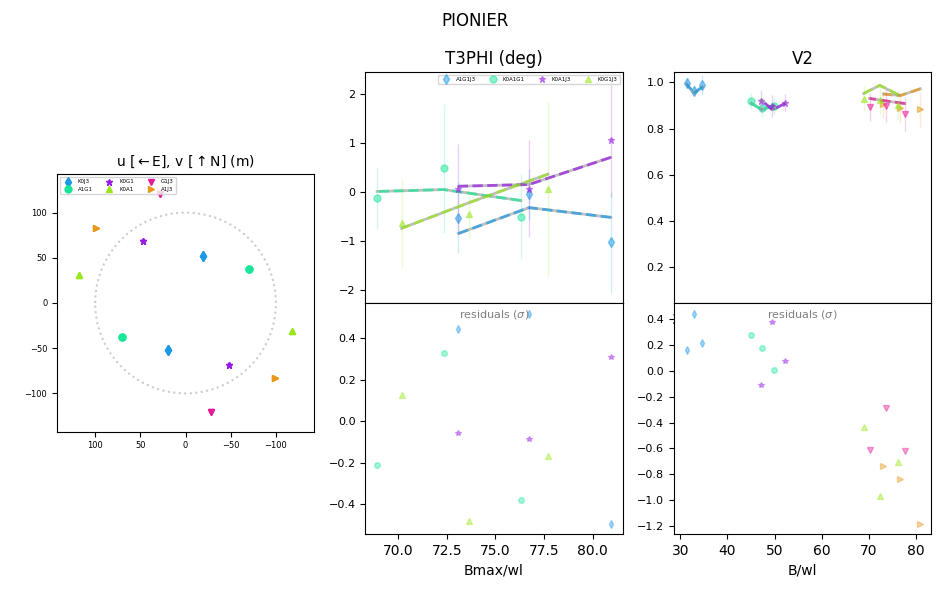}
   \caption{The best-fit data for HD\,132058. On the left, the $u-v$ coverage from the observations with PIONIER can be seen. The different colours correspond to the telescope pair with which that particular data was acquired at the VLTI. In the middle, the closure phase (T3PHI) fit and residuals are shown in terms of the spatial frequency (B$_{avg}$/$\lambda$, written as Bavg/wl on the axes). On the right, the fit to the squared visibilities (V2) and the associated residuals are shown, again in terms of spatial frequency. Across the two fits, the data are represented as points whilst the fit as a continuous line.}
    \end{figure*}
    
    \begin{figure*}
   \centering
   \includegraphics[width=140mm]{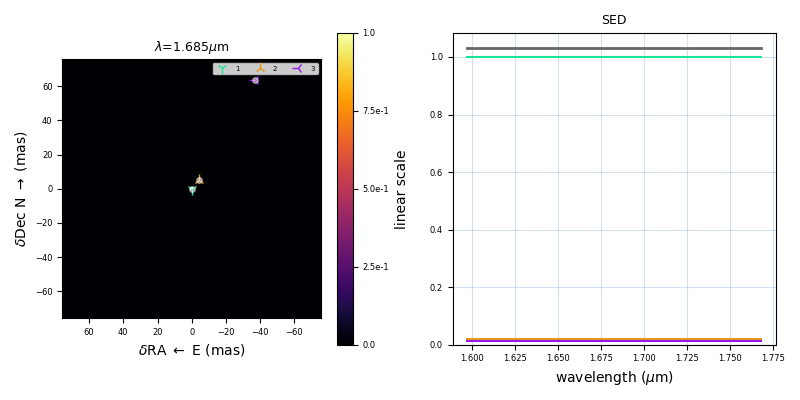}
   \caption{Model image created based on the best-fitting model of HD\,132058.}
    \end{figure*}

    \begin{figure*}
   \centering
   \includegraphics[width=120mm]{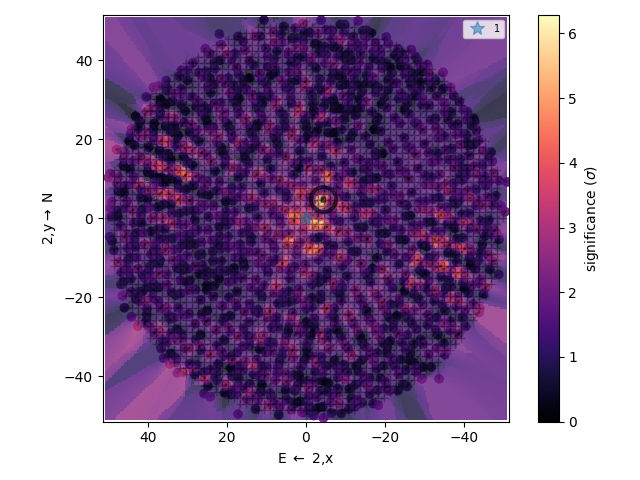}
   \caption{Plot showing the grid search used to search for the secondary companion around HD\,132058 and the companion's significance.}
    \end{figure*}
    
    \begin{figure*}
   \centering
   \includegraphics[width=120mm]{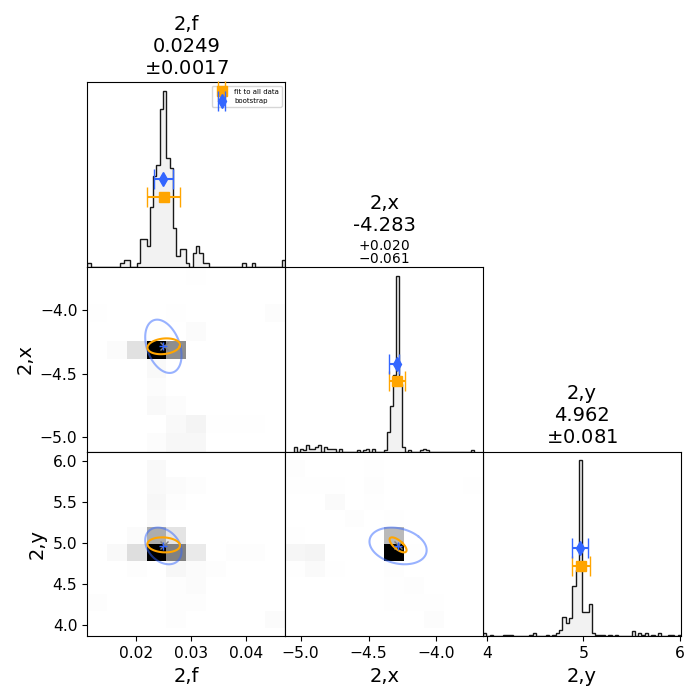}
   \caption{Bootstrapping plot showing the error determination for the parameters of the secondary star of HD\,132058.}
    \end{figure*}

    \begin{figure*}
   \centering
   \includegraphics[width=120mm]{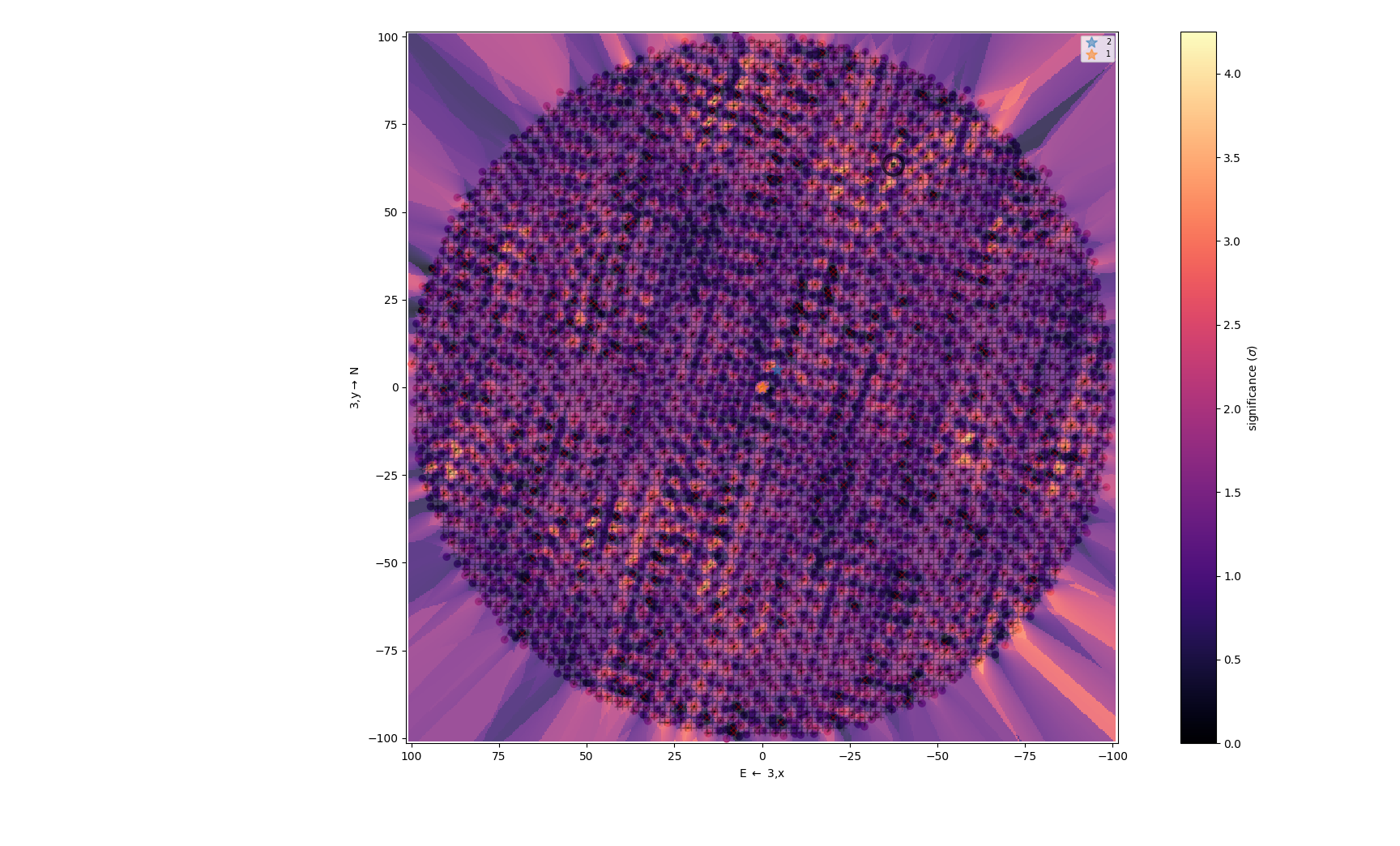}
   \caption{Plot showing the grid search used to search for the companion around HD\,132058 and the companion's significance.}
    \end{figure*}
    
    \begin{figure*}
   \centering
   \includegraphics[width=150mm]{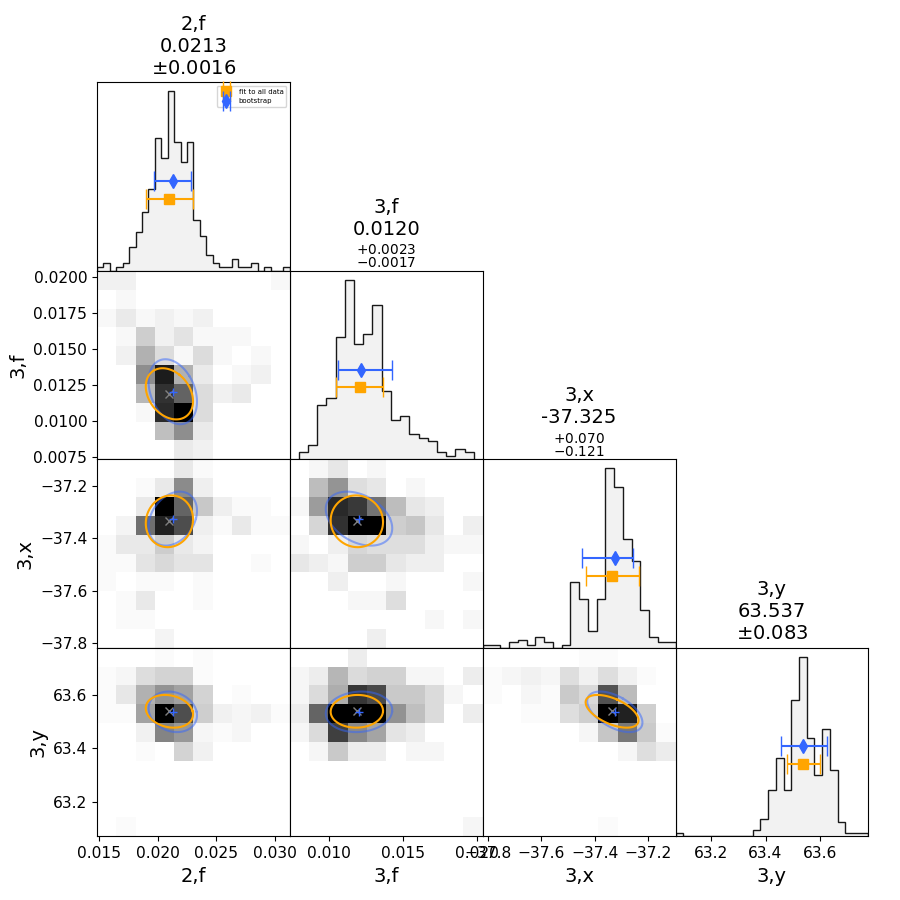}
   \caption{Bootstrapping plot showing the error determination for the dataset of HD\,132058.}
    \end{figure*}

                        \begin{figure*}
   \centering
   \includegraphics[width=170mm]{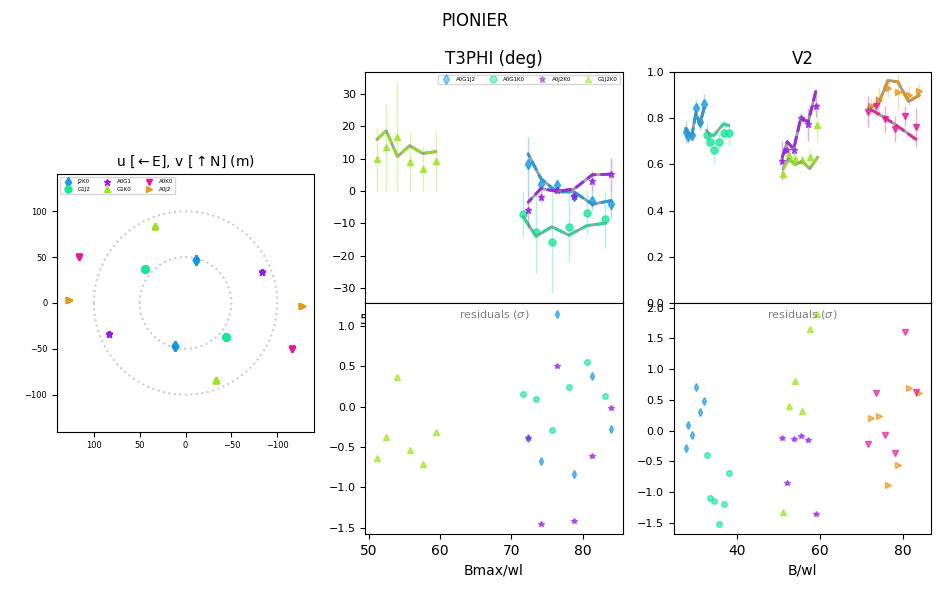}
   \caption{The best-fit data for HD\,147932. On the left, the $u-v$ coverage from the observations with PIONIER can be seen. The different colours correspond to the telescope pair with which that particular data was acquired at the VLTI. In the middle, the closure phase (T3PHI) fit and residuals are shown in terms of the spatial frequency (B$_{avg}$/$\lambda$, written as Bavg/wl on the axes). On the right, the fit to the squared visibilities (V2) and the associated residuals are shown, again in terms of spatial frequency. Across the two fits, the data are represented as points whilst the fit as a continuous line.}
    \end{figure*}
    
    \begin{figure*}
   \centering
   \includegraphics[width=140mm]{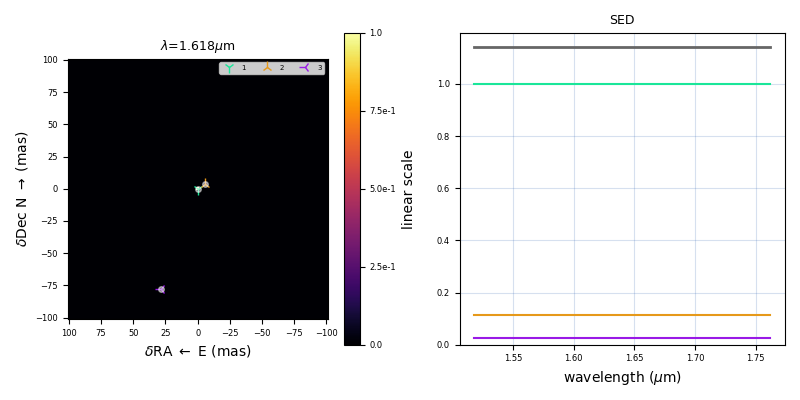}
   \caption{Model image created based on the best-fitting model of HD\,147932.}
    \end{figure*}

    \begin{figure*}
   \centering
   \includegraphics[width=120mm]{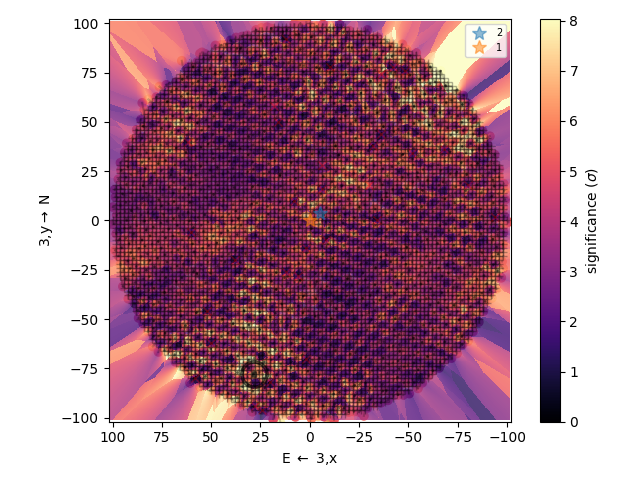}
   \caption{Plot showing the grid search used to search for the secondary companion around HD\,147932 and the companion's significance.}
    \end{figure*}
    
    \begin{figure*}
   \centering
   \includegraphics[width=120mm]{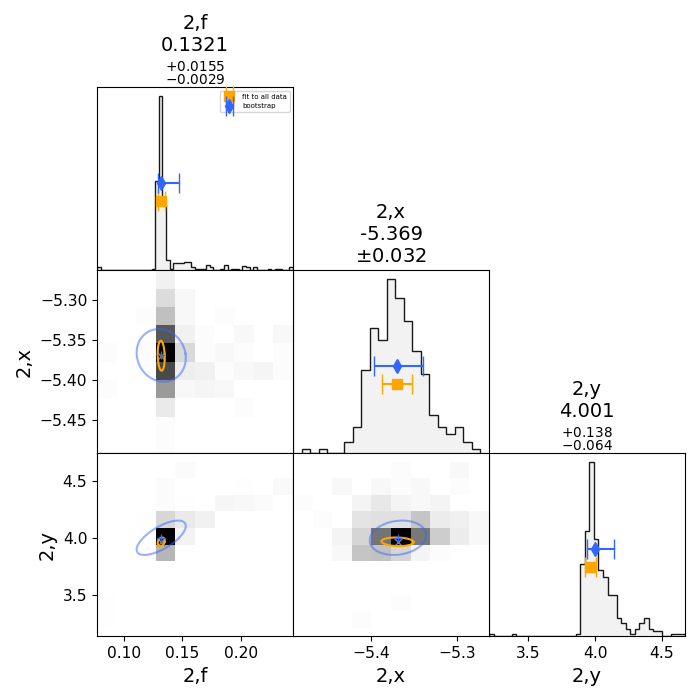}
   \caption{Bootstrapping plot showing the error determination for the parameters of the secondary star of HD\,147932.}
    \end{figure*}

    \begin{figure*}
   \centering
   \includegraphics[width=120mm]{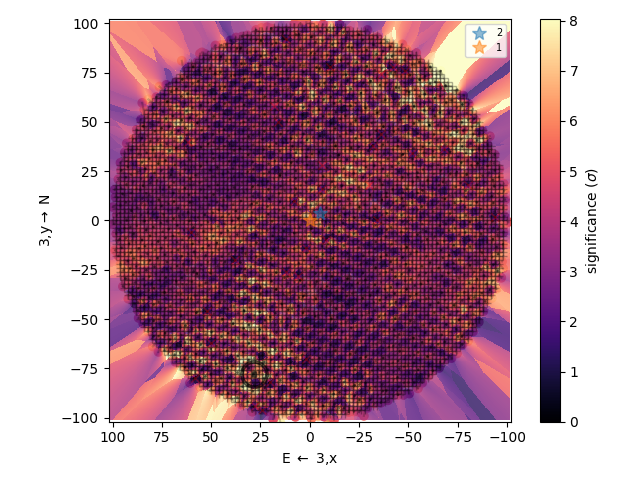}
   \caption{Plot showing the grid search used to search for the tertiary companion around HD\,147932 and the companion's significance.}
    \end{figure*}
    
    \begin{figure*}
   \centering
   \includegraphics[width=120mm]{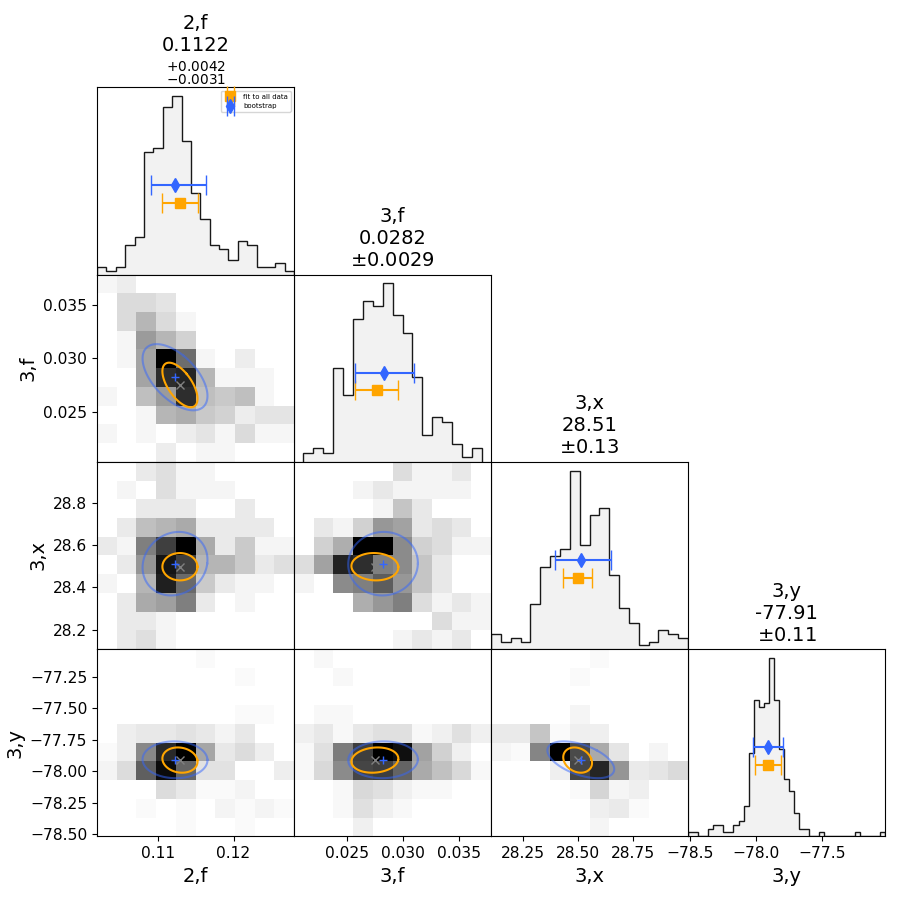}
   \caption{Bootstrapping plot showing the error determination for tertiary of HD\,147932.}
    \end{figure*}

                        \begin{figure*}
   \centering
   \includegraphics[width=170mm]{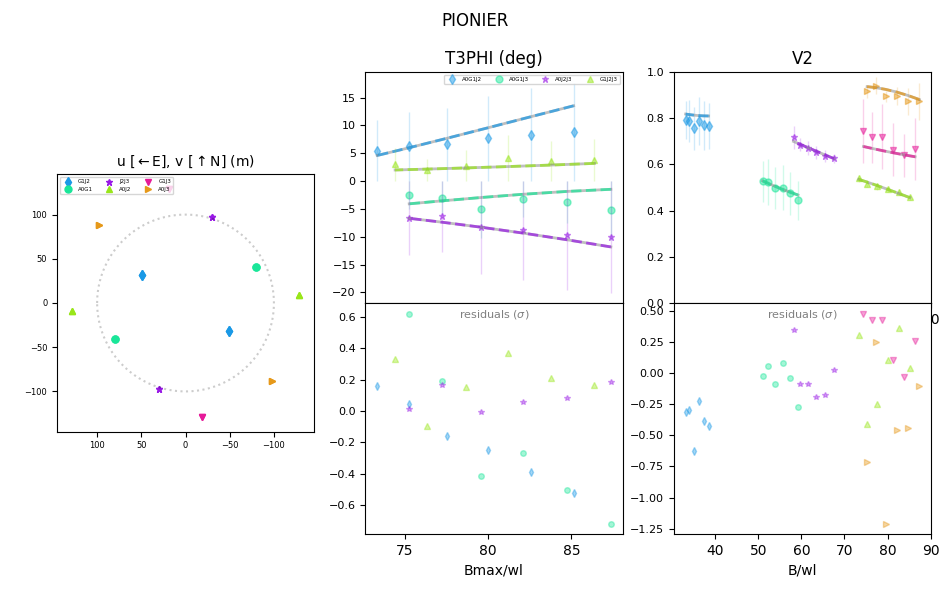}
   \caption{The best-fit data for HD\,161701. On the left, the $u-v$ coverage from the observations with PIONIER can be seen. The different colours correspond to the telescope pair with which that particular data was acquired at the VLTI. In the middle, the closure phase (T3PHI) fit and residuals are shown in terms of the spatial frequency (B$_{avg}$/$\lambda$, written as Bavg/wl on the axes). On the right, the fit to the squared visibilities (V2) and the associated residuals are shown, again in terms of spatial frequency. Across the two fits, the data are represented as points whilst the fit as a continuous line.}
    \end{figure*}
    
    \begin{figure*}
   \centering
   \includegraphics[width=140mm]{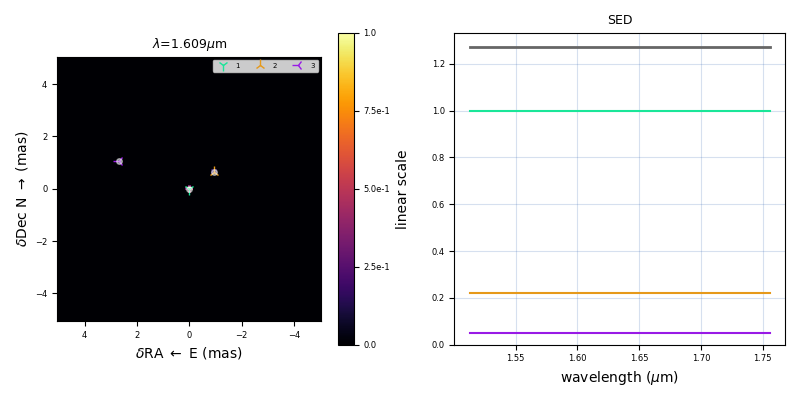}
   \caption{Model image created based on the best-fitting model of HD\,161701.}
    \end{figure*}

        \begin{figure*}
   \centering
   \includegraphics[width=120mm]{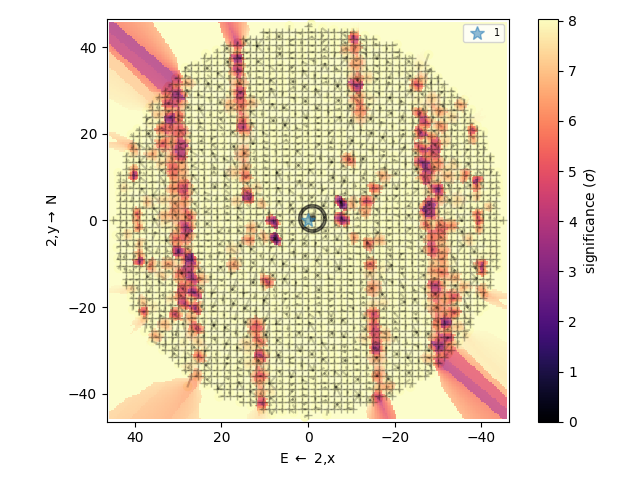}
   \caption{Plot showing the grid search used to search for the secondary companion around HD\,161701 and the companion's significance.}
    \end{figure*}
    
    \begin{figure*}
   \centering
   \includegraphics[width=120mm]{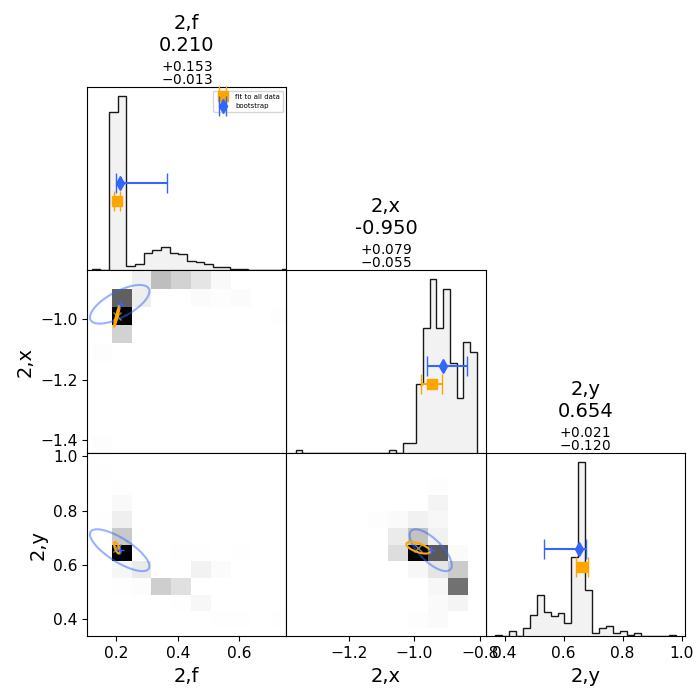}
   \caption{Bootstrapping plot showing the error determination for the parameters of the secondary star of HD\,161701.}
    \end{figure*}

    \begin{figure*}
   \centering
   \includegraphics[width=120mm]{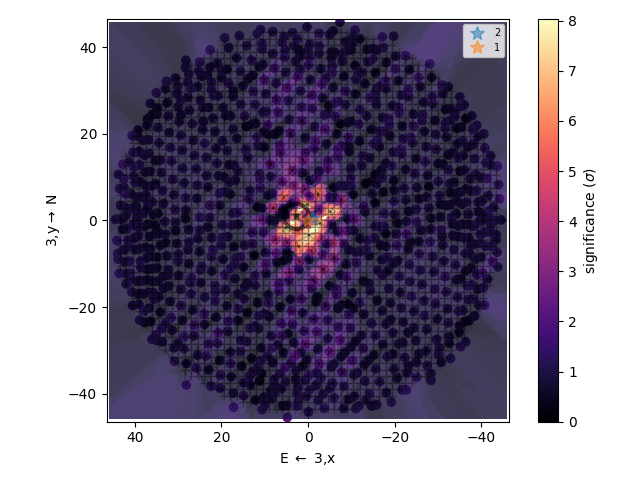}
   \caption{Plot showing the grid search used to search for the tertiary companion around HD\,161701 and the companion's significance.}
    \end{figure*}
    
    \begin{figure*}
   \centering
   \includegraphics[width=120mm]{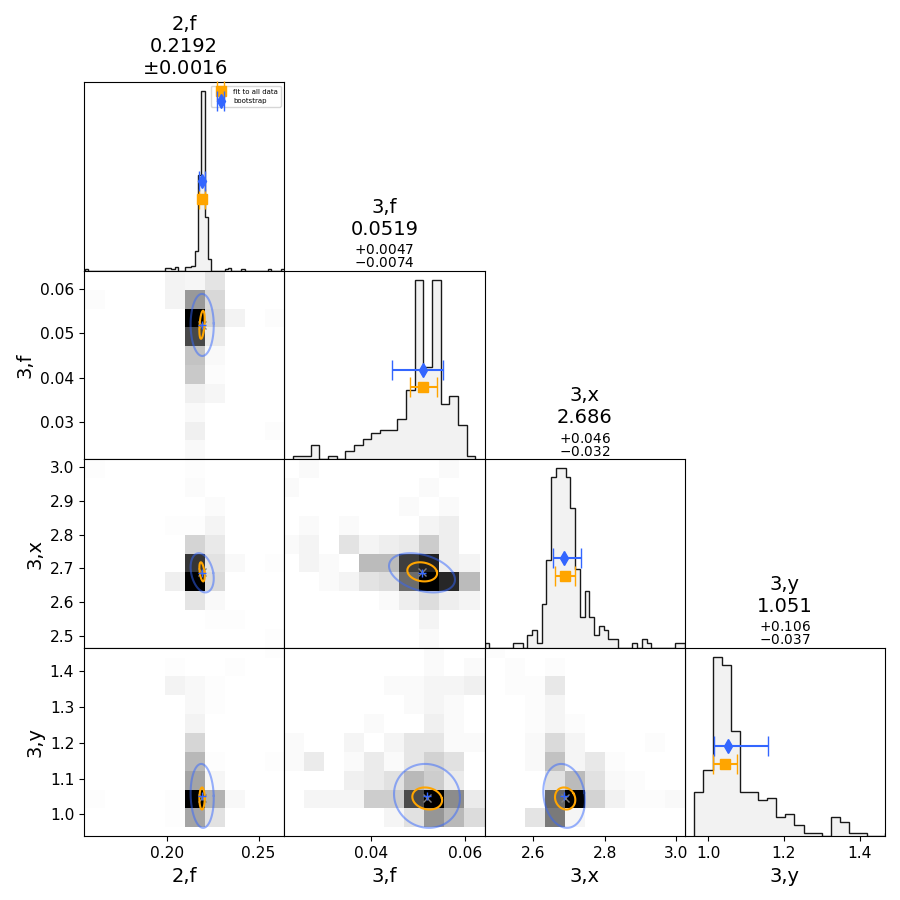}
   \caption{Bootstrapping plot showing the error determination for the tertiary HD\,161701.}
    \end{figure*}

                        \begin{figure*}
   \centering
   \includegraphics[width=170mm]{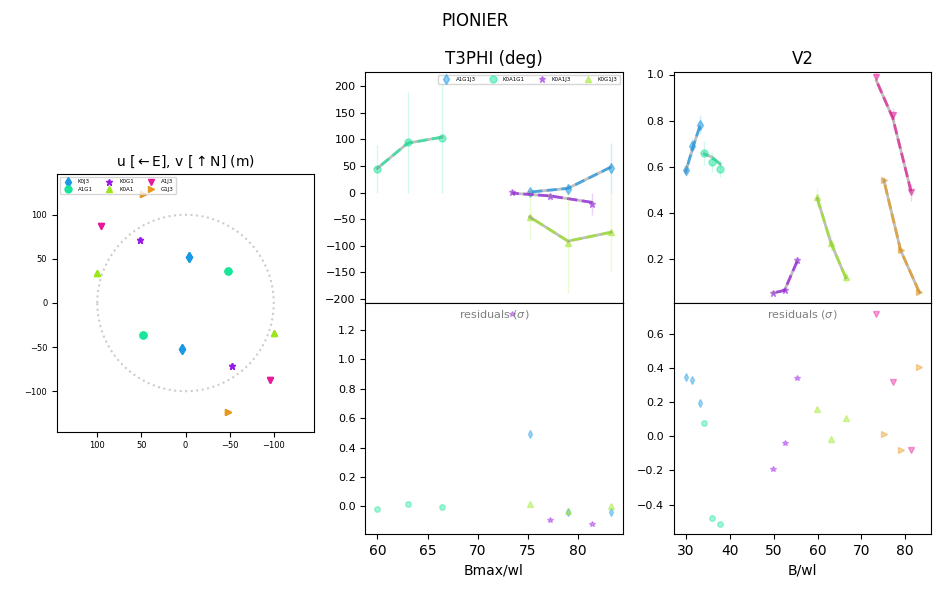}
   \caption{The best-fit data for HD\,193933. On the left, the $u-v$ coverage from the observations with PIONIER can be seen. The different colours correspond to the telescope pair with which that particular data was acquired at the VLTI. In the middle, the closure phase (T3PHI) fit and residuals are shown in terms of the spatial frequency (B$_{avg}$/$\lambda$, written as Bavg/wl on the axes). On the right, the fit to the squared visibilities (V2) and the associated residuals are shown, again in terms of spatial frequency. Across the two fits, the data are represented as points whilst the fit as a continuous line.}
    \end{figure*}
    
    \begin{figure*}
   \centering
   \includegraphics[width=140mm]{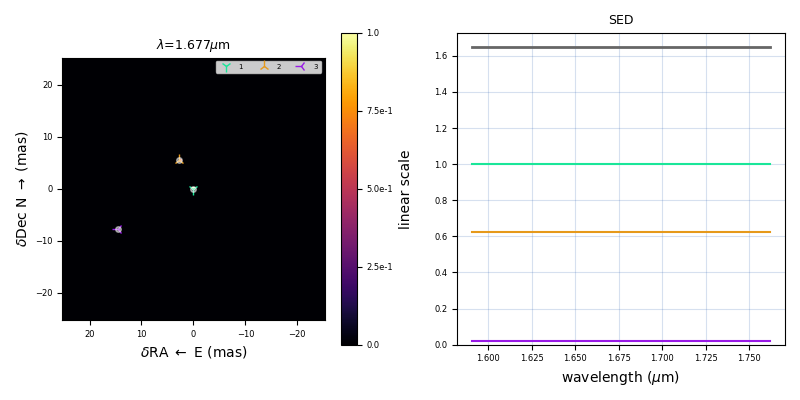}
   \caption{Model image created based on the best-fitting model of HD\,193933.}
    \end{figure*}

        \begin{figure*}
   \centering
   \includegraphics[width=120mm]{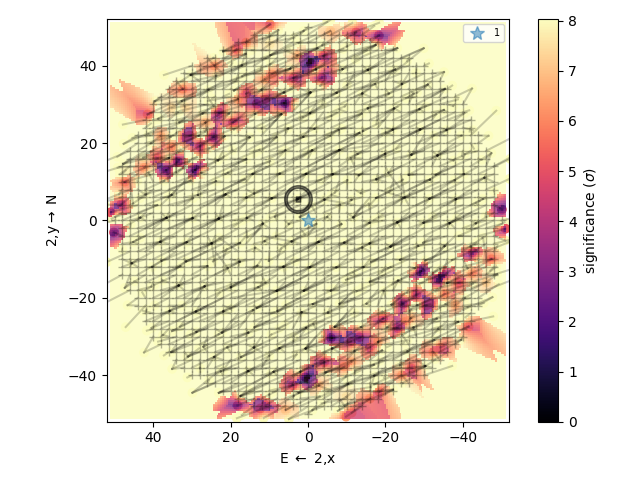}
   \caption{Plot showing the grid search used to search for the secondary companion around HD\,193933 and the companion's significance.}
    \end{figure*}
    
    \begin{figure*}
   \centering
   \includegraphics[width=120mm]{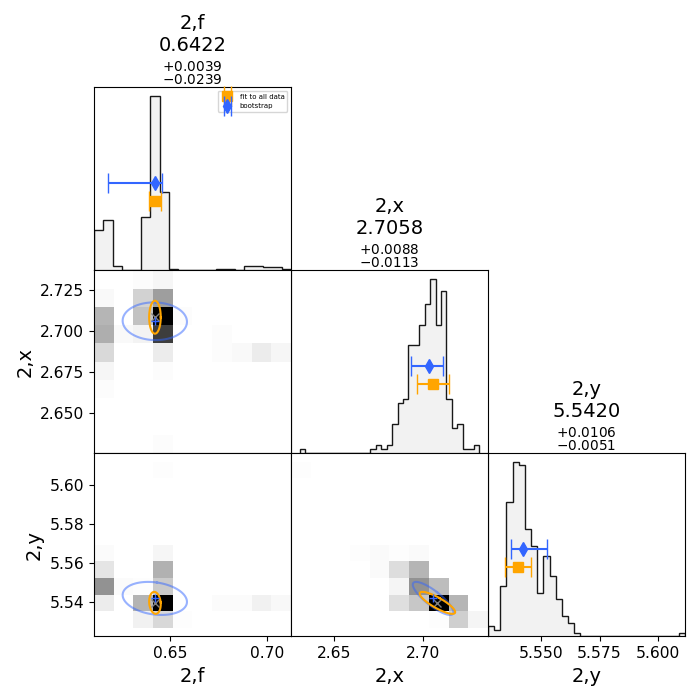}
   \caption{Bootstrapping plot showing the error determination for the parameters of the secondary star of HD\,193933.}
    \end{figure*}

    \begin{figure*}
   \centering
   \includegraphics[width=120mm]{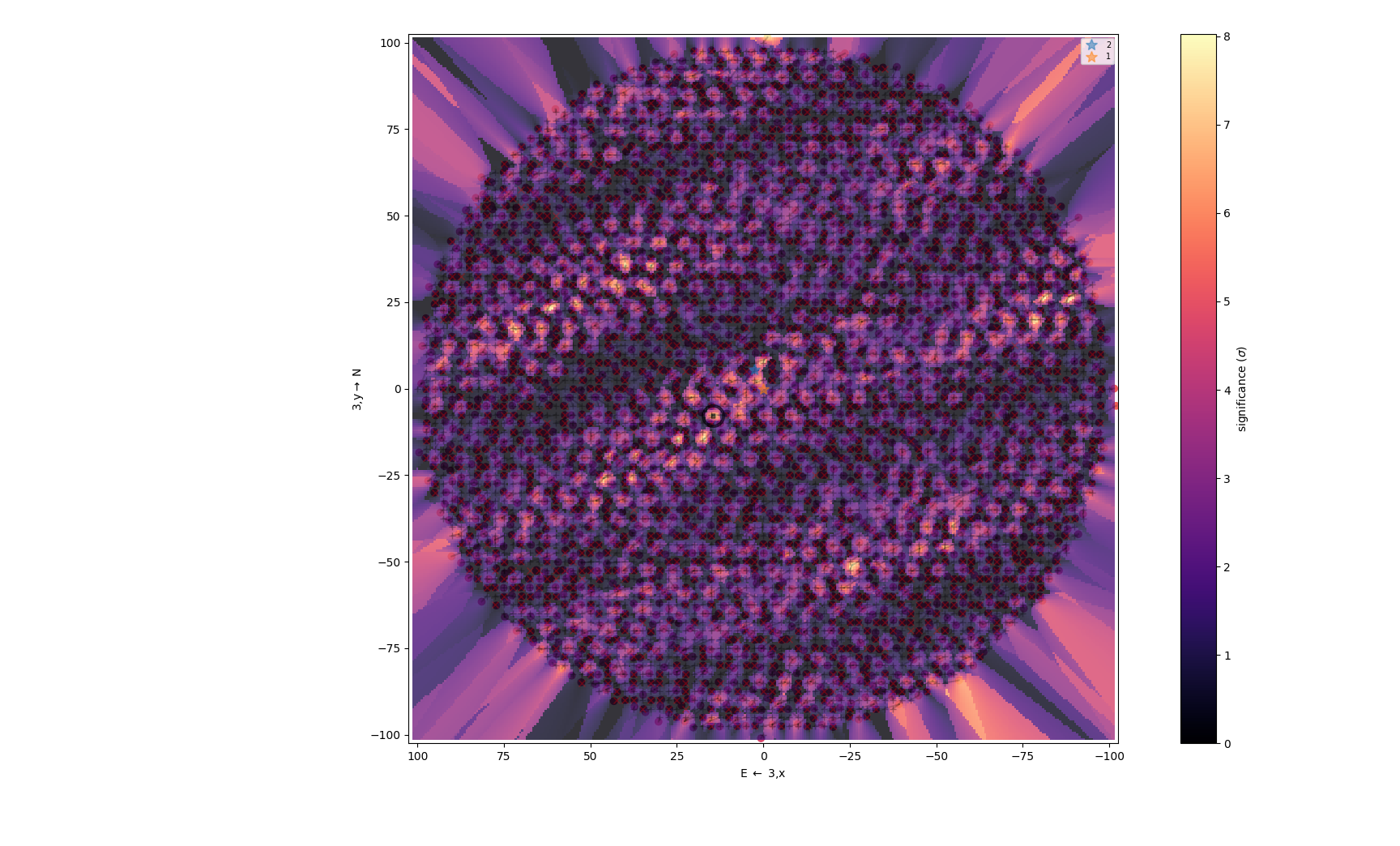}
   \caption{Plot showing the grid search used to search for the tertiary companion around HD\,193933 and the companion's significance.}
    \end{figure*}
    
    \begin{figure*}
   \centering
   \includegraphics[width=120mm]{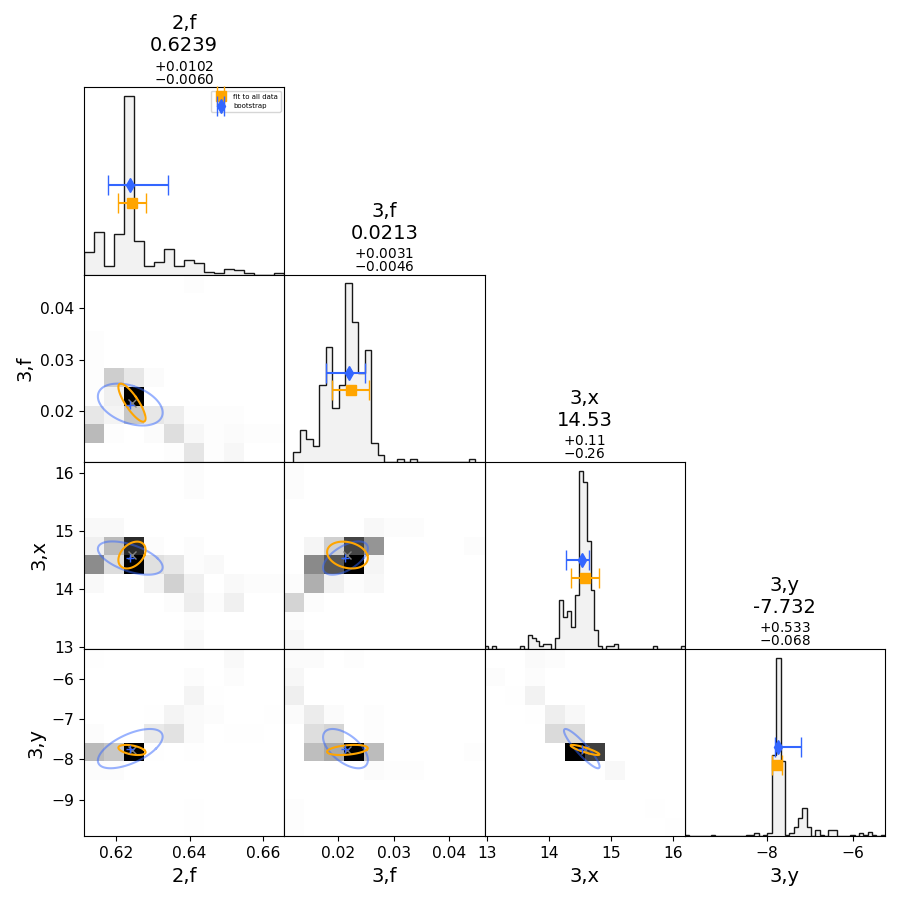}
   \caption{Bootstrapping plot showing the error determination for the tertiary of HD\,193933.}
    \end{figure*}

\end{appendix}

\end{document}